\documentclass[journal=chreay,manuscript=review,layout=twocolumn]{achemso}

\usepackage{graphicx}
\usepackage{dcolumn}
\usepackage{bm}
\usepackage[hidelinks]{hyperref}
\usepackage{color}
\usepackage{glossaries} 
\usepackage{amsmath,amssymb}
\usepackage{microtype} 
\usepackage{textcomp}
\usepackage{soul}
\usepackage{booktabs}
\usepackage{multirow}

\newcommand\here[1]{\fcolorbox{red}{red}{\rule{0pt}{6pt}\rule{6pt}{0pt}}\quad}

\newcommand\tran{\mkern-2mu\raise1.25ex\hbox{$\scriptscriptstyle\top$}\mkern-3.5mu}

\author{Oliver T.\ Unke$^1$}
\affiliation{Machine Learning Group, Technische Universit\"at Berlin, 10587 Berlin, Germany}
\alsoaffiliation{DFG Cluster of Excellence ``Unifying Systems in Catalysis'' (UniSysCat), Technische Universit\"at Berlin, 10623 Berlin, Germany}

\author{Stefan Chmiela$^1$}
\affiliation{Machine Learning Group, Technische Universit\"at Berlin, 10587 Berlin, Germany}

\author{Huziel E.\ Sauceda}
\affiliation{Machine Learning Group, Technische Universit\"at Berlin, 10587 Berlin, Germany}
\alsoaffiliation{
 BASLEARN, BASF-TU joint Lab, Technische Universit\"at Berlin, 10587 Berlin, Germany
}

\author{Michael Gastegger}
\affiliation{Machine Learning Group, Technische Universit\"at Berlin, 10587 Berlin, Germany}
\alsoaffiliation{DFG Cluster of Excellence ``Unifying Systems in Catalysis'' (UniSysCat), Technische Universit\"at Berlin, 10623 Berlin, Germany}

\author{Igor Poltavsky}
\affiliation{Department of Physics and Materials Science, University of Luxembourg, L-1511 Luxembourg City, Luxembourg.}

\author{Kristof T. Sch\"utt}
\affiliation{Machine Learning Group, Technische Universit\"at Berlin, 10587 Berlin, Germany}

\author{Alexandre Tkatchenko}
\affiliation{Department of Physics and Materials Science, University of Luxembourg, L-1511 Luxembourg City, Luxembourg.}
\email{alexandre.tkatchenko@uni.lu}

\author{Klaus-Robert M\"uller}
\affiliation{Machine Learning Group, Technische Universit\"at Berlin, 10587 Berlin, Germany}
\alsoaffiliation{BIFOLD--Berlin Institute for the Foundations of Learning and Data, Berlin, Germany}
\alsoaffiliation{Department of Artificial Intelligence,
	Korea University, Anam-dong, Seongbuk-gu, Seoul 02841, Korea}
\alsoaffiliation{Max Planck Institute for Informatics, Stuhlsatzenhausweg, 66123 Saarbr\"ucken, Germany}
\alsoaffiliation{Google Research, Brain team, Berlin, Germany}
\email{klaus-robert.mueller@tu-berlin.de}

\title{Machine Learning Force Fields}

\abbreviations{ML,FF,PES}

\keywords{Machine Learning, Force Field, Potential Energy Surface}

\begin{document}
\thanks{$^1$these authors contributed equally.}


\begin{abstract}
In recent years, the use of Machine Learning (ML) in computational chemistry has enabled numerous advances previously out of reach due to the computational complexity of traditional electronic-structure methods. One of the most promising applications is the construction of ML-based force fields (FFs), with the aim to narrow the gap between the accuracy of \textit{ab initio} methods and the efficiency of classical FFs. 
The key idea is to learn the statistical relation between chemical structure and potential energy without relying on a preconceived notion of fixed chemical bonds or knowledge about the relevant interactions. Such universal ML approximations are in principle only limited by the quality and quantity of the reference data used to train them. This review gives an overview of applications of ML-FFs and the chemical insights that can be obtained from them. The core concepts underlying ML-FFs are described in detail and a step-by-step guide for constructing and testing them from scratch is given. The text concludes with a discussion of the challenges that remain to be overcome by the next generation of ML-FFs.
\end{abstract}

\glsaddall
\printglossary[title=Machine Learning Terminology, toctitle=Glossary,style=list,nonumberlist]

\setcounter{tocdepth}{2} 
\tableofcontents

\section{Introduction}
\label{sec:introduction}
In 1964, physicist Richard Feynman famously remarked ``\textit{that all things are made of atoms and that everything that living things do can be understood in terms of the jigglings and wigglings of atoms}''.~\cite{feynman1963feynman} As such, an atomically resolved picture can provide invaluable insights on biological (and other) processes. The first molecular dynamics (MD) study of a protein in 1977 by \citet{mccammon1977dynamics} did not consider explicit solvent molecules and was limited to less than 10~ps of simulation.
Still, it challenged the (at that time) common belief that proteins are essentially rigid structures\cite{phillips1981biomolecular} and, instead, suggested that the interior of proteins behaves more fluid-like. 
Since then, systems consisting of more than a million atoms have been studied\cite{schulz2009scaling}, simulation times extended to the millisecond regime\cite{shaw2008anton}, and the study of entire viruses in atomic detail made possible\cite{freddolino2006molecular,zhao2013mature}. Recently, a distributed computing effort even allowed to study the viral proteome of SARS-CoV-2 for a total of 0.1 seconds of simulation time \cite{zimmerman2020citizen}.

To perform MD simulations, typically, the Newtonian equations of motion are integrated numerically, which requires knowledge of the forces acting on individual atoms at each time step of the simulation\cite{karplus2002molecular}. In principle, the most accurate way to obtain these forces is by solving the Schr\"odinger equation (SE), which describes the physical laws underlying most chemical phenomena and processes\cite{dirac1929quantum}. Unfortunately, an analytic solution is only possible for two-body systems such as the hydrogen atom. For larger chemical structures, the SE can only be solved approximately. However, even with approximations, an accurate numerical solution is a computationally demanding task. For example, the CCSD(T) method (coupled cluster with singles, doubles and perturbative triples), which is widely regarded as the ``gold standard'' of chemistry (as its predictions compare well with experimental results)\cite{gordon2005theory}, scales $\propto N^7$ with the number of atoms $N$. \footnote{Strictly speaking, the true scaling of the CCSD(T) method is $\mathcal{O}(n^7)$, where $n$ is the number of basis functions used for the wave function ansatz. Depending on the desired accuracy and which atoms are present (more precisely, how many electrons are in their shells), $n$ can vary greatly. However, the number of atoms is usually a good proxy.} Because of this, it is unfeasible to calculate the forces for many different configurations of large chemical systems, which is required for running MD simulations, with accurate methods. Instead, simple empirical functions are commonly used to model the relevant interactions. From these so-called force fields (FFs), atomic forces can be readily derived analytically. 

\begin{figure*}[t]
	\centering
	\includegraphics[width=\textwidth]{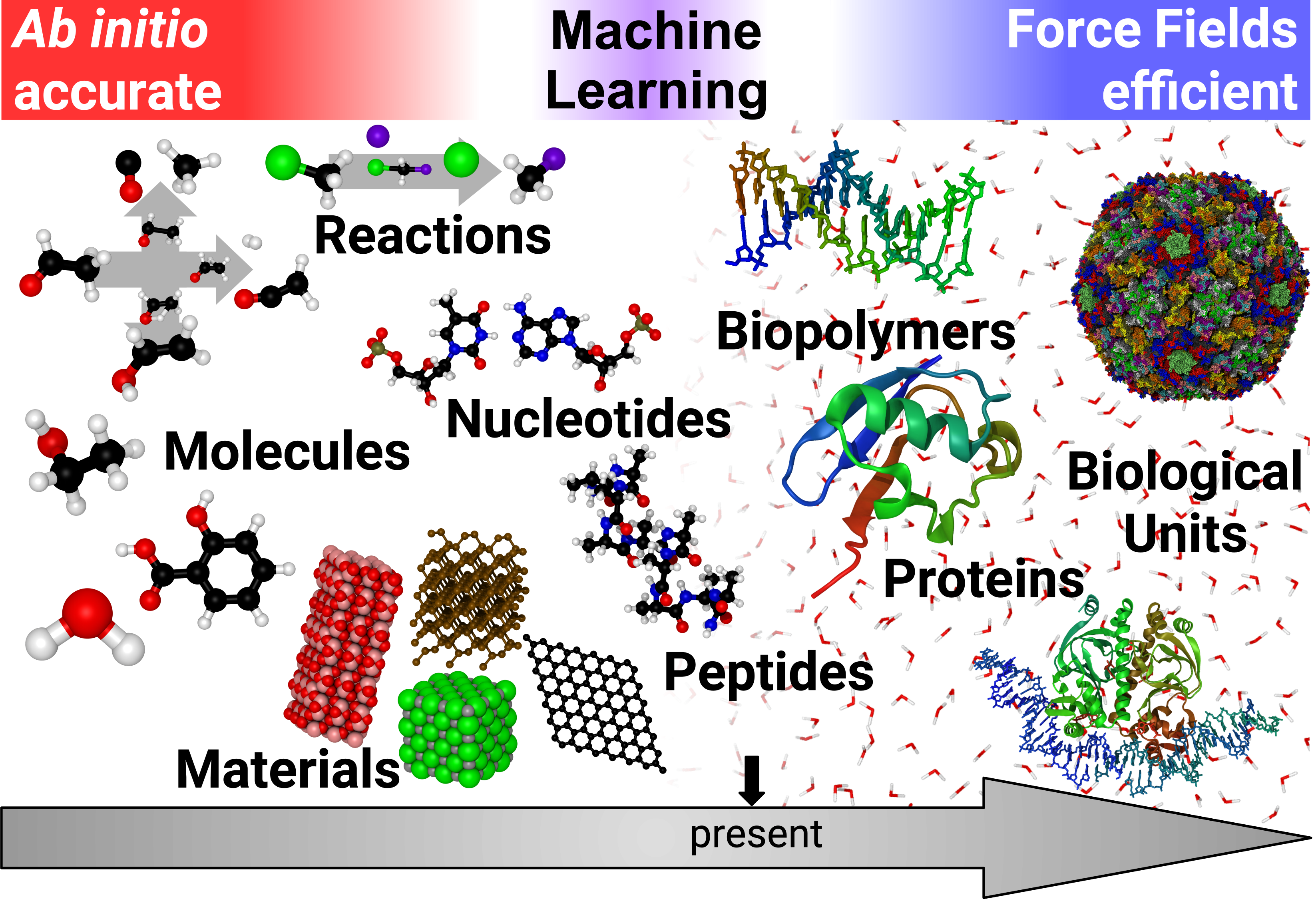}
	\caption{Accurate \textit{ab initio} methods are computationally demanding and can only be used to study small systems in gas phase or regular periodic materials. Larger molecules in solution, such as proteins, are typically modeled by force fields, empirical functions which trade accuracy for computational efficiency. Machine learning methods are closing this gap and allow to study increasingly large chemical systems at \textit{ab initio} accuracy with force field efficiency.}
	\label{fig:overview}
\end{figure*}

Most conventional FFs model chemical interactions as a sum over bonded and non-bonded terms\cite{gonzalez2011force,unke2020high}. The former can be described with simple functions of the distances between directly bonded atoms, or angles and dihedrals between atoms sharing some of their bonding partners. The non-bonded terms consider pair-wise combinations of atoms, typically by modeling electrostatics with Coulomb's law (assuming a point charge at each atom's position) and dispersion with a Lennard-Jones potential\cite{jones1924determination}. Due to the computational efficiency of these terms, such classical FFs allow to study systems consisting of many thousands of atoms. However, while offering a qualitatively reasonable description of chemical interactions, the quality of MD simulations, and hence the insights that can be obtained from them, are ultimately limited by the accuracy of the underlying FF.\cite{vitalini2015dynamic} This is particularly problematic when polarization, or many-body interactions in general, are of significant importance, as these effects are not adequately modeled by the terms described above. While it is possible to construct polarizable FFs\cite{halgren2001polarizable,warshel2007polarizable,shi2013polarizable,lopes2013polarizable} or account for other important effects, e.g.\ anisotropic charge distributions\cite{rasmussen2007force,unke2017minimal}, to improve accuracy (in exchange for computational efficiency), it is not always clear \textit{a priori} when such modifications are necessary. Beyond that, conventional FFs require a preconceived notion of bonding patterns and thus cannot describe bond breaking or bond formation. While there exist reactive FFs offering an approximated description of reactions\cite{warshel1980empirical,van2001reaxff,nagy2014multisurface}, they are often not sufficiently accurate for quantitative studies or restricted to specific types of reactions. Mixed quantum mechanics/molecular mechanics (QM/MM) treatments\cite{senn2009qm} pose an alternative solution: Here, the SE is only solved for a small part of the system where high accuracy is required or reactions may take place, e.g.\ the active site of an enzyme. Meanwhile, all remaining atoms are treated at the FF level of accuracy. Although this is more efficient than a pure quantum-mechanical approach, it is often necessary to model a large number of atoms at the QM level for converged results\cite{kulik2016large}, which is still highly computationally demanding.

In a ``dream scenario'' for computational chemists and biologists, it would be possible to treat even large systems at the highest levels of theory, which would require prohibitively large computational resources in the real world. Machine learning (ML) methods could help to achieve this dream in a principled manner, thus closing the gap between the accuracy of \textit{ab initio} methods and the efficiency of classical FFs (Fig.~\ref{fig:overview}). ML methods aim to learn the functional relationship between inputs (chemical descriptors) and outputs (properties) from patterns or structure in the data. Ideally, a trained learning machine would then reflect the underlying effective ``rules'' of quantum mechanics.\cite{schutt2020machine} Practically, ML models can take a shortcut by not having to solve any equations that follow from the physical laws governing the structure-property relation. Because of this unique ability, ML methods have been enjoying growing popularity in the chemical sciences in recent years. They allow to explore chemical space and predict the properties of compounds with both unprecedented efficiency and high accuracy.\cite{rupp2012fast,montavon2013machine,hansen2013assessment,hansen2015machine,von2020exploring,schutt2020machine} ML has also been used to augment and accelerate traditional methods used in molecular simulations, e.g.\ for sampling equilibrium states\cite{noe2019boltzmann,kohler2020equivariant} and rare events\cite{zhang2019targeted}, computing reaction rates\cite{koner2019exhaustive}, exploring protein folding dynamics\cite{noe2020machine} and other biophysical processes\cite{sonderby2015convolutional,almagro2017deeploc,botlani2018machine,boninsegna2018sparse,senior2020improved}, Markov state modeling\cite{scherer2015pyemma,mardt2018vampnets,wehmeyer2018time,wu2018deep,chen2019nonlinear,klus2019kernel,sidky2019high}, and coarse-graining\cite{chen2018learning,wang2019machine,nuske2019coarse,wang2020ensemble} (for a recent review on applications of ML in molecular simulatons, see Ref.~\citenum{noe2020machinemolsim}). Recent advances made it even possible to predict molecular wave functions, which can act as an interface between ML and quantum chemistry\cite{schutt2019unifying,gastegger2020deep}, as knowledge of the wave function allows to deduce many different quantum mechanical observables at once. ML can also be combined with more traditional semi-empirical methods, for example by predicting accurate repulsive potentials for density functional tight-binding approaches.\cite{stohr2020accurate} Instead of circumventing equations, ML methods can also help when solving them: They have been used to find novel density functionals\cite{snyder2012finding,brockherde2017bypassing,bogojeski2019density} and solutions of the Schr\"odinger equation\cite{hermann2019deep,carleo2017solving}. Other promising applications include the generation of molecular structures to tackle inverse design problems,\cite{gebauer2018generating,gebauer2019symmetry,hoffmann2019generating,winter2019efficient,simm2020reinforcement} or planning chemical syntheses\cite{strieth2020machine}.

\begin{figure*}
	\centering
	\includegraphics[width=\textwidth]{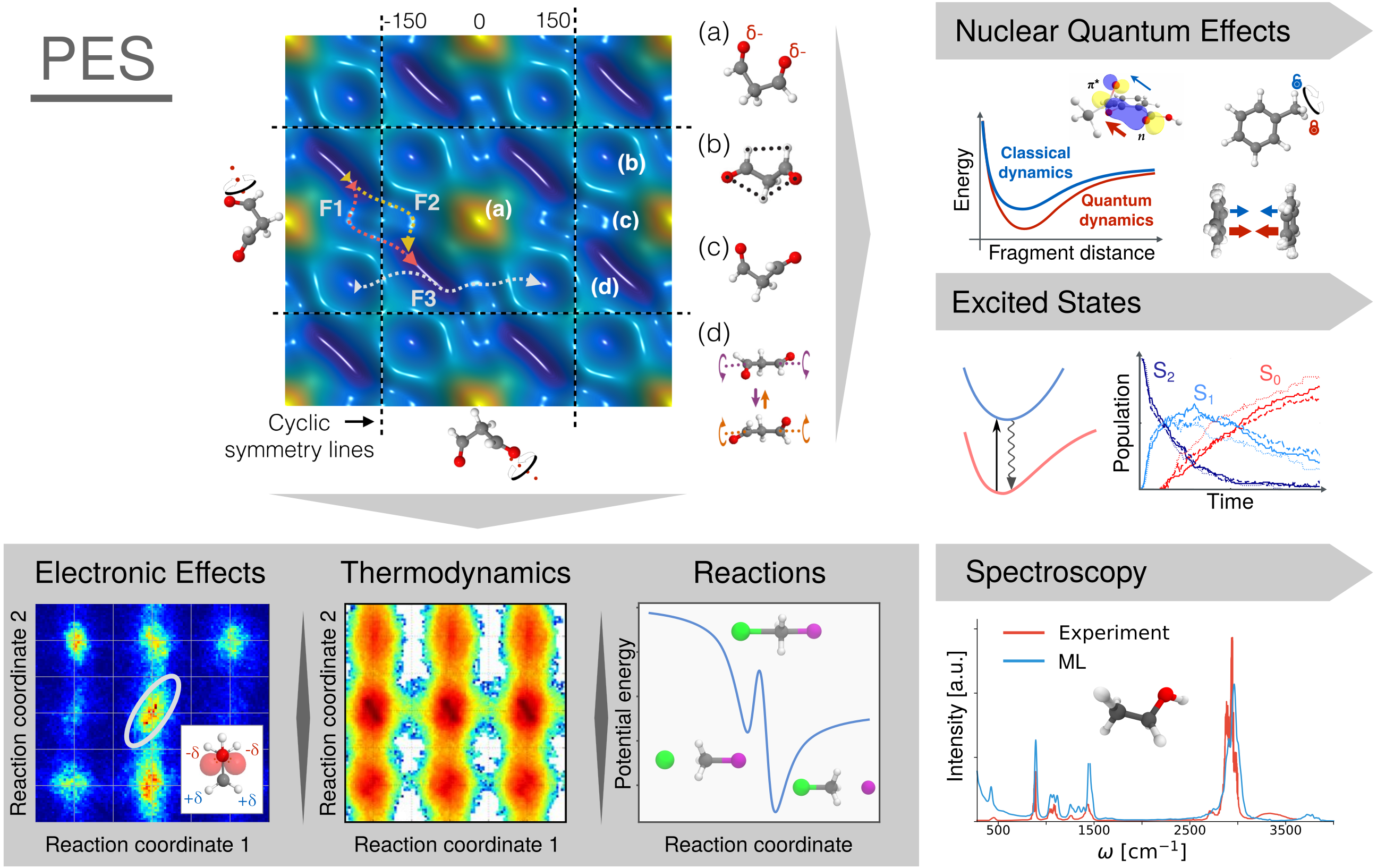}
	\caption{ML-FFs combine the accuracy of \textit{ab initio} methods and the efficiency of classical FFs. They provide easy access to a system's potential energy surface (PES), which can in turn be used to derive a plethora of other quantities.
		By using them to run MD simulations on a single PES, ML-FFs allow chemical insights inaccessible to other methods (see gray box). For example, they accurately model electronic effects and their influence on thermodynamic observables and allow a natural description of chemical reactions, which is difficult or even impossible with conventional FFs. 
		Their efficiency also allows them to be applied in situations where the Born--Oppenheimer approximation begins to break down and a single PES no longer provides an adequate description. An example is the study of nuclear quantum effects and electronically excited states (upper right).
		Finally, ML-FFs can be further enhanced by modeling additional properties. This provides direct access to a wide range of molecular spectra, building a bridge between theory and experiment (lower right).
		In general, such studies would be prohibitively expensive with \textit{ab initio} methods.}
	\label{fig:achieved}
\end{figure*}

For constructing ML-FFs, suitable reference data to learn the relevant structure-property relation include energy, forces, or a combination of both, obtained from \textit{ab initio} calculations. Contrary to conventional FFs, no preconceived notion of bonding patterns needs to be assumed. Instead, all chemical behavior is learned from the reference data. This allows to reconstruct the important interactions purely from atomic positions without imposing a restricted analytical form on the interatomic potential and enables a natural description of chemical reactions. For example, it is now possible to construct ML-FFs for small molecules from CCSD(T) reference data close to spectroscopic accuracy and with computational efficiency similar to conventional FFs\cite{chmiela2018, sauceda2019molecular}. This has enabled studies that would be prohibitively expensive with conventional methods of computational chemistry and allowed to obtain novel chemical insights (see Fig.~\ref{fig:achieved}).

Other properties than energies and forces can be predicted as well: For example, dipole moments, which are a measure for how polar molecules are, can be used to calculate infrared spectra from MD simulations\cite{gastegger2017machine, christensen2019operators,kaser2020reactive} and allow a comparison to experimental measurements. Other prediction targets could be used to screen large compound databases for molecules with desirable properties several orders of magnitude faster than with \textit{ab initio} methods. The HOMO/LUMO gap, which is important for materials used in electronic devices such as solar cells\cite{agnihotri2014computational}, is only one prominent example of many potentially interesting prediction targets.

This review will focus on the construction of ML-FFs for the usage in MD simulations and other applications (for details on how to set up such simulations or how to extract physical insights from them, refer to Refs.~\citenum{payne1986molecular,adcock2006molecular,paquet2015molecular}). The text is structured as follows:  Section~\ref{sec:mathematical_and_coceptual_framework} reviews fundamental concepts of chemistry (\ref{subsec:chemistry_foundations}) and machine learning (\ref{subsec:machine_learning_foundations}) relevant to the construction of ML-FFs and discusses special considerations when the two are combined (\ref{subsec:combining_ml_and_chemistry}). As this article is intended for both chemists and machine learning practitioners, these sections provide all readers with the necessary background to understand the remainder of the review. Experts in either field may want to skip over the respective sections, as they discuss fundamentals. The next part (Section~\ref{sec:best_practices_and_pitfalls}) serves as a step-by-step guide and reference for readers that want to apply ML-FFs in their own research. Here, the best practices for constructing ML-FFs are outlined, possible problems that may be encountered along the way (and how to avoid them) are discussed and an overview of several software packages, which may be used to accelerate the construction of ML-FFs, is provided. Section~\ref{sec:applications_of_machine_learned_force_fields} lists several applications of ML-FFs described in the literature and highlights physical and chemical insights made possible through the use of ML. The review is concluded in Section~\ref{sec:challenges} by a discussion of obstacles that still need to be overcome to extend the applicability of ML-FFs to an even broader context.

\section{Mathematical and conceptual framework}
\label{sec:mathematical_and_coceptual_framework}
Section~\ref{subsec:chemistry_foundations} reviews important chemical concepts such as the potential energy surface and invariance properties of physical systems, which are essential for constructing physically meaningful models. It is meant as a short summary of the most important physical principles and fundamental chemical knowledge for readers with a primarily ML-focused background who are interested in constructing ML-FFs. On the other hand, to offer readers with a chemical background a first orientation, an overview of two important methodologies in Machine Learning, namely kernel-based learning approaches and artificial neural networks, is given in Section~\ref{subsec:machine_learning_foundations}. Finally, Section~\ref{subsec:combining_ml_and_chemistry} lists constraints related to the physical invariances mentioned earlier and gives examples of models for constructing ML-FFs and how they implement these constraints in practice.

\subsection{Chemistry foundations}
\label{subsec:chemistry_foundations}
The Schr\"odinger equation (SE)\cite{schrodinger1926undulatory}, which describes the interaction of atomic nuclei and electrons, is sufficient for understanding most chemical phenomena and processes\cite{dirac1929quantum}. Unfortunately, it can only be solved analytically for very simple systems, such as the hydrogen atom. For more complex systems like molecules, exact numerical solutions are often impractical due to a steep increase of computational costs as a function of system size. For this reason, numerous approximation schemes have been devised to enable insights into more complicated chemical systems. Virtually all of these are based on the Born-Oppenheimer (BO) approximation\cite{born1927quantentheorie}, which decouples electronic and nuclear motion so that the latter can be neglected. It is assumed that electrons adjust instantaneously to changes in the nuclear positions, which is motivated by the observation that atomic nuclei are heavier than electrons by several orders of magnitude, thus moving on a vastly different timescale. Hence, the nuclear positions appear almost stationary to the electrons and therefore enter the resulting ``electronic SE'' only parametrically: The energy of the electrons depends on the external potential caused by the nuclei, which in turn is fully determined by their positions and nuclear charges. By summing electronic energy and Coulomb repulsion between nuclei, the total potential energy of the system is obtained, which is one of the most important properties of molecules. Alongside entropic contributions, it determines the relative stability of different compounds, whether reactions are exothermic or endothermic, and can even serve as proxy for more complex properties. For example, the potency of some drugs can be estimated from their binding energy to biomolecules\cite{meng2011molecular}.

\subsubsection{The potential energy surface}
\label{subsubsec:the_potential_energy_surface}
\begin{figure}
\centering
\includegraphics[width=\columnwidth]{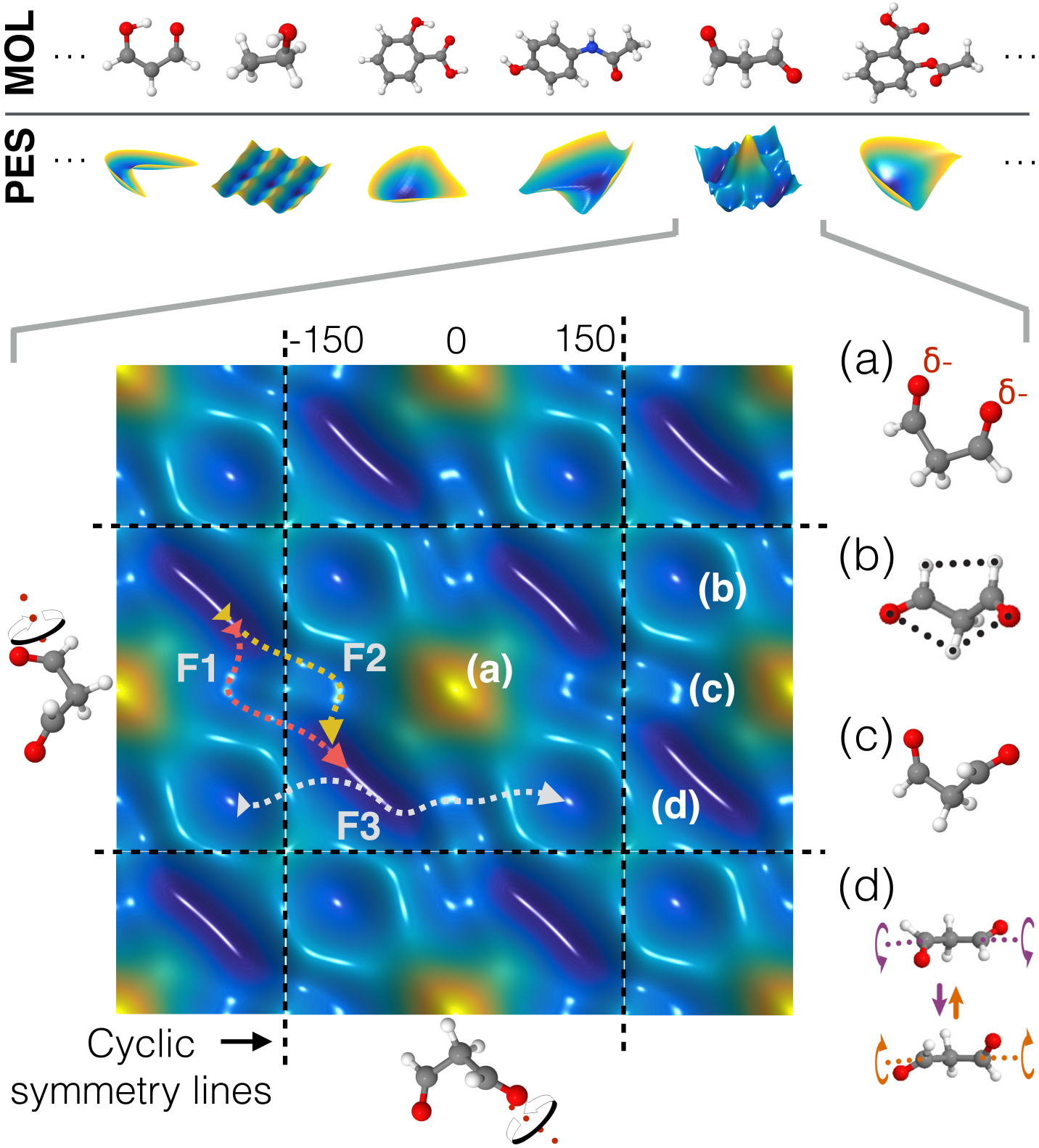}
\caption{Top: Two-dimensional projections of the PESs of different molecules, highlighting rich topological differences and various possible shapes. Bottom: Cut through the PES of keto-malondialdehyde for rotations of the two aldehyde groups. Note that the shape repeats periodically for full rotations. Regions with low potential energy are drawn in blue and high energy regions in yellow. Structure (a) leads to
a steep increase in energy due to the proximity of the two oxygen
atoms carrying negative partial charges. Local minima of the PES are shown in (b) and (c), whereas (d) displays structural fluctuations around the global minimum. By running molecular dynamics simulations, the most common transition paths (F1, F2, and F3) between the different minima could be revealed.
}
\label{fig:various_PES} 
\end{figure}

By introducing a parametric dependency between energy and nuclei, the BO approximation implies the existence of a functional relation $f:\{Z_i,\mathbf{r}_i\}_{i=1}^{N}\mapsto E$, which maps the nuclear charges $Z_i$ and positions $\mathbf{r}_i$ of $N$ atoms directly to their potential energy $E$. This function, called the potential energy surface (PES), governs the dynamics of a chemical system, similar to a ball rolling on a hilly landscape. Minima (``valleys'') on the PES correspond to stable molecules and significant structural changes (or even chemical reactions) occur when a system crosses over a transition state (``ridge'') from one minimum into another (Fig.~\ref{fig:various_PES}).

Knowledge of the PES therefore also allows to predict how a system evolves over time. For example, by studying a thermal ensemble of molecules starting from the same minimum on the PES, it is possible to determine which fraction of them will reach different minima and in what time frame, allowing to assess their reactivity and which products are formed. It is also possible to deduce the macroscopic thermodynamic properties of a system by studying how it behaves at an atomic level. In such molecular dynamics (MD) simulations, a classical treatment of nuclear dynamics is sometimes sufficiently accurate. In case of significant nuclear delocalization, which may occur in systems with light atoms, strong bonds, or for shallow potential energy landscapes\cite{tuckerman1993efficient}, nuclear quantum effects (NQEs) must be included as well. Even then, methods like path-integral MD establish a one-to-one correspondence between the properties of a quantum object and a classical system with an extended phase space, eliminating the need to solve the nuclear SE\cite{chandler1981exploiting, habershon2013ring, poltavsky2016modeling}.

At each time step of a dynamics simulation, the forces $\mathbf{F}_i$ acting on each atom~$i$ must be known so that the equations of motion can be integrated numerically (e.g.\ using the Verlet algorithm\cite{verlet1967computer}). They can be derived from the PES by using the relation $\mathbf{F}_i=-\nabla_{\mathbf{r}_i} E$, i.e.\ the forces are the negative gradient of the potential energy $E$ with respect to the atomic positions $\mathbf{r}_i$ (see also Section~\ref{subsubsec:invariances_of_physical_systems}). Forces can also be used to perform geometry optimizations, e.g.\ to find special configurations of atoms which correspond to critical points on the PES. For example, the height of a reaction barrier can be computed from the energy difference between the saddle point (transition state) and either of the two minima (equilibrium structures) which are connected by it.

Although the BO approximation simplifies the SE, even approximate solutions can be computationally demanding, in particular for large systems containing many degrees of freedom. Thus, it is often unfeasible to derive \textit{ab initio} energies and forces for each time step of an MD simulation. For this reason, analytical functions, i.e.\ force fields (FFs), are typically used to represent the PES, circumventing the problem of solving an equation altogether. The difficulty is then shifted to finding an appropriate functional form and parametrization of the FF. ML methods automate this demanding and time-consuming process by learning an appropriate function from reference data.

\subsubsection{Invariances of physical systems}
\label{subsubsec:invariances_of_physical_systems}
Closed physical systems are governed by various conservation laws that describe invariant properties. They are fundamental principles of nature that characterize symmetries that must not be violated. As such, conservation laws provide strong constraints that can be used as guiding principles in search of physically plausible ML models. The basic invariances of molecular systems are directly derived from Noether's theorem\cite{noether1918invarianten}, which states that each conserved quantity is associated with a differentiable symmetry of the action of a physical system. Typical conserved quantities include the total energy (following from temporal invariance), as well as angular and linear momentum (roto-translational invariance). 
Energy conservation imposes a particular structure on vector fields in order for them to be valid force fields with corresponding potentials. Namely, forces must be the negative gradient of the potential energy with respect to atomic positions. This relation ensures that when atoms move, they always acquire the same amount of kinetic energy as they lose in potential energy (and \textit{vice versa}), i.e.\ the total energy is constant (the work done along closed paths is zero). The conservation of linear and angular momentum implies that the potential energy of a molecule only depends on the relative position of its atoms to each other, i.e.\ it does not change with rigid rotations and/or translations.
Another invariance (not derived from Noether's theorem) follows from the fact that, from the perspective of the electrons, atoms with the same nuclear charge appear identical to each other. They can thus be exchanged without affecting the energy and forces, which makes the PES symmetric with respect to permutations of some of its arguments. To ensure physically meaningful predictions, ML-FFs must be made invariant under the same transformations as the true PES by introducing appropriate constraints.

\subsection{Machine learning foundations}
\label{subsec:machine_learning_foundations}
A question that frequently arises for researchers new to the field of ML concerns the difference of ML modeling to plain interpolation in the noise free regression case. After all, the Shannon sampling theorem gives bounds for the number of ``training samples'' needed to reconstruct a band-limited signal exactly\cite{shannon1949communication}. Since the regression tasks considered in this review use \textit{ab initio} data as reference, they can be considered practically noise-free. Furthermore, PESs are usually smooth, i.e.\ there is a well-defined frequency cutoff in the spectrum of this ``signal''. Thus, both requirements for Shannon interpolation are satisfied and it should in principle be possible to reconstruct FFs via interpolation of the training samples without error, provided there are enough of them.

This is where ML diverges from signal interpolation theory. In practice, there is often not enough data available to fully capture all the necessary information for a perfect reconstruction. In that case, the goal of ML methods is not to recover the training data, but rather to estimate the true process with its underlying regularities that also describes all new and unseen data -- this is often denoted as generalization.
The key to generalization is selecting a model based on the well known principle of Occam's razor, i.e.\ the notion that simpler hypotheses are more likely to be correct\cite{schaffer2015not}. The capacity of the model can be controlled using the bias--variance trade-off\cite{geman1992neural} (a compromise between expressiveness and complexity) and is practically done by exercising model selection techniques (see Section~\ref{subsubsec:model_selection_how_to_choose_hyperparameters}) such as cross-validation that leave out part of the  data  from the ML training process and use it later to obtain a valid estimate of the generalization error.\cite{muller2001introduction,hansen2013assessment}
The reason why regularization is often needed is that ML algorithms are \textit{universal approximators}, i.e.\ they can approximate any continuous function on a closed interval arbitrarily close. Since for a finite amount of reference data infinitely many such functions are thinkable, a regularization mechanism is often needed to select a preferably simple function from the vast space of possibilities.

\begin{figure}
	\includegraphics[width=\columnwidth]{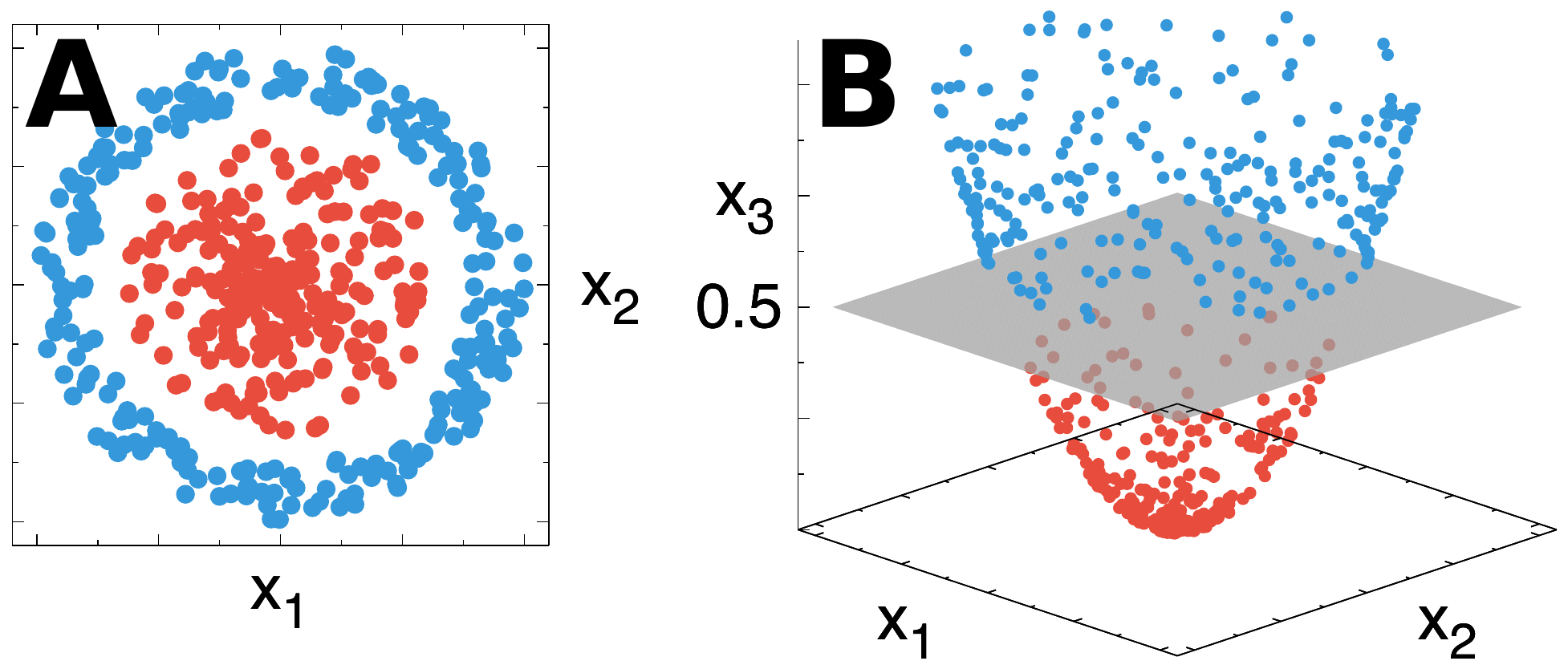}
	\caption{\textbf{A}: The blue and red points with coordinates
		($x_1$, $x_2$) are linearly
		inseparable. \textbf{B}: By defining a suitable mapping from the
		input space ($x_1$, $x_2$) to a higher-dimensional feature
		space ($x_1$, $x_2$, $x_3$), blue and red points become
		linearly separable (gray plane at $x_3=0.5$).}
	\label{fig:linear_separability_in_feature_space}
\end{figure}

ML methods typically rely on the fact that nonlinear problems, such as predicting energy from nuclear positions, can be ``linearized'' by mapping the input
to a (often higher-dimensional) ``feature space'' (see
Fig.~\ref{fig:linear_separability_in_feature_space}).\cite{scholkopf1997kernel,scholkopf1998nonlinear,scholkopf1999input,muller2001introduction} 
Note that such feature spaces are explicitly constructed for kernel-based learning methods (see Section~\ref{subsubsec:kernel_based_methods}) or learned respectively for deep learning models \cite{bishop1995neural} (see Section~\ref{subsubsec:artificial_neural_networks}).
Kernel-based methods achieve this by taking advantage of the so-called \textit{kernel trick}\cite{muller1997predicting,boser1992training,scholkopf1998nonlinear,theodoridis2008pattern,theodoridis2020machine} which allows implicitly operating in a high-dimensional feature space without explicitly performing any computation there. In contrast, artificial neural networks (NNs) decompose a complex non-linear function into a composition of linear transformations with learnable parameters connected by nonlinear activation functions. With increasingly many of such nonlinear transformations organized in ``layers'' (\emph{deep} NNs), it is possible to efficiently  learn highly  complex feature spaces.

\begin{figure}
	\centering
	\includegraphics[width=1.0\columnwidth]{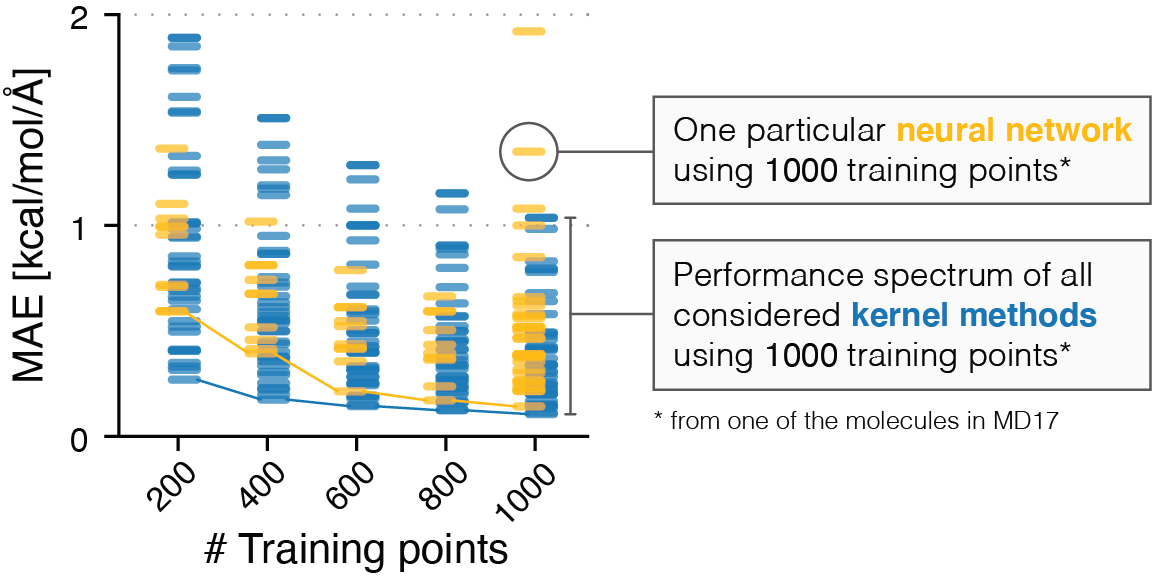}
	\caption{Mean absolute force prediction errors (MAEs) of different ML models trained on molecules in the MD17 dataset\cite{chmiela2017}, colored by model type. Overall, kernel methods (GDML\cite{chmiela2017}, sGDML\cite{chmiela2018}, FCHL18/19\cite{faber2018alchemical,christensen2020fchl}) are slightly more data efficient, i.e.\ they produce more accurate predictions with smaller training datasets, but neural network architectures (PhysNet\cite{unke2019}, SchNet\cite{schutt2017schnet}, DimeNet\cite{klicpera_dimenet_2020}, EANN\cite{zhang2019embedded}, DeePMD\cite{zhang2018deep}, DeepPot-SE\cite{zhang2018end}, ACSF\cite{behler2007generalized}, HIP-NN\cite{lubbers2018hierarchical}) catch up quickly with increasing training set size and continue to improve when more data for training is available.}
	\label{fig:md17_scatter}
\end{figure}

While NNs tend to require more training data to reach the same accuracy as kernel methods (see Fig.~\ref{fig:md17_scatter}),\cite{kamath2018neural} they typically scale better to larger data sets. In general, neither method is strictly superior over the other\cite{wolpert1996lack} and both have advantages and disadvantages that must be weighed against each other for a specific application. Recently, it has even been discovered that in the limit of infinitely wide layers, deep NNs are equivalent to kernel methods, which shifts the main differentiating factor between both methodologies to how they are constructed and trained\cite{lee2017deep, matthews2018gaussian} and makes deep NNs accessible to kernel-based analysis methods \cite{braun2008relevant,montavon2011kernel}.

In the following, kernel methods and neural networks are described in more detail to highlight the most important properties that differentiate both methodologies.

\subsubsection{Kernel-based methods}
\label{subsubsec:kernel_based_methods}

\begin{figure*}
    \centering
    \includegraphics[width=0.95\textwidth]{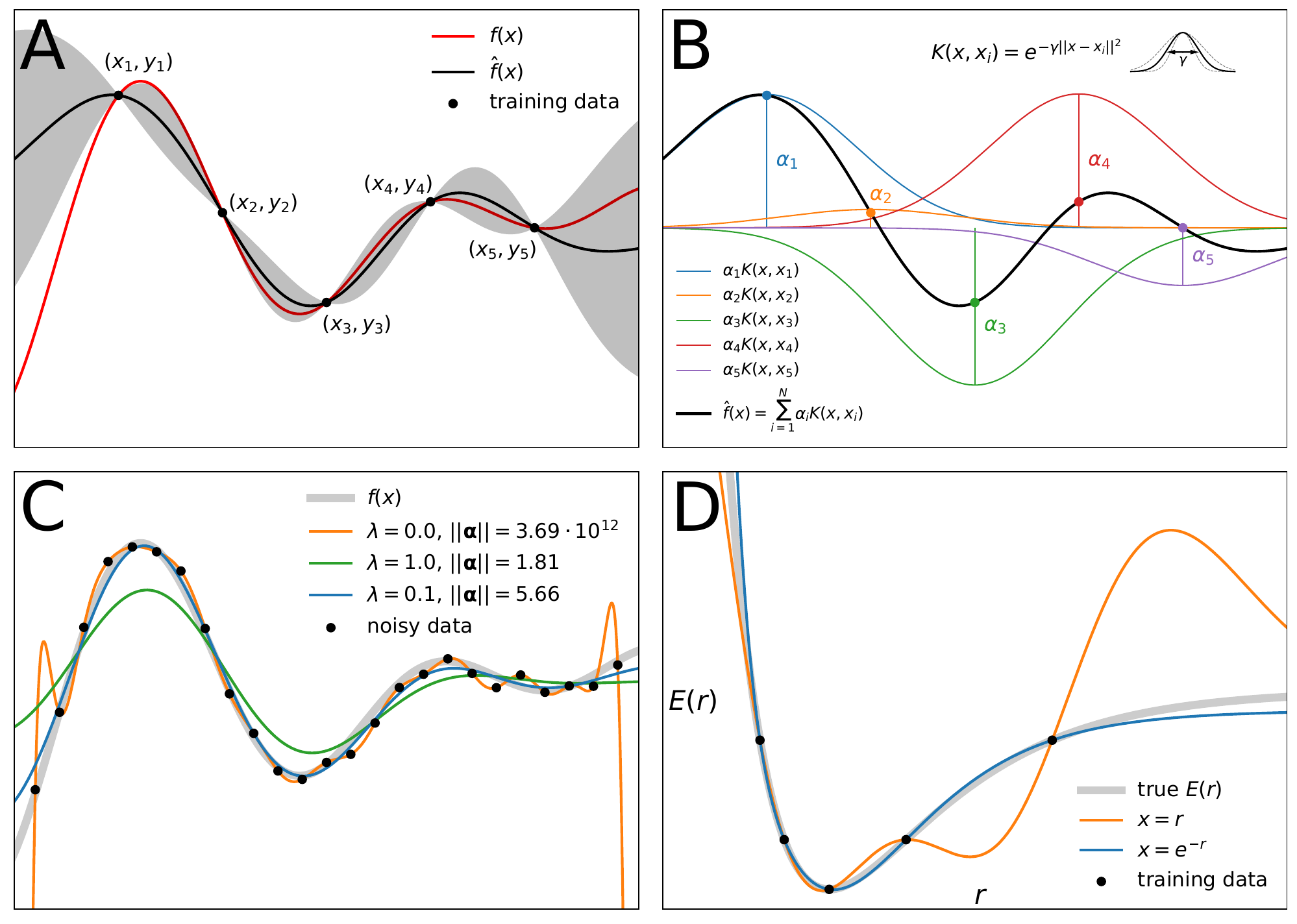}
    \caption{Overview of the mathematical concepts that form the basis of kernel methods.
    \textbf{A}: Gaussian process regression of a one-dimensional function $f(x)$ (red line) from $M=5$ data samples $(x_i,y_i)$. The black line $\hat{f}(x)$ depicts the mean (Eq.~\ref{eq:gaussian_process_mean}) of the conditional probability $p(y_*|\mathbf{y})$ (see Eq.~\ref{eq:gaussian_process_distribution}), whereas the gray area depicts two standard deviations from its mean (see Eq.~\ref{eq:gaussian_process_variance}). Note that predictions are most confident in regions where training data is present. \textbf{B}: The function $\hat{f}(x)$ can be expressed as a linear combination of $M$ kernel functions $K(x,x_i)$ weighted with regression coefficients $\alpha_i$ (see Eq.~\ref{eq:kernel_regression}). In this example, the Gaussian kernel (Eq.~\ref{eq:gaussian_kernel}) is used (the hyperparameter $\gamma$ controls its width). \textbf{C}: Influence of noise on prediction performance. Here, the function $f(x)$ (thick gray line) is learned from $M=25$ samples, however, each data point $(x_i,y_i)$ contains observational noise (see Eq.~\ref{eq:gaussian_process_noise_model}). When the coefficients $\alpha_i$ are determined without regularization, i.e.\ no noise is assumed to be present, the model function reproduces the training samples faithfully, but undulates wildly between data points (orange line, $\lambda=0$). The regularized solution (blue line, $\lambda=0.1$, see Eq.~\ref{eq:krr_coefficient_relation_regularized}) is much smoother and stays closer to the true function $f(x)$, but individual data points are not reproduced exactly. When the regularization is too strong (green line, $\lambda=1.0$), the model function becomes unable to fit the data. Note how regularization shrinks the magnitude of the coefficient vectors  $\lVert\boldsymbol{\alpha}\rVert$. \textbf{D}: For constructing force fields, it is necessary to encode molecular structure with a representation $\mathbf{x}$. The choice of this structural descriptor may strongly influence model performance. Here, the potential energy $E$ of a diatomic molecule (thick gray line) is learned from $M=5$ data points by two kernel machines using different structural representations (both models use a Gaussian kernel). When the interatomic distance $r$ is used as descriptor (orange line, $x=r$), the predicted potential energy oscillates between data points, leading to spurious minima and qualitatively wrong behavior for large $r$. A model using the descriptor $x=e^{-r}$ (blue line) predicts a physically meaningful potential energy curve that is qualitatively correct even when the model extrapolates.}
    \label{fig:kernel_basics}
\end{figure*}

\begin{figure}
    \centering
    \includegraphics[width=\columnwidth]{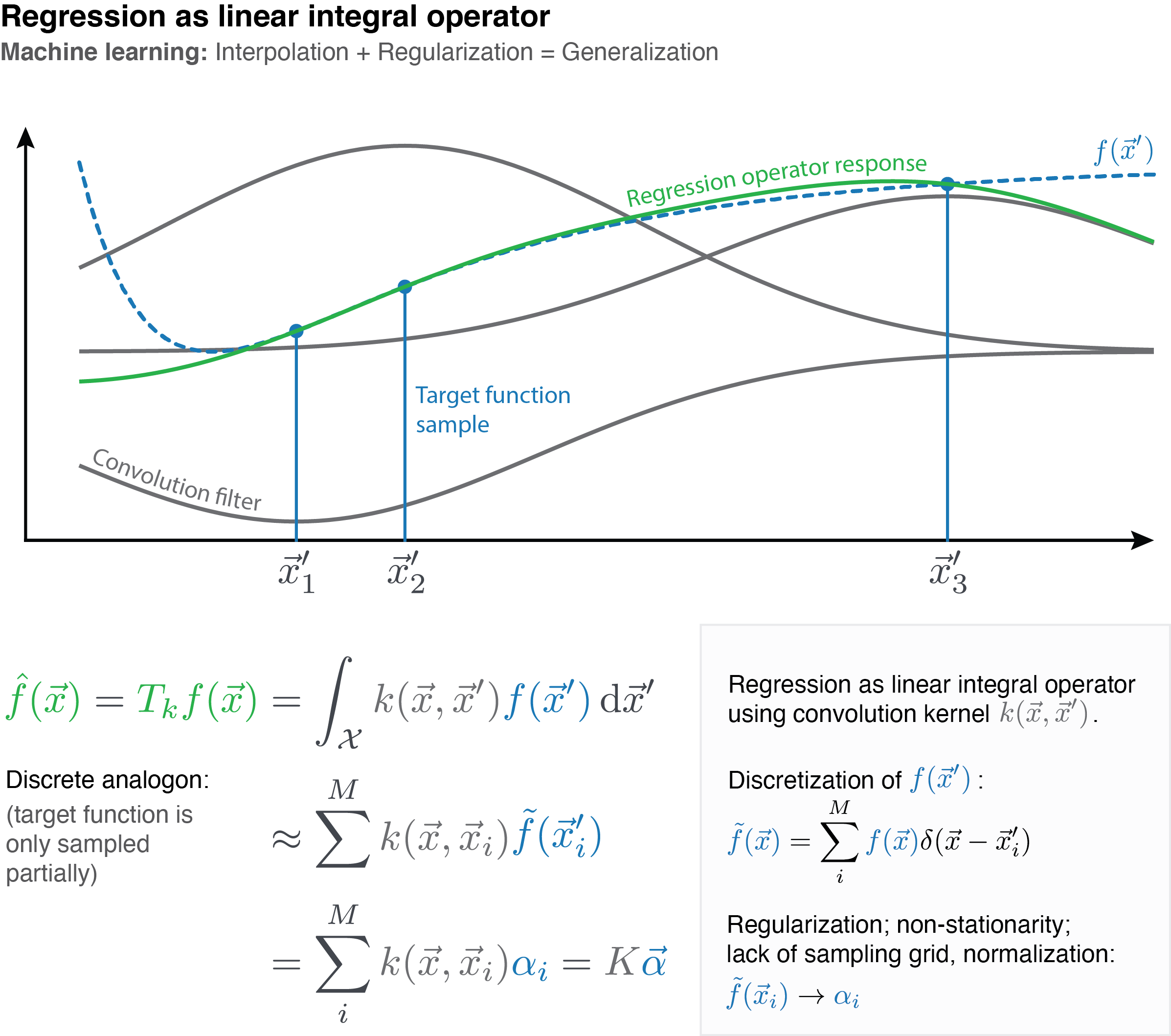}
    \caption{Kernel ridge regression can be understood as a linear integral operator $T_k$ that is applied to the (only partially known) target function of interest $f(\mathbf{x})$. Such operators are defined as convolutions with a continuous kernel function $K$, whose response is the regression result. Because the training data is typically not sampled on a grid, this convolution task transforms to a linear system that yields the regression coefficients $\mathbf{\alpha}$. Because only $T_k f(\mathbf{x})$ and not the true $f(\mathbf{x})$ is recovered, the challenge is to find a kernel that defines an operator that leaves the relevant parts of its original function invariant. This is why the Gaussian kernel (Eq.~\ref{eq:gaussian_kernel}) is a popular choice: Depending on the chosen length scale $\gamma$, it attenuates high frequency components, while passing through the low frequency components of the input, therefore making only minimal assumptions about the target function. However, stronger assumptions (e.g.\ by combining kernels with physically-motivated descriptors) increase the sample efficiency of the regressor.}
    \label{fig:regression_operators}
\end{figure}

Given a data set $\{(y_i; \mathbf{x}_i)\}_{i=1}^{M}$ of $M$ reference values $y_i \in \mathbb{R}$ for inputs $\mathbf{x}_i \in
\mathbb{R}^D$, kernel regression aims to estimate $y_*$ for unknown inputs $\mathbf{x}_*$. For example, for PES construction, $y$ is the potential energy and $\mathbf{x}$ encodes structural information about the atoms, i.e.\ their nuclear charges and relative positions in space. Popular choices for such ``descriptors'' are vectors of internal coordinates, Coulomb matrices\cite{rupp2012fast}, representations of atomic environments (e.g.\ symmetry functions\cite{behler2011atom},
SOAP\cite{bartok2013representing} or
FCHL\cite{faber2018alchemical,christensen2020fchl}), or an encoding of crystal structure\cite{schutt2014represent,faber2015crystal,faber2016machine}. 

The representer theorem states that the functional relation 
\begin{equation}
    y = f(\mathbf{x}) + \epsilon\,,
    \label{regressionwithnoise}
\end{equation}
where $\epsilon$ denotes measurement noise,  can be optimally approximated as a linear combination
\begin{equation}
f(\mathbf{x}) \approx \hat{f}(\mathbf{x}) = \sum_{i = 1}^{M}
\alpha_i K(\mathbf{x},\mathbf{x}_i)\,,
\label{eq:kernel_regression}
\end{equation}
where $\alpha_i$ are coefficients and $K(\mathbf{x},\mathbf{x'})$ is a (typically nonlinear) symmetric and positive semi-definite function\cite{wahba1990spline,scholkopf2001generalized,argyriou2009there} that measures the similarity of two compound descriptors $\mathbf{x}$ and $\mathbf{x'}$ (see Fig.~\ref{fig:regression_operators}).\footnote{The function $K(\mathbf{x},\mathbf{x'})$ computes the inner product of two points $\phi(\mathbf{x})$ and $\phi(\mathbf{x'})$ in some Hilbert space $\mathcal{H}$ (the feature space) without the need to evaluate (or even know) the mapping $\phi: \mathbb{R}^D \mapsto \mathcal{H}$ explicitly, i.e.\ $K(\mathbf{x},\mathbf{x'})$ is a reproducing kernel of $\mathcal{H}$.\cite{berlinet2011reproducing,muller2001introduction}}
Examples for such functions $K$ are the polynomial kernel \begin{equation}
K(\mathbf{x},\mathbf{x'}) = (\langle\mathbf{x},\mathbf{x'}\rangle+ c)^d\,,
\label{eq:polynomial_kernel}
\end{equation}
where hyperparameter $d$ is the degree of the polynomial and $\langle \cdot, \cdot \rangle$ is the dot product, or the Gaussian kernel given by
\begin{equation}
K(\mathbf{x},\mathbf{x'}) = e^{-\gamma\lVert
	\mathbf{x}-\mathbf{x'}\rVert^2}
\label{eq:gaussian_kernel}
\end{equation}
with hyperparameter $\gamma$ controlling its width/scale and $\lVert \cdot \rVert$ denoting the $L^2$-norm (see Refs.~\citenum{muller2001introduction, hansen2013assessment, scholkopf2002learning, theodoridis2008pattern} for more examples of kernel functions).

The structure and number of dimensions of the associated Hilbert space $\mathcal{H}$ depends on the choice of $K(\mathbf{x},\mathbf{x'})$ and dimension of the inputs $\mathbf{x}$ and $\mathbf{x'}$. As an example, consider the polynomial kernel (Eq.~\ref{eq:polynomial_kernel}) with degree $d=2$ and two-dimensional inputs. The corresponding homogeneous $(c=0)$ polynomial mapping is given by $\phi:(x_1,x_2)\mapsto(x_1^2, \sqrt{2}x_1x_2, x_2^2)$, so the associated $\mathcal{H}$ is three-dimensional. While in this case, it is still possible to compute $\phi$ and evaluate the inner product of two points $\phi(\mathbf{x})$ and $\phi(\mathbf{x'})$ explicitly, the advantage of using kernels becomes apparent when the Gaussian kernel (Eq.~\ref{eq:gaussian_kernel}) is considered. Rewriting Eq.~\ref{eq:gaussian_kernel} as 
\begin{equation}
K(\mathbf{x},\mathbf{x'}) = e^{-\gamma\lVert \mathbf{x}\rVert^2}e^{-\gamma\lVert \mathbf{x'}\rVert^2}e^{2\gamma \langle\mathbf{x},\mathbf{x'}\rangle}
\label{eq:gaussian_kernel_rewritten}
\end{equation}
and expanding the third factor in a Taylor series
$e^{2\gamma\langle\mathbf{x},\mathbf{x'}\rangle}=\sum_{d=0}^\infty\frac{1}{d!}\left(
2\gamma\langle\mathbf{x},\mathbf{x'}\rangle\right)^d$ reveals that the Gaussian kernel is equivalent to an infinite sum over (scaled) polynomial kernels (see Eq.~\ref{eq:polynomial_kernel}) and the associated $\mathcal{H}$ is $\infty$-dimensional. Fortunately, by using the kernel function $K(\mathbf{x},\mathbf{x'})$, it is possible to operate in
$\mathcal{H}$ implicitly and evaluate $\hat{f}(\mathbf{x})$ (Eq.~\ref{eq:kernel_regression}) without computing the mapping $\phi$. This is often referred to as the
\textit{kernel trick}\cite{muller1997predicting,boser1992training,scholkopf1998nonlinear,muller2001introduction,scholkopf2002learning,theodoridis2008pattern}. 

It remains the question how the coefficients $\alpha_i$ in Eq.~\ref{eq:kernel_regression} are determined. One way to do so is by adopting a Bayesian, or probabilistic, point of view.\cite{rasmussen2003gaussian,murphy2012machine} Here, it is assumed that the reference data $\{(y_i; \mathbf{x}_i)\}_{i=1}^{M}$ are generated by a Gaussian process (GP), i.e.\ drawn from a multivariate Gaussian distribution. For simplicity, it can be assumed that this distribution has a mean of zero, as other values can be generated by simply adding a constant term. Further, the possibility that the reference data might be contaminated by noise (for example due to uncertainties in measuring $y_i$) is accounted for explicitly. Typically, Gaussian noise is assumed, i.e.\
\begin{equation}
y_i =  f(\mathbf{x}_i) + \mathcal{N}(0,\lambda)\,,
\label{eq:gaussian_process_noise_model}
\end{equation}
where $\lambda$ is the variance of the normally distributed noise $\mathcal{N}$. In the GP picture, the choice of $K(\mathbf{x},\mathbf{x'})$ expresses an assumption about the underlying function class. For example, choosing the Gaussian kernel implies that $f(\mathbf{x})$ does not change drastically over a length scale controlled by $\gamma$ (see Eq.~\ref{eq:gaussian_kernel}). As such, a particular kernel function $K$ corresponds to an implicit regularization, i.e.\ an assumption about the underlying smoothness properties of the function to be estimated.\cite{smola1998connection} The challenge lies in finding a kernel that represents the structure in the data that is being modeled as good as possible.\cite{smola1998connection,braun2008relevant} Many kernels are able to approximate continuous functions on a compact subset arbitrarily well,\cite{smola1998connection,micchelli2006universal} but a strong prior has the advantage of restricting the hypothesis space, which drastically improves the convergence of the learning task with respect to the available training data.\cite{huang2016communication} 

Under these conditions, it is now possible to rigorously answer the question
``given the data $\mathbf{y}=\left[y_1 \cdots y_M\right]^{\top}$,
how likely is it to observe the value $y_*$ for input
$\mathbf{x}_*$?'' As $y_*$ is generated by the same GP as the reference data, the conditional probability $p(y_*|\mathbf{y})$ can be expressed as
\begin{equation}
\begin{bmatrix}
\mathbf{y}\textbf{}\\
y^*
\end{bmatrix}
\sim
\mathcal{N}\left(0,
\begin{bmatrix}
\mathbf{K} + \lambda\mathbf{I}_M & \mathbf{K}_*^{\top}\\
\mathbf{K}_* &  K(\mathbf{x}_*,\mathbf{x}_*)
\end{bmatrix}
\right)\,,
\label{eq:gaussian_process_distribution}
\end{equation}
where $\mathbf{I}_M$ is the identity matrix of size $M$, $\mathbf{K}$ is the $M\times M$ kernel matrix\cite{hofmann2008kernel,muller2001introduction} with entries $K_{ij}=K(\mathbf{x}_i,\mathbf{x}_j)$ and $\mathbf{K}_*=\left[K(\mathbf{x}_*,\mathbf{x}_1) \cdots
K(\mathbf{x}_*,\mathbf{x}_M)\right]$. In other words, Eq.~\ref{eq:gaussian_process_distribution} expresses a probability distribution over possible predictions, where its mean value
\begin{equation}
\bar{y}_* = \mathbf{K}_* \left(\mathbf{K} + \lambda\mathbf{I}_M\right)^{-1}\mathbf{y}
\label{eq:gaussian_process_mean}
\end{equation}
is the most likely estimate for $y_*$ (given the reference data) and its variance 
\begin{equation}
\mathrm{var}(y_*) = K(\mathbf{x}_*,\mathbf{x}_*) - \mathbf{K}_* \left(\mathbf{K} + \lambda\mathbf{I}_N\right)^{-1} \mathbf{K}_*^{\top}
\label{eq:gaussian_process_variance}
\end{equation}
provides information about how strongly other likely predictions vary from the mean. Note that while Eq.~\ref{eq:gaussian_process_variance} can be used as uncertainty estimate for a particular prediction, it should not be confused with error bars. The optimal coefficients $\boldsymbol{\alpha} = \left[\alpha_1 \cdots
\alpha_M\right]^{\top}$ in Eq.~\ref{eq:kernel_regression} are thus given by 
\begin{equation}
\boldsymbol{\alpha} = \left(\mathbf{K}+\lambda\mathbf{I}_M\right)^{-1}\mathbf{y}
\label{eq:krr_coefficient_relation_regularized}
\end{equation}
or simply $\boldsymbol{\alpha} = \mathbf{K}^{-1}\mathbf{y}$ in the noise-free case ($\lambda=0$). However, even in the absence of noise, it can be beneficial to choose a non-zero $\lambda$ to obtain a regularized solution. The addition of $\lambda > 0$ to the diagonal of $\mathbf{K}$ increases numerical stability and has the effect of damping the magnitude of the coefficients, thereby increasing the smoothness of $\hat{f}(\mathbf{x})$. The downside is that the known reference values $y_i$ are only approximately reproduced. This, however, also decreases the chance of overfitting and can lead to better generalization, i.e.\ increased accuracy when predicting unknown values. 

Matrix factorization methods like Cholesky decomposition\cite{golub2012matrix} are typically used to efficiently solve the linear problem in Eq.~\ref{eq:krr_coefficient_relation_regularized} in closed form. However, this type of approach scales as $\mathcal{O}(M^3)$ with the number of reference data and may become problematic for extremely large data sets. Iterative, e.g.\ gradient-based, solvers can reduce the complexity to $\mathcal{O}(M^2)$.\cite{raykar2007fast} Once the coefficients have been determined, the value $y_*$ for an arbitrary input $\mathbf{x}_*$ can be estimated according to
Eq.~\ref{eq:kernel_regression} with $\mathcal{O}(M)$ complexity (a sum over all $M$ reference data points is required). 

Alternatively, a variety of approximation techniques exploit that kernel matrices usually have a small \emph{numerical} rank, i.e.\ a rapidly decaying eigenvalue spectrum. This enables approximate factorizations $\mathbf{R}\mathbf{R}^T \approx \mathbf{K}$, where $\mathbf{R}$ is either a rectangular matrix $\in \mathbb{R}^{M \times L}$ with $L < M$ or sparse. As a result Eq.~\ref{eq:krr_coefficient_relation_regularized} becomes easier to solve, albeit the result will not be exact\cite{williams2001using, quinonero2005unifying, snelson2006sparse, rahimi2008random, rudi2017falkon}. 

A straightforward approach to approximate a linear system is to pick a representative or random subset of $L$ points $\tilde{\mathbf{x}}$ from the dataset (in principle, even arbitrary $\tilde{\mathbf{x}}$ could be chosen) and construct a rectangular kernel matrix $\mathbf{K}
_{LM}\in\mathbb{R}
^{L\times M}$ with entries $K_{LM,ij} = K(\tilde{\mathbf{x}}_i, \mathbf{x}_j)$. Then, the corresponding coefficients can be obtained via the Moore-Penrose pseudoinverse\cite{moore1920reciprocal,penrose1955generalized}
\begin{equation}
\tilde{\boldsymbol{\alpha}} = (1+\lambda)^{-1}\left(\mathbf{K}_{LM}\mathbf{K}_{LM}^{\top}\right)^{-1}\mathbf{K}_{LM}\mathbf{y}\,.
\label{eq:krr_coefficient_relation_approximated}
\end{equation}
Solving Eq.~\ref{eq:krr_coefficient_relation_approximated} scales as $\mathcal{O}(ML^2)$ and is much less computationally demanding than inverting the original matrix in Eq.~\ref{eq:krr_coefficient_relation_regularized}. Once the $L$ coefficients $\tilde{\boldsymbol{\alpha}}$ are obtained, the model can be evaluated with $\hat{f}(\mathbf{x})=\sum_L \tilde{\alpha}_iK(\mathbf{x},\tilde{\mathbf{x}}_i)$, i.e.\ an additional benefit is that evaluation now scales as $\mathcal{O}(L)$ instead of $\mathcal{O}(M)$ (see Eq.~\ref{eq:kernel_regression}).

However, the approximation above gives rise to an over-determined system with fewer parameters than training points and therefore reduced model capacity. Strictly speaking, the involved matrix does not satisfy the properties of a kernel matrix anymore, as it is neither symmetric nor positive semi-definite. To obtain a kernel matrix that still maintains these properties, the Nystr\"om~\cite{williams2001using} approximation
\begin{equation}
\mathbf{K} \approx \tilde{\mathbf{K}}=\mathbf{K}_{L M}^{\top} \mathbf{K}_{L L}^{-1} \mathbf{K}_{L M}
\label{eq:nystroem_approx}
\end{equation}
can be used instead. Here, the sub-matrix $\mathbf{K}_{L L}$ is a true kernel matrix between all inducing points $\tilde{\mathbf{x}}_i$. Using the Woodbury matrix identity\cite{cutajar2016preconditioning}, the regularized inverse is given by
\begin{equation}
\begin{aligned}
&(\tilde{\mathbf{K}}+\lambda \mathbf{I}_M)^{-1} = \lambda^{-1}[\mathbf{I}_M-\mathbf{K}_{L M}^{\top} \\ 
& \: \left(\lambda \mathbf{K}_{L L}+\mathbf{K}_{L M} \mathbf{K}_{L M}^{\top}\right)^{-1} \mathbf{K}_{L M}]
\end{aligned}
\label{eq:nystroem_regularized_inverse}
\end{equation}
and $\tilde{\boldsymbol{\alpha}} = (\tilde{\mathbf{K}}+\lambda \mathbf{I}_M)^{-1} \mathbf{y}$. The computational complexity of solving the Nystr\"om approximation is $\mathcal{O}(L^3 + ML^2)$.

It should be mentioned that kernel regression methods are known under different names in the literature of different communities. Due to their relation to GPs, some authors prefer the name Gaussian process regression (GPR). Others favor the term kernel ridge regression (KRR), since determining the coefficients with Eq.~\ref{eq:krr_coefficient_relation_regularized} corresponds to solving a least-squares objective with $L^2$-regularization in the kernel feature space $\phi$ and is similar to ordinary ridge regression\cite{tikhonov1977solutions}. Sometimes, the method is also referred to as reproducing kernel Hilbert space (RKHS) interpolation, since Eq.~\ref{eq:kernel_regression} ``interpolates'' between known reference values (when coefficients are determined with $\lambda=0$, all known reference values are reproduced exactly). All these methods are formally equivalent and essentially differ only in the manner the relevant equations are derived. There are small philosophical differences, however: For example, in the KRR and RKHS pictures, $\lambda$ in Eq.~\ref{eq:krr_coefficient_relation_regularized} is a regularization hyperparameter that has to be introduced \textit{ad hoc}, whereas in the GPR picture, $\lambda$ is directly related to the Gaussian noise in Eq.~\ref{eq:gaussian_process_noise_model}. The expansion coefficients obtained from Eq.~\ref{eq:krr_coefficient_relation_regularized} can change drastically depending on the choice of $\lambda$, so this is an important detail. Further, while Eq.~\ref{eq:gaussian_process_variance} can be used to compute uncertainty estimates for all kernel regression methods, the GPR picture allows to relate it directly to the variance of a Gaussian process.

The most important concepts discussed in this section are summarized visually in Fig~\ref{fig:kernel_basics}.

\subsubsection{Artificial neural networks}
\label{subsubsec:artificial_neural_networks}

\begin{figure*}
    \centering
    \includegraphics[width=\textwidth]{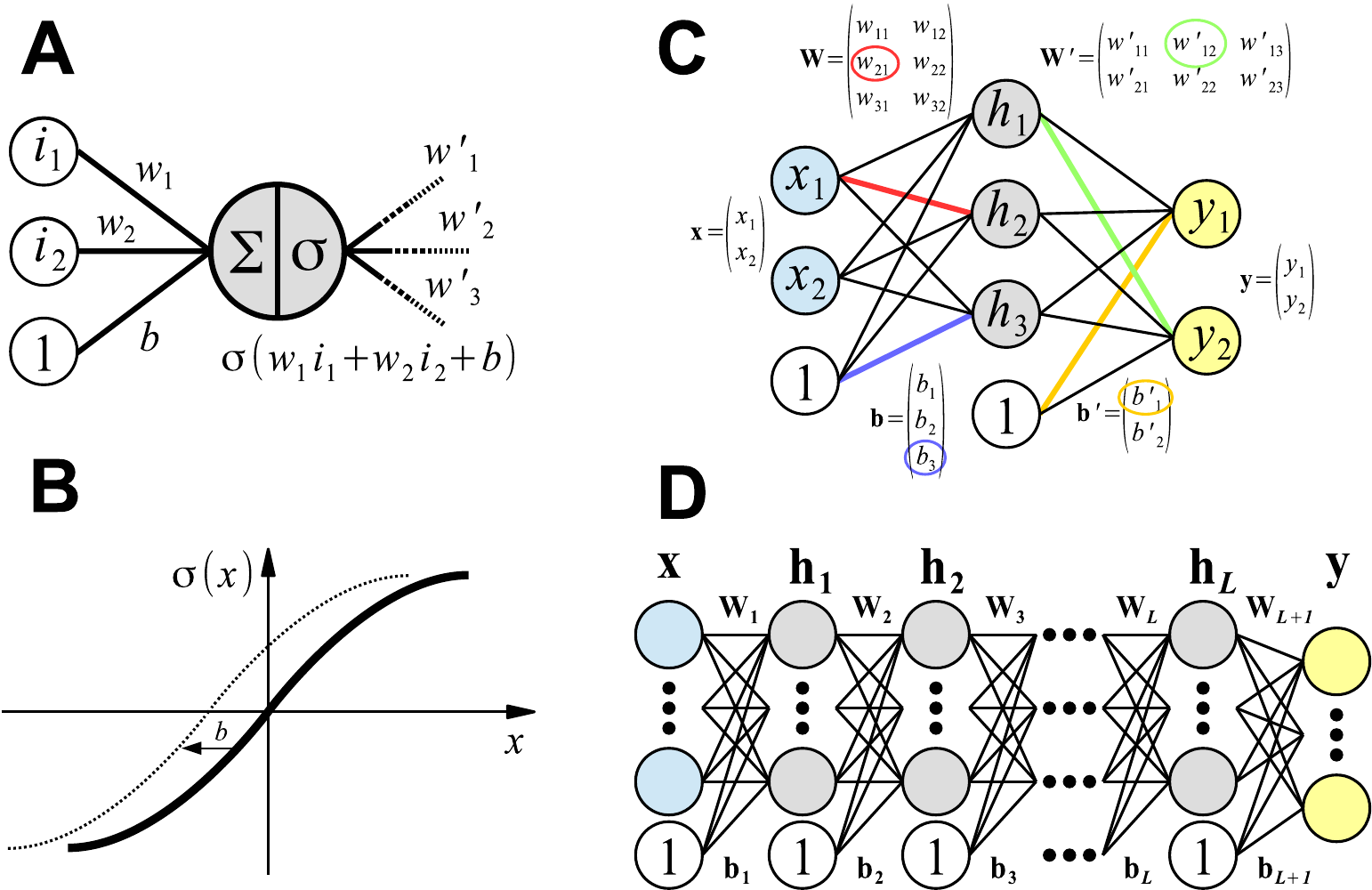}
    \caption{Schematic representation of the mathematical concepts underlying artificial (feed-forward) neural networks.
    \textbf{A}: A single artificial neuron can have an arbitrary number of inputs and outputs. Here, a neuron that is connected to two inputs $i_1$ and $i_2$ with ``synaptic weights'' $w_1$ and $w_2$ is depicted. The bias term $b$ can be thought of as the weight of an additional input with a value of 1. Artificial neurons compute the weighted sum of their inputs and pass this value through an activation function $\sigma$ to other neurons in the neural network (here, the neuron has three outputs with connection weights $w'_1$, $w'_2$, and $w'_3$). \textbf{B}: A possible activation function $\sigma(x)$. The bias term $b$ effectively shifts the activation function along the $x$-axis. Many non-linear functions are valid choices, but the most popular are sigmoid transformations such as $\tanh(x)$ or (smooth) ramp functions, e.g.\ $\mathrm{max}(0,x)$ or $\ln(1+e^x)$. \textbf{C}: Artificial neural network with a single hidden layer of three neurons (gray) that maps two inputs $x_1$ and $x_2$ (blue) to two outputs $y_1$ and $y_2$ (yellow), see Eq.~\ref{eq:single_layer}. For regression tasks, the output neurons typically use no activation function. Computing the weighted sums for the neurons of each layer can be efficiently implemented as a matrix vector product (Eq.~\ref{eq:dense_layer}). Some entries of the weight matrices ($\mathbf{W}$ and  $\mathbf{W'}$) and bias vectors ($\mathbf{b}$ and  $\mathbf{b'}$) are highlighted in color with the corresponding connection in the diagram. \textbf{D}: Schematic depiction of a \emph{deep} neural network with $L$ hidden layers (Eq.~\ref{eq:deep_feedforward_nn}). Compared to using a single hidden layer with many neurons, it is usually more parameter-efficient to connect multiple hidden layers with fewer neurons sequentially.}
    \label{fig:nn_basics}
\end{figure*}
Originally, artificial neural networks (NNs) were, as suggested by their name, intended to model the intricate networks formed by biological neurons\cite{mcculloch1943logical}. Since then, they have become a standard ML algorithm\cite{mcculloch1943logical,kohonen1988introduction,abdi1994neural,bishop1995neural,clark1999neural,ripley2007pattern,haykin2009neural,lecun2012efficient,theodoridis2020machine} only remotely related to their original biological inspiration. In the simplest case, the fundamental building blocks of NNs are 
dense (or ``fully-connected'') layers -- linear transformations from input vectors
$\mathbf{x}\in \mathbb{R}^{n_{\rm in}}$ to output vectors
$\mathbf{y}\in \mathbb{R}^{n_{\rm out}}$ according to
\begin{equation}
\mathbf{y} = \mathbf{W}\mathbf{x} + \mathbf{b}\,,
\label{eq:dense_layer}
\end{equation}
where both weights $\mathbf{W}\in \mathbb{R}^{n_{\rm out} \times n_{\rm in}}$ and biases $\mathbf{b}\in \mathbb{R}^{n_{\rm out}}$ are parameters, and $n_{\rm in}$ and $n_{\rm out}$ denote the
number of dimensions of $\mathbf{x}$ and $\mathbf{y}$, respectively. Evidently, a single dense layer can only express linear functions. Non-linear relations between inputs and outputs can only be modeled when at least two dense layers are stacked and combined with a non-linear \emph{activation function} $\sigma$:
\begin{equation}
\label{eq:single_layer}
\begin{aligned}
\mathbf{h} &= \sigma\left(\mathbf{W}\mathbf{x} + \mathbf{b}\right)\,,\\
\mathbf{y} &= \mathbf{W'}\mathbf{h} + \mathbf{b'}\,.
\end{aligned}
\end{equation}
Provided that the number of dimensions of the ``hidden layer'' $\mathbf{h}$ is large enough, this arrangement can approximate any mapping between inputs $\mathbf{x}$ and outputs $\mathbf{y}$ to arbitrary precision, i.e.\ it is a general function approximator\cite{gybenko1989approximation,hornik1991approximation}.

In theory, \emph{shallow} NNs as shown above are sufficient to approximate any functional relationship.\cite{hornik1991approximation} However, \emph{deep} NNs with multiple hidden layers are often superior and were shown to be more
parameter-efficient.\cite{eldan2016power,cohen2016expressive,telgarsky2016benefits,lu2017expressive} To construct a deep NN, $L$ hidden layers are combined sequentially
\begin{equation}
\label{eq:deep_feedforward_nn}
\begin{aligned}
\mathbf{h}_1 &= \sigma\left(\mathbf{W}_1\mathbf{x} + \mathbf{b}_1\right)\,,\\
\mathbf{h}_2 &= \sigma\left(\mathbf{W}_2\mathbf{h}_1 + \mathbf{b}_2\right)\,,\\
&\vdots\\
\mathbf{h}_L &= \sigma\left(\mathbf{W}_L\mathbf{h}_{L-1} + \mathbf{b}_L\right)\,,\\
\mathbf{y} &= \mathbf{W}_{L+1}\mathbf{h}_L + \mathbf{b}_{L+1}\,,
\end{aligned}
\end{equation}
mapping the input $\mathbf{x}$ to several intermediate feature representations $\mathbf{h}_l$, until the output $\mathbf{y}$ is obtained by a linear regression on the features $\mathbf{h}_L$ in the final layer. For PES construction, typically, the NN maps a representation of chemical structure $\mathbf{x}$ to a one-dimensional output representing the energy. Contrary to the coefficients $\boldsymbol{\alpha}$ in kernel methods (see Eq.~\ref{eq:krr_coefficient_relation_regularized}), the parameters $\{\mathbf{W}_l, \mathbf{b}_l\}_{l=1}^{L+1}$ of an NN cannot be fitted in closed form. Instead, they are initialized randomly and optimized (usually using a variant of stochastic gradient descent) to minimize a loss function that measures the discrepancy between the output of the NN and the reference data.\cite{montavon2012neural} A common choice is the mean squared error (MSE), which is also used in kernel methods. During training, the loss and its gradient are estimated from randomly drawn batches of training data, making each step independent of the number of training data $M$. On the other hand, finding the coefficients for kernel methods scales as $\mathcal{O}(M^3)$ due to the need of inverting the $M\times M$ kernel matrix. Evaluating an NN according to
Eq.~\ref{eq:deep_feedforward_nn} for a single input $\mathbf{x}$ scales linearly with respect to the number of model parameters. The same is true for kernel methods, but here the number of model parameters is tied to the number of reference data $M$ used for training the model (see Eq.~\ref{eq:kernel_regression}), which means that evaluating kernel methods scales $\mathcal{O}(M)$. As the evaluation cost of NNs is independent of $M$ and only depends on the chosen architecture, they are typically the method of choice for learning large datasets. A schematic overview of the mathematical concepts behind NNs is given in Fig.~\ref{fig:nn_basics}.

\subsubsection{Model selection: How to choose hyperparameters}
\label{subsubsec:model_selection_how_to_choose_hyperparameters}
\begin{figure*}[t]
	\centering
	\includegraphics[width=\textwidth]{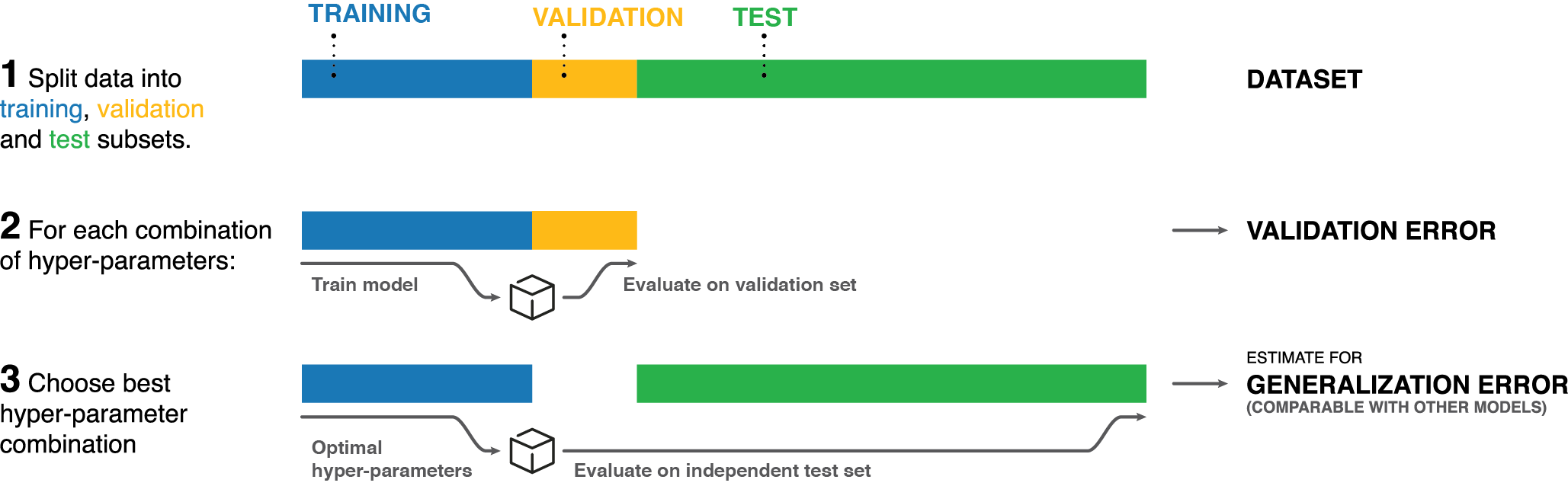}
	\caption{Overview of model selection process.}
	\label{fig:cross_validation} 
\end{figure*}

In addition to the parameters that are determined when learning an ML model for a given dataset, e.g.\ the weights $\mathbf{W}$ and biases $\mathbf{b}$ in NNs or the regression coefficients $\boldsymbol{\alpha}$ in kernel methods, many models contain hyperparameters that need to be chosen before training. They allow to tune a given model to the prior beliefs about the dataset/underlying physics and thus play a significant role in how a model generalizes to different data patterns. Two types of hyperparameters can be distinguished: The first kind influences the composition of the model itself, such as the type of kernel or the NN architecture, whereas the second kind affects the training procedure and thus the final parameters of the trained model. Examples for hyperparameters are the width (number of neurons per layer) and depth (number of hidden layers) of an NN, the kernel width $\gamma$ (see Eq.~\ref{eq:gaussian_kernel}), or the strength of regularization terms (e.g.\ $\lambda$ in Eq.~\ref{eq:krr_coefficient_relation_regularized}). 

The range of valid values is strongly dependent on the hyperparameter in question. For example, certain hyperparameters might need to be selected from the positive real numbers (e.g.\ $\gamma$ and $\lambda$, see above), while others are restricted to positive integers or have interdependencies (such as depth and width of an NN). This is why hyperparameters are often optimized with primitive exhaustive search schemes like grid search or random search in combination with educated guesses for suitable search ranges, or more sophisticated Bayesian approaches\cite{snoek2012practical}. Common gradient-based optimization methods can typically not be applied effectively. Fortunately, for many hyperparameters, model performance is fairly robust to small changes and good default values can be determined which work across many different datasets.

Before any hyperparameters may be optimized, a so-called test set must be split off from the available reference data and kept strictly separate. The remainder of the data is further divided into a training and a validation set. This is done because the performance of ML models is  not judged by how well they predict the data they were trained on, as it is often possible to achieve arbitrarily small errors in this setting. Instead, the generalization error, i.e.\ how well the model is able to predict unseen data, is taken as indicator for the quality of a model. For this reason, for every trial combination of hyperparameters, a model is trained on the training data and its performance measured on the validation set to estimate the generalization error. Finally, the best performing model is selected. To get better statistics for estimates of the generalization error, instead of splitting the remaining data (reference data excluding test set) into just two parts, it is also possible to divide it into $k$ parts (or folds). Then, a total of $k$ models is trained, each using $k-1$ folds as the training set and the last fold as validation set. This method is known as $k$-fold cross validation\cite{hastie2009elements,hansen2013assessment}.

As the validation data influences model selection (even though it is not used directly in the training process), the validation error may give too optimistic estimates and is no reliable way to judge the true generalization error of the final model. A more realistic value can be obtained by evaluating the model on the held-out test set, which has neither direct nor indirect influence on model selection. To not invalidate this estimate, it is crucial not to further tweak any parameters or hyperparameters in response to test set performance. More details on how to construct ML models (including the selection of hyperparameters and the importance of keeping an independent test set) can be found in Section~\ref{sec:best_practices_and_pitfalls}. The model selection process is summarized in Fig.~\ref{fig:cross_validation}.

\subsection{Combining machine learning and chemistry}
\label{subsec:combining_ml_and_chemistry}
The need for ML methods often arises from the lack of theory to describe a desired mapping between input and output. A classical example for this is image classification: It is not clear how to distinguish between pictures of different objects, as it is unfeasible to formulate millions of rules by hand to solve this task. Instead, the best results are currently achieved by learning statistical image characteristics from hundreds of thousands of examples that were extracted from a large dataset representing a particular object class. From that, the classifier learns to estimate the distribution inherent to the data in terms of feature extractors with learned parameters like convolution filters that reflect different scales of the image statistics.\cite{lee2017deep,theodoridis2020machine,bishop1995neural} This working principle represents the best approach known to date to tackle this particular challenge in the field of computer vision.

On the other hand, the benchmark for solving molecular problems is set by rigorous physical theory that provides essentially exact descriptions of the relationships of interest. While the introduction of approximations to exact theories is common practice and essential to reduce their complexity to a workable level, those simplifications are always physical or mathematical in nature. This way, the generality of the theory is only minimally compromised, albeit with the inevitable consequence of a reduction in predictive power. In contrast, statistical methods can be essentially exact, but only in a potentially very narrow regime of applicability. Thus, a main role of ML algorithms in the chemical sciences has been to shortcut some of the computational complexity of exact methods by means of empirical inference, as opposed to providing some mapping between input and output at all (as is the case for image classification). Notably, recent developments could show that machine learning can provide novel insight beyond providing efficient shortcuts of complex physical computations.\cite{schutt2017quantum,carleo2017solving,brockherde2017bypassing,chmiela2017,schutt2019unifying,noe2019boltzmann,sauceda2019molecular,sauceda2020} 

Force field construction poses unique challenges that are absent from traditional ML application domains, as much more stringent demands on accuracy are placed on ML approaches that attempt to offer practical alternatives to established methods. Additionally, considerable computational cost is associated with the generation of high-level \textit{ab initio} training data, with the consequence that practically obtainable datasets with sufficiently high quality are typically not very large. This is in stark contrast with the abundance of data in traditional ML application domains, such as computer vision, natural language processing etc. The challenge in chemistry, however, is to retain the generality, generalization ability and versatility of ML methods, while making them accurate, data-efficient, transferable, and scalable.

\subsubsection{Physical constraints}
\label{subsubsec:physical_constraints}
To increase data efficiency and accuracy, ML-FFs can (and should) exploit the invariances of physical systems (see Section~\ref{subsubsec:invariances_of_physical_systems}), which provide additional information in ways that are not directly available for other ML problems. Those invariances can be used to reduce the function space from which the model is selected, in this manner effectively reducing the degrees of freedom for learning,\cite{chmiela2018,anselmi2016invariance} i.e. making the learning problem easier and thus also solvable with a fraction of data. As ML algorithms are universal approximators with virtually no inherent flexibility restrictions, it is important that physically meaningful solutions are obtained. In the following, important physical constraints of such solutions and possible ways of their realization are discussed in detail. Furthermore, existing kernel-based methods and neural network architectures tailored for the construction of FFs and how they implement these physical constraints in practice are described.

\paragraph{Energy conservation} A necessary requirement for ML-FFs is that, in the absence of external forces, the total energy of a chemical system is conserved (see Section~\ref{subsubsec:invariances_of_physical_systems}). When the potential energy is predicted by any differentiable method and forces derived from its gradient, they will be conservative by construction. However, when forces are predicted directly, this is generally not true, which makes deriving energies from force samples slightly more complicated. The main challenge to overcome is that not every vector field is necessarily a valid gradient field. Therefore, the learning problem cannot simply be cast in terms of a standard multiple output regression task, where the output variables are modeled without enforcing explicit correlations. 
\begin{figure*}
	\centering
	\includegraphics[width=\textwidth]{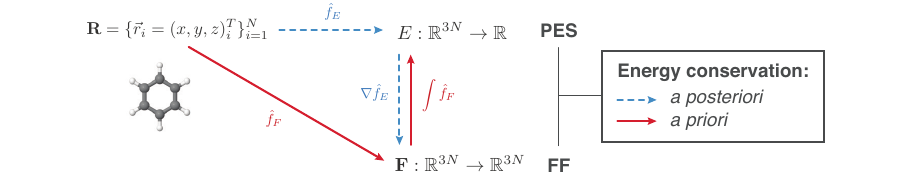}
	\caption{Differentiation of an energy estimator (blue) versus direct force reconstruction (red). The law of energy conservation is trivially obeyed in the first case, but requires explicit \emph{a priori} constraints in the latter scenario. The challenge in estimating forces directly lies in the complexity arising from their high $3N$-dimensionality (three force components for each of the $N$ atoms) in contrast to predicting a single scalar for the energy.}
	\label{fig:e_vs_f_learning}
\end{figure*}
A big advantage of predicting forces directly is that they are true quantum-mechanical observables within the BO approximation by virtue of the Hellmann-Feynman theorem\cite{hellman1937einfuhrung,feynman1939forces}, i.e.\ they can be calculated \emph{analytically} and therefore at a relatively low additional cost when generating \textit{ab initio} reference data. As a rough guideline, the computational overhead for analytic forces scales with a factor of only around $\sim$1--7 on top of the energy calculation.\cite{chmiela2019}. In contrast, at least $3N + 1$ energy evaluations would be necessary for a numerical approximation of the forces by using finite differences. For example, at the PBE0/DFT (density functional theory with the Perdew-Burke-Ernzerhof hybrid functional) level of theory\cite{adamo1999toward}, calculating energy and analytical forces for an ethanol molecule takes only $\sim$1.5 times as long as calculating just the energy (the exact value is implementation-dependent), whereas for numerical gradients, a factor of at least $\sim$10 would be expected. 

As forces provide additional information about how the energy changes when an atom is moved, they offer an efficient way to sample the PES, which is why it is desirable to formulate ML models that can make \emph{direct} use of them in the training process. Another benefit of a direct reconstruction of the forces is that it avoids the amplification of estimation errors due to the derivative operator that would otherwise be applied to the PES reconstruction (see Fig.~\ref{fig:e_vs_f_learning}).\cite{chmiela2017,sauceda2019molecular,snyder2012finding}

\paragraph{Roto-translational invariance}
A crucial requirement for ML-FFs is the rotational and translational invariance of the potential energy, i.e.\ $E\left(\mathbf{R}\right) = E\left(\mathcal{R}\mathbf{R} +\mathcal{T}\right)$, where $\mathcal{R}$ and $\mathcal{T}$ are rigid rotations and translations and $\mathbf{R}$ are the Cartesian coordinates of the atoms. As long as the representation $\mathbf{x}(\mathbf{R})$ of chemical structure chosen as input for the ML model itself is roto-translationally invariant, ML-FFs inherit its desired properties and even the gradients will automatically behave in the correct equivariant way due to the outer derivative $\partial (\mathbf{x}(\mathcal{R}\mathbf{R}+\mathcal{T}) / \partial \mathbf{R} = \mathcal{R}\partial\mathbf{x}(\mathbf{R})/\partial\mathbf{R}$. One example of appropriate features to construct a representation $\mathbf{x}$ with the desired properties are pairwise distances. For a system with $N$ atoms, there are $\binom{N}{2}$ different pairwise distances, which results in reasonably sized feature sets for systems with a few dozen atoms. Apart from very few pathological cases, this representation is complete, in the sense that any possible configuration of the system can be described exactly and uniquely\cite{bartok2013representing}. However, while pairwise distances serve as an efficient parametrization of some geometry distortions like bond stretching, they are relatively inefficient in describing others, e.g.\ rotations of functional groups. In the latter case, many distances are affected even for slight angular changes, which can pose a challenge when trying to learn the geometry-energy mapping. Complex transition paths or reaction coordinates are often better described in terms of bond and torsion angles in addition to pairwise distances. The problem is that the number of these features grows rather quickly, with $\binom{N}{3}$ and $\binom{N}{4}$, respectively. At that rate, the size of the feature set quickly becomes a bottleneck, resulting in models that are slow to train and evaluate. While an expert choice of relevant angles would circumvent this issue, it reduces some of the 
``data-driven'' flexibility that ML models are typically appreciated for. Note that models without roto-translational invariance are practically unusable, as they will start to generate spurious linear and/or angular momentum during dynamics simulations.

\begin{figure}
	\centering
	\includegraphics[width=1.0\columnwidth]{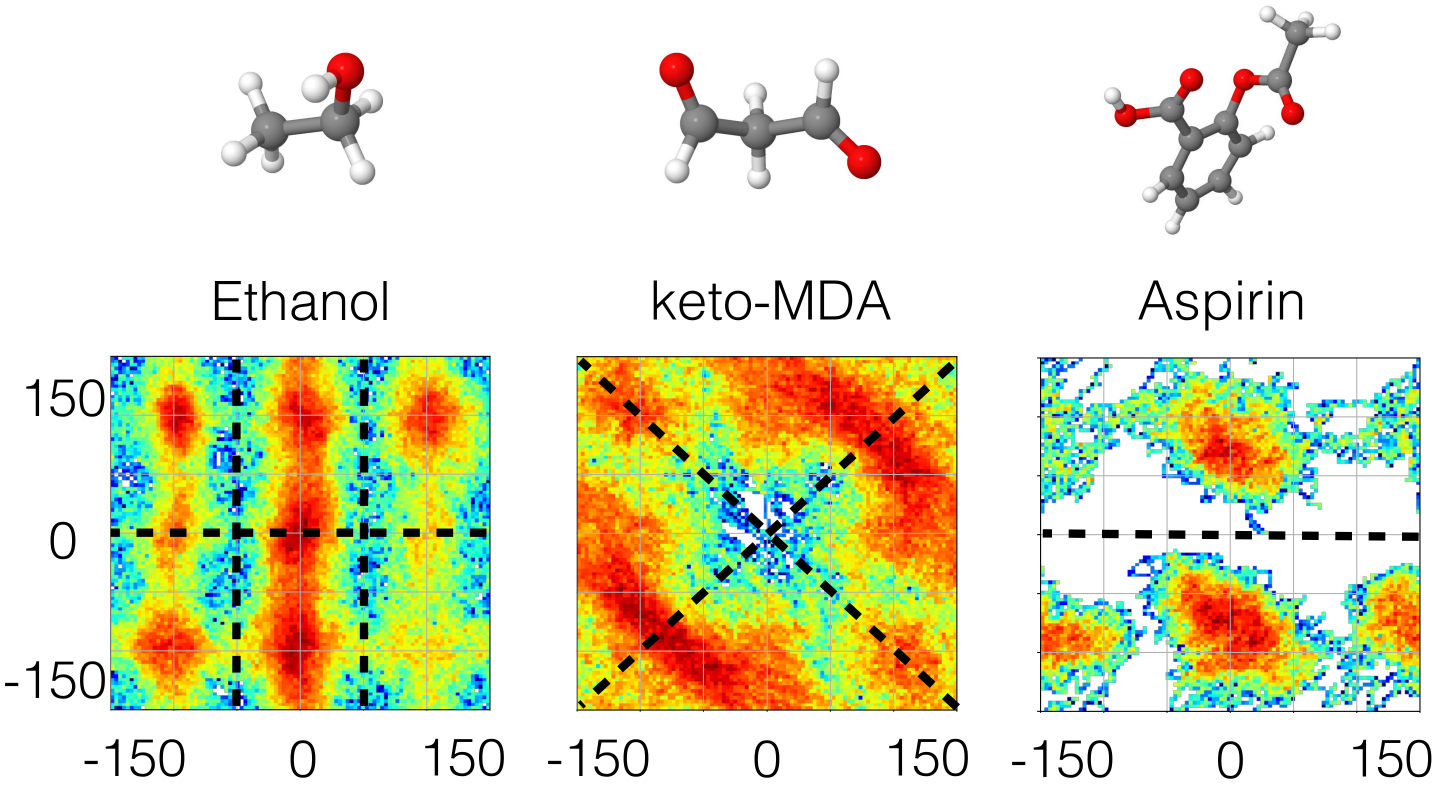}
	\caption{Regions of the PESs for ethanol, keto-malondialdehyde and aspirin visited during a 200~ps \textit{ab initio} MD simulation at 500~K using the PBE+TS/DFT level of theory~\cite{PBE1996,tkatchenko2009accurate} (density functional theory with the Perdew-Burke-Ernzerhof functional and Tkatchenko-Scheffler dispersion correction). The black dashed lines indicate the symmetries of the PES. Note that regions related by symmetry are not necessarily visited equally often.}
	\label{fig:FigBiasedSampl}
\end{figure}

\paragraph{Indistinguishability of identical atoms}
In the BO approximation, the potential energy of a chemical system only depends on the charges and positions of the nuclei. As a consequence, the PES is symmetric under permutation of atoms with the same nuclear charge. However, symmetric regions are not necessarily sampled in an unbiased way during MD simulations (see Fig.~\ref{fig:FigBiasedSampl}). Consequently, ML-FFs that are not constrained to treat all symmetries equivalently may (due to the uneven sampling) predict different results when permuting atoms.
While it is in principle possible to arrive at a ML-FF that is symmetric with respect to permutations of same-species atoms indirectly via data augmentation\cite{montavon2012learning,montavon2013machine} or by simply using datasets that naturally include all relevant symmetric configurations in an unbiased way, there are obvious scaling issues with this approach. It is much more efficient to impose the right constraints onto the functional form of the ML-FF, such that all relevant symmetric variants of a certain atomic configuration appear equivalent automatically. Such symmetric functions can be constructed in various ways, each of which has advantages and disadvantages.

Assignment-based approaches do not symmetrize the ML-FF \textit{per se}, but instead aim to normalize its input, such that all symmetric variants of a configuration are mapped to the same internal representation. In its most basic realization, this assignment is done heuristically, i.e.\ by using inexact, but computationally cheap criteria. Examples for this approach are the Coulomb matrix\cite{rupp2012fast} or the Bag-of-Bonds\cite{hansen2015machine} descriptors, that use simple sorting schemes for that purpose. Histograms\cite{huo2017unified, christensen2020fchl} and some density-based\cite{bartok2013representing,hirn2017wavelet,eickenberg2017solid} approaches follow that same principle, although not explicitly. All of those schemes have in common that they compare the features in aggregate as opposed to individually. A disadvantage is that dissimilar features are likely to be compared to each other or treated as the same, which limits the accuracy of the prediction. Such weak assignments are better suited for datasets with diverse conformations rather than gathered from MD trajectories that contain many similar geometries. In the latter case, the assignment of features might change as the geometry evolves, which would lead to discontinuities in the prediction and would effectively be treated by the ML model as noise (see $\epsilon$ in Eq.~\ref{regressionwithnoise}).

An alternative path is to recover the true correspondence of molecular features using a graph matching approach\cite{kriege2016valid,Vert2008}. Each input $\mathbf{x}$ is matched to a canonical permutation of atoms $\tilde{\mathbf{x}} = \mathbf{P}\mathbf{x}$ before generating the prediction. This procedure effectively compresses the PES to one of its symmetric subdomains (see dashed black lines in Fig.~\ref{fig:FigBiasedSampl}), but in an exact way. Note that graph matching is in all generality an NP-complete problem which can only be solved approximately. In practice, however, several algorithms exist to ensure at least consistency in the matching process if exactness can not be guaranteed\cite{pachauri2013solving}. A downside of this strategy is that any input must pass through a matching process, which is relatively costly, despite being approximate. Another issue is that the boundaries of the symmetric subdomains of the PES will necessarily lie in the extrapolation regime of the reconstruction in which prediction accuracy tends to degrade. As the molecule undergoes symmetry transformations, these boundaries are frequently crossed, to the detriment of prediction performance.

Arguably the most universal way of imposing symmetry, especially if the functional form of the model is already given, is via invariant integration over the relevant symmetry group $f_\text{sym}(\mathbf{x}) = \int_{\pi \in \mathcal{S}} f(\mathbf{P}_\pi \mathbf{x})$. Typically, $\mathcal{S}$ would be the permutation group and $\mathbf{P}_\pi$ the corresponding permutation matrix that transforms each vector of atom positions $\mathbf{x}$. Some approaches
\cite{bartok2013representing, bartok2015g, de2016comparing} avoid this implicit ordering of atoms in $\mathbf{x}$ by adopting a three-dimensional density representation of the molecular geometry defined by the atom positions, albeit at the cost of losing rotational invariance, which then must be recovered by integration. 
Invariant integration gives rise to functional forms that are truly symmetric and do not require any pre- or post-processing of the in- and output data. A significant disadvantage is however, that the cardinality of even basic symmetry groups is exceedingly high, which affects both training and prediction times. 

This combinatorial challenge can be solved by limiting the invariant integral to the physical point group and fluxional symmetries that actually occur in the training dataset. Such a sub-group of meaningful symmetries can be automatically recovered and is often rather small\cite{chmiela2019}. For example, each of the molecules benzene, toluene and azobenzene have only 12 physically relevant symmetries, whereas their full symmetric groups have orders $6!6!$, $7!8!$ and $12!10!2!$ symmetries respectively.

\subsubsection{(Symmetric) Gradient Domain Machine Learning ((s)GDML)}
\label{subsubsec:symmetric_gradient_domain_machine_learning}
Gradient domain machine learning (GDML) is a kernel-based method introduced as a data efficient way to obtain accurate reconstructions of flexible molecular force fields from small reference datasets of high-level \textit{ab initio} calculations\cite{chmiela2017}. Contrary to most other ML-FFs, instead of predicting the energy and obtaining forces by derivation with respect to nuclear coordinates, GDML predicts the forces directly. As mentioned in Section~\ref{subsubsec:physical_constraints}, forces obtained in this way may violate energy conservation. To ensure conservative forces, the key idea is to use a kernel $\mathbf{K}\left(\mathbf{x},\mathbf{x}^{\prime}\right) = \nabla_{\mathbf{x}}K_{E}\left(\mathbf{x},\mathbf{x}^{\prime}\right)\nabla_{\mathbf{x}^{\prime}}^{\top}$ that models the forces $\mathbf{F}$ as a transformation of an unknown potential energy surface $E$ such that
\begin{equation}
\begin{aligned}
\boldsymbol{\mathbf{F}}&=-\nabla E\\&\sim\mathcal{G}\mathcal{P}\left[-\nabla\mu_{E},(\mathbf{x}),\nabla_{\mathbf{x}}K_{E}\left(\mathbf{x},\mathbf{x}^{\prime}\right)\nabla_{\mathbf{x}^{\prime}}^{\top}\right]\,.
\end{aligned}
\label{eq:gdml_key_idea}
\end{equation}
Here, $\mu_{E}: \mathbb{R}^{D} \rightarrow \mathbb{R}$ and $K_{E}: \mathbb{R}^{D} \times \mathbb{R}^{D} \rightarrow \mathbb{R}$ are the prior mean and covariance functions of the latent energy-based Gaussian process $\mathcal{G}\mathcal{P}$, respectively. The descriptor of chemical structure $\mathbf{x}\in \mathbb{R}^{D}$ consists of the inverse of all $D$ pair-wise distances, which guarantees roto-translational invariance of the energy. Training on forces is motivated by the fact that they are available analytically from electronic structure calculations, with only moderate computational overhead atop energy evaluations. The big advantage is that for a training set of size $M$, only $M$ reference energies are available, whereas there are three force components for each of the $N$ atoms, i.e.\ a total of $3NM$ force values. This means that a kernel-based model trained on forces contains more coefficients (see Eq.~\ref{eq:kernel_regression}) and is thus also more flexible than an energy-based variant. Additionally, the amplification of noise due to the derivative operator is avoided.

A limitation of the GDML method is that the structural descriptor $\mathbf{x}$ is not permutationally invariant, because the values of its entries (inverse pairwise distances) change when atoms are re-ordered. An extension of the original approach, sGDML\cite{chmiela2018,chmiela2019} (symmetric GDML), additionally incorporates all relevant rigid space group symmetries, as well as dynamic non-rigid symmetries of the system at hand into the kernel, to further improve its efficiency and ensure permutational invariance. Usually, the identification of symmetries requires chemical and physical intuition about the system at hand, which is impractical in an ML setting. Here, however, a data-driven multi-partite matching approach is employed to automatically recover permutations of atoms that appear within the training set\cite{chmiela2019}. A matching process finds permutation matrices $\mathbf{P}$ that realize the assignment between adjacency matrices $(\mathbf{A})_{ij} = \|\vec{r}_i - \vec{r}_j\|$ of molecular graph pairs $G$ and $H$ in different energy states
\begin{equation}
\operatorname*{arg\,min}_{\tau} \mathcal{L}(\tau) = \|\mathbf{P}(\tau)\mathbf{A}_G\mathbf{P}(\tau)\tran - \mathbf{A}_H\|^2
\label{eq:matching_objective}
\end{equation}
and thus between symmetric transformations\cite{Umeyama1988}. The resulting approximate local pairwise assignments are subsequently globally synchronized using transitivity as the consistency criterion\cite{pachauri2013solving} to eliminate impossible assignments. By limiting this search to the training set, combinatorially feasible, but physically irrelevant permutations $\tau$ are automatically excluded (ones that are inaccessible without crossing impassable energy barriers). Such hard symmetry constraints (derived from the training set) greatly reduce the intrinsic complexity of the learning problem without biasing the estimator, since no additional approximations are introduced.

\subsubsection{Gaussian approximation potentials (GAPs)}
\label{subsubsec:gaussian_approximation_potentials}
Gaussian approximation potentials (GAPs)\cite{bartok2010gaussian} were originally developed for materials such as bulk crystals, but were later also applied to molecules\cite{bartok2017machine}.
They scale linearly with the number of atoms of a system and can accommodate for periodic boundary conditions.
Similar to high-dimensional neural network potentials\cite{behler2007generalized} (see Section~\ref{subsubsec:high_dimensional_neural_network_potentials}), GAPs decompose each system into atom-centered environments $i$ such that its energy can be written as the sum of atomic contributions
\begin{equation}
E=\sum_{i=1}^N E_i(\{\mathbf{r}_{ij}\}_{j \in [1, N]})\,,
\label{eq:gap_local_contributions}
\end{equation}
with $\mathbf{r}_{ij} = \mathbf{r}_{j} - \mathbf{r}_{i}$ and $\mathbf{r}_{i}$ being the position of atom $i$. A smooth cutoff function is applied to the pairwise distances $\lVert \mathbf{r}_{ij} \rVert$ to ensure that the contributions $E_i$ are local and no discontinuities are introduced when atoms enter or leave the cutoff radius. Even though such a decomposition is inherently non-unique and no labels for atom-wise energies are available in the reference data, they can still be approximated 
by a Gaussian process: The sum over atomic environments can be moved into the kernel function, yielding a kernel for systems $\mathbf{x}$ and $\mathbf{x}'$ with $N$ and $N'$ atoms, respectively:
\begin{equation}
K(\mathbf{x}, \mathbf{x}') = \sum_{i=1}^N \sum_{j=1}^{N'} K_\text{local}(x_i, x'_j)\,.
\label{eq:gap_kernel}
\end{equation}
Thus, reference energies for the whole system are sufficient for the model to learn a suitable energy decomposition into atomic environments.

Several descriptors and kernels for GAPs have been developed based on a local ``atomic density'' $\rho(\mathbf{r}) = \sum_j \delta(\mathbf{r} - \mathbf{r_j})$.
Initially, \citet{bartok2010gaussian} proposed to employ local atomic coordinates projected onto a 4D hyper sphere. 
Since this projection can represent the volume of a 3D sphere, the introduction of an additional radial basis can be avoided.
To achieve rotational invariance, the bispectrum of 4D spherical harmonics of these coordinates was used as a descriptor. Alternatively, the SOAP (smooth overlap of atomic positions) kernel\cite{bartok2013representing} is defined as the integral over rotations $\mathcal{R}$ of atomic densities
\begin{equation}
K(\rho, \rho') = \int d\mathcal{R} \left| \int \rho(\mathbf{r}) \rho'(\mathcal{R}\mathbf{r}) \mathrm{d}\mathbf{r} \right|^n\,.
\label{eq:soap_kernel}
\end{equation}
Given smoothed local densities $\rho(\mathbf{r})=\sum_j\exp(-\gamma\|\mathbf{r}-\mathbf{r_j}\|^2)$, it has been shown that the SOAP kernel is equivalent to the linear kernel over the SO(3) power spectrum and bispectrum for $n=2$ and $n=3$, respectively\cite{bartok2013representing}.
Both approaches are invariant to permutation of neighboring atoms as well as the rotation of the local environment.
Further representations include best matches of the atomic densities over rotations\cite{de2016comparing} and kernels for symmetry-adapted prediction of tensorial properties\cite{grisafi2018symmetry,csanyi2020machine}.

\subsubsection{Neural Network Potentials}
\label{subsubsec:high_dimensional_neural_network_potentials}
\begin{figure*}[t]
    \centering
    \includegraphics[width=0.99\textwidth]{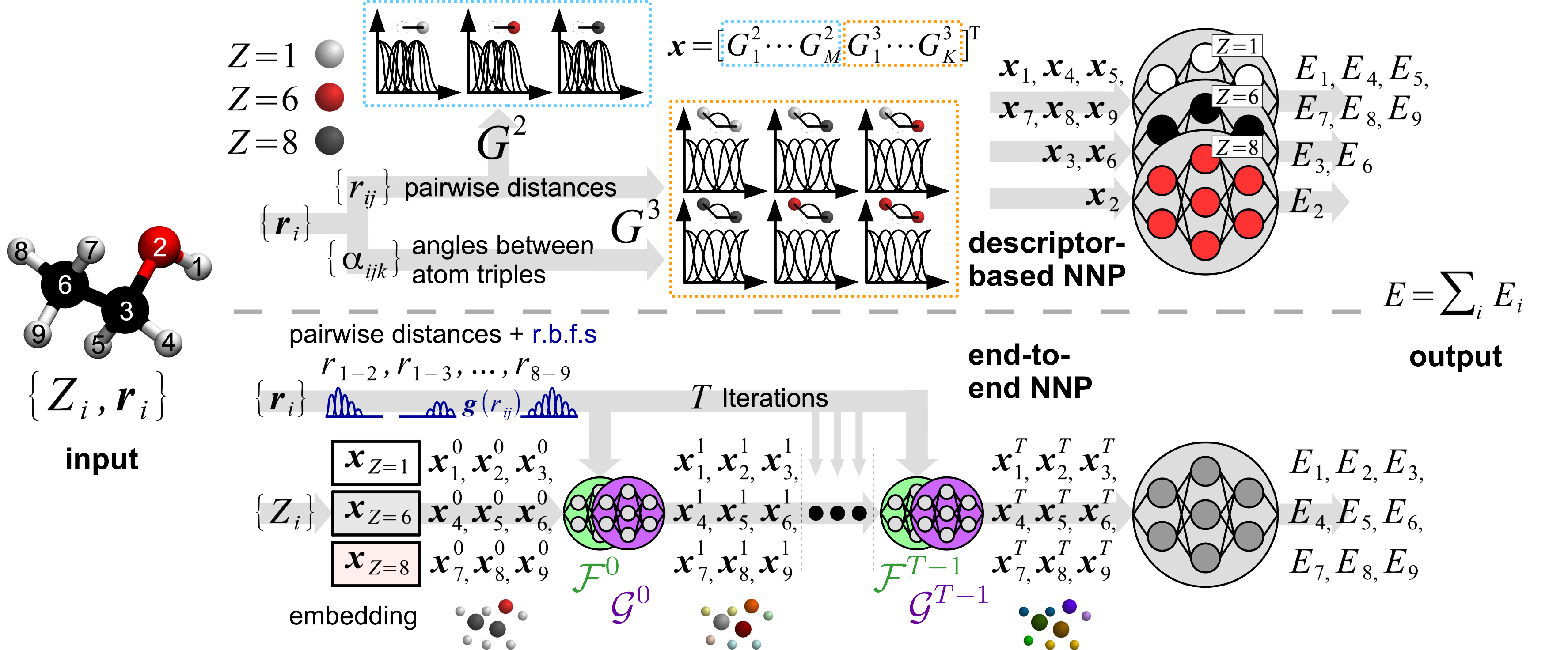}
    \caption{Overview of descriptor-based (top) and end-to-end (bottom) NNPs. Both types of architecture take as input a set of $N$ nuclear charges $Z_i$ and Cartesian coordinates $\mathbf{r}_i$ and output atomic energy contributions $E_i$, which are summed to the total energy prediction $E$ (here $N=9$, an ethanol molecule is used as example). In the descriptor-based variant, pairwise distances $r_{ij}$ and angles $\alpha_{ijk}$ between triplets of atoms are calculated from the Cartesian coordinates and used to compute hand-crafted two-body ($G^2$) and three-body ($G^3$) atom-centered symmetry functions (ACSFs) (see Eqs.~\ref{eq:two_body_symmetry_function}~and~\ref{eq:three_body_symmetry_function}). For each atom $i$, the values of $M$ different $G^2$ and $K$ different $G^3$ ACSFs are collected in a vector $\mathbf{x}_i$, which serves as a fingerprint of the atomic environment and is used as input to an NN predicting $E_i$. Information about the nuclear charges is encoded by having separate NNs and sets of ACSFs for all (combinations of) elements. In end-to-end NNPs, $Z_i$ is used to initialize the vector representation $\mathbf{x}_i^0$ of each atom to an element-dependent (learnable) embedding (atoms with the same $Z_i$ start from the same representation). Geometric information is encoded by iteratively passing these descriptors (along with pairwise distances $r_{ij}$ expanded in radial basis functions $\mathbf{g}(r_{ij})$) in $T$ steps through NNs representing interaction functions $\mathcal{F}^t$ and atom-wise refinements $\mathcal{G}^t$ (see Eq.~\ref{eq:general_message_passing}). The final descriptors $\mathbf{x}^T_i$ are used as input for an additional NN predicting the atomic energy contributions (typically, a single NN is shared among all elements).}
    \label{fig:nnp_overview}
\end{figure*}

The first neural network potentials (NNPs) used a set of internal coordinates, e.g.\ distances and angles, as structural representation to model the PES\cite{blank1995neural,brown1996combining,tafeit1996neural,no1997description,prudente1998fitting}. While being roto-translationally invariant, internal coordinates impose an arbitrary order on the atoms and are thus not reflecting the equivalence of permuted inputs. As a result, the NNP might assign different energies to symmetrically equivalent structures.
Beyond that, the number of atoms influences the dimensionality of the input $\mathbf{x}$, limiting the applicability of the PES to chemical systems of the same size.
Decomposing the energy prediction in the spirit of a many-body expansion circumvents these issues\cite{manzhos2006random,manzhos2007using,malshe2009development}, however, it scales unfavorably with system size and number of chemical species, because each term in the many-body expansion has to be modeled by a separate NN. 

\citet{behler2007generalized} were the first to propose so-called high-dimensional neural networks potentials (HDNNPs), where the total energy of a chemical system is expressed as a sum over atomic contributions $E=\sum_{i}E_i$, predicted by the same NN (or one for each element). The underlying assumption is that the energetic contribution $E_i$ of each atom depends mainly on its local chemical environment. As all atoms of the same type are treated identically and summation is commutative, the output does not change when the input is permuted.
Due to the decomposition into atomic contributions, systems with varying numbers of atoms can be predicted by the same NNP. In principle, this framework also enables transferability between system sizes, e.g.\ a model can be trained on small systems, but applied to predict energies and forces for larger systems. 
However, this requires sufficient sampling of the local environments to remove spurious correlations caused by the training data distribution, as well as corrections for long-range effects.

The introduction of HDNNPs inspired many NN architectures that can be broadly categorized into two types. \emph{Descriptor-based} NNPs\cite{behler2011atom,khorshidi2016amp,artrith2017efficient,unke2018reactive} rely on fixed rules to encode the environment of an atom in a vector $\mathbf{x}$, which is then used as input for an ordinary feed-forward NN (see Eq.~\ref{eq:deep_feedforward_nn}). These architectures include many variants of the original Behler-Parrinello network, such as ANI\cite{smith2017ani} and TensorMol\cite{yao2018tensormol}. 
On the other hand, \textit{end-to-end} NNPs\cite{duvenaud2015convolutional,kearnes2016molecular,schutt2017quantum,gilmer2017neural}, take nuclear charges and Cartesian coordinates as input and learn a suitable representation from the data. 

Many end-to-end NNPs have been inspired by the graph neural network by \citet{scarselli2008graph} and were later collectively cast as message-passing neural networks (MPNNs)\cite{gilmer2017neural}. In this type of model, molecules are regarded as undirected graphs, where atoms are represented by nodes and interactions between them as edges. By exchanging information between nodes along edges (message-passing), complex chemical interactions can be modeled. A prominent example is the Deep Tensor Neural Network (DTNN)\cite{schutt2017quantum}. Since its introduction, this approach has been refined to create new architectures, such as
SchNet\cite{schutt2017schnet, schutt2018quantum},
HIP-NN\cite{lubbers2018hierarchical} or
PhysNet\cite{unke2019}.
End-to-end NNPs that do not directly fall into the category of MPNNs are covariant compositional networks that are able to employ features of higher angular momentum\cite{hy2018predicting,anderson2019cormorant,weiler20183d} as well as models using a pseudo-density as input\cite{eickenberg2017solid}.

Because no fixed rule is used to construct descriptors, end-to-end NNPs are able to automatically adapt the environment representations $\mathbf{x}$ to the reference data (in contrast to the descriptor-based variant). However, as long as $\mathbf{x}$ is invariant with respect to translation, rotation, and permutation of symmetry equivalent atoms, both types of NNPs adhere to all physical constraints outlined in Section~\ref{subsubsec:physical_constraints}. NNPs are commonly used to predict energies, while conservative forces are obtained by derivation. Despite being energy-based, it is still possible to incorporate information from \textit{ab initio} forces by including them in the loss term optimized during training. 
At this point, it should be noted that the requirement for continuously differentiable models excludes the use of certain activation functions, for example the popular ReLU activation\cite{nair2010rectified}, when constructing ML-FFs based on neural networks. To avoid discontinuities in the forces, activation functions used for NNPs must always be smooth.

\paragraph{Descriptor-based NNPs}
The first descriptor-based NNP introduced by \citet{behler2007generalized} uses atom-centered symmetry functions (ACSFs)\cite{behler2011atom} consisting of two-body terms
\begin{equation}
G^2_i = \sum_{j \neq i}^{} e^{-\eta(r_{ij}-r_s)^2}f_{\rm cut}(r_{ij})
\label{eq:two_body_symmetry_function}
\end{equation}
and three-body terms
\begin{equation}
\begin{split}
G^3_i = 2^{1-\zeta}\sum_{j,k \neq i}^{} \left(1+\lambda\cos\theta_{ijk}\right)^\zeta e^{-\eta\left(r_{ij}^2+r_{ik}^2+r_{jk}^2\right)} \\ 
\times f_{\rm cut}(r_{ij})f_{\rm cut}(r_{ik})f_{\rm cut}(r_{jk})
\end{split}
\label{eq:three_body_symmetry_function}
\end{equation}
 to encode information about the chemical environment of each atom $i$. Here, $r_{ij}$ is the distance between atoms $i$ and $j$, $\theta_{ijk}$ the angle spanned by atoms $i$, $j$ and $k$ centered around $i$, and the summations run over all atoms within a cutoff distance $r_{\rm cut}$. As the atom order is irrelevant for the values of $G^2_i$ and $G^3_i$ and only internal coordinates are used to calculate them, all physical invariants are satisfied. A cutoff function such as
\begin{equation}
f_{\rm cut}(r) = \begin{cases} 
\dfrac{1}{2}\left[\cos\left(\dfrac{\pi r}{r_{\rm cut}}\right)+1\right]\,, & r\leq r_{\rm cut} \\
0\,, & r > r_{\rm cut}
\end{cases}
\label{eq:behler_parrinello_cutoff}
\end{equation}
ensures that $G^2_i$ and $G^3_i$ vary smoothly when atoms enter or leave the cutoff sphere and the parameters $\eta$, $r_s$, $\zeta$,  and $\lambda(=\pm1)$ determine to which distances, or combinations of angles and distances, the ACSFs are most sensitive. When sufficiently many $G^2_i$ and $G^3_i$ with different parameters are combined and stored in a vector $\mathbf{x}_i$, they form a ``fingerprint'' of the local environment of atom $i$. This environment descriptor is then used as input for a neural network for predicting the energy contributions $E_i$ of atoms~$i$ and the total energy $E=\sum_{i}E_i$ is obtained by summation.

Since the ACSFs only use geometric information, they work best for systems containing only atoms of one element, for example crystalline silicon\cite{behler2007generalized}. To describe multi-component systems, typically, the symmetry functions are duplicated for each combination of elements and separate NNs are used to predict the energy contributions for atoms of the same type\cite{behler2015constructing}. Since the combinatorial explosion can lead to a large number of ACSFs for systems containing many different elements, an alternative is to modify the ACSFs with element-dependent weighting functions\cite{gastegger2018wacsf}. Most descriptor-based NNPs, such as ANI\cite{smith2017ani} and TensorMol\cite{yao2018tensormol}, use variations of Eqs.~\ref{eq:two_body_symmetry_function}~and~\ref{eq:three_body_symmetry_function} (sometimes allowing parameters of ACSFs to be optimized during training) to construct the environment descriptors $\mathbf{x}_i$. Different ways to encode the structural information are possible, for example using three-dimensional Zernike functions\cite{khorshidi2016amp}, or the coefficients of a spherical harmonics expansion\cite{unke2018reactive}, but the general principle remains the same. Also, while most descriptor-based NNPs use separate parametrizations for different elements, it is also possible to use a single NN to predict all atomic energy contributions\cite{unke2018reactive}. The common feature for all variations of this approach is that the functional form of the environment descriptor is predetermined and manually designed.

\paragraph{End-to-end NNs}
A potential drawback of the previously introduced ACSFs is that they must be chosen by an expert before training the neural network. If the choice of symmetry functions is poor, for example when the resulting descriptor is (nearly) identical for two very different structures, the expressive power of the neural network and the achievable accuracy are limited \textit{a priori}.
Additionally, a growing number of input dimensions can quickly become computationally expensive, both for calculating the descriptors and for evaluating the NN. This is especially the case when modeling multi-component systems, where commonly orthogonality is assumed between different elements (which increases the number of input dimensions) or the descriptors are simply weighted by an element-dependent factor (which may limit the structural resolution of the descriptor).

In contrast, end-to-end NNPs directly take atomic types and positions as inputs to learn suitable representations from the reference data. Similar to descriptor-based NNPs, many end-to-end NNPs obtain the total energy $E$ as a sum of atomic contributions $E_i$. However, those are predicted from learned features $\mathbf{x}_i$ encoding information about the local chemical environment of each atom $i$. This allows them to adapt the features based on the size and distribution of the training set as well as the chemical property of interest during the training process. The idea is to learn a mapping to a high-dimensional feature space, so that structurally (and energetically) similar atomic environments lie close together and dissimilar ones far apart.

Within the deep tensor neural network framework\cite{schutt2017quantum}, this is achieved by iteratively refining the atomic features $\mathbf{x}_i$ based on neighboring atoms. The features are initialized to $\mathbf{x}_i^0 = \mathbf{e}_{Z_i}$, where $\mathbf{e}_{Z}$ are learnable element-dependent representations that are updated for $T \in [3, 6]$ steps. This procedure is inspired by diffusion graph kernels\cite{kondor2002diffusion} as well as the graph neural network model by \citet{scarselli2008graph}.
Many end-to-end networks have adapted this approach which can be written in general as
\begin{equation}
\mathbf{x}_{i}^{t+1} = \mathcal{G}^t\left(\mathbf{x}_{i}^{t} + \sum_{j \neq i} \mathcal{F}^{t}\left(\mathbf{x}_{i}^{t}, \mathbf{x}_{j}^{t},\mathbf{g}(r_{ij})\right)f_{\rm cut}(r_{ij})\right)\,,
\label{eq:general_message_passing}
\end{equation}
where the summation runs over all atoms within a distance $r_{\rm cut}$ and a cutoff function $f_{\rm cut}$ ensures smooth behavior when atoms cross the cutoff.
Here, the ``atom-wise'' function $\mathcal{G}^{t}$ is used to refine the atomic features after they have been updated with information from neighboring atoms through the interaction-function $\mathcal{F}^{t}$. Usually, the interatomic distance $r_{ij}$ is not used directly as input to $\mathcal{F}^{t}$, but expanded in a set of uniformly spaced radial basis functions\cite{schutt2017quantum,schutt2017schnet,unke2019} to form a vectorial input $\mathbf{g}(r_{ij})$.
Both $\mathcal{F}^{t}$ and $\mathcal{G}^{t}$ functions are NNs with the specific implementations varying between different end-to-end NNP architectures. As only pair-wise distances are used and the order of atoms is irrelevant due to the commutative property of summation, the features $\mathbf{x}_i$ obtained by Eq.~\ref{eq:general_message_passing} are automatically roto-translationally and permutationally invariant (and thus also the energy predictions).

\citet{gilmer2017neural} have cast graph networks of this structure as message-passing neural networks and proposed a variant that uses a set2set decoder\cite{vinyals2015order} instead of a sum over energy contributions to achieve permutational invariance of the energy.
SchNet\cite{schutt2017schnet} takes an alternative view of the problem and models interactions between atoms with convolutions. The convolution filters need to be continuous (to have smooth predictions) but are evaluated at finite points, i.e.\ the positions of neighboring atoms. To ensure rotational invariance, only radial convolution filters are used, leading again to an interaction function that is a special case of Eq.~\ref{eq:general_message_passing}.

While the previously introduced approaches aim to learn as much as possible from the reference data, several models have been proposed to better exploit chemical domain knowledge.
The hierarchical interacting particle neural network (HIP-NN)\cite{lubbers2018hierarchical} obtains the prediction as a sum over atom-wise contributions $E_i^t$ that are predicted after every update step $t$. A regularizer penalizes larger energy contributions in deeper layers, i.e.\ enforcing a declining, hierarchical prediction of the energy.
PhysNet\cite{unke2019} modified the energy function to include explicit terms for electrostatic and dispersion interactions,
\begin{equation}
E = \sum_{i=1}^N E_i + k_e \sum_{i=1}^N \sum_{j>i}^N \tilde{q}_i \tilde{q}_j \chi(r_{ij}) + E_\text{D3}\,,
\end{equation}
where $E_\text{D3}$ is Grimme's D3 dispersion correction\cite{grimme2010consistent}, $k_e$ is Coulomb's constant and $\tilde{q}_i$ are corrected partial charges predicted by the network that are guaranteed to sum to the total charge of the molecule. In an ablation study on a dataset of S$_\text{N}$2 reactions\cite{unke2019sn2}, it was shown that the inclusion of long-range terms improves prediction accuracy for energies and forces while models without these terms show qualitatively wrong asymptotic behavior.\cite{unke2019}

\section{Best practices and Pitfalls}
\label{sec:best_practices_and_pitfalls}
\begin{figure*}[t]
    \centering
    \includegraphics[width=\textwidth]{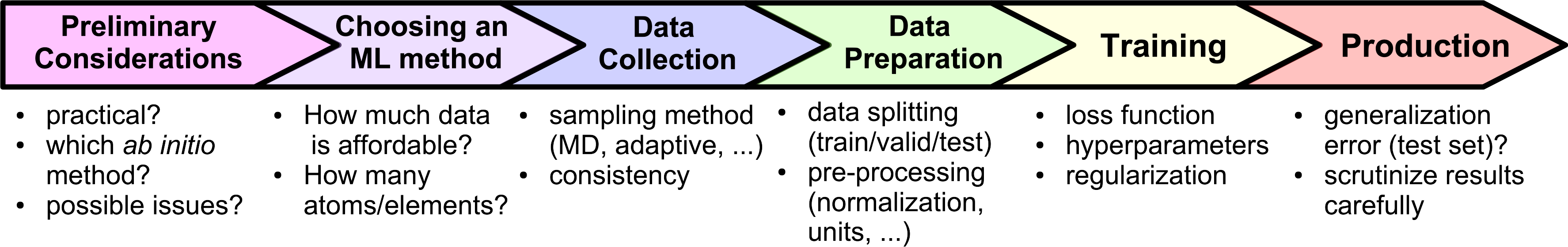}
    \caption{Overview of the most important steps when constructing and using ML-FFs.}
    \label{fig:best_practices}
\end{figure*}

A number of careful modeling steps are necessary to construct an ML-FF for a particular problem of interest (Fig.~\ref{fig:best_practices}). Even before starting this process, some forethought is appropriate due to certain limitations of \textit{ab initio} methods themselves. This section gives an overview about all steps necessary to construct an ML-FF from scratch and highlights possible ``pitfalls'', i.e.\ issues that may occur along the way, in particular when the recommended practices are not followed. First, some preliminary considerations, which should be taken before starting with the construction of an ML-FF, are discussed (\ref{subsec:preliminary_considerations}). Next, basic principles for choosing an appropriate ML method for a specific task are given (\ref{subsec:choosing_an_appropriate_ml_method}). Then, the importance of high quality reference data, different strategies to collect it (\ref{subsec:data_collection}), and how the data has to be prepared (\ref{subsec:data_preparation}), are outlined. This is followed by an overview of how to train an ML model on the collected data (\ref{subsec:training_the_model}) and guidelines for using the trained ML-FF in a production setting, e.g.\ for running MD simulations (\ref{subsec:using_the_model_in_production}). Finally, popular software packages for constructing ML-FFs are briefly described and code examples are given (\ref{subsec:example_code_and_software_packages}).

\subsection{Preliminary considerations}
\label{subsec:preliminary_considerations}
Before running any \textit{ab initio} calculations to collect data for training ML models, it is advisable to think about the limitations of the chosen level of theory itself. The issues discussed here are problem-specific and often not unique to ML-FFs, but PES reconstruction in general. As such, a comprehensive list is not possible, but a few examples are given below.

\paragraph{Practicability}
On the spectrum of quantum chemistry methods, ML-FFs fit into the niche between highly efficient conventional FFs\cite{monticelli2013force} and accurate, but computationally expensive \textit{ab initio} methods.\cite{friesner2005ab} Efficiency-wise, they are still inferior to classical FFs, because their functional forms are considerably more complex and thus more expensive to evaluate. Even the fastest ML-FFs are still one to three orders of magnitude slower\cite{chmiela2019,brickel2019reactive,sauceda2020molecular}. On the other end, ML-FFs are lower bounded by the accuracy of the reference data used for training, which means that the underlying \textit{ab initio} method will always be at least equally accurate. In practical terms, this means that in order to be useful, ML-FFs need to offer time savings over directly running \textit{ab initio} calculations and an improved accuracy compared to conventional FFs. For this purpose, the full procedure of data generation, training and inference must be taken into account, as opposed to just regarding inference speed, which will be much quicker than \textit{ab initio} methods. While this consideration sounds trivial at first, it is still advisable to think about whether constructing an ML-FF really is economical. For example, if the goal is to run just a single short MD trajectory, the question is how much data is necessary for the model to reach the required accuracy. Some models may require several thousands of training points to produce accurate enough predictions, even for fairly small molecules. Then, when factoring in the overall time required for going through the process of creating the ML model, testing it, and running the MD simulation, it might be more efficient to simply run an \textit{ab initio} MD simulation in the first place. Further, not every ML method is equally applicable or appropriate for all systems due to methodical and/or conceptual constraints. Such limitations are discussed in greater detail in Section~\ref{subsec:choosing_an_appropriate_ml_method}.

\paragraph{Multireference effects}
Many \textit{ab initio} methods use a single Slater determinant to express the wave function of a system. The problem with this approach is that different determinants may be dominant in different regions of the PES, leading to a poor description of the wave function if the wrong determinant is chosen. Especially when many calculations are performed for various strongly distorted geometries, for example when a reaction is studied and bonds need to be broken, it may happen that the solution ``jumps'' discontinuously from one electronic state to another, leading to inconsistent reference data. When an ML model is trained on such a data set, it will try to find a compromise between the inconsistencies and its performance typically be unsatisfying. It is therefore advisable to check for possible multireference effects prior to generating data and, if necessary, switch to a multireference method (for a comprehensive review on multireference methods, see Ref.~\citenum{szalay2012multiconfiguration}).

\paragraph{Strong delocalization}
The models discussed in this review all assume that energy contributions are local to some degree. This assumption is either introduced explicitly by a cutoff radius, or it enters the model through the use of a specific structural descriptor. For example, by using inverse distances to encode chemical structures for kernel methods (as is done e.g.\ in GDML, see Section~\ref{subsubsec:symmetric_gradient_domain_machine_learning}), relative changes between close atoms are weighed more strongly when comparing two conformations. While assuming locality is valid in many practical applications, there exist many cases where this assumption breaks down. An example are extensive conjugated $\pi$-systems, where a rotation around certain bonds might break the favorable interaction between electrons, leading to a ``non-local'' energy contribution. If such effects exist, an appropriate model should be chosen, for example the cutoff radius may need to be larger than usual or a different structural descriptor must be picked. 

\subsection{Choosing an appropriate ML method}
\label{subsec:choosing_an_appropriate_ml_method}
Several different variants of ML-FFs have been discussed in Section~\ref{subsec:combining_ml_and_chemistry} and many more are described in the literature. Although all these methods can be applied to construct ML-FFs for any chemical system, some methods might be more promising than others for certain tasks. For researchers who want to apply ML methods to a specific problem for the first time, the abundance of different models to choose from may be overwhelming and it might be difficult to find an appropriate choice. 

In the following, possible applications of ML-FFs are broadly categorized based on simple questions about the task at hand. For each case, advantages and disadvantages of individual models are discussed to provide help and guidance to the reader for identifying an appropriate model for their use case.

\paragraph{How much reference data can be used for training?} \textit{
	When in doubt which method to use, a rule of thumb could be to prefer kernel methods when there are less than $\sim10^3$--$10^4$ training points and NN-based approaches otherwise (but this may also be a matter of preference).
}

Depending on the desired accuracy, the amount of \emph{ab initio} reference data which can be collected within a reasonable time frame may be vastly different. For example, if reference calculations are performed at the DFT level of theory, it is often feasible to collect several thousands of data points, even for relatively large molecules. On the other hand, if CCSD(T) accuracy and a large basis set is required, already a few hundred reference calculations for small molecules can require a considerable amount of computing time. Although it is of course always desirable to perform as few reference calculations as possible, for some tasks, collecting a large data set is unavoidable. For example, if a model should be able to predict a variety of different molecules containing many different elements, the relevant chemical space must be sampled sufficiently.

In general, kernel-based models tend to achieve good prediction accuracies even with few training points, whereas NNs often need more data to reach their full potential (although there may be exceptions for both model variants, see also Fig.~\ref{fig:md17_scatter}). Further, the optimal model parameters for kernel models can be determined analytically (see Eq.~\ref{eq:krr_coefficient_relation_regularized}), which, at least for small datasets, is typically faster than training a NN via (a variant of) stochastic gradient descent. However, when the data set size $M$ is very large, solving Eq.~\ref{eq:krr_coefficient_relation_regularized} analytically can become prohibitively expensive as it scales $\mathcal{O}(M^3)$ (and requires $\mathcal{O}(M^2)$ memory to store the kernel matrix). Further, evaluating kernel models scales with $\mathcal{O}(M)$ (see Eq.~\ref{eq:kernel_regression}), whereas the cost of evaluating NN-based methods has (as long as the number of parameters does not have to be increased for larger datasets) constant complexity. For this reason, NNs tend to be more suitable for large datasets. Note that there are approximations which improve the scaling of kernel methods (so they can be applied even to very large datasets) at the cost of accuracy (see Eqs.~\ref{eq:krr_coefficient_relation_approximated}-\ref{eq:nystroem_regularized_inverse}).

\paragraph{Should the model be able to predict a single type of chemical system or multiple different ones?}
\textit{
	To be applicable to multiple systems, a model must decompose its prediction into atomic contributions. Models that use no such decomposition must either use a fixed size descriptor or several separate models need to be trained.
}

Some ML-FFs only need to be able to predict systems with a fixed composition and number of atoms, for example to study the dynamics of a single molecule, whereas other applications require the ability to predict different systems with varying size, e.g.\ when clusters consisting of a different number and kind of molecules are studied with the same model.

While all ML-FFs can be applied in the first case, the latter requires either that the length of chemical descriptors is independent of the number of atoms, or that model predictions are decomposed into local contributions based on fixed-size fingerprints of atomic environments (which naturally makes them extensive). Most NNPs (see Section~\ref{subsubsec:artificial_neural_networks}) and many kernel methods, e.g.\ GAPs (see Section~\ref{subsubsec:gaussian_approximation_potentials}) or FCHL\cite{faber2018alchemical,christensen2020fchl}, use such a decomposition and can be applied to differently sized chemical systems without issues. An exception are e.g.\ (s)GDML models (see Section~\ref{subsubsec:symmetric_gradient_domain_machine_learning}), which encode chemical structures as vectors of inverse distances between atomic pairs. Consequently, the length of the descriptor changes with the number of atoms and the model can only be applied to a single type of system. In some special cases, it may be possible to choose a maximum descriptor length and pad descriptors of smaller molecules with zeros, but this may introduce other problems and/or reduce the accuracy.

\paragraph{Will the model be applied to single or multi-component systems?}
\textit{
	If only a handful of elements is relevant, all models are equally suitable. When a large number of elements needs to be considered, the model should be able to encode and use information about atom types efficiently.
}

As long as an ML-FF is only applied to single-component systems (consisting of a single element), for example elemental carbon or silicon, all relevant information is contained in the relative arrangement of atoms and nuclear charges need not be encoded explicitly. However, as soon as there are multiple atom types (as is common for most applications of ML-FFs), the model must have some way to distinguish between them. A notable exception are some models such as (s)GDML, which use inverse pairwise distances as structural descriptor. Here, information about atom types is implicitly contained, because specific entries always correspond to the same combination of atom types. 

Many local descriptors of atomic environments only use geometric information in the form of distances and angles between pairs and triplets of atoms (see Eqs.~\ref{eq:two_body_symmetry_function}~and~\ref{eq:three_body_symmetry_function}). To include information about atomic types,  geometric features have to be included separately for every possible combination of elements, leading to a drastic increase of descriptor size (descriptors for kernel machines based e.g.\ on SOAP\cite{bartok2013representing} or FCHL\cite{faber2018alchemical,christensen2020fchl} also grow in size when the number of atom types is increased). Many descriptor-based NNPs further use separate NNs to predict atomic contributions of different elements (see Fig.~\ref{fig:nnp_overview}). A disadvantage of these approaches is that the number of terms in the descriptor increases combinatorially with the number of elements covered by the model (in particular if three-body or even four-body terms are used), which impacts the computational cost of training and evaluating the model. Also, larger amounts of training data may become necessary for good results. As long as only a few elements need to be considered, these downsides are not an issue, but if a model for a significant fraction of the periodic table is required, a more efficient representation is desirable. Most end-to-end NNPs employ so-called element embeddings (see Fig.~\ref{fig:nnp_overview}), which do not become more complex when the number of elements is increased. This has the additional benefit of  potentially increasing the data efficiency of the model by utilizing alchemical information. Another alternative is to introduce element-dependent weighting functions (instead of duplicating terms in ACSF descriptors)\cite{gastegger2018wacsf}. 

\paragraph{Will the model be applied to small or large systems?} \textit{Models for very large target systems should be able to exploit chemical locality, so that reference calculations for fragments can be used as training data. Additionally, this allows trivial parallelization of predictions over multiple machines.}

Often, ML-FFs are used to study small or medium-sized molecules. In such cases, all models are equally applicable. For very large systems containing many atoms however, some methods have particularly advantageous properties. For example, it might be infeasible to run \textit{ab initio} calculations for the full target system. In this case, being able to fragment the system into smaller parts, for which reference calculations are affordable, is very useful. 

To be trainable on such fragments, ML-FFs must introduce an explicit assumption about chemical locality by introducing a cutoff radius. Every method that decomposes predictions into a sum of local atomic contributions can thus be trained in this way. ML-FFs without cutoffs on the other hand need reference data for the complete system (see above). Another advantage of local models is that their predictions are \emph{embarrassingly parallel}: The contributions of individual atoms can be calculated on separate machines (storing a copy of the model), each requiring only information about neighboring atoms within the cutoff radius. Apart from possible efficiency benefits, this may even become necessary if the computations to handle all atoms do not fit into the memory of a single machine (for example when the system of interest consists of millions of atoms\cite{jia2020pushing}). Note that while not all ML methods to construct FFs can be parallelized in this way, most models contain mostly linear operations, which are amenable to other parallelization methods, e.g.\ by utilizing GPUs (graphics processing units).

At this point, a subtle difference between cutoffs used in NNPs of the message-passing type (see Section~\ref{subsubsec:high_dimensional_neural_network_potentials}) and descriptor-based NNPs (as well as kernel machines based on local atomic environments) should be pointed out. In message-passing schemes, information between all atoms within the cutoff radius is exchanged over $T$ iterations, thus the \emph{effective} cutoff radius increases by a factor of $T$. This means that in order to distribute the computation over multiple machines, it is either necessary to communicate updates to other machines after each iteration, or a sufficiently large subdomain needs to be stored on all machines.

\paragraph{Are long-range interactions expected to be important for the system of interest?} \textit{If strong long-range contributions to the energy are present, it is advisable to either use a model without cutoffs, or augment the pure ML approach by explicitly including physical interaction terms.}

As described earlier, many ML-FFs introduce cutoffs to exploit chemical locality. An obvious downside of this approach is that all interactions beyond the cutoff cannot be represented. For uncharged molecules without strong dipole moments, relevant interactions are usually sufficiently short-ranged that this is not problematic. However, when strong long-ranged (e.g.\ charge-dipole) interactions are important, cutoffs may introduce significant errors. Models such as (s)GDML, which consider the whole chemical structure without introducing cutoffs, do not suffer from this issue in principle.  

While it is possible to simply increase the cutoff distance until more long-ranged contributions can be neglected, this decreases the computational and data efficiency of models which were designed with cutoffs in mind. A better alternative could be to include the relevant physical interaction terms explicitly in the model. For example, TensorMol\cite{yao2018tensormol} and PhysNet\cite{unke2019} include such correction terms by default, but other models can be augmented in a similar fashion. Although not strictly necessary, even models without cutoffs may profit from such terms by an increased data efficiency.

\subsection{Data collection}
\label{subsec:data_collection}
A fundamental component of any ML model is the reference data. While its architecture and other technical details are responsible for the potential accuracy of a model, the choice of reference data and its quality defines the reliability and range of applicability of the final model. Any deficiencies that are present in the data will inevitably also lead to artifacts in models trained on it, a principle often colloquially stated as ``garbage in, garbage out''\cite{sanders2017garbage}. As such, the reference data is one of the most important components of an ML-FF. The generation of datasets in computational chemistry and physics are challenges on their own. First of all, each reference point is a result of computationally expensive and often nontrivial calculations (see Section~\ref{subsec:chemistry_foundations}), which limits the amount of data that can be collected. Furthermore, the dimensionality of the configurational space of molecules, solids, or liquids is so vast that -- except for trivial cases -- it is not apparent how to identify the representative geometries in the ocean of possibilities. The optimal choice of reference data might even need to be adapted to the individual properties of the respective ML model that consumes it and/or its intended application. In the following, several strategies for sampling the PES and generating reference datasets are outlined (multiple of these approaches can be combined). Afterwards, problems that may occur due to insufficient sampling are highlighted and general remarks about the importance of a consistent reference dataset are given.

\begin{figure*}[t]
	\centering
	\includegraphics[width=1.0\textwidth]{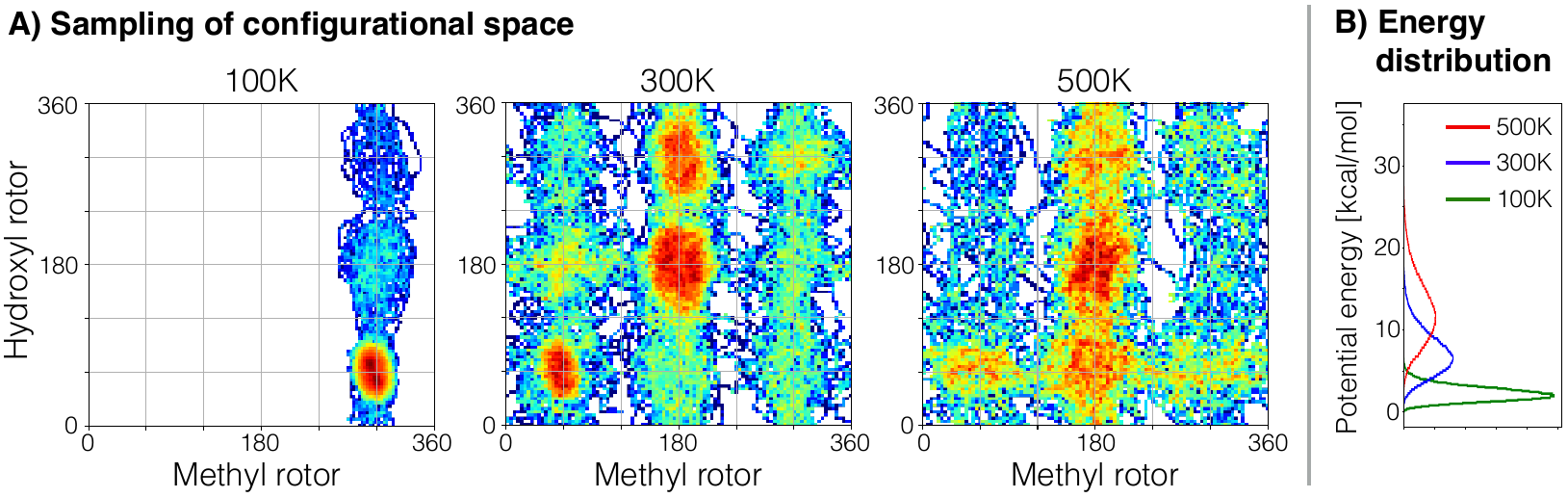}
	\caption{\textbf{A}: Two-dimensional projection of the sampled regions of the PES of ethanol at 100~K, 300~K and 500~K from running AIMD simulations with FHI-aims\cite{FHIaims2009} (Fritz Haber Institute \textit{ab initio} molecular simulations) at the PBE+TS/DFT level of theory\cite{PBE1996,tkatchenko2009accurate}. The length of the simulation was 500 ps. \textbf{B}: Distribution of sampled potential energies for the three different temperatures.}
	\label{fig:FigEthSampl}
\end{figure*}

\paragraph{AIMD sampling}
A good starting point to assemble the reference dataset is by sampling the PES using \textit{ab initio} molecular dynamics (AIMD) simulations. Albeit expensive in terms of the amount of necessary reference calculations, this technique constitutes a straightforward way to explore configurational space. Here, the temperature of the simulation determines which regions of the PES and what energy ranges (according to the Boltzmann distribution) are explored (see Fig.~\ref{fig:FigEthSampl}). For example, if the aim is to construct an ML-FF for calculating the vibrational spectrum of ethanol at 300~K, generating the database at 500~K is a safe option since the subspace of configurations relevant at 300~K is contained in the resulting database (see Fig.~\ref{fig:FigEthSampl}A). Sampling at higher temperatures ensures that the model does not enter the extrapolation regime during production runs, which is practically guaranteed to happen when a lower temperature is used for sampling. In general, the resulting dataset will be biased towards lower energy regions of the PES, where the system spends most of the simulation time. For this reason, pure AIMD sampling is only advisable when the intended application of the final ML model involves MD simulations for equilibrium or close to equilibrium properties, where rare events do not play a major role. Examples of this are the study of vibrational spectra, minima population, or thermodynamic properties.

\begin{figure}
	\centering
	\includegraphics[width=1.0\columnwidth]{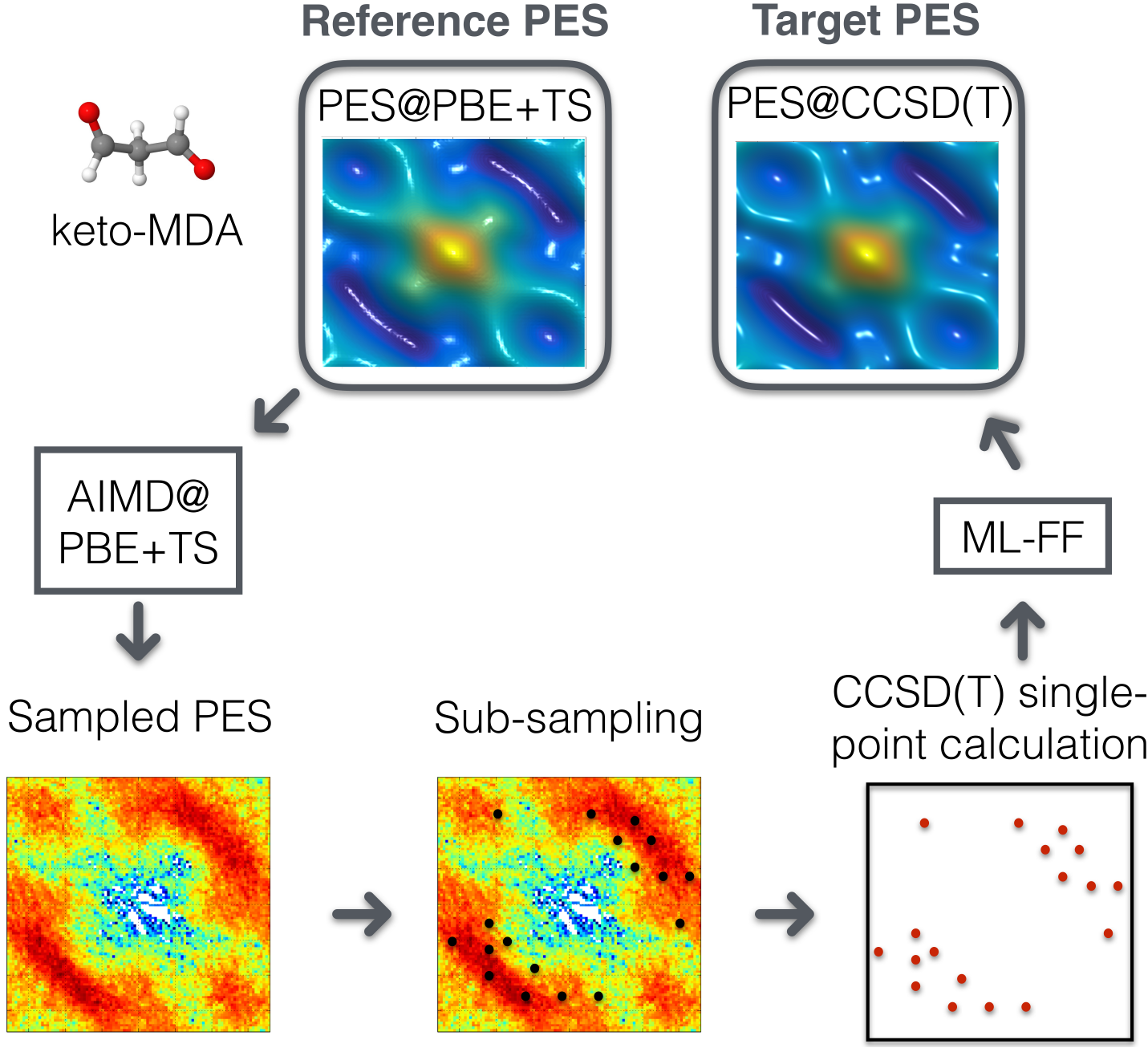}
	\caption{Procedure followed to generate a database at the CCSD(T) level of theory for keto-malondialdehyde using sampling by proxy. An AIMD simulation at 500~K computed at the PBE+TS/DFT level of theory is used to sample the molecular PES. Afterwards, the trajectory is sub-sampled (black dots) to generate a subset of representative geometries, for which single-point calculations at the CCSD(T) level of theory are performed (red dots). This highly accurate reference data is then used to train an ML-FF.}
	\label{fig:FigSubSampl}
\end{figure}

\paragraph{Sampling by proxy}
Constructing reliable reference datasets from AIMD simulations can be computationally expensive. While system size plays a major role, other phenomena, such as the presence of intramolecular interactions and fluxional groups, can also influence how quickly the PES is explored. Because of this, long simulation times may be required to visit all relevant regions. For example, generating $2\times10^5$ conformations from AIMD using a relatively affordable level of theory (e.g.\ PBE+TS/DFT with a small basis set) can take between a few days to several weeks (depending on the size of the molecule). With higher levels of theory, the required computation time may increase to months, or, when  highly accurate methods such as CCSD(T) are required, even become prohibitively long (several years).

To resolve this issue, a possible strategy is to sample the PES at a lower level of theory to generate a long trajectory that covers many regions on the PES. The collected dataset is then subsampled to generate a small, but representative set of geometries, which serve as input for performing single-point calculations at a higher level of theory (see Fig.~\ref{fig:FigSubSampl}).
This strategy works best when the PES has a similar topology at both levels of theory, so it can be expected that configurations generated at the lower level are representative of configurations that would be visited in an AIMD simulation at the higher level (see the two-dimensional projections of the PES in Fig.~\ref{fig:FigSubSampl}). When the two PESs are topologically very different, e.g.\ when a semi-empirical method or even a conventional FF is used to generate the initial trajectory, it may happen that the relevant regions of the PES at the higher level of theory are not covered sufficiently. Then, when an ML-FF is trained on the collected dataset and used for running an MD simulation, the trajectory may enter the extrapolation regime and the model might give unphysical predictions. Thus, extra care should be taken when two very different levels of theory are used for sampling by proxy.

\paragraph{Adaptive sampling}
Another method to minimize the amount of expensive \textit{ab initio} calculations is called adaptive sampling or \textit{on-the-fly} ML\cite{csanyi2004learn}. Here, a preliminary ML-FF is trained on only a small initial set of reference data and then used to run an MD simulation. During the dynamics, additional conformations are collected whenever the model predictions become unreliable according to an uncertainty criterion. Then, new reference calculations are performed for the collected structures and the training of the ML model is continued or started from scratch on the augmented dataset. The process is repeated until no further unreliable regions can be discovered during MD simulations.

When following this strategy, the quality of the uncertainty estimate is crucial for an efficient sampling of the PES: If the estimate is overconfident, deviations from the reference PES might be missed. If the estimate is overly cautious, many redundant \textit{ab initio} calculations have to be performed. There exist several ways to estimate the uncertainty of an ML-FF. For example, Bayesian methods learn a probability distribution over models, which enables straightforward uncertainty estimates (see the predictive variance of a Gaussian process, Eq.~\ref{eq:gaussian_process_variance}). For models where an explicit uncertainty estimate is not available, e.g.\ neural networks, a viable alternative is \textit{query-by-committee}\cite{seung1992query,behler2015constructing}. Here, an ensemble of models is trained, for example on different subsets of the reference data and each starting from a different parameter initialization. Then, the discrepancy between their predictions can be used as uncertainty estimate. Query-by-committee has been successfully employed to sample PESs using neural networks for water dimers\cite{morawietz2012neural}, organic molecules\cite{gastegger2017machine,unke2019} as well as across chemical compound space\cite{smith2018less}. Other alternatives, for example using dropout\cite{srivastava2014dropout} as a Bayesian approximation\cite{gal2016dropout}, could also be used.

Collecting data ``on-the-fly'' is even possible without uncertainty estimates. Instead, additional reference calculations are performed at fixed intervals during the MD simulation\cite{csanyi2004learn,li2015molecular}. This relies on the assumption that the probability of reaching the extrapolation regime of an ML model rises with increasing length of the MD trajectory.
While performing \textit{ab initio} calculations in regular intervals will discover all deviations of the model eventually, this variant of on-the-fly ML does not exploit any information about the already collected reference set and may thus lead to many redundant data points. More detailed reviews on uncertainty estimation and active sampling of PESs can be found in Refs.~\citenum{gastegger2020molecular}
~and~\citenum{shapeev2020active}.

\paragraph{Metadynamics sampling} Similar to adaptive sampling, metadynamics sampling\cite{barducci2011metadynamics,herr2018metadynamics} uses a preliminary ML-FF to run MD simulations to find structures for which to run reference calculations. However, the dynamics are biased to increase the probability for visiting unexplored regions on the PES. This is achieved by placing ``Gaussian bump functions'' on the PES in regions that have already been visited, i.e.\ the potential energy of already known structures is artificially raised. It is possible to combine metadynamics with the uncertainty estimates used in adaptive sampling to only select the most relevant structures.

\paragraph{Normal mode sampling}
It is also possible to sample the PES without running any kind of MD simulation. In normal mode sampling\cite{smith2017ani}, the idea is to start from a minimum on the PES and generate distorted structures by randomly displacing atoms along the normal modes. They are the eigenvectors of the mass-weighted Hessian matrix obtained at the minimum position, i.e.\ a harmonic approximation of the molecular vibrations. From the associated force constants (related to the eigenvalues), the increase in potential energy for displacements along individual normal modes can be estimated. Since they are orthogonal to each other, it is straightforward to combine multiple random displacements along different normal modes such that the resulting structures are sampled from a Boltzmann distribution at a certain temperature. In other words, structures generated like this are drawn from the same distribution as if an ``approximated PES'' was sampled with a (sufficiently long) MD simulation. This approximated PES is equivalent to a Taylor expansion of the original PES around the minimum position, truncated after the quadratic term (the contribution of the linear term vanishes at extrema). 

Structures generated from random normal mode sampling are not correlated, in contrast to those obtained from adjacent time steps in MD simulations, which makes this approach an efficient way to explore the PES.  However, the disadvantage is that only regions close to minima can be sampled. Additionally, the harmonic approximation is only valid for small distortions, i.e.\ the larger the temperature, the more the sampled distribution diverges from the Boltzmann distribution on the true PES. Because of these limitations, it is best to combine normal mode sampling with other sampling methods, for example to generate an initial reference dataset, which is later expanded by adaptive sampling.

\begin{figure}[t]
	\includegraphics[width=\columnwidth]{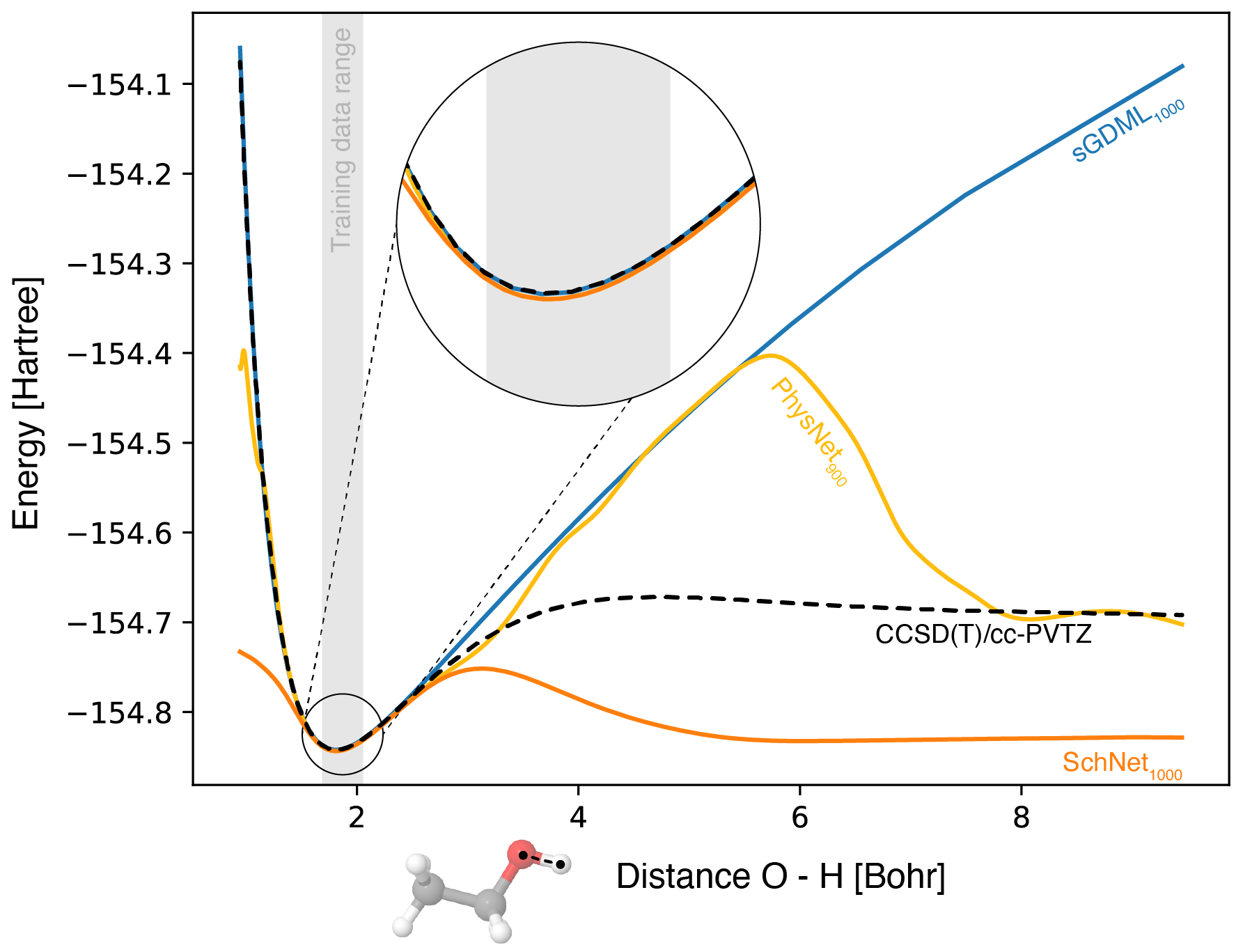}
	\caption{One-dimensional cut through the PES of ethanol along the O--H bond distance for different ML-FFs (solid blue, yellow and orange lines) compared to \textit{ab initio} reference data (dashed black line). Close to the region sampled by the training data (range highlighted in gray), all model predictions are virtually identical to the reference method (see zoomed view). When extrapolating far from the sampled region, the different models have increasingly large prediction errors and behave unphysically.}
	\label{fig:etoh_dissociation_curve}
\end{figure}

\paragraph{Problems due to insufficient sampling} Because their extrapolation capabilities are limited, ML methods only give reliable predictions in regions where training data is present.\cite{sugiyama2007covariate} When generating reference data, it is therefore important that all regions of the PES that may be relevant for a later study are sampled sufficiently. For example, when studying a reaction, the data should not only cover configurations corresponding to educt and product structures, but also the region around the transition state and along the transition pathway. When the reaction coordinate defining the transition process is already known, a straightforward way to generate the reference data would be to sample the transition path region. However, even when an ML model can reproduce the entire reference dataset with the required accuracy, it is still possible to run into issues when the model is used to study the reaction. If the rare transition process was not sampled sufficiently, it is not guaranteed that MD simulations with the ML-FF reproduce it correctly. The reference data may be restricted to a specific subset of molecular configurations along the transition pathway. Hence, the model can enter the extrapolation regime somewhere between the boundary states and the transition pathways generated by an MD simulation might be unreasonable. Another potential issue is that after passing the transition state region, typically, a large amount of potential energy is converted to internal motions such as bond vibrations. As a result, the effective temperature defined by the kinetic energy exceeds the ambient conditions by orders of magnitude. Even when using a thermostat in the simulation, thermal energy increases so rapidly that it may not be able to handle the increase in temperature immediately. As a consequence, the trajectory visits high-energy configurations, which may not be included in the reference data, and the model again has to extrapolate.

When ML-FFs enter the extrapolation regime, i.e.\ they are used to predict values outside the sampled regions of the PES, unphysical effects may be observed. Consider for example the dissociation of the O\nobreakdash--H bond in the hydroxyl group of ethanol (Fig.~\ref{fig:etoh_dissociation_curve}). 
Here, different models were trained on data gathered from an MD simulation of ethanol at 500~K and used to predict how the energy changes when the O\nobreakdash--H distance of the hydroxyl group is shortened or elongated to extreme values well outside the range sampled during the dynamics. In this example, while the sGDML model is able to accurately extrapolate to much shorter distances than are present in the training data, it still fails to predict the bond dissociation. The NNP models (PhysNet and SchNet) exhibit qualitatively wrong short-range behavior and spurious minima on the PES, which may trap trajectories during MD simulations. Because of these limited extrapolation capabilities, it is advisable to sample larger regions of the PES than are expected to be visited during MD simulations, so that there is a ``buffer'' and models never enter the unreliable extrapolation regime during production runs. For example, when an ML-FF is to be used for a study at a temperature of 300~K, the PES should be sampled around 500~K or higher.

\paragraph{Importance of data consistency} Although it may appear trivial, it is crucial that all data used for training a model is internally consistent: A single level of theory (method and basis set) should be used to calculate the reference data. When multiple quantum chemical codes (or even different versions of the same code) are used for data generation, it should be checked that their output is numerically identical when given the same input geometry (if they do not then this will effectively manifest itself like noisy outputs, severely deteriorating the precision of the ML model). Further, many \textit{ab initio} codes automatically re-orient the input geometry such that the principal moments of inertia are aligned with the $x$-, $y$- and $z$\nobreakdash-axes, so extra care should be taken when forces or other orientation-dependent quantities (i.e.\ electric moments) are extracted to verify they are consistent with the input geometry. When some calculation settings need to be adapted for a subset of the data, e.g.\ for cases with difficult convergence, it is important to check that values computed with the modified settings are consistent with the rest of the data. Additionally, for training some ML models, it may be essential that atoms are ordered in a particular way throughout the data set. For example, the permutational symmetry of (s)GDML models is limited to the transformations recovered from the training set, whereas the NN models discussed in this review are fully agnostic with respect to atom indexing.

\subsection{Data preparation}
\label{subsec:data_preparation}
After the reference data is collected, it has to be prepared for the training procedure. This includes splitting the data into different subsets, which are reserved for separate purposes. Some models may also require that the data is preprocessed in some way before the training can start. In the following, important aspects of these preparation steps are highlighted.

\paragraph{Splitting the data}
Prior to training any ML model, it is necessary to split the reference data into disjoint subsets for training/validation and testing (see Section~\ref{subsubsec:model_selection_how_to_choose_hyperparameters}). While the training/validation set is used for fitting the model, the test set is only ever used after a model is trained to estimate its generalization error, i.e.\ to judge how well the model performs on unseen data.\cite{hansen2013assessment,hastie2009elements} It is very important to keep the two splits separate, as it is easily possible to achieve training errors that are several orders of magnitude lower than the true generalization error when the model is not properly regularized. Many models also feature hyperparameters, such as kernel widths, regularization terms or learning rates, that must be tuned by comparing several trained model variants on a third dataset used purely for validation (a subset of the training/validation set). Note that information from the validation set will still enter the model indirectly, i.e.\ it also participates in the training process. This is why a strict separation of the training/validation set from the test set is crucial. Undetected duplicates in the dataset can complicate splitting, as the contamination of the test set with training data (``data leakage'') might go unnoticed. In this case, the model is effectively trained on part of the test set and estimates of the generalization error might be too optimistic and unreliable. Such a scenario can occur even when no obvious mistakes were made, e.g.\ when the structures for a dataset are sampled by running a long MD simulation where snapshots are written very frequently. Structures collected from adjacent time steps may be highly correlated in this case and when splitting the data randomly into training and test sets, a large portion of both sets will be almost identical. In such a case, instead of using a random split, a better approach would be to use a time-split of the dataset,\cite{lemm2011introduction} e.g.\ using the first 80\% of the MD trajectory as the training/validation set and reserving the last 20\% for testing. 

\paragraph{Data preprocessing}
Prior to training a model, the raw data is often processed in some way to improve the numerical stability of the ML algorithm. For example, a common practice is normalization, where inputs (or prediction targets) are scaled and shifted to lie in the range $-1\dots1$ or to have a mean of zero and unit variance. The constants required for such transformations must never be extracted from the complete dataset. Instead, only the training set may be used to obtain this information.\cite{muller2001introduction,lemm2011introduction,hansen2013assessment} Otherwise, estimates of the generalization error on the test set may be overconfident (this is another form of data leakage). While normalization may be less common for the purpose of constructing ML-FFs, any ``data-dependent`` transformation must be done carefully. For example, it may be desirable to subtract the mean energy of structures from the energy labels in order to obtain numbers with smaller absolute values (for numerical reasons). This mean energy should be calculated only from the structures in the training set.

If a model is trained using a hybrid loss that incorporates multiple interdependent properties, such as energy and forces, it is important to consider the effects of the normalization procedure on the functional relationship of those values. For example, multiplying the energy labels by a factor requires that the forces are treated in the same way, because the factor carries over to the derivative (scaling energies and forces by different factors would therefore introduce inconsistencies in the data). Also, while subtracting the mean value from energy labels is valid, it is not correct to add any constant to the force labels, because that would translate into a linear term in the energy domain (the energy is related to the forces through integration). Consequently, the consistency between both label types would be broken and an energy conserving model would be incapable of learning. Even when doing simple unit transformations, care should be taken not to introduce any inconsistencies. For example, when energy labels are converted from $E_{\rm h}$ to kcal~mol$^{-1}$ and atom coordinates from $a_0$ to \AA, force labels have to also be converted to kcal~mol$^{-1}$~\AA$^{-1}$ so that all data is consistent. Depending on which code was used to obtain the reference data, it is even possible that units for some labels \emph{must} be converted, because they may be given in different unit systems in the raw data (\textit{ab initio} codes often report energy and forces in atomic units, whereas for coordinates, angstroms are popular).

\subsection{Training the model}
\label{subsec:training_the_model}
After the data has been collected and prepared, the next step is training the ML-FF. During the training process, the parameters of the model are tuned to minimize a loss function, which measures the discrepancy between the training data and the model predictions. In some cases, e.g.\ most kernel methods, the optimal solution can be found analytically. When this is not possible, e.g.\ when training neural networks, the parameters are typically optimized iteratively by gradient descent or a similar algorithm. Because standard gradient descent tends to converge very slowly, some authors have proposed to augment it with terms mimicking momentum\cite{nesterov1983method,qian1999momentum} or adaptive step sizes\cite{duchi2011adaptive,zeiler2012adadelta}. Not only training times, but also the achievable accuracy varies greatly with different optimization algorithms, so it is best to try different schemes (see Ref.~\citenum{ruder2016overview} for an overview over different popular methods). A good default choice is the Adam optimizer\cite{kingma2015adam}, which converges quickly and gives good results for many different NN architectures. The hyperparameters of a model (e.g.\ the number of layers or their width in the case of NNs) can also be selected in this step, albeit by checking the model performance on the validation set after training (instead of optimizing them directly). This section details the training process and highlights important points to consider, e.g.\ the choice of loss function or how to prevent overfitting of the model to the training data.

\paragraph{Choosing the loss function}
For regression tasks, a standard choice for the loss function is the mean squared error (MSE) given by $\mathcal{L}=\frac{1}{N}\sum_{i=1}^{N} (y_i-\hat{y}_i)^2$, because it punishes outliers disproportionately. Here, the index $i$ runs over all $M$ samples of the training data, $y_i$ is the reference value for data point $i$ and $\hat{y}_i$ is the corresponding model prediction. When the MSE is used as loss function, it is implicitly assumed that any noise present in the reference data is distributed normally, which without additional information, is a sensible guess for most data. Further, the MSE loss allows finding the optimal parameters analytically (due to convexity) for linear ML algorithms, such as kernel ridge regression (see Eq.~\ref{eq:krr_coefficient_relation_regularized} in Section~\ref{subsubsec:kernel_based_methods}). However, the MSE is not necessarily the best choice for all cases. For example, to make the model less sensitive to outliers, a common alternative is to use a mean absolute error (MAE) loss given by $\mathcal{L}=\frac{1}{N}\sum_{i=1}^{N} \lvert y_i-\hat{y}_i \rvert$. Other functional forms, such as Huber loss\cite{huber1992robust} or even an adaptive loss\cite{barron2019general}, are also possible, provided they are a meaningful measure of model performance.

After deciding on the general form of the loss function, the question remains which labels $y$ to use as a reference. 
While the potential energy is an obvious choice, in classical MD, the PES is explored via integration of Newton’s second law of motion, which exclusively involves atomic forces. Since an important objective of ML-FFs is to reproduce the dynamical behavior of molecules in MD simulations as well as possible, it could even be argued that accurate force predictions should take priority over energy predictions in MD applications. However, since energy labels are usually available as a byproduct of force calculations, it seems reasonable to include both label types in the hope that this will help improve the overall prediction performance for both quantities. This gives rise to models based on hybrid loss functions that simultaneously penalize force $\mathbf{F}$ and energy $E$ training errors. Assuming an MSE loss, it generally takes the form
\begin{equation}
\mathcal{L} = \frac{1}{N}\sum_{i=1}^{N} \underbrace{\lVert \hat{\mathbf{F}}_i-\mathbf{F}_{i}\rVert^{2}}_{\mathcal{L}_{\mathbf{F}_{i}}}+\eta\underbrace{(\hat{E}_i-E_{i})^{2}}_{\mathcal{L}_{E_{i}}}\,,
\label{eq:multi_objective_loss}
\end{equation}
where the hyperparameter $\eta$ determines the relative weighting between both loss terms to account for differences in units, information content, and noise level of the label types. A bilateral reduction of both loss terms is only possible if the objectives are non-competing, i.e.\ when the optimal parameter set is equally effective across both tasks. For this to be true, it must hold that 
\begin{equation}
\begin{aligned}
\lVert\hat{\mathbf{F}}_i - \mathbf{F}_i\rVert^2 =& \eta(\hat{E}_i - E_i)^2\,,\\
\Rightarrow \lVert-\nabla_{\mathbf{R}}(\hat{E}_i - E_i)\rVert^2 =&  \eta(\hat{E}_i - E_i)^2
\end{aligned}
\label{eq:loss_relation}
\end{equation}
 at every training point $i$ (here, the relation $\mathbf{F}=-\nabla_{\mathbf{R}}E$ was substituted). Otherwise, the objectives $\mathcal{L}_{\mathbf{F}_{i}}$ and $\mathcal{L}_{E_{i}}$ are necessarily minimized by a different set of model parameters. Eq.~\ref{eq:loss_relation} is only true in general when $\mathcal{L}_{E_{i}}=\mathcal{L}_{\mathbf{F}_{i}}=0$ for all~$i$, which is not fulfilled in practice, because both labels may contain noise and they can usually not be fitted perfectly. A model trained using a hybrid loss (Eq.~\ref{eq:multi_objective_loss}) will thus have to compromise between fulfilling both objectives on the training data, as opposed to joining energy and force labels for a performance gain on both. For this reason, the use of hybrid loss functions (or how to weight different contributions) warrants careful consideration depending on the intended application of the final model. Some models, e.g.\ (s)GDML (see \ref{subsubsec:symmetric_gradient_domain_machine_learning}), do not even include energy constraints in their loss function at all and are trained on forces only. The energy can still be recovered via integration, but it does not participate in the training procedure except for determining the integration constant. In the end, the ultimate measure of a model's quality should not be how well it minimizes a particular loss function, but instead how well it is able to reproduce the experimental observables of interest. Also, it is important to keep in mind that the loss function measured on the training data is only a proxy for the true objective of any model, which is to generalize to unseen data. Compromising between the energy and force labels of the training data can even improve prediction accuracy for \emph{both} label types on unseen data. For a more thorough discussion on the role of gradient reference data and how it can improve prediction performance, see Ref.~\cite{chmiela2019towards,christensen2020on,meyer2020machine}.

\paragraph{Tuning hyperparameters}
Hyperparameters, such as kernel widths or the depth and width of a neural network, are typically optimized independently of the parameters that determine the model fit to the data: A hyperparameter configuration is chosen, the model is trained, and its performance is measured on the validation set. This process is repeated for as many trials as are affordable or until the desired accuracy is reached. Here it is crucial that no test data is used to measure model performance when tuning hyperparameters, so the ability to estimate the generalization error on the test set is not compromised. Choosing good values for the hyperparameter regimes requires some experience and intuition of the problem at hand. Fortunately, many models are quite robust and good default hyperparameters exist, which do not require any further tuning to arrive at good results. In other cases, hyperparameter tuning can be automated (for example via grid or random search\cite{lecun2012efficient,muller2001introduction,bergstra2012random,hansen2013assessment,chmiela2019}) and does not need to be performed manually. See also Section~\ref{subsubsec:model_selection_how_to_choose_hyperparameters} for a more detailed discussion on tuning hyperparameters.

\begin{figure}[t]
	\centering
	\includegraphics[width=\columnwidth]{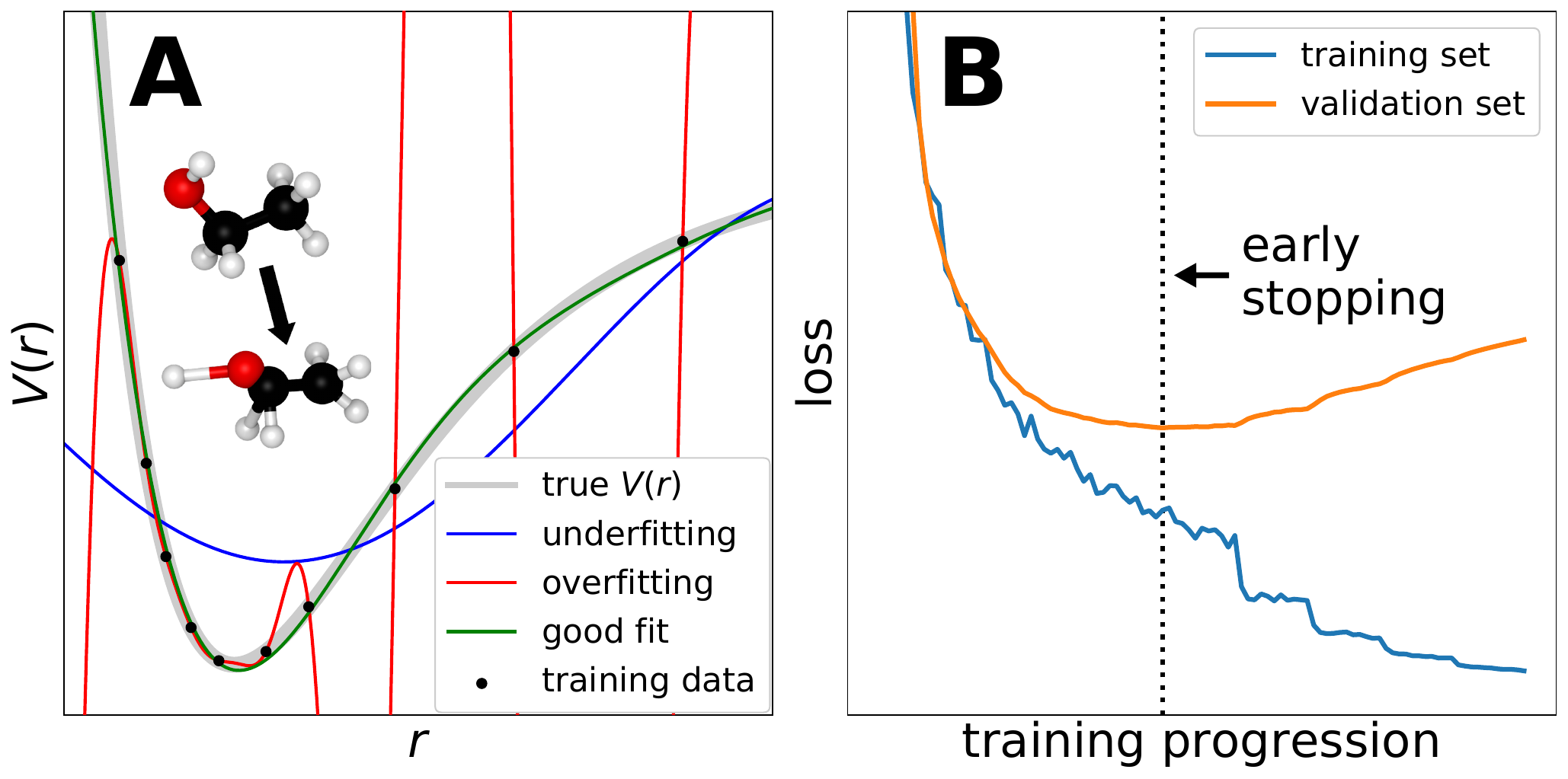}
	\caption{\textbf{A}: One-dimensional cut through a PES predicted by different ML models. The overfitted model (red line) reproduces the training data (black dots) faithfully, but oscillates wildly in between reference points, leading to ``holes'' (spurious minima) on the PES. During an MD simulation, trajectories may become trapped in these regions and produce unphysical structures (inset). The properly regularized model (green line) may not reproduce all training points exactly, but fits the true PES (gray line) well, even in regions where no training data is present. However, too much regularization may lead to underfitting (blue line), i.e.\ the model becomes unable to reproduce the training data at all. \textbf{B}: Typical progress of the loss measured on the training set (blue) and on the validation set (orange) during the training of a neural network. While the training loss decreases throughout the training process, the validation loss saturates and eventually increases again, which indicates that the model starts to overfit. To prevent overfitting, the training can be stopped early once the minimum of the validation loss is reached (dotted vertical line).}
	\label{fig:overfitting} 
\end{figure}

\paragraph{Regularization}
Because ML models contain many parameters (sometimes even more than the number of data points used for training), it is possible or even likely that they ``overfit'' to the training data. An overfitted model achieves low prediction errors on the training set, but performs significantly worse on unseen data (Fig.~\ref{fig:overfitting}A). The aim of regularization methods is to prevent this unwanted effect by limiting or decreasing the complexity of a model. 

When the loss function is minimized iteratively by gradient descent or similar algorithms, as is common practice for training NNs, one of the most simple methods to prevent overfitting is \emph{early stopping}\cite{prechelt1998early}): In the beginning of the training process, prediction errors typically decrease on both training and validation data. At some point however, because the validation set is not used to directly optimize parameters, the performance on the training data will continue to improve, whereas the loss measured on the validation set will stagnate at a constant value or even begin to increase again. This indicates that the model starts overfitting. Early stopping simply halts the training process as soon as the validation error converges (instead of waiting for convergence of the training error), see Fig.~\ref{fig:overfitting}B. Early stopping also limits the size of the neural network weights and thus implicitly limits the complexity of the underlying function class. Similar to tuning hyperparameters, only the validation set, but never the test set, must be used for determining the stopping point.

Another method of regularization is the introduction of penalty terms to the loss function.  Since overfitted models often are characterized by high variance in the prediction (see Fig.~\ref{fig:overfitting}A), the idea is to penalize large model parameters. For example, $L^2$ regularization (adding the squared magnitude of parameters to the loss) shrinks the $L^2$-norm of the parameter vectors towards zero and prevents very large parameter values. On the other hand, $L^1$ regularization (adding the absolute values of parameters to the loss) shrinks their $L^1$-norm, i.e.\ it favors sparse parameter combinations. Typically, the regularization term is weighted by an additional hyperparameter $\lambda$ that determines its strength (like all hyperparameters, $\lambda$ has to be tuned on the validation set). Note that solving Eq.~\ref{eq:krr_coefficient_relation_regularized} to determine the parameters of a kernel method will result in an $L^2$-regularized model trained on the MSE loss function.

\subsection{Using ML-FFs in production}
\label{subsec:using_the_model_in_production}
The main motivation for training an ML-FF is to use it for some production task, such as running an MD simulation. Before doing so however, it is advisable to verify that it fulfills the accuracy requirements for its intended application. At this point in time, the test set becomes important: Since it was neither used directly nor indirectly during the training process, the data in the test set allows to estimate the performance of a model on truly unseen data, i.e.\ how well it generalizes. For this, it is common practice to compute summary errors on the test set, for example the mean absolute error (MAE) or root mean squared error (RMSE), as a measure of the overall accuracy of a model. In general, such a way of quantifying accuracy gives an overview of the ML model's performance on the given dataset and provides a simple way to benchmark. 

However, summary errors are biased towards the densely sampled regions of the PES, whereas much larger errors can be expected for less populated regions. Therefore, while summary errors measured on the test set are typically a good indicator for the quality of a model, they are not necessarily the best way to judge how well an ML-FF performs at its primal objective, namely capturing the relevant quantum interactions present in the original molecular system. In other words, performance measures evaluated on the test set should not be trusted blindly. They are only reliable when the test set is representative of the new data encountered during production tasks, i.e.\ when they are drawn from the same distribution. When a model has to extrapolate, it might give unreliable predictions, even when its performance on the test set is satisfactory. When in doubt, especially when an ML-FF is used for a different task than it was originally constructed for, it is better to collect a few new reference data points to verify that a model is still valid for its use case. Because of the generally limited extrapolation capabilities of ML models, results obtained from studies with ML-FFs should always be scrutinized more carefully than e.g.\ results obtained with conventional FFs. For example, it is advisable to randomly select a few trajectories and verify that the sampled structures look ``physically sensible'', e.g.\ no extremely short or long bonds are present and atoms have no unusual valencies. Since the PES is a high-dimensional object, rare events, where a trajectory visits a part of configurational space that is not sampled in the reference data, are always possible, even when the PES was carefully sampled. If any questionable model predictions are found, it is advisable to double-check their accuracy with additional reference calculations. 

\subsection{Example code and software packages}
\label{subsec:example_code_and_software_packages}
While many modern ML-FFs are conceptually simple, their implementation is often not straightforward, involving many intricate details that can not be exhaustively covered in  publications. Instead, those details are best conveyed by a reference implementation of the respective model. Publicly available well maintained codes allow to replicate numerical experiments and to build on top of existing models with minimal effort. 

In this section, example code snippets for training and evaluating kernel- and NN-based ML-FFs with the \texttt{sGDML}\cite{chmiela2019} (\ref{subsubsec:sgdml_package}) and \texttt{SchNetPack}\cite{schutt2018schnetpack} (\ref{subsubsec:schnetpack_package}) software packages are given. This is followed by a short description of other popular software packages for the construction of ML-FFs (\ref{subsubsec:other_software_packages}) as a first orientation for interested readers. Note that the list is not comprehensive and many other similar packages exist.

\subsubsection{The \texttt{sGDML} package}
\label{subsubsec:sgdml_package}
A reference implementation of the (s)GDML model is available as Python software package at \url{http://www.sgdml.org}~\cite{chmiela2019}. It includes a command-line interface that guides the user through the complete process of model creation and testing, in an effort to make this ML approach accessible to broad practitioners.  Interfaces to the Atomic Simulation Environment (ASE)\cite{ase17} or i-PI\cite{ipiv2} make it straightforward to perform MD simulations, vibrational analyses, structure optimizations, nudged elastic band computations, and more. 

To get started, only user-provided reference data is needed, specifically a set of Cartesian geometries with corresponding total energy and atomic-force labels. Force labels are necessary, because sGDML implements energy conservation as an explicit linear operator constraint by modeling the FF reconstruction as the transformation of an underlying energy model (see Section~\ref{subsubsec:symmetric_gradient_domain_machine_learning}). The trained model will give predictions at the accuracy of the reference data and can  be queried like any other FF.

\paragraph{Dataset preparation} The \texttt{sGDML} package uses a proprietary format for its datasets, but scripts to import and export to all file types supported by the ASE package\cite{ase17}, which covers most popular standards, are included. To convert a \texttt{<dataset>}, simply call
\begin{verbatim}
	sgdml_dataset_via_ase.py <dataset>
\end{verbatim}
and follow the instructions.

\paragraph{Training} The most convenient way to reconstruct a FF is via the command line interface:
\begin{verbatim}
sgdml all <dataset> <ntrain> <nvalid>
\end{verbatim}
This command will automatically generate a fully trained and cross-validated model and save it to a file, i.e.\ model selection and hyperparameter tuning (see Section~\ref{subsubsec:model_selection_how_to_choose_hyperparameters}) are performed automatically. The parameters \texttt{<ntrain>} and \texttt{<nvalid>} specify the sample sizes for the training and validation subsets, respectively. All remaining points are reserved for testing. Each subset is sampled from the provided reference \texttt{<dataset>} without overlap.

\paragraph{Using the model} To use the trained model, the sGDML predictor is instantiated from the \texttt{<model>} file  generated above and energy and forces are queried for a given geometry (for example stored in an XYZ file \texttt{<xyz>}):
\begin{verbatim}
import numpy
from sgdml.predict import GDMLPredict
from sgdml.utils import io

# Load model from file
parameters = numpy.load("<model>")
gdml = GDMLPredict(parameters)

# Load structure from xyz file
r,_ = io.read_xyz("<xyz>")

# Evaluate model 
# (energies e and forces f)
e,f = gdml.predict(r)
\end{verbatim}

It is also possible to run MD simulations using ASE and the \texttt{Calculator} interface included with the \texttt{sGDML} package:

\begin{verbatim}
import ase
from sgdml.intf.ase_calc import \
    SGDMLCalculator
from ase.md.velocitydistribution import \
    MaxwellBoltzmannDistribution

# Load sGDML model as ASE calculator
calc = SGDMLCalculator("<model>")

# Load structure and attach calculator
atoms = ase.io.read("<xyz>")
atoms.set_calculator(calc)

# Initialize momenta at 300 K
MaxwellBoltzmannDistribution(
    atoms, 300*units.kB)

# Setup MD using the velocity Verlet 
# integrator and a time step of 0.2 fs 
dyn = ase.md.verlet.VelocityVerlet(
    atoms, 0.2*ase.units.fs,
    trajectory="<trajectory>")

# Simulate for 1000 steps
dyn.run(1000)
\end{verbatim}

To run this script, a trained model (\texttt{<model>}) and an initial geometry (\texttt{<xyz>}) are needed. The resulting MD trajectory is stored in a file \texttt{<trajectory>}. For more details and applications examples, please visit the documentation at \url{www.sgdml.org/doc/}.
	
\subsubsection{The \texttt{SchNetPack} package}
\label{subsubsec:schnetpack_package}
\texttt{SchNetPack}\cite{schutt2018schnetpack} is a toolbox for developing and applying deep neural networks to the atomistic modeling of molecules and materials available from \url{https://schnetpack.readthedocs.io/}. It offers access to models based on (weighted) atom-centered symmetry functions and the deep tensor neural network SchNet, which can be coupled to a wide range of output modules to predict potential energy surfaces and forces, as well as a growing number of other quantum-chemical properties. \texttt{SchNetPack} is designed to be readily extensible to other neural network potentials such as the DTNN~\cite{schutt2017quantum} or PhysNet~\cite{unke2019}. It provides extensive functionality for training and deploying these models, including access to common benchmark datasets. It also provides an Atomic Simulation Environment (ASE)\cite{ase17} \texttt{Calculator} interface, which can be used for performing a wide variety of tasks implemented in ASE. Moreover, \texttt{SchNetPack} includes a fully functional MD suite, which can be used to perform efficient MD and PIMD simulations in different ensembles.

As it is based on the PyTorch deep learning framework\cite{paszke2019pytorch}, \texttt{SchNetPack} models are highly efficient and can be applied to large datasets and across multiple GPUs. Combined with the modular design paradigm of the code package, these features also allow for a straightforward implementation and evaluation of new models. Similar to the \texttt{sGDML} package, the central commodity for training models in \texttt{SchNetPack} is a dataset containing the Cartesian geometries (including unit cells and periodic boundary conditions, if applicable) and atom types, as well as the target properties to be modeled (e.g.\ energies, forces, dipole moments, etc.). More information can be found in Ref.~\citenum{schutt2018schnetpack}.

\paragraph{Dataset preparation} \texttt{SchNetPack} uses an adapted version of the ASE database format to handle reference data. The package provides several routines for preparing custom datasets, as well as a range of pre-constructed dataset classes for popular benchmarks (e.g.\ QM9\cite{ramakrishnan2014quantum} and MD17\cite{chmiela2017}), which will automatically download and format the data. For example, molecular data from the MD17 dataset can be loaded via
\begin{verbatim}
spk_load.py md17 <molecule> <path>
\end{verbatim}
where \texttt{<molecule>} indicates the molecule for which data should be loaded (e.g.\ \texttt{ethanol}), while the second argument specifies where the data is stored locally.

\texttt{SchNetPack} also provides a utility script for converting data files in the extended XYZ format, which is able to handle a wide variety of properties, to the database format used internally. Conversion can be invoked with the command
\begin{verbatim}
spk_parse.py <input> <target>
\end{verbatim}
where the arguments specify the file paths to the \texttt{<input>} data file and \texttt{<target>} database in \texttt{SchNetPack} format, respectively.

\paragraph{Training} 
As for the \texttt{sGDML} package, training and evaluating ML models in \texttt{SchNetPack} can be performed via a command line interface. For example, a basic model can be trained with the script:
\begin{verbatim}
spk_run.py train [model_type]
    [dataset_type] <dataset> <model>
    --split <ntrain> <nvalid>
\end{verbatim}
Here, \texttt{[model\_type]} specifies which kind of NNP to use (\texttt{wacsf} for a descriptor-based NNP using wACSFs\cite{gastegger2018wacsf}, or \texttt{schnet} for the SchNet\cite{schutt2017schnet} end-to-end NNP architecture) and
 \texttt{[dataset\_type]} specifies either a preexisting dataset (e.g.\ \texttt{qm9} or \texttt{md17}), or a \texttt{custom} dataset provided by the user. The next two arguments are the paths to the reference \texttt{<dataset>} and the file the trained \texttt{<model>} will be written to. The arguments
\texttt{<ntrain>} and \texttt{<nvalid>} specify the sample sizes for the training and validation subsets, while the remaining points are reserved for testing. \texttt{SchNetPack} offers a wide range of additional settings to modify the training process (e.g.\ model composition, use of GPU, how different properties should be treated etc.), see \url{https://schnetpack.readthedocs.io/}.

\paragraph{Using the model} 
Once a model has been trained, it can be evaluated in several different ways. The most basic method is to perform predictions via:
\begin{verbatim}
import torch
import ase
from schnetpack.data.atoms \
	import AtomsConverter

# Load model from file
spk_model = torch.load("<model>")

# Set up converter for ASE atoms
converter = AtomsConverter()

# Load structure from xyz file
atoms = ase.io.read("<xyz>")
inputs = converter(atoms)

# Evaluate model and collect
# predictions in a dictionary
results = spk_model(inputs)
\end{verbatim}

It is also possible to use the \texttt{SchNetPack} MD suite to perform various simulations with the trained model. 
Continuing the above example, a basic MD run can be carried out as:
\begin{verbatim}
import ase
import schnetpack.md as md

# Set up the system using 1 replica
# (use n_replicas > 1 for PIMD)
md_system = md.System(n_replicas=1)

# Load structure from xyz file
atoms = ase.io.read("<xyz>")
md_system.load_molecules(atoms)

# Initialize momenta at 300 K
init = md.MaxwellBoltzmannInit(300)
init.initialize_system(md_system)

# Set up integrator 
tstep = 0.2 # fs
integrator = md.VelocityVerlet(tstep)

# Prepare model for MD (specify 
# units, required properties, etc.)
calculator = md.SchnetPackCalculator(
    spk_model,
    required_properties=["forces"],
    force_handle="forces",
    position_conversion='A',
    force_conversion='kcal/mol/A')

# Combine everything
simulator = md.Simulator(
    md_system, integrator, calculator)

# Simulate for 1000 steps
simulator.simulate(1000)
\end{verbatim}

Simulations can be further modified via hooks, which introduce temperature and pressure control, as well as various sampling schemes.
Further documentation of the code package and usage tutorials can be found at \url{https://schnetpack.readthedocs.io/}.

\subsubsection{Other software packages}
\label{subsubsec:other_software_packages}
\paragraph{AMP: Atomistic Machine-learning Package}
AMP is a Python package designed to integrate closely with the Atomistic Simulation Environment\cite{ase17} (ASE) and aims to be as intuitive as possible. Its modular architecture allows many different combinations of structural descriptors and model types. The main idea of AMP is to construct ML-FFs \emph{on-demand}, i.e.\ simulations are first started with an \textit{ab initio} method and later switched to the ML-FF once the model is sufficiently accurate. The package is described in greater detail in Ref.~\citenum{khorshidi2016amp} and on its official website \url{https://amp.readthedocs.io/}.

\paragraph{\ae net} The Atomic Energy NETwork (\ae net) package includes tools for constructing and applying neural network-based ML-FFs. It is written in Fortran 95/2003 and utilizes efficient BLAS (Basic Linear Algebra Subprograms) and LAPACK (Linear Algebra PACKage) routines for performing linear algebra. A Python interface is also included. More details can be found on \url{https://github.com/atomisticnet/aenet/}.

\paragraph{DeePMD-kit}
The DeePMD-kit is a package written in Python/C++ aiming to minimize the effort required to build deep NNPs with different structural descriptors. It is based on the TensorFlow deep learning framework\cite{abadi2016tensorflow} and offers interfaces to the high-performance classical and path-integral MD packages LAMMPS\cite{plimpton1993fast} and i-PI\cite{ipiv2}. More details on the DeePMD-kit can be found in Ref.~\citenum{wang2018deepmd} or on \url{https://github.com/deepmodeling/deepmd-kit/}.

\paragraph{Dscribe} Dscribe is a Python package for transforming atomic structures into fixed-size numerical fingerprints.\cite{dscribe} These descriptors can then be used as input for neural networks or kernel machines to construct ML-FFs. Supported representations include the standard Coulomb matrix\cite{rupp2012fast} and variants for the description of periodic systems\cite{faber2015crystal}, ACSFs\cite{behler2011atom}, SOAP\cite{bartok2013representing}, and MBTR\cite{huo2017unified}. More details can be found on the official website \url{https://singroup.github.io/dscribe/} or in Ref.~\citenum{dscribe}.

\paragraph{n2p2} The neural network potential package (n2p2) allows to use existing parametrizations of Behler-Parinello NNPs to predict energies and forces (either with standalone tools or with the LAMMPS MD package\cite{plimpton1993fast}), but it also provides training tools for generating new potentials. It is mainly written in C++. For further details, refer to \url{https://compphysvienna.github.io/n2p2/}.

\paragraph{PROPhet} The PROPerty Prophet (or short: PROPhet) package uses neural networks to predict the relationship between chemical structure and material properties. As such, it can also be used to generate NN-based ML-FFs. It includes tools to automatically extract properties of interest from the output files of several \textit{ab initio} codes and an interface to the LAMMPS MD package\cite{plimpton1993fast}. More details can be found on \url{https://biklooost.github.io/PROPhet/}.

\paragraph{QML}
QML is a toolkit for learning properties of molecules and solids written in Python.\cite{christensen2017qml} It supplies building blocks to construct efficient and accurate kernel-based ML models, such as different kernel functions and premade implementations of many different structural representations, e.g.\ Coulomb matrix\cite{rupp2012fast}, SLATM\cite{huang2016communication}, and FCHL\cite{christensen2020fchl}. The package is primarily intended for the general prediction of chemical properties, but can also be used for the construction of ML-FFs. For further details, refer to the official website \url{https://www.qmlcode.org} or the github repository \url{https://github.com/qmlcode/qml/}.

\paragraph{RKHS toolkit}
The RKHS toolkit is mainly intended for constructing highly accurate and efficient PESs for studying scattering reactions of small molecules. As described in section~\ref{subsubsec:kernel_based_methods}, the evaluation of kernel-based methods scales linearly with the number of training points $M$ (see Eq.~\ref{eq:kernel_regression}). By using special kernel functions and precomputed lookup tables, the RKHS toolkit allows to bring this cost down to $\mathcal{O}(\log N)$. However, it requires that the training data has grid structure, which limits its applicability to small systems, where it is meaningful to sample the PES by scanning a list of values for each internal coordinate. The implemented kernel functions also allow to encode physical knowledge about the long-range decay behavior of certain coordinates, which enables accurate extrapolation well beyond the range covered in the training data. A Fortran90 implementation of the toolkit can be downloaded from \url{https://github.com/MMunibas/RKHS/} and the algorithmic details are described in Ref.~\citenum{unke2017toolkit}.

\paragraph{RuNNer} The RuNNer Code was the first implementation of high-dimensional neural network potentials and the source code is freely available. Details on how to obtain access can be found on \url{https://www.uni-goettingen.de/de/560580.html}.

\paragraph{TensorMol} The TensorMol package allows to train NNPs that explictly account for electrostatic interactions. It is based on the TensorFlow deep learning framework\cite{abadi2016tensorflow} and includes an interface to i-PI\cite{ipiv2} for performing path integral simulations. For further information, refer to Ref.~\citenum{yao2018tensormol} or \url{https://github.com/jparkhill/TensorMol}.

\section{Physical and Chemical Insights from Machine Learned Force Fields}
\label{sec:applications_of_machine_learned_force_fields}

In nature, the atoms in chemical systems are in constant motion, giving rise to various configurations and reactive events. A large number of experimental observations are not based on a single molecule or atom, but instead on ensembles of various species subject to external conditions, such as temperature or pressure. Consequently, properties associated with individual structures are not sufficient to characterize macroscopic systems. One way to compute ensemble averages are molecular dynamics (MD) simulations, where the time evolution of a system is governed by the atomic forces derived from its associated potential energy surface (PES). From the ergodic hypothesis\cite{boltzmann2017vorlesungen} it is known that the expected value of an observable $A$ can also be obtained from the time average $\langle A\rangle_{t}=T^{-1}\sum_{t=1}^{T}{A_t}$, where $A_t$ is the value of $A$ corresponding to the structure at time step $t$ of the dynamics trajectory and $T$ is their total number. Of course, this relation is valid only when the dynamics is long enough to visit all configurations of the system accessible under the simulation conditions. 

In order to obtain meaningful statistics with MD simulations, many thousands (or millions) of successive PES evaluations are necessary. Due to their high computational cost, accurate electronic structure PESs quickly become intractable for such simulations, which is why highly efficient classical force fields (FFs) are usually employed for running MD simulations. However, this efficiency comes at a cost: Conventional FFs completely neglect or misrepresent some potentially relevant contributions to the potential energy, such as polarization, charge transfer, or electronic effects, which limits their usefulness in modeling complex chemical phenomena. Machine-learned FFs (ML-FFs) offer a unique combination of computational efficiency and high accuracy, opening up tantalizing new possibilities in the simulation of the dynamics of molecules, surfaces, materials and condensed phases. They are able to model \emph{all} chemical interactions -- including those that are typically neglected by conventional FFs. The high accuracy of ML-FFs allows to obtain qualitatively different and novel insights, which would otherwise only be accessible from computationally infeasible \emph{ab initio} MD (AIMD) simulations. 
In the following, some chemical insights made possible by ML-FFs, which could not have been obtained with conventional FFs, are highlighted in greater detail. A brief overview is given in Table~\ref{tab:overview_applications}. Note that the given examples represent only a tiny fraction of the published literature, an exhaustive list is beyond the scope of the current review. Interested readers can find further examples in other review articles, e.g.\ in Refs.~\citenum{botu2017machine,behler2017first,deringer2019machine}.

\begin{table*}
	\resizebox{0.9\textwidth}{!}{%
		\begin{tabular}{lllrl}
			\toprule
			Category                         & Reference                                                      & ML-FF  & Max. $N_\mathrm{atoms}$ & Reference theory \\
			
			\midrule
			\multirow{3}{*}{Electronic Effects}      & Sauceda \emph{et. al.}\cite{sauceda2019molecular}      & sGDML       & 21      & CCSD(T), CCSD      \\	
			& Sauceda \emph{et. al.}\cite{sauceda2020molecular}      & sGDML       & 21      & CCSD(T), CCSD      \\		
			& Sauceda \emph{et. al.}\cite{chmiela2018}               & sGDML       & 21      & PBE, CCSD(T), CCSD \\
			\midrule
			\multirow{8}{*}{Thermodynamics}          & Morawietz \emph{et. al.}\cite{morawietz2016van}        & BP-NNP      & 6\,912  & RPBE, BLYP \\
			& Andrade \emph{et. al.}\cite{andrade2020free}           & DeepMD      & 426     & SCAN \\                              
			& Deringer \emph{et. al.}\cite{deringer2017machine}      & GAP         & 1\,000  & LDA \\
			& Behler \emph{et. al.}\cite{behler2008pressure}         & BP-NNP      & 64      & LDA \\
			& Bartok \emph{et. al.}\cite{bartok2018machine}          & GAP         & 23\,496 & PW91 \\
			& Deringer \emph{et. al.}\cite{deringer2018realistic}    & GAP         & 4\,096  & PW91 \\                                
			& Bonati \emph{et. al.}\cite{bonati2018silicon}          & DeepMD      & 680     & SCAN \\
			& Brickel \emph{et. al.}\cite{brickel2019reactive}       & PhysNet     & 6       & MP2 \\
			\midrule
			\multirow{7}{*}{Reactions}               & Unke \emph{et. al.}\cite{unke2016collision}            & RKHS        & 3       & UCCSD(T) \\
			& Denis \emph{et. al.}\cite{denis2017quantum}            & RKHS        & 3       & UCCSD(T)-F12a \\
			& Lu \emph{et. al.}\cite{lu2020comprehensive}            & PIP-NN      & 7       & UCCSD(T)-F12a \\
			& Sweeny \emph{et. al.}\cite{sweeny2020thermal}          & PhysNet     & 7       & MP2 \\
			& K{\"a}ser \emph{et. al.}\cite{kaser2020isomerization}  & PhysNet     & 7       & MP2 \\
			& Rivero \emph{et. al.}\cite{rivero2019reactive}         & PhysNet     & 19      & M06-2X \\
			& Liu \emph{et. al.}\cite{liu2018constructing}           & BP-NNP      & 38      & RPBE \\
			\midrule
			\multirow{5}{*}{Nuclear Quantum Effects} & Chmiela \emph{et. al.}\cite{chmiela2017}               & GDML        & 21      & PBE \\
			& Chmiela \emph{et. al.}\cite{chmiela2018}               & sGDML       & 21      & CCSD, CCSD(T) \\
			& Sch{\"u}tt \emph{et. al.}\cite{schutt2018}             & SchNet      & 20      & PBE \\
			& Sauceda \emph{et. al.}\cite{sauceda2020}               & sGDML       & 21      & CCSD, CCSD(T) \\
			& Hellstr{\"o}m \emph{et. al.}\cite{hellstrom2018nuclear}& BP-NNP      & 1\,700  & RPBE \\
			\midrule
			\multirow{3}{*}{Excited States}          & Chen \emph{et. al.}\cite{chen2018deep}                 & HDNN        & 5       & CASSCF \\
			& Westermayr \emph{et. al.}\cite{westermayr2019machine}  & NN          & 6       & MR-CISD \\
			& Westermayr \emph{et. al.}\cite{westermayr2020combining}& SchNet      & 6       & MR-CISD, CASSCF \\
			\midrule   
			\multirow{4}{*}{Spectroscopy}            & Gastegger \emph{et. al.}\cite{gastegger2017machine}    & BP-NNP      & 209     & BLYP, BP86, B2PLYP \\
			& Yao \emph{et. al.}\cite{yao2018tensormol}              & BP-NNP      & 60      & $\omega$B97X-D \\
			& Raimbault \emph{et. al.}\cite{raimbault2019using}      & SOAP        & 80      & PBE \\
			& Sommers \emph{et. al.}\cite{sommers2020raman}          & DeepMD      & 512     & SCAN \\
			\bottomrule
		\end{tabular}
	}
	\caption{Overview of different topics and applications of ML-FFs discussed in this section. In all cases, the type of employed ML-FF is given along with the number of atoms of the largest system used to study the respective phenomenon. The basic level of reference theory (neglecting basis sets and dispersion corrections for clarity) are also reported.}
	\label{tab:overview_applications}
\end{table*}

\subsection{Electronic Effects}
\begin{figure}[t]
	\centering
	\includegraphics[width=\columnwidth]{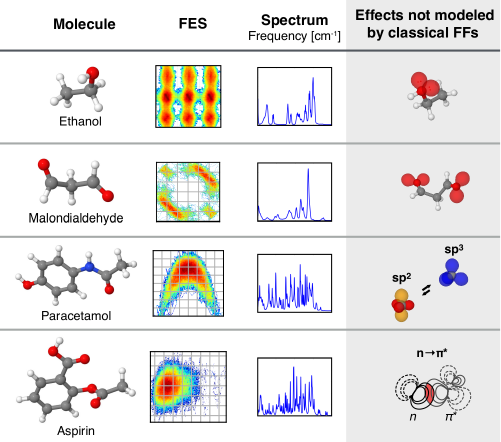}
	\caption{Visualization of electronic effects, which are accurately modeled by ML-FFs, but neglected by conventional FFs. Electron lone pairs, hybridization changes and orbital donation effects all influence the dynamics of molecules and hence the properties that are computed from MD simulations. When predicting, e.g.\ Gibbs/Helmholtz free energy surfaces (FESs) or molecular spectra, neglecting them will lead to qualitatively different results.}
	\label{fig:notmodeledbyffs}
\end{figure}

A good example for the power of ML-FFs is a recent study of the dynamics of small molecules (malondialdehyde, ethanol, salicylic acid, paracetamol, aspirin) with atomic forces at CCSD(T) quality.\cite{sauceda2019molecular} AIMD simulations were run at 500~K at the PBE+TS/DFT level of theory\cite{PBE1996,tkatchenko2009accurate} and the collected configurations randomly subsampled to calculate energies and forces at the CCSD(T) level of theory (reference data for aspirin was calculated at CCSD accuracy). For each molecule, an ML-FF was constructed from 1000 data points with the sGDML\cite{chmiela2019} method (see Section~\ref{subsubsec:symmetric_gradient_domain_machine_learning}) and used to run MD simulations at 300~K. Running simulations of this quality with \textit{ab initio} methods is impossible, as they would require up to a billion times more computation time. Conventional FFs were shown to be no viable alternative to ML-FFs, as they do not adequately describe, or even completely neglect, effects which strongly influence the dynamics -- and hence the properties -- of the studied molecules (Fig.~\ref{fig:notmodeledbyffs}).

For example, in ethanol, the lone pairs of the oxygen atom interact with the partially positively charged hydrogen atoms of the methyl group. Due to this attraction, the configuration where both lone pairs are adjacent to a hydrogen atom is visited most frequently during a dynamics simulation. Any derived property, e.g.\ the Gibbs/Helmholtz free energy surface (FES) or the infrared spectrum, is only accurate when this effect is properly described. Conventional FFs do not account for lone pairs and are thus unable to predict the molecular properties correctly. 

A similar effect can be observed in malondialdehyde. Here, the lone pairs of the two oxygen atoms strongly repel each other, which drives the dynamics away from configurations where they are close. While conventional FFs can crudely model electrostatic repulsion between the oxygen atoms with negative partial charges, the steric contributions from the overlap of the electron clouds is not described, causing a qualitatively different dynamics.

Paracetamol is another molecule where lone pairs influence the stability of specific configurations: The partially positively charged phenyl-hydrogen adjacent to the oxygen atom of the acetamide group interacts with its lone pairs and favors a specific dihedral angle. Additionally, the nitrogen atom of the acetamide group is $sp^2$ hybridized, which allows conjugation to the electrons in the phenyl system and leads to the planar geometry of paracetamol. When the nitrogen hybridization state is changed to $sp^3$, the energetically favorable interaction is broken and corresponding configurations are thus rarely visited during room temperature dynamics. However, at higher temperatures, the hybridization state may switch frequently -- conventional FFs are unable to describe this.

Another important electronic effect can be observed in aspirin. Here, an occupied (lone pair) $n$ orbital of the carbonyl group overlaps with an unoccupied antibonding $\pi^*$ orbital in the ester group. This $n\rightarrow\pi^*$ interaction dictates the relative arrangement of these functional groups in the global minimum structure.\cite{sauceda2020molecular} The effect is even amplified during dynamics, since thermal fluctuations enhance the overlap.\cite{chmiela2018} 

These and many other electronic effects, e.g.\  $n\rightarrow\sigma^*$ interactions, hyperconjugation, and Jahn-Teller distortions are captured automatically by ML-FFs. In contrast, including them in conventional FFs would require additional terms, whose functional form (and even which effects need to be modeled) are typically unknown \textit{a priori}.

\subsection{Thermodynamics}
A typical application of classical FFs is the study of thermodynamic properties of bulk systems, such as enthalpies, entropies, and phase diagrams. However, their limited accuracy is a major obstacle for quantitative predictions, as small inaccuracies in the interaction of a few particles will inevitably lead to big discrepancies when studying many particles. A good example for this are van der Waals (vdW) interactions. They are weak contributions to the total potential energy for small molecules in gas phase, but they add up in large condensed systems and bulk materials and can strongly influence their properties and dynamics.\cite{hermann2017first} While conventional FFs account for vdW interactions, they typically do so with a relatively crude model based on the Lennard-Jones potential\cite{jones1924determination}, which is insufficient for quantitative predictions in many cases. A prime example is water: It is the most studied liquid in literature and many different conventional FFs for water (some with additional special-purpose terms) have been proposed in the last decades, yet none of them is able to reproduce all experimentally measured properties of water in MD simulations\cite{guillot2002reappraisal}.

Here, ML-FFs offer a promising alternative. \citet{morawietz2016van} trained a descriptor-based NNP on periodic configurations of liquid and crystalline water, for which reference data was calculated with different DFT functionals.
MD simulations with the ML-FF revealed that the thermodynamic anomalies of water, such as its density maximum and negative volume of melting, are due to a delicate balance of weak vdW forces. The study was able to accurately predict experimentally measured radial distribution functions, as well as temperature dependent shear viscosities and diffusion coefficients. As ML-FFs are naturally able to describe bond breaking and formation, the study could even investigate proton transfer between different water molecules. 

The ability to analyze thermodynamic properties of reactive events is a major advantage of ML-FFs over conventional methods. For example, a recent study investigated the Gibbs free energy of proton transfer in liquid water at a titanium oxide surface.\cite{andrade2020free} A descriptor-based NNP was trained using reference data collected through an adaptive sampling approach and used to run MD simulations. The study revealed that a significant fraction of water molecules forms short-lived hydroxyl groups on the titanium oxide surface, which strongly influence its surface chemistry. Such insights are key to understanding phenomena such as surface functionalization and photocatalytic processes.

Another application where the flexibility of ML-FFs is a major advantage is the modeling of bulk materials. For example, Gaussian approximation potentials (GAPs, see Section~\ref{subsubsec:gaussian_approximation_potentials}) and NNPs have been constructed for elemental carbon\cite{deringer2017machine} and silicon\cite{behler2008pressure,bartok2018machine,deringer2018realistic,bonati2018silicon}. They allow to investigate a wide range of phenomena of liquid, crystalline, and amorphous solid phases, including defects and crack propagation. Modeling these effects accurately is only possible with ML-FFs or prohibitively expensive AIMD simulations. It is even possible to predict accurate phase diagrams of such systems with ML-FFs\cite{behler2008pressure,bartok2018machine}. Since this requires a correct model of bond formation and breaking, as well as changes of bonding patterns, such insights could not be obtained from conventional FFs. 

\subsection{Reactions}
One of the most significant advantages of ML-FFs over conventional FFs is their natural ability to model chemical reactions. Even in cases where it is possible to construct special purpose classical FFs that are able to describe reactions, they are typically much less accurate than their ML-FF counterparts. For example, a recent study compared an ML-FF constructed with a message-passing NNP with two classical methods to obtain a reactive FF for the Cl--CH$_3$--Br transformation.\cite{brickel2019reactive} Here, the ML-FF achieved up to three orders of magnitude lower errors and yielded qualitatively and quantitatively different predictions for the Helmholtz free energy surface along the reaction path.  
It is therefore no surprise that one of the first fields where ML-FFs were employed with great success are reaction dynamics. Here, the chemical transformations associated with molecular collisions over short time and length scales are studied. These simulations offer detailed atomistic insights into the reaction mechanism, providing access to rate constants and scattering cross sections, as well as insights on how the molecular energy is distributed between different modes, all of which can be directly related to experiments. In order to yield quantitative predictions, sufficient statistics and highly accurate PESs are required, making them an excellent application for ML-FFs. Studies typically involve small molecular systems, which are treated at high levels of accuracy, such as the collision of N$_2^+$ and Ar\cite{unke2016collision,denis2017quantum} or the Cl+CH$_3$OH~$\rightarrow$~HCl+CH$_3$O/CH$_2$OH reaction\cite{lu2020comprehensive}. Typical conventional FFs require fixed bonding patterns and are thus intrinsically unsuited for studying chemical reactions. While there also exist reactive variants of classical FFs, they do not reach the accuracy of ML-FFs. For example, a recent study investigated the thermal activation of methane by MgO$^+$ with a message-passing NNP (see Section~\ref{subsubsec:high_dimensional_neural_network_potentials}) and a reactive classical FF.\cite{sweeny2020thermal} Here, the ML-FF achieved prediction errors up to two orders of magnitude lower than the classical variant compared to \textit{ab initio} data. In addition, the disagreement between experimental rate constants and those predicted from MD simulations was lower by a factor of two with the ML-FF compared to the values obtained from the classical FF. The remaining discrepancy between prediction and experiment was further investigated and it was determined that the deviation was not due to inaccuracies of the ML-FF \textit{per se}, but instead could be traced back to the multireference character of the transition state, i.e.\ problems with the \textit{ab initio} reference data itself.

Even though it is possible to construct classical reactive FFs for specific reactions, there are cases where this is exceedingly difficult. A good example is a recent study where the photo-tautomerization reaction of acetaldehyde was investigated, which is speculated to be a major pathway for formic acid formation in the atmosphere\cite{kaser2020isomerization}. After being photo-excited, acetaldehyde contains enough energy that it may not only tautomerize to ethenol, but also dissociate into carbon monoxide and methane, or into hydrogen and ethenone. An accurate description of all three possible reaction pathways with the same FF is extremely difficult to achieve with conventional methods. The NNP used for the study on the other hand was trained on MP2/aug-cc-pVTZ\cite{moller1934note,dunning1989gaussian} reference data and allowed an unbiased description of all relevant processes at \textit{ab initio} quality. Analyzing a total of 12000 individual trajectories, the study concluded that the formation of ethenol from photo-tautomerization of acetaldehyde is unlikely under atomospheric conditions. This insight could not have been obtained by running AIMD simulations in a reasonable time frame: The combined simulation time of 1~\textmu s would amount to ten billion single point calculations (a time step of $\Delta t=0.1$~fs was used due to the large excitation energies). In contrast, less than 500k structures were used for training the ML-FF, i.e.\ the time spent for running \textit{ab initio} calculations was reduced by more than five orders of magnitude by employing an ML-FF.

Due to the efficiency of ML-FFs, scattering simulations can now even be extended to involve larger organic molecules. For example, a study of the minimum dynamic path\cite{unke2019sampling} of Diels-Alder reactions of 1,3-dibromo-1,3-butadiene and maleic anhydride with an end-to-end NNP has revealed that molecular rotations are a major driving force for the formation of products\cite{rivero2019reactive}, an effect which had not been described previously in the literature for this type of reaction. ML-FFs can even be applied to reactions between molecules and surfaces. For example, a study by \citet{liu2018constructing} investigated (reactive) HCl scattering on a gold surface using a descriptor-based NNP.

For a recent review on neural network-based PESs for small molecules and reactions, see Ref.~\citenum{manzhos2020neural}.

\subsection{Nuclear Quantum Effects}

\begin{figure*}
	\centering
	\includegraphics[width=\textwidth]{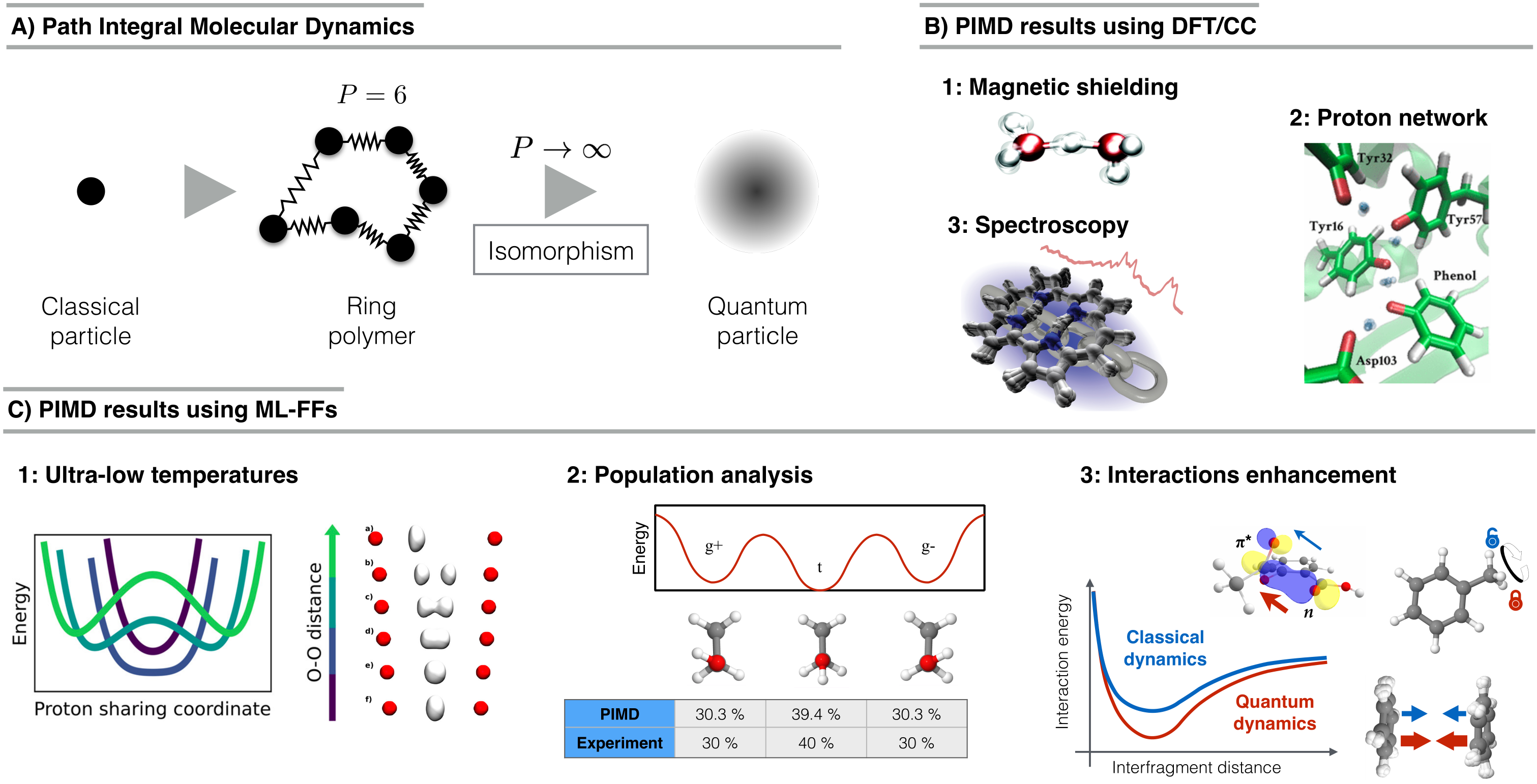}
	\caption{
		\textbf{A)} Schematic description of the path integral (ring polymer) molecular dynamics (PIMD) method, where quantum particles are approximated by a classical ring polymer with $P$ beads. There is an exact isomorphism between these two systems when $P\to\infty$, i.e.\ their statistical properties become equivalent. \textbf{B)} PIMD simulations using DFT or coupled cluster calculations. 1: Coupled cluster PIMD simulations of the Zundel model to compute $^1H$ magnetic shielding tensor (adapted with permission from Ref.~\citenum{Spura_CC-PIMD_PCCP2015}. Copyright \citeyear{Spura_CC-PIMD_PCCP2015} published by the PCCP Owner Societies under CC~BY-NC~3.0 \url{https://creativecommons.org/licenses/by-nc/3.0/}.). 2: Example of hydrogen-bond networks and their NQE implications on biological functions and enzyme catalysis (adapted with permission from Ref.~\citeyear{Wang_Ketosteroid-PIMD_JPCB2017}. Copyright \citeyear{Wang_Ketosteroid-PIMD_JPCB2017} American Chemical Society.). 3: IR spectrum of the porphycen molecule computed from PIMD simulations (adapted with permission from Ref.~\citenum{Litman_Porphycen-PIMD_JACS2019}. Copyright \citeyear{Litman_Porphycen-PIMD_JACS2019} American Chemical Society.). \textbf{C)} PIMD simulations using ML-FFs trained on DFT or coupled cluster data. 1: Ultra-low temperature dynamics of the Zundel model obtained from PIMD simulations (adapted with permission from Ref.~\citenum{Schran_Zundel-PIMD_JCTC2018}. Copyright \citeyear{Schran_Zundel-PIMD_JCTC2018} American Chemical Society.). 2: Comparison of the statistical sampling of different conformers of ethanol between experiment and simulations (adapted with permission from Ref.~\citenum{chmiela2018}. Copyright \citeyear{chmiela2018} \citeauthor{chmiela2018}). 3: Schematic description of the enhancement in intra- and inter-molecular interactions due to NQEs (adapted with permission from Ref.~\citenum{sauceda2020}. Copyright \citeyear{sauceda2020} \citeauthor{sauceda2020}).
	}
	\label{fig:pimd}
\end{figure*}

Predictive simulations of molecular systems and materials require not only highly accurate representations of the potential energy surface (PES), but also appropriate statistical sampling of the PES. While classical MD simulations are sufficient for this in some cases, the quantum nature of nuclei plays an important role in many systems. Nuclear quantum fluctuations are a fundamental phenomenon in nature resulting from Heisenberg's uncertainty principle\cite{heisenberg1985anschaulichen}, hence physical and chemical properties of molecular or biological systems, as well as nano- and bulk-materials, may be affected by them up to certain extent. In particular, light elements, such as protons and atoms in the first row of the periodic table, are prone to display nuclear quantum effects (NQEs) even at room temperature. Furthermore, materials or molecules formed by heavier atoms, but having strong bonds or being at low temperatures, exhibit significant NQEs\cite{Merchant1973,Kirchner2000,NanotubeCTE2004,ThermalExpSi2018,ThermalExpZnSb2015,Poltavsky-graphene,NQEreview2018,Freitas_disloc-PIMD_npjCM2018}.

Consequently, in order to generate predictive simulations of many physical properties, NQEs must be incorporated. A widely used methodology to perform quantum dynamics is path integral molecular dynamics (PIMD). This method is based on the isomorphism between a quantum particle and a classical harmonic ring polymer of $P$ beads (i.e.\ $P$ harmonically-coupled copies of the particle), where the equality holds for $P\to\infty$ (see Fig.~\ref{fig:pimd}A)\cite{chandler1981exploiting}. In practice, convergence of thermodynamical properties can be achieved using only a small number of beads. For light atoms at room temperature for example, $P\sim$16--32 is often sufficient to converge mechanical properties\cite{chmiela2019,Spura_CC-PIMD_PCCP2015,Litman_Porphycen-PIMD_JACS2019,Poltavsky-graphene}. This number can be reduced even further by using more sophisticated thermodynamic estimators\cite{pPI_2018}.

Given that PIMD simulations require energies and forces for $P$ copies of the system of interest, it is infeasible to use \textit{ab initio} methods to derive them in most cases. There are some exceptions: For example, PIMD simulations to study the IR spectrum of the porphycen molecule have been performed using DFT with the B3LYP functional and it was shown that the correct Helmholtz free energy and vibrational spectrum can only be recovered by considering NQEs (see Fig.~\ref{fig:pimd}B:3)\cite{Litman_Porphycen-PIMD_JACS2019}. Another example are PIMD simulations of the Zundel model at the CCSD level of theory to study the impact of NQEs on its structure and the $^1$H magnetic shielding tensor\cite{Spura_CC-PIMD_PCCP2015}. However, 
both of these studies required supercomputers to make the calculations possible in a reasonable time frame. On the other hand, ML-FFs can replicate the same results at a fraction of the computational cost, i.e.\ speed-ups by a factor of 10$^5$--10$^7$ (depending on the reference level of theory) can be achieved\cite{schutt2017schnet,chmiela2018}.
This gain in computational efficiency makes it possible to run PIMD simulations for a wide range of systems and offers the chance to reveal new chemical and physical insights.

For example, \citet{chmiela2017} performed room temperature PIMD simulations of aspirin using a GDML model (see Section~\ref{subsubsec:sgdml_package}) trained on PBE+TS\cite{PBE1996,tkatchenko2009accurate} reference data to investigate the paths followed between different minima on its PES. In a followup study, \citet{chmiela2018} compared free energies and vibrational density of states of a variety of medium-sized molecules obtained from PIMD simulations with a model trained on CCSD or CCSD(T) reference data to the same quantities obtained from a model trained on PBE+TS/DFT data. The authors found that even though the PESs at the two different levels of theory are very similar, tiny differences may still lead to largely different free energies. Additionally, it was shown that the experimentally determined populations for different conformations of ethanol can only be recovered from simulations when including NQEs (see Fig.~\ref{fig:pimd}C:2).

In another study, \citet{schutt2018} investigated the dynamics of C$_{20}$ fullerene using a NNP trained on PBE+TS/DFT reference data. Here, including NQEs broadens the radial distribution function significantly, which also increases the molecular polarizability\cite{sauceda2020}. A change in the distribution of interatomic distances also influences electronic effects: A recent study of Ref.~\citenum{sauceda2019molecular} (mentioned earlier in the paragraph on electronic effects) investigated NQEs in small organic molecules\cite{sauceda2020}. The study revealed that NQEs can dynamically strengthen molecular interactions by enhancing $n\rightarrow\pi^*$ donation through increasing orbital overlap, or by strengthening electrostatic interactions between neighboring charge densities (see Fig.~\ref{fig:pimd}C:3). 
Another interesting observed effect is a temporary change of bond orders, which can lead to emerging localized transient states of methyl rotors. The study also showed that vdW interactions are strengthened by NQEs: Since interatomic distances expand on average due to thermal and quantum dilations, the molecular polarizability is also increased (see Fig.~\ref{fig:pimd}C:3). Other observed implications of NQEs include ``bonding'' between hydroxyl groups and hindered rotor dynamics, which leads to molecular stiffening and smoother Helmholtz free energy surfaces. 

ML-FFs also make it possible to go far beyond the system size accessible with standard electronic structure methods. In Ref.~\citenum{hellstrom2018nuclear}, a descriptor-based NNP was used to study the influence of NQEs on aqueous NaOH solutions of different concentrations ($\sim$1000 atoms).  It could be shown that NQEs exert a subtle influence on the solvation structure in the Na$^+$ environment and significantly increase the proton transfer rates and hence diffusion coefficients of the different species. The accuracy of the ML-FF also made it possible to identify error cancellation effects in the reference method, leading to artificially good agreement with experiment in the absence of NQEs.

\subsection{Excited States}
The Born--Oppenheimer approximation breaks down when modeling the dynamics of molecular excited states, which are essential for understanding photochemical processes. An extension to classical MD, which allows for the simulation of such phenomena, is quantum-classical surface hopping MD.
In this approach, the excited state dynamics of a molecule are simulated by letting it evolve on a set of PESs associated with the different electronic states.
To describe the distribution of the molecule between the different states, the effective PES governing the time evolution changes according to stochastic criteria, e.g.\ based on coupling terms between the relevant states.
The correct quantum statistics are then recovered from multiple independent simulations.
These simulations are computationally intensive, as they do not only require the computation of multiple PES, but also different coupling terms.
This is further amplified by the need for a large number of trajectories in order to obtain reliable statistics. As such, quantum-classical surface hopping simulations can profit greatly from the efficiency and versatility of ML-FFs.

In Ref.~\citenum{chen2018deep}, for example, the authors used descriptor-based NNPs to study the excited state dynamics of the methylene imine molecule, as well as regions close to the conical intersection between the singlet ground and excited states. It could be shown that the NNPs are able to recover the effective PES with high accuracy and allow for efficient simulations to estimate the state populations of the system. Here, the coupling between the different surfaces was computed based on the Zhu–Nakamura approximation\cite{zhu1993two}, which relies on the energy differences between states. 
More accurate quantum mechanical descriptions of the inter-state couplings rely on so-called non-adiabatic coupling vectors (NACs), which introduce several additional challenges from an ML perspective. First, NACs exhibit the same rotational equivariance as molecular forces. Second, they grow rapidly for states lying close in energy. And finally, as a quantity computed between different states, they are determined only up to an arbitrary phase. The latter property in particular complicates the construction of ML models, as the random nature of the phase factor needs to be compensated during training. Early works relied on a costly preprocessing of the reference data\cite{westermayr2019machine}. Ref.~\citenum{westermayr2020combining}, however, demonstrated that the phase problem can be overcome by introducing phase-less loss functions during the training procedure. Using a modified end-to-end NNP to describe the excited state dynamics of the methylenimmonium cation, it could be shown that using such loss terms completely eliminates the need for a preprocessing step. In addition, the work modeled the NACs as derivatives of a proxy potential, thus accounting for their transformation under rotations of the molecule. The combination of these approaches not only made it possible to obtain accurate population statistics for the studied system, but could also greatly extend the time scales accessible by the simulation beyond the limits of conventional electronic structure approaches. 

For a recent review on machine learning for electronically excited sates, see Ref.~\citenum{westermayr2020machine2}.

\subsection{Spectroscopy}
As stated  at the beginning of this section, MD simulations are an excellent tool to model the temporal autocorrelation functions of various quantities, which can in turn be used to predict experimental observables, such as diffusion coefficients. These quantities need not be restricted to properties derived from the PES, but encompass other electronic properties such as dipole moments or polarizabilities. Access to the corresponding time autocorrelation functions enables the simulation of a wide range of molecular spectra, which can be directly related to experiment. The most prominent examples are infrared and Raman spectra derived from the autocorrelation functions of dipole moments and polarizabilities, respectively. Both types of vibrational spectra are of great practical interest, since they can be measured accurately via experiment and provide insights into the atomic structure of molecules and materials. However, these spectra can be subject to a series of complex quantum mechanical effects, such as vibrational anharmonicities. 
Hence, high level electronic structure treatments are required in order to obtain quantitatively accurate predictions of experimental results. Unfortunately, computing the required autocorrelation functions based purely on electronic structure calculations quickly grows prohibitively expensive, as simulations covering sufficient time scales are required in order to yield reliable spectra. In addition, if the influence of temperature or other phenomena should be studied in detail, a large number of such simulations is required. 
Recently, ML-FFs  have emerged as invaluable tools for obtaining reliable molecular spectra. A growing number of ML-FFs now provide access to quantities beyond the PES, e.g.\ dipole moments or polarizabilities. As such, they offer the possibility to perform these simulations in only a fraction of the time required by an \textit{ab initio} approach or even make them possible at all.

Ref.~\citenum{gastegger2017machine} demonstrates the potential inherent to ML-FFs based on the prediction of infrared spectra for organic molecules including the protonated alanine tripeptide. By combining a descriptor-based NNP model of the PES with a dipole moment model based on latent NN-predicted atomic charges, highly accurate infrared spectra could be obtained for all studied systems. The efficiency of such an approach was demonstrated based on an alkane containing more than 200 atoms, where it was possible to reduce a projected computation time of 9000 years with the original \textit{ab initio} method to only a few days (including the reference calculations needed for training the ML models). Moreover, the high accuracy of the predictions made it possible to identify shortcomings of the original reference methods and study how they influence the infrared spectrum of the tripepdtide. A similar latent charge based approach was employed in Ref.~\citenum{yao2018tensormol} to model infrared spectra of various amino acids. This study could not only obtain accurate spectra, but also demonstrated that the latent charges predicted by the dipole model constitute a valid ML driven scheme for deriving atomic partial charges, which can be used to model long-range electrostatic interactions explicitly. This scheme has since been employed in many physically augmented models (e.g.\ TensorMol\cite{yao2018tensormol} or PhysNet\cite{unke2019}).

In a similar manner, ML models capable of predicting polarizability tensors offer access to Raman spectra. Ref.~\citenum{raimbault2019using} introduces a symmetry adapted approach for modeling polarizability tensors using Gaussian process regression (GPR) based on the SOAP\cite{bartok2013representing} kernel. The authors use this model to study the Raman spectra of paracetamol in gas phase and various molecular crystals and achieve excellent agreement with electronic structure methods in both cases. 
Not only is the proposed approach highly data efficient, requiring only a small number ($<$1000) of reference data, but it could also be shown that the resulting model is transferable between different polymorphic forms of the crystal.
Ref.~\citenum{sommers2020raman} models Raman spectra of liquid water using descriptor-based NNPs to predict molecular polarizabilities. The computational efficiency of the approach made it possible to obtain Raman spectra for a system containing 416 water molecules based on two nanosecond trajectories at DFT level accuracy, a feat which would be infeasible with the original reference method. 
As a consequence, the influence of temperature effects on the Raman spectra of water and heavy water could be studied in detail. The atomic resolution of the employed ML approach made it possible to decompose the simulated spectra into intramolecular and intermolecular contributions, offering insights into the mechanisms governing the temperature dependence of the different spectral features.

\section{Challenges}
\label{sec:challenges}
Following the best practices outlined in the previous section, the current generation of ML-FFs is applicable to a wide range of problems in  chemistry that involve small- to medium-sized systems. While this space of chemical compounds is already significant in size, the ``dream scenario'' of chemists and biologists referenced in the introduction can only be realized with access to larger system sizes. Not only does the number of stable structures increase exponentially with added atomistic degrees of freedom,\cite{blum2009970,ruddigkeit2012enumeration} many interesting phenomena play out at nanoscale resolution, which is inaccessible to ML methods as of yet. This is because some steps involved in the construction of ML-FFs,  like sampling the reference data, which are solvable at small scale, become seemingly insurmountable obstacles at larger scales due to unfeasible computing times. The complexity of interactions, e.g.\ the non-classical behavior of nuclei, as well as significant contributions from large fluctuations, increase the space of conformations that need to be learned. To further complicate things, the cost of accurate \textit{ab initio} calculations increases steeply with expanding system size, limiting the amount of reference data that can be collected within a reasonable time frame. This also means that a growing number of atom correlations need to be represented by a model in order to capture the full scope of interactions present in the real system. Below, some considerations in reconciling the somewhat contradicting demands of scalability, transferability, data efficiency and accuracy in large-scale ML-FFs are outlined.

\subsection{Locality and smoothness assumptions}
\label{subsec:locality_and_smoothness_assumptions}
A fundamental challenge that must be faced by \textit{ab initio} methods, conventional FFs, and ML models alike is the many-body problem. Most properties of a physical system are determined by the interaction of many particles, whether those are electrons or, on a higher abstraction level, atoms. In fact, the reason that \textit{ab initio} calculations are expensive to obtain is due to the challenging computational scaling properties of high-dimensional many-body problems. As a result, the hierarchy of different levels of theory is directly defined by the level of correlation treatment in the respective wave function parametrization. Because the number of electronic degrees of freedom of a system is much higher than the number of atoms, the computational limitations of \textit{ab initio} methods become evident very quickly, even for small systems. Atomistic approximations scale more favorably, because they need to correlate less particles, but they are subject to the same scaling laws. The only escape is to neglect some correlations in favor of a reduced problem size. Unfortunately, it is to date impossible to reliably determine which interactions can be removed with minimal impact, without compromising the full many-body solution. Thus, the ideal of a local model is in conflict with the very nature of many-body systems. Although it is possible to recover some effects such as non-local charge transfer by means of a charge equilibration scheme,\cite{ghasemi2015interatomic,ko2020fourth} a general solution for this problem does not exist. While not fully justified from a physics perspective, assuming locality is still a useful inductive bias, which can help generalization and computational efficiency. It also helps when collecting reference data, as it implies that larger systems can be predicted using the information learned from smaller systems. Another assumption, which all ML-FFs discussed in this review make, is that the PES is smooth. This is a necessary requirement for most practical applications, since a non-smooth PES implies force discontinuities, which would lead to instabilities during MD simulations. Smoothness is also a requirement from the ML perspective, as only regular signals can be reconstructed from limited observations. 

For most commonly used NNPs and many kernel-based ML-FFs, locality is built into the design explicitly through the introduction of a cutoff radius. The global interactions between atoms are modeled by accumulating individual local atomic contributions. In this ``mean-field approximation'', the interaction of a particle with its surroundings is reduced to an effective one-body problem, i.e.\ an interaction of that particle with the average effect of its neighbors. As similar neighborhoods can be identified in different compounds across chemical space, these assumptions allow to build models from reference calculations of small molecules, which are transferable to much larger structures.\cite{huang2017dna,huang2020quantum} However, the lack of explicit higher-order terms comes at the cost of potentially loosing some important interaction effects, similar to the Hartree-Fock method and Kohn-Sham DFT in \textit{ab initio} calculations.

On the other hand, some models (e.g.\ (s)GDML) capture global correlations in the sense that a single prediction is obtained for the whole structure. Of course, this relies on reference calculations that are accurate enough to contain the relevant information. Global interactions of large systems can not be accurately inferred from a training set of small molecules or molecular fragments, which is why reference calculations for the exact target structure are necessary. It can therefore become difficult to collect enough reference data for large structures. In addition, even models that use no atom-wise decomposition might still implicitly assume that interactions are local to some degree due to their chemical descriptor. For example, in (s)GDML models, systems are encoded as a vector of inverse pair-wise distances. Therefore, structural changes between distant atoms contribute less strongly to changes in the overall descriptor than proximal atoms.
 
\begin{figure*}
\includegraphics[width=\textwidth]{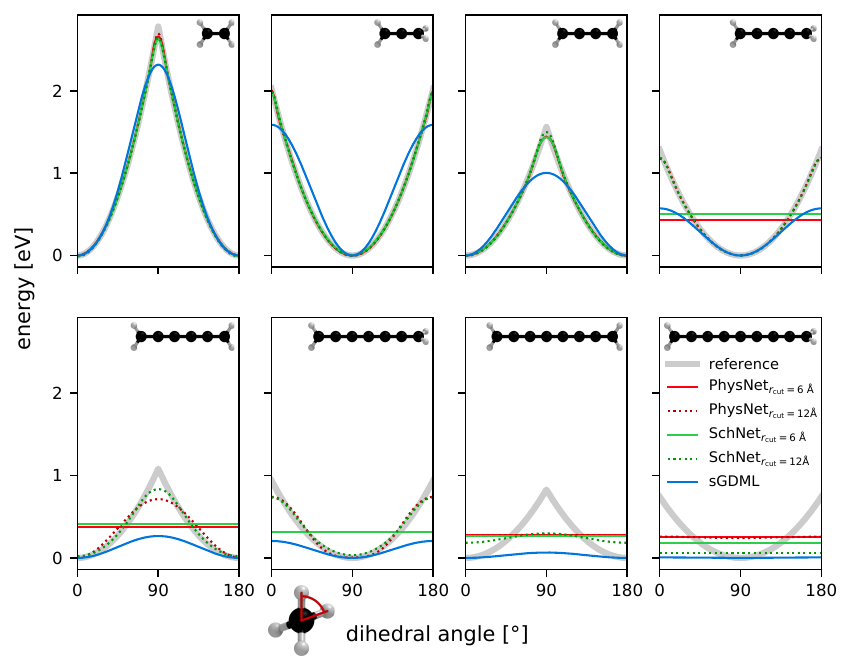}
\caption{Energy profiles of different ML-based PESs for a rotation of the dihedral angle between the terminal methylene groups of cumulenes (C$_{2+n}$H$_4$) of different sizes ($0\le n \le 7$). All reference calculations were performed with the semi-empirical MNDO method\cite{dewar1977ground} and models were trained on 4500 structures (with an additional 450 structures used for validation) collected from MD simulations at 1000~K. Because rotations of the dihedral angle are not sufficiently sampled at this temperature, the dihedral was rotated randomly before performing the reference calculations. Instead of a sharp cusp at the maximum of the rotation barrier, all models predict a smooth curve. Predictions become worse for increasing cumulene sizes with the cusp region being over-smoothed more strongly. For $n=7$, all models fail to predict the angular energy dependence. Note that NNP models (such as PhysNet and SchNet) may already fail for smaller cumulenes when the cutoff distance is chosen too small ($r_{\rm cut} = 6$~\AA), as they are unable to encode information about the dihedral angle in the environment-descriptor. However, it is possible to increase the cutoff ($r_{\rm cut} = 12$~\AA) to counter this effect.}
\label{fig:cumulene_scans} 
\end{figure*}

While locality and smoothness are valid assumptions for the majority of chemical systems, there are pathological cases where they break down and ML models that rely on them perform poorly. As an example, consider cumulenes -- hydrocarbons of the form C$_{2+n}$H$_4$ ($n \ge 0$) with $n+1$ cumulative double bonds. These molecules have a rigid linear geometry with the two terminal methylene groups forming an equilibrium dihedral angle of 0$^{\circ}$ (when $n$ is even) or 90$^{\circ}$ (when $n$ is odd). Rotating the dihedral angle out of its equilibrium position results in a sharp increase in potential energy even though the methylene groups may be separated by several angstroms when $n$ is large. This is due to the energetically favorable overlap of $\pi$-orbitals along the carbon chain (a highly non-local interaction), which is broken when the methylene groups are rotated against each other. Additionally, the potential energy exhibits a sharp ``cusp'' at the maximum energy (i.e.\ it is not smooth), because the ground state electronic configuration switches abruptly from one state to another (strictly speaking, multi-reference calculations would be necessary here). One-dimensional projections of the PESs predicted by ML-FFs along the rotation of the dihedral angle reveal several problems (Fig.~\ref{fig:cumulene_scans}). For example, all models predict smooth approximations by design, which is beneficial for running MD simulations, but results in large prediction errors around the cusp. Further, when the number of double bonds ($n+1$), i.e.\ the ``non-locality'' of relevant interactions, is increased, the quality of predictions decreases dramatically, until all models are unable to reproduce the energy profile. 

Note that by design, NNPs relying on message-passing are unable to resolve information about the dihedral angle if information between hydrogen atoms on opposite ends of the molecule cannot be exchanged directly (i.e.\ $r_{\rm cut}$ is too small) and predict constant energies in this case. The same is true for descriptor-based NNPs, as fingerprints of chemical environments also only consider atoms up to a cutoff (see Eqs.~\ref{eq:two_body_symmetry_function}~and~\ref{eq:three_body_symmetry_function}). Any kernel method taking as input local structural descriptors relying on cutoff radii (e.g.\  SOAP\cite{bartok2013representing} or  FCHL19\cite{christensen2020fchl}) will suffer from the same problems. Even when a ``global'' descriptor of chemical structure such as inverse pair-wise distances is chosen (e.g.\ Coulomb matrix\cite{rupp2012fast}), changes in the dihedral angle between distant groups of atoms are not resolved sufficiently for accurate predictions (see sGDML model in Fig.~\ref{fig:cumulene_scans}). The only way to fix this problem in general is to drop the locality assumption completely, for example by including all $\binom{N}{4}$ possible dihedral angles in the structural descriptor (without introducing additional factors that decrease the weight of these features with increasing distance between atoms). However, due to the combinatorial explosion of the number of possible dihedral angles, this would lead to extremely large descriptors whenever the number of atoms $N$ is not very small. The resulting models would be slow to evaluate and require a lot of reference data to give robust predictions (to prevent them from entering the extrapolation regime). An expert choice, i.e.\ including only a single relevant dihedral angle in the descriptor, is a possible way around this issue, but requires prior knowledge of the problem at hand and goes somewhat against ML philosophy. 

As a final remark, it should be mentioned that conventional FFs only include terms for dihedral angles between directly bonded atoms, so they are equally unable to predict the energy profiles of the larger cumulenes shown in Fig.~\ref{fig:cumulene_scans}. As such, relying on chemical locality is an assumption made by virtually all methods for approximating PESs and is not specific to just ML methods.

\subsection{Transferability, scalability and long-range interactions}
\label{subsec:transferability_scalability_and_long_range_interactions}
The concept of chemical locality discussed above also plays a central role in the transferabilty and scalability of ML models for atomistic systems. Transferability indicates how well models can adapt to compounds varying in their chemical composition, while scalability indicates how efficiently these models scale with respect to the size of systems modeled. Both concepts are closely related and inherently rooted in chemical locality. The assumption that interactions between atoms are local implies that similar structural motifs will give rise to comparable interactions and hence similar contributions to the properties of a molecule or material. In an ML context, chemical locality allows a model to reuse the information learned for different parts of a molecule for similar features in different systems. In this manner, a large atomistic system could in principle be assembled from smaller components like a jigsaw puzzle.\cite{huang2017dna} The former aspect is crucial in order to make models transferable, while the latter allows for the development of architectures whose evaluation cost scales linearly with system size.

ML-FFs exploiting chemical locality offer several advantages compared to other models. If trained properly, they can be applied to systems of different size and composition. The training procedure benefits in a similar manner, as local models can be trained on structures containing different numbers of atoms. Moreover, it is also possible to use only fragments of the original system during construction of a model. This property is very attractive in situations where accurate reference computations for the whole system are infeasible due to system size and/or scaling of the computational method. Local chemical environments are also less diverse than complete structures, potentially reducing the need for extensive sampling and decreasing the chances that models enter the extrapolation regime in a production setting.
In addition, local models scale linearly with system size, as interactions are limited to the cutoff radius and can be evaluated efficiently. In contrast, models without cutoffs are typically more limited in their practical applicability for extended systems. They always require reference computations to be performed for the whole system and, once trained, can only be reused for this particular molecule or material.

Despite these advantages, local ML models suffer from several inherent problems. In order to construct models which exploit locality, a chemical system needs to be partitioned in one way or another. This can for example be achieved by limiting interactions to terms involving only a certain number of atoms (similar to conventional FFs) or by restricting them to local atom-centered environments.
These approximations place strong limitations on which kind of interactions can be described. As a result, local ML models have difficulty when dealing with the situations where non-local effects are important, such as strongly conjugated systems and excited states (see Section~\ref{subsec:locality_and_smoothness_assumptions}). For standard simulations, the presence of long-range interactions, such as electrostatic and dispersion effects, are much more common phenomena. These are particularly important for modeling extended systems, where ML models are typically believed to offer a significant advantage over more conventional FFs. Since the structure and dynamical behavior of such systems is influenced greatly by long-range interactions, ML models need to be able to account for them in a satisfying manner.

Recovering long-range effects necessitates a balancing act between physical accuracy and computational efficiency, as the scalability of local models hinges on there being a limited number of interactions which need to be evaluated. This feat is further complicated by the typical energy scales of these interactions, which are small compared to local contributions such as bond energies. 
For these reasons, it is not advisable to account for long-range interactions by simply increasing the size of local environments. While local models with sufficiently large cutoffs are able to learn the relevant effects in principle, it may require a disproportionately large amount of data to reach an acceptable level of accuracy for an interaction with a comparably simple functional form. The reason is that average gradients and curvature in different regions of the PES may differ by several orders of magnitude, which makes it difficult to achieve uniformly low prediction errors across all regions. Hence, an optimal description would require to employ different characteristic scales. 

\begin{figure}[t]
\centering
\includegraphics[width=\columnwidth]{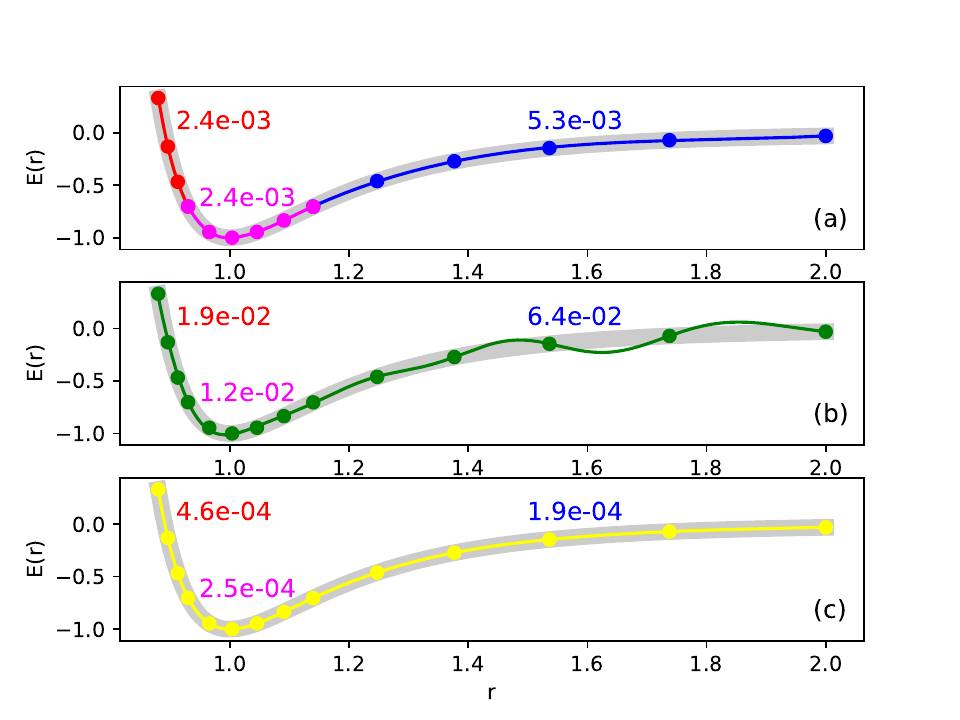}
\caption{Lennard-Jones potential (thick gray line) predicted by KRR with a Gaussian kernel. In (a), the potential energy is decomposed into short- (red), middle- (magenta) and long-range (blue) parts, which are learned by separate models (symbols show the training data and solid lines the model predictions). The mean squared prediction errors (in arbitrary units) for the respective regions are shown in the corresponding colors. In (b), the entire potential is learned by a single model using the same training points (green). All models in (a) and (b) use $r$ as the structural descriptor. Panel (c) shows a single model learning the potential, but using $r^{-1}$ as structural descriptor (yellow). The mean squared errors (a.u.) for different parts of the potential in (b) and (c) are reported independently to allow direct comparison with the values reported in (a).}
\label{fig:LJ}
\end{figure}

For illustration, consider the following toy examples: In the first variant, a Lennard-Jones (LJ) potential\cite{jones1924determination} is separated into a region around its minimum, a repulsive short-range, and an attractive long-range part. The task is to learn each of the three regions with a separate model (see Fig.~\ref{fig:LJ}a). In the second variant, a single model is trained on all regions at once (see Fig.~\ref{fig:LJ}b). Here, all models are kernel-based and use a Gaussian kernel (Eq.~\ref{eq:gaussian_kernel}). The kernel hyper-parameter $\gamma$ is optimized by a grid-search and cross-validation. Compared to the models trained on individual regions, the prediction errors of the model for all regions increase by around an order of magnitude. Further, it shows spurious oscillations between training points in the long-range region. When the optimal values of $\gamma$ for the different models are compared, the reason for failure when training on all regions at once becomes apparent: The optimal values of $\gamma$ are $198.88$, $75.47$, $0.08$ for the short-, middle-, and long-range models, respectively, which highlights the multi-scale nature of the PES. On the other hand, when training on all regions at once, the model necessarily has to compromise, which leads to an optimal value of $\gamma=22.12$. In this toy example, the multi-scale problem can be solved by switching from using $r$ as a structural descriptor to the more appropriate inverse distance $r^{-1}$ (Fig.~\ref{fig:LJ}c). Unfortunately, for realistic (high-dimensional) PESs with multiple minima, it can be difficult to find an appropriate descriptor to address the multi-scale nature of the PES, which leads to data-inefficient models. As a result, more training data is needed to reach an acceptable accuracy, which is problematic considering the computational cost of high-quality reference calculations.

One possibility of overcoming these limitations is by instead partitioning the energy into contributions modeled entirely via ML (short-range) and contributions described via explicit physical relations based on local quantities predicted via ML (long-range). A prime example for such an approach is the treatment of electrostatics, as was first introduced in Ref.~\citenum{morawietz2012neural}. Here, an ML model is used to predict partial charges for each atom based on their local environment. These charges can then be used in standard Coulomb and Ewald summation to compute the long-range electrostatic energy of a system. While such schemes initially relied on point charge reference data obtained from (arbitrary) partitioning methods of the \textit{ab initio} electron density (e.g.\ Hirshfeld charges\cite{hirshfeld1977bonded}), they have since been extended to operate on charges derived from an ML model for dipole moments (a true quantum mechanical observable).\cite{gastegger2017machine,yao2018tensormol,unke2019} Here, scalar partial charges $q_i$ are predicted for each atom $i$ and the molecular dipole moment is constructed as $\boldsymbol{\mu}=\sum_i q_i \mathbf{r}_i$, where $\mathbf{r}_i$ are the atomic positions (the predicted $q_i$ can be corrected to guarantee charge conservation\cite{unke2019}). The discrepancy between reference and predicted dipole moments is included in the loss function used for training the model (see Section~\ref{subsec:training_the_model}) and the partial charges consequently derived in a purely data-driven manner.

Contrary to electrostatics, accounting for dispersion interactions is not as straightforward, because the exact physical form of dispersion interactions is still debated and a variety of approximate schemes have been proposed.\cite{hermann2017first} In addition, dispersion corrections typically depend on coefficients computed from atomic polarizabilities as local properties. The corresponding quantum mechanical observable is the molecular polarizability tensor. In contrast to charges (scalars) derived from dipole moments (a vector quantity), predicting molecular polarizabilities requires rotationally equivariant ML models.\cite{wilkins2019accurate} Because of this, many ML approaches rely on the same empirical pair-wise dispersion potentials employed for correcting density functional theory computations.\cite{morawietz2013density,uteva2017interpolation,unke2019}

To summarize, local ML architectures are a promising approach towards transferable and scalable models, but they have a number of drawbacks which will still need to be addressed in the future. Promising alternative approaches to achieve transferability are ML models based directly on electronic structure methods, i.e.\ ``semi-empirical ML''\cite{li2018density,zubatyuk2019machine,stohr2020accurate} and models for electron density and Hamiltonians\cite{schutt2019unifying}.
These approaches express fundamental quantum chemical quantities in a local representation, e.g.\ Hamiltonian matrix elements in an atomic orbital basis. Non-locality can then be introduced via the ``correct'' mathematical mechanism, e.g.\ matrix diagonalization in the case of Hamiltonians. This physically motivated structure allows such models to recover a wide range of interactions while still being transferable. They are also better suited to predict intensive properties of molecules (whose magnitude is independent of system size), where assuming additive atomic contributions is not valid. A downside of such models compared to conventional ML-FFs is the increased computational cost due to the additional matrix operations. 

With respect to scalability, hybrid approaches similar to QM/MM\cite{senn2009qm} might constitute valid alternatives to pure ML models. Although several orders of magnitude more efficient than electronic structure theory, even local ML models encounter problems when faced with systems containing tens of thousands of atoms. Compared to conventional FFs, the more complex functional form underlying ML-FFs leads to an increased computational cost. In such cases, partitioning the system into regions treated at different levels of approximation can lead to a significant speedup. ML models can for example be embedded into regions modeled by classical force fields, yielding ML/MM like simulation protocols. Restricting elaborate ML approaches to only a subset of a chemical systems would make it possible to employ more accurate approximations in a manner analogous to conventional QM/MM. For example, in Ref.~\citenum{lahey2020simulating}, the authors study protein--ligand binding with a ML/MM approach: The ligand is described by an NN-based ML-FF and treated as if it was in gas phase. Coupling to the protein environment (described by a conventional FF) is achieved solely through non-bonded dispersion and electrostatic interactions. The disadvantage of such a simple embedding is that the ``quantum region'' cannot be polarized by the ``classical region''.  A more sophisticated embedding was recently proposed by \citet{gastegger2020machine}. Here, the region described by the ML-FF is explicitly polarized by the electric field induced by surrounding point charges, i.e.\ the electric field is an additional input to the model.
Alternative approaches, describe the effect of the classical environment by augmenting structural descriptors such as ACSFs by additional terms explicitly depending on the MM point charges.\cite{zhang2018solvation}  A similar approach is followed in Ref.~\citenum{boselt2020machine}, where the classical environment is described by auxiliary atom types.

\section{Concluding remarks}
\label{sec:concluding_remarks}
The last decades have witnessed significant advances in statistical learning that allowed ML techniques to enter our daily lives, industrial practice and scientific research. 

Classically, automation in industry and scientific fields relied on hand-crafted rules that represented human knowledge.\cite{nilsson1982principles} Not only is the creation of rule-based systems laborious and may require to cover an excessive number of cases, it often leads to rigid structures that are unable to adapt well to new situations. Even worse, some concepts are difficult or impossible to formalize, such as human perception for image classification. 

Modern statistical ML algorithms\cite{murphy2012machine,theodoridis2020machine} such as deep learning\cite{bishop1995neural,lecun2015deep, schmidhuber2015deep,goodfellow2016deep} or kernel-based learning\cite{cortes1995support,vapnik1995nature,scholkopf1998nonlinear,muller2001introduction,scholkopf2002learning} enable models that freely adapt to knowledge that is implicitly contained in datasets (in an abstract form) and thus offer a more robust way of solving problems than rule-based reasoning. 
For the field of molecular simulations, the potential of ML methods may help to bridge the accuracy-efficiency gap between first-principles electronic structure methods and conventional (rule-based) FFs. Bringing both fields together has raised many questions and still poses some fundamental challenges for new generations of ML-FFs. At this point in time, ML-FFs have already become a successful and practical tool in computational chemistry. 

Starting from a broad perspective, this review has focused on the role of ML for constructing force fields and assessed what can be achieved with these new techniques at the current stage of development. This has been contrasted with problems that are (so far) beyond the reach of present methods. Illustrative examples of the relevant chemistry and ML concepts have been discussed to demonstrate the practical usefulness that modern ML techniques can bring to chemistry and physics. This includes an overview of the most important considerations behind the construction of modern ML-FFs, such as the incorporation of physical invariances, choice of ML algorithms, and loss functions. Special attention has been given to the topic of validating ML-FFs, which requires particular care in scientific applications.\cite{lapuschkin2019unmasking} Furthermore, a comprehensive list of best practices, pitfalls, and challenges has been provided, which will serve as a useful guideline for practitioners standing on either side of this growing interdisciplinary field. These ``tricks of the trade''\cite{montavon2012neural} are often assumed to be obvious and thus omitted from publications -- here they have been deliberately spelled out to avoid unnecessary barriers to enter the field. Additionally, a small catalog of software tools that can enable and accelerate the implementation of ML-FFs has been provided as a pointer for readers wishing to adopt  ML methods in their own research.

While routinely performing computational studies of condensed phase systems (e.g.\ proteins in solution) at the highest levels of theory is still beyond reach, ML methods have already made other ``smaller dreams'' a reality: Just a decade ago, it would have been unthinkable to study the dynamics of molecules like aspirin at coupled cluster accuracy. Today, a couple hundred \textit{ab initio} reference calculations are enough to construct ML-FFs that reach this accuracy within a few tens of wavenumbers.\cite{sauceda2020construction}  In the past, even if suitable reference data was available, constructing accurate force fields was labor-intensive and required human effort and expertise. Nowadays, by virtue of automatic ML methods, the same task is as effortless as the push of a button. Thanks to the speed-ups offered by ML methods over conventional approaches, studies that previously required supercomputers to be feasible in a realistic time frame\cite{Spura_CC-PIMD_PCCP2015,Litman_Porphycen-PIMD_JACS2019} can now be performed on a laptop computer\cite{schutt2017schnet,chmiela2018}. 

In addition to enabling studies that were prohibitively expensive in the past, ML methods have also led to new chemical insights on systems that were thought to be already well understood. Even relatively small molecules were shown to display non-trivial electronic effects, influencing their dynamics and allowing a better understanding of experimental observations.\cite{sauceda2020} Many other unknown chemical effects potentially wait to be discovered by studies now possible with ML-FFs. At the speed at which improvements to existing ML-FFs are published, it is not unreasonable to expect significant advances that will make similar studies possible for larger systems and help realize many more ``dreams'' in the near future. 

Concluding, ML-FFs are a highly active line of research with many unexplored avenues and attractive applications in chemistry, with possibilities to contribute to a better understanding of fundamental quantum chemical properties and ample opportunity for novel theoretical, algorithmic and practical improvement. Given the success of this relatively young interdisciplinary field, it is to be expected that ML-FFs will become a fundamental part of modern computational chemistry. 

\begin{acknowledgement}
OTU acknowledges funding from the Swiss National Science
Foundation (Grant No. P2BSP2\_188147). AT was supported by the European Research Council (ERC-CoG ``BeStMo'').
KRM was supported in part  by the Institute of Information \& Communications Technology Planning \& Evaluation (IITP) grants funded by the Korea Government (No. 2017-0-00451, Development of BCI based Brain and Cognitive Computing Technology for Recognizing User’s Intentions using Deep Learning) and funded by the Korea Government (No. 2019-0-00079,  Artificial Intelligence Graduate School Program, Korea University), and was partly supported by the German Ministry for Education and Research (BMBF) under Grants 01IS14013A-E, 01GQ1115, 01GQ0850, 01IS18025A, 031L0207D and 01IS18037A; the German Research Foundation (DFG) under Grant Math+, EXC 2046/1, Project ID 390685689. We would like to thank Stefan Ganscha for his valuable input to the manuscript. Correspondence to AT and KRM. 
\end{acknowledgement}

\section*{Biographies}
\noindent
Oliver T.\ Unke is an SNSF postdoctoral research fellow in the Machine Learning Group at Technische Universit\"at Berlin. He received his Ph.D.\ in Chemistry from the University of Basel in 2019. His research has focused on developing methods for constructing accurate potential energy surfaces and their application in molecular dynamics simulations.\\

\noindent
Stefan Chmiela is a senior researcher at the Berlin Institute for the Foundations of Learning and Data (BIFOLD). He received his Ph.D.\ from Technische Universit\"at Berlin in 2019. His research interests include Hilbert space learning methods for applications in quantum chemistry, with particular focus on data efficiency and robustness.\\

\noindent 
Huziel E.\ Sauceda is a postdoctoral researcher in the Machine Learning Group at Technische Universit\"at Berlin and part of the BASLEARN project at the same institution. He obtained his bachelor degree in Physics at the Universidad Aut\'onoma de Sinaloa in Mexico, and his master and Ph.D.\ at the Institute of Physics of the Universidad Nacional Aut\'onoma de M\'exico (UNAM) in Mexico City. Between 2016 and 2019, he was a postdoctoral researcher at the Fritz Haber Institute of the Max Planck Society in Berlin. His research interests include \textit{ab initio} simulations, nuclear quantum effects, and thermodynamics of (nano)materials, as well as development and applications of machine learning methods to quantum chemistry and materials science.\\

\noindent
Michael Gastegger is a postdoctoral researcher in the BASLEARN project of the Machine Learning Group at Technische Universit\"at Berlin. He received his Ph.D.\ in Chemistry from the University of Vienna in Austria in 2017. His research interests include the development of machine learning methods for quantum chemistry and their application in simulations.\\

\noindent
Igor Poltavsky is a senior researcher at the University of Luxembourg. He received his Ph.D.\ from B.\ Verkin Institute for Low Temperature Physics \& Engineering in 2009. His research interests include statistical physics, imaginary-time path integral methods, nuclear quantum effects, \textit{ab initio} simulations, and machine learning.\\

\noindent
Kristof T.\ Sch\"utt is a senior researcher at the Berlin Institute for the Foundations of Learning and Data (BIFOLD). He received his master's degree in computer science in 2012 and his PhD at the machine learning group of Technische Universit\"at Berlin in 2018. Until September 2020, he worked at the Audatic company developing neural networks for real-time speech enhancement. His research interests include interpretable neural networks, representation learning, generative models, and machine learning applications in quantum chemistry.\\

\noindent
Alexandre Tkatchenko is a Professor of Theoretical Chemical Physics at the University of Luxembourg and Visiting Professor at Technische Universit\"at Berlin. He obtained his bachelor degree in Computer Science and a Ph.D.\ in Physical Chemistry at the Universidad Autonoma Metropolitana in Mexico City. Between 2008 and 2010, he was an Alexander von Humboldt Fellow at the Fritz Haber Institute of the Max Planck Society in Berlin. Between 2011 and 2016, he led an independent research group at the same institute. Tkatchenko serves on editorial boards of two society journals: Physical Review Letters (APS) and Science Advances (AAAS). He received a number of awards, including elected Fellow of the American Physical Society, the 2020 Dirac Medal from WATOC, the Gerhard Ertl Young Investigator Award of the German Physical Society, and two flagship grants from the European Research Council: a Starting Grant in 2011 and a Consolidator Grant in 2017. His group pushes the boundaries of quantum mechanics, statistical mechanics, and machine learning to develop efficient methods to enable accurate modeling and obtain new insights into complex materials.\\

\noindent
Klaus-Robert M\"uller 
has been a professor of computer science at Technische Universit{\"a}t Berlin since 2006; at the same time he is directing rsp.\ co-directing the Berlin Machine Learning Center and the Berlin Big Data Center. He studied physics in Karlsruhe from 1984 to 1989 and obtained his Ph.D.\ degree in computer science at Technische Universit{\"a}t Karlsruhe in 1992. After completing a postdoctoral position at GMD FIRST in Berlin, he was a research fellow at the University of Tokyo from 1994 to 1995. In 1995, he founded the Intelligent Data Analysis group at GMD-FIRST (later Fraunhofer FIRST) and directed it until 2008. From 1999 to 2006, he was a professor at the University of Potsdam. He was awarded the Olympus Prize for Pattern Recognition (1999), the SEL Alcatel Communication Award (2006), the Science Prize of Berlin by the Governing Mayor of Berlin (2014), the Vodafone Innovations Award (2017). In 2012, he was elected member of the German National Academy of Sciences-Leopoldina, in 2017 of the Berlin Brandenburg Academy of Sciences and also in 2017 external scientific member of the Max Planck Society. In 2019 and 2020 he became Highly Cited researcher in the cross-disciplinary area. His research interests are intelligent data analysis and Machine Learning in the sciences (Neuroscience (specifically Brain-Computer Interfaces), Physics, Chemistry) and in  industry.

\bibliography{references}

\providecommand{\latin}[1]{#1}
\makeatletter
\providecommand{\doi}
  {\begingroup\let\do\@makeother\dospecials
  \catcode`\{=1 \catcode`\}=2 \doi@aux}
\providecommand{\doi@aux}[1]{\endgroup\texttt{#1}}
\makeatother
\providecommand*\mcitethebibliography{\thebibliography}
\csname @ifundefined\endcsname{endmcitethebibliography}
  {\let\endmcitethebibliography\endthebibliography}{}
\begin{mcitethebibliography}{327}
\providecommand*\natexlab[1]{#1}
\providecommand*\mciteSetBstSublistMode[1]{}
\providecommand*\mciteSetBstMaxWidthForm[2]{}
\providecommand*\mciteBstWouldAddEndPuncttrue
  {\def\EndOfBibitem{\unskip.}}
\providecommand*\mciteBstWouldAddEndPunctfalse
  {\let\EndOfBibitem\relax}
\providecommand*\mciteSetBstMidEndSepPunct[3]{}
\providecommand*\mciteSetBstSublistLabelBeginEnd[3]{}
\providecommand*\EndOfBibitem{}
\mciteSetBstSublistMode{f}
\mciteSetBstMaxWidthForm{subitem}{(\alph{mcitesubitemcount})}
\mciteSetBstSublistLabelBeginEnd
  {\mcitemaxwidthsubitemform\space}
  {\relax}
  {\relax}

\bibitem[Feynman \latin{et~al.}(1963)Feynman, Leighton, and
  Sands]{feynman1963feynman}
Feynman,~R.~P.; Leighton,~R.~B.; Sands,~M. \emph{The Feynman Lectures On
  Physics. Vol. 1}; Addison-Wesley, 1963\relax
\mciteBstWouldAddEndPuncttrue
\mciteSetBstMidEndSepPunct{\mcitedefaultmidpunct}
{\mcitedefaultendpunct}{\mcitedefaultseppunct}\relax
\EndOfBibitem
\bibitem[McCammon \latin{et~al.}(1977)McCammon, Gelin, and
  Karplus]{mccammon1977dynamics}
McCammon,~J.~A.; Gelin,~B.~R.; Karplus,~M. Dynamics Of Folded Proteins.
  \emph{Nature} \textbf{1977}, \emph{267}, 585--590\relax
\mciteBstWouldAddEndPuncttrue
\mciteSetBstMidEndSepPunct{\mcitedefaultmidpunct}
{\mcitedefaultendpunct}{\mcitedefaultseppunct}\relax
\EndOfBibitem
\bibitem[Phillips(1981)]{phillips1981biomolecular}
Phillips,~D. \emph{Biomolecular Stereodynamics}; Adenine Press: Guilderland,
  NY, 1981\relax
\mciteBstWouldAddEndPuncttrue
\mciteSetBstMidEndSepPunct{\mcitedefaultmidpunct}
{\mcitedefaultendpunct}{\mcitedefaultseppunct}\relax
\EndOfBibitem
\bibitem[Schulz \latin{et~al.}(2009)Schulz, Lindner, Petridis, and
  Smith]{schulz2009scaling}
Schulz,~R.; Lindner,~B.; Petridis,~L.; Smith,~J.~C. Scaling Of
  Multimillion-atom Biological Molecular Dynamics Simulation On A Petascale
  Supercomputer. \emph{J. Chem. Theory Comput.} \textbf{2009}, \emph{5},
  2798--2808\relax
\mciteBstWouldAddEndPuncttrue
\mciteSetBstMidEndSepPunct{\mcitedefaultmidpunct}
{\mcitedefaultendpunct}{\mcitedefaultseppunct}\relax
\EndOfBibitem
\bibitem[Shaw \latin{et~al.}(2008)Shaw, Deneroff, Dror, Kuskin, Larson, Salmon,
  Young, Batson, Bowers, Chao, \latin{et~al.} others]{shaw2008anton}
Shaw,~D.~E.; Deneroff,~M.~M.; Dror,~R.~O.; Kuskin,~J.~S.; Larson,~R.~H.;
  Salmon,~J.~K.; Young,~C.; Batson,~B.; Bowers,~K.~J.; Chao,~J.~C.,
  \latin{et~al.}  Anton, A Special-purpose Machine For Molecular Dynamics
  Simulation. \emph{Commun. ACM} \textbf{2008}, \emph{51}, 91--97\relax
\mciteBstWouldAddEndPuncttrue
\mciteSetBstMidEndSepPunct{\mcitedefaultmidpunct}
{\mcitedefaultendpunct}{\mcitedefaultseppunct}\relax
\EndOfBibitem
\bibitem[Freddolino \latin{et~al.}(2006)Freddolino, Arkhipov, Larson,
  McPherson, and Schulten]{freddolino2006molecular}
Freddolino,~P.~L.; Arkhipov,~A.~S.; Larson,~S.~B.; McPherson,~A.; Schulten,~K.
  Molecular Dynamics Simulations Of The Complete Satellite Tobacco Mosaic
  Virus. \emph{Structure} \textbf{2006}, \emph{14}, 437--449\relax
\mciteBstWouldAddEndPuncttrue
\mciteSetBstMidEndSepPunct{\mcitedefaultmidpunct}
{\mcitedefaultendpunct}{\mcitedefaultseppunct}\relax
\EndOfBibitem
\bibitem[Zhao \latin{et~al.}(2013)Zhao, Perilla, Yufenyuy, Meng, Chen, Ning,
  Ahn, Gronenborn, Schulten, Aiken, \latin{et~al.} others]{zhao2013mature}
Zhao,~G.; Perilla,~J.~R.; Yufenyuy,~E.~L.; Meng,~X.; Chen,~B.; Ning,~J.;
  Ahn,~J.; Gronenborn,~A.~M.; Schulten,~K.; Aiken,~C., \latin{et~al.}  Mature
  {HIV-1} Capsid Structure By Cryo-electron Microscopy And All-atom Molecular
  Dynamics. \emph{Nature} \textbf{2013}, \emph{497}, 643--646\relax
\mciteBstWouldAddEndPuncttrue
\mciteSetBstMidEndSepPunct{\mcitedefaultmidpunct}
{\mcitedefaultendpunct}{\mcitedefaultseppunct}\relax
\EndOfBibitem
\bibitem[Zimmerman \latin{et~al.}(2020)Zimmerman, Porter, Ward, Singh, Vithani,
  Meller, Mallimadugula, Kuhn, Borowsky, Wiewiora, \latin{et~al.}
  others]{zimmerman2020citizen}
Zimmerman,~M.~I.; Porter,~J.~R.; Ward,~M.~D.; Singh,~S.; Vithani,~N.;
  Meller,~A.; Mallimadugula,~U.~L.; Kuhn,~C.~E.; Borowsky,~J.~H.;
  Wiewiora,~R.~P., \latin{et~al.}  Citizen Scientists Create An Exascale
  Computer To Combat COVID-19. \emph{BioRxiv} \textbf{2020}, \relax
\mciteBstWouldAddEndPunctfalse
\mciteSetBstMidEndSepPunct{\mcitedefaultmidpunct}
{}{\mcitedefaultseppunct}\relax
\EndOfBibitem
\bibitem[Karplus and McCammon(2002)Karplus, and McCammon]{karplus2002molecular}
Karplus,~M.; McCammon,~J.~A. Molecular Dynamics Simulations Of Biomolecules.
  \emph{Nat. Struct. Biol} \textbf{2002}, \emph{9}, 646--652\relax
\mciteBstWouldAddEndPuncttrue
\mciteSetBstMidEndSepPunct{\mcitedefaultmidpunct}
{\mcitedefaultendpunct}{\mcitedefaultseppunct}\relax
\EndOfBibitem
\bibitem[Dirac(1929)]{dirac1929quantum}
Dirac,~P. A.~M. Quantum Mechanics Of Many-electron Systems. \emph{Proc. R. Soc.
  Lond. A} \textbf{1929}, \emph{123}, 714--733\relax
\mciteBstWouldAddEndPuncttrue
\mciteSetBstMidEndSepPunct{\mcitedefaultmidpunct}
{\mcitedefaultendpunct}{\mcitedefaultseppunct}\relax
\EndOfBibitem
\bibitem[Gordon and Schmidt(2005)Gordon, and Schmidt]{gordon2005theory}
Gordon,~M.~S.; Schmidt,~M.~W. Theory And Applications Of Computational
  Chemistry: The First Forty Years. \emph{Dykstra, CE} \textbf{2005},
  1167--1189\relax
\mciteBstWouldAddEndPuncttrue
\mciteSetBstMidEndSepPunct{\mcitedefaultmidpunct}
{\mcitedefaultendpunct}{\mcitedefaultseppunct}\relax
\EndOfBibitem
\bibitem[Gonz{\'a}lez(2011)]{gonzalez2011force}
Gonz{\'a}lez,~M. Force Fields And Molecular Dynamics Simulations. {\'E}cole
  th{\'e}matique de la Soci{\'e}t{\'e} Fran{\c{c}}aise de la Neutronique. 2011;
  pp 169--200\relax
\mciteBstWouldAddEndPuncttrue
\mciteSetBstMidEndSepPunct{\mcitedefaultmidpunct}
{\mcitedefaultendpunct}{\mcitedefaultseppunct}\relax
\EndOfBibitem
\bibitem[Unke \latin{et~al.}(2020)Unke, Koner, Patra, K{\"a}ser, and
  Meuwly]{unke2020high}
Unke,~O.~T.; Koner,~D.; Patra,~S.; K{\"a}ser,~S.; Meuwly,~M. High-dimensional
  Potential Energy Surfaces For Molecular Simulations: From Empiricism To
  Machine Learning. \emph{Mach. Learn.: Sci. Technol.} \textbf{2020}, \emph{1},
  13001\relax
\mciteBstWouldAddEndPuncttrue
\mciteSetBstMidEndSepPunct{\mcitedefaultmidpunct}
{\mcitedefaultendpunct}{\mcitedefaultseppunct}\relax
\EndOfBibitem
\bibitem[Lennard-Jones(1924)]{jones1924determination}
Lennard-Jones,~J.~E. On The Determination Of Molecular Fields. -- {II}. From
  The Equation Of State Of A Gas. \emph{Proc. R. Soc. Lond. A} \textbf{1924},
  \emph{106}, 463--477\relax
\mciteBstWouldAddEndPuncttrue
\mciteSetBstMidEndSepPunct{\mcitedefaultmidpunct}
{\mcitedefaultendpunct}{\mcitedefaultseppunct}\relax
\EndOfBibitem
\bibitem[Vitalini \latin{et~al.}(2015)Vitalini, Mey, No{\'e}, and
  Keller]{vitalini2015dynamic}
Vitalini,~F.; Mey,~A.~S.; No{\'e},~F.; Keller,~B.~G. Dynamic Properties Of
  Force Fields. \emph{J. Chem. Phys.} \textbf{2015}, \emph{142},
  02B611\_1\relax
\mciteBstWouldAddEndPuncttrue
\mciteSetBstMidEndSepPunct{\mcitedefaultmidpunct}
{\mcitedefaultendpunct}{\mcitedefaultseppunct}\relax
\EndOfBibitem
\bibitem[Halgren and Damm(2001)Halgren, and Damm]{halgren2001polarizable}
Halgren,~T.~A.; Damm,~W. Polarizable Force Fields. \emph{Curr. Opin. Struct.
  Biol.} \textbf{2001}, \emph{11}, 236--242\relax
\mciteBstWouldAddEndPuncttrue
\mciteSetBstMidEndSepPunct{\mcitedefaultmidpunct}
{\mcitedefaultendpunct}{\mcitedefaultseppunct}\relax
\EndOfBibitem
\bibitem[Warshel \latin{et~al.}(2007)Warshel, Kato, and
  Pisliakov]{warshel2007polarizable}
Warshel,~A.; Kato,~M.; Pisliakov,~A.~V. Polarizable Force Fields: History, Test
  Cases, And Prospects. \emph{J. Chem. Theory Comput.} \textbf{2007}, \emph{3},
  2034--2045\relax
\mciteBstWouldAddEndPuncttrue
\mciteSetBstMidEndSepPunct{\mcitedefaultmidpunct}
{\mcitedefaultendpunct}{\mcitedefaultseppunct}\relax
\EndOfBibitem
\bibitem[Shi \latin{et~al.}(2013)Shi, Xia, Zhang, Best, Wu, Ponder, and
  Ren]{shi2013polarizable}
Shi,~Y.; Xia,~Z.; Zhang,~J.; Best,~R.; Wu,~C.; Ponder,~J.~W.; Ren,~P.
  Polarizable Atomic Multipole-based {AMOEBA} Force Field For Proteins.
  \emph{J. Chem. Theory Comput.} \textbf{2013}, \emph{9}, 4046--4063\relax
\mciteBstWouldAddEndPuncttrue
\mciteSetBstMidEndSepPunct{\mcitedefaultmidpunct}
{\mcitedefaultendpunct}{\mcitedefaultseppunct}\relax
\EndOfBibitem
\bibitem[Lopes \latin{et~al.}(2013)Lopes, Huang, Shim, Luo, Li, Roux, and
  MacKerell~Jr]{lopes2013polarizable}
Lopes,~P.~E.; Huang,~J.; Shim,~J.; Luo,~Y.; Li,~H.; Roux,~B.;
  MacKerell~Jr,~A.~D. Polarizable Force Field For Peptides And Proteins Based
  On The Classical Drude Oscillator. \emph{J. Chem. Theory Comput.}
  \textbf{2013}, \emph{9}, 5430--5449\relax
\mciteBstWouldAddEndPuncttrue
\mciteSetBstMidEndSepPunct{\mcitedefaultmidpunct}
{\mcitedefaultendpunct}{\mcitedefaultseppunct}\relax
\EndOfBibitem
\bibitem[Rasmussen \latin{et~al.}(2007)Rasmussen, Ren, Ponder, and
  Jensen]{rasmussen2007force}
Rasmussen,~T.~D.; Ren,~P.; Ponder,~J.~W.; Jensen,~F. Force Field Modeling Of
  Conformational Energies: Importance Of Multipole Moments And Intramolecular
  Polarization. \emph{Int. J. Quantum Chem.} \textbf{2007}, \emph{107},
  1390--1395\relax
\mciteBstWouldAddEndPuncttrue
\mciteSetBstMidEndSepPunct{\mcitedefaultmidpunct}
{\mcitedefaultendpunct}{\mcitedefaultseppunct}\relax
\EndOfBibitem
\bibitem[Unke \latin{et~al.}(2017)Unke, Devereux, and Meuwly]{unke2017minimal}
Unke,~O.~T.; Devereux,~M.; Meuwly,~M. Minimal Distributed Charges: Multipolar
  Quality At The Cost Of Point Charge Electrostatics. \emph{J. Chem. Phys.}
  \textbf{2017}, \emph{147}, 161712\relax
\mciteBstWouldAddEndPuncttrue
\mciteSetBstMidEndSepPunct{\mcitedefaultmidpunct}
{\mcitedefaultendpunct}{\mcitedefaultseppunct}\relax
\EndOfBibitem
\bibitem[Warshel and Weiss(1980)Warshel, and Weiss]{warshel1980empirical}
Warshel,~A.; Weiss,~R.~M. An Empirical Valence Bond Approach For Comparing
  Reactions In Solutions And In Enzymes. \emph{J. Am. Chem. Soc.}
  \textbf{1980}, \emph{102}, 6218--6226\relax
\mciteBstWouldAddEndPuncttrue
\mciteSetBstMidEndSepPunct{\mcitedefaultmidpunct}
{\mcitedefaultendpunct}{\mcitedefaultseppunct}\relax
\EndOfBibitem
\bibitem[Van~Duin \latin{et~al.}(2001)Van~Duin, Dasgupta, Lorant, and
  Goddard]{van2001reaxff}
Van~Duin,~A.~C.; Dasgupta,~S.; Lorant,~F.; Goddard,~W.~A. {ReaxFF}: A Reactive
  Force Field For Hydrocarbons. \emph{J. Phys. Chem. A} \textbf{2001},
  \emph{105}, 9396--9409\relax
\mciteBstWouldAddEndPuncttrue
\mciteSetBstMidEndSepPunct{\mcitedefaultmidpunct}
{\mcitedefaultendpunct}{\mcitedefaultseppunct}\relax
\EndOfBibitem
\bibitem[Nagy \latin{et~al.}(2014)Nagy, Yosa~Reyes, and
  Meuwly]{nagy2014multisurface}
Nagy,~T.; Yosa~Reyes,~J.; Meuwly,~M. Multisurface Adiabatic Reactive Molecular
  Dynamics. \emph{J. Chem. Theory Comput.} \textbf{2014}, \emph{10},
  1366--1375\relax
\mciteBstWouldAddEndPuncttrue
\mciteSetBstMidEndSepPunct{\mcitedefaultmidpunct}
{\mcitedefaultendpunct}{\mcitedefaultseppunct}\relax
\EndOfBibitem
\bibitem[Senn and Thiel(2009)Senn, and Thiel]{senn2009qm}
Senn,~H.~M.; Thiel,~W. {QM/MM} Methods For Biomolecular Systems. \emph{Angew.
  Chem. Int. Ed.} \textbf{2009}, \emph{48}, 1198--1229\relax
\mciteBstWouldAddEndPuncttrue
\mciteSetBstMidEndSepPunct{\mcitedefaultmidpunct}
{\mcitedefaultendpunct}{\mcitedefaultseppunct}\relax
\EndOfBibitem
\bibitem[Kulik \latin{et~al.}(2016)Kulik, Zhang, Klinman, and
  Martinez]{kulik2016large}
Kulik,~H.~J.; Zhang,~J.; Klinman,~J.~P.; Martinez,~T.~J. How Large Should The
  {QM} Region Be In {QM/MM} Calculations? The Case Of Catechol
  O-methyltransferase. \emph{J. Phys. Chem. B} \textbf{2016}, \emph{120},
  11381--11394\relax
\mciteBstWouldAddEndPuncttrue
\mciteSetBstMidEndSepPunct{\mcitedefaultmidpunct}
{\mcitedefaultendpunct}{\mcitedefaultseppunct}\relax
\EndOfBibitem
\bibitem[Sch{\"u}tt \latin{et~al.}(2020)Sch{\"u}tt, Chmiela, von Lilienfeld,
  Tkatchenko, Tsuda, and M{\"u}ller]{schutt2020machine}
Sch{\"u}tt,~K.~T.; Chmiela,~S.; von Lilienfeld,~O.~A.; Tkatchenko,~A.;
  Tsuda,~K.; M{\"u}ller,~K.-R. \emph{Machine Learning Meets Quantum Physics};
  Springer, 2020\relax
\mciteBstWouldAddEndPuncttrue
\mciteSetBstMidEndSepPunct{\mcitedefaultmidpunct}
{\mcitedefaultendpunct}{\mcitedefaultseppunct}\relax
\EndOfBibitem
\bibitem[Rupp \latin{et~al.}(2012)Rupp, Tkatchenko, M{\"u}ller, and
  Von~Lilienfeld]{rupp2012fast}
Rupp,~M.; Tkatchenko,~A.; M{\"u}ller,~K.-R.; Von~Lilienfeld,~O.~A. Fast and
  Accurate Modeling of Molecular Atomization Energies with Machine Learning.
  \emph{Phys. Rev. Lett.} \textbf{2012}, \emph{108}, 058301\relax
\mciteBstWouldAddEndPuncttrue
\mciteSetBstMidEndSepPunct{\mcitedefaultmidpunct}
{\mcitedefaultendpunct}{\mcitedefaultseppunct}\relax
\EndOfBibitem
\bibitem[Montavon \latin{et~al.}(2013)Montavon, Rupp, Gobre,
  Vazquez-Mayagoitia, Hansen, Tkatchenko, M{\"u}ller, and
  Von~Lilienfeld]{montavon2013machine}
Montavon,~G.; Rupp,~M.; Gobre,~V.; Vazquez-Mayagoitia,~A.; Hansen,~K.;
  Tkatchenko,~A.; M{\"u}ller,~K.-R.; Von~Lilienfeld,~O.~A. Machine Learning of
  Molecular Electronic Properties in Chemical Compound Space. \emph{New J.
  Phys.} \textbf{2013}, \emph{15}, 095003\relax
\mciteBstWouldAddEndPuncttrue
\mciteSetBstMidEndSepPunct{\mcitedefaultmidpunct}
{\mcitedefaultendpunct}{\mcitedefaultseppunct}\relax
\EndOfBibitem
\bibitem[Hansen \latin{et~al.}(2013)Hansen, Montavon, Biegler, Fazli, Rupp,
  Scheffler, Von~Lilienfeld, Tkatchenko, and M\"{u}ller]{hansen2013assessment}
Hansen,~K.; Montavon,~G.; Biegler,~F.; Fazli,~S.; Rupp,~M.; Scheffler,~M.;
  Von~Lilienfeld,~O.~A.; Tkatchenko,~A.; M\"{u}ller,~K.-R. Assessment and
  Validation of Machine Learning Methods for Predicting Molecular Atomization
  Energies. \emph{J. Chem. Theory Comput.} \textbf{2013}, \emph{9},
  3404--3419\relax
\mciteBstWouldAddEndPuncttrue
\mciteSetBstMidEndSepPunct{\mcitedefaultmidpunct}
{\mcitedefaultendpunct}{\mcitedefaultseppunct}\relax
\EndOfBibitem
\bibitem[Hansen \latin{et~al.}(2015)Hansen, Biegler, Ramakrishnan, Pronobis,
  Von~Lilienfeld, Mu\"ller, and Tkatchenko]{hansen2015machine}
Hansen,~K.; Biegler,~F.; Ramakrishnan,~R.; Pronobis,~W.; Von~Lilienfeld,~O.~A.;
  Mu\"ller,~K.-R.; Tkatchenko,~A. Machine Learning Predictions Of Molecular
  Properties: Accurate Many-body Potentials And Nonlocality In Chemical Space.
  \emph{J. Phys. Chem. Lett.} \textbf{2015}, \emph{6}, 2326--2331\relax
\mciteBstWouldAddEndPuncttrue
\mciteSetBstMidEndSepPunct{\mcitedefaultmidpunct}
{\mcitedefaultendpunct}{\mcitedefaultseppunct}\relax
\EndOfBibitem
\bibitem[von Lilienfeld \latin{et~al.}(2020)von Lilienfeld, M{\"u}ller, and
  Tkatchenko]{von2020exploring}
von Lilienfeld,~O.~A.; M{\"u}ller,~K.-R.; Tkatchenko,~A. Exploring Chemical
  Compound Space With Quantum-based Machine Learning. \emph{Nat. Rev. Chem.}
  \textbf{2020}, \emph{4}, 347–--358\relax
\mciteBstWouldAddEndPuncttrue
\mciteSetBstMidEndSepPunct{\mcitedefaultmidpunct}
{\mcitedefaultendpunct}{\mcitedefaultseppunct}\relax
\EndOfBibitem
\bibitem[No{\'e} \latin{et~al.}(2019)No{\'e}, Olsson, K{\"o}hler, and
  Wu]{noe2019boltzmann}
No{\'e},~F.; Olsson,~S.; K{\"o}hler,~J.; Wu,~H. Boltzmann Generators: Sampling
  Equilibrium States Of Many-body Systems With Deep Learning. \emph{Science}
  \textbf{2019}, \emph{365}, eaaw1147\relax
\mciteBstWouldAddEndPuncttrue
\mciteSetBstMidEndSepPunct{\mcitedefaultmidpunct}
{\mcitedefaultendpunct}{\mcitedefaultseppunct}\relax
\EndOfBibitem
\bibitem[K{\"o}hler \latin{et~al.}(2020)K{\"o}hler, Klein, and
  No{\'e}]{kohler2020equivariant}
K{\"o}hler,~J.; Klein,~L.; No{\'e},~F. Equivariant Flows: Exact Likelihood
  Generative Learning For Symmetric Densities. \emph{arXiv preprint
  arXiv:2006.02425} \textbf{2020}, \relax
\mciteBstWouldAddEndPunctfalse
\mciteSetBstMidEndSepPunct{\mcitedefaultmidpunct}
{}{\mcitedefaultseppunct}\relax
\EndOfBibitem
\bibitem[Zhang \latin{et~al.}(2019)Zhang, Yang, and No{\'e}]{zhang2019targeted}
Zhang,~J.; Yang,~Y.~I.; No{\'e},~F. Targeted Adversarial Learning Optimized
  Sampling. \emph{J. Phys. Chem. Lett.} \textbf{2019}, \emph{10},
  5791--5797\relax
\mciteBstWouldAddEndPuncttrue
\mciteSetBstMidEndSepPunct{\mcitedefaultmidpunct}
{\mcitedefaultendpunct}{\mcitedefaultseppunct}\relax
\EndOfBibitem
\bibitem[Koner \latin{et~al.}(2019)Koner, Unke, Boe, Bemish, and
  Meuwly]{koner2019exhaustive}
Koner,~D.; Unke,~O.~T.; Boe,~K.; Bemish,~R.~J.; Meuwly,~M. Exhaustive
  State-to-state Cross Sections For Reactive Molecular Collisions From
  Importance Sampling Simulation And A Neural Network Representation. \emph{J.
  Chem. Phys.} \textbf{2019}, \emph{150}, 211101\relax
\mciteBstWouldAddEndPuncttrue
\mciteSetBstMidEndSepPunct{\mcitedefaultmidpunct}
{\mcitedefaultendpunct}{\mcitedefaultseppunct}\relax
\EndOfBibitem
\bibitem[No{\'e} \latin{et~al.}(2020)No{\'e}, De~Fabritiis, and
  Clementi]{noe2020machine}
No{\'e},~F.; De~Fabritiis,~G.; Clementi,~C. Machine Learning For Protein
  Folding And Dynamics. \emph{Curr. Opin. Struc. Biol.} \textbf{2020},
  \emph{60}, 77--84\relax
\mciteBstWouldAddEndPuncttrue
\mciteSetBstMidEndSepPunct{\mcitedefaultmidpunct}
{\mcitedefaultendpunct}{\mcitedefaultseppunct}\relax
\EndOfBibitem
\bibitem[S{\o}nderby \latin{et~al.}(2015)S{\o}nderby, S{\o}nderby, Nielsen, and
  Winther]{sonderby2015convolutional}
S{\o}nderby,~S.~K.; S{\o}nderby,~C.~K.; Nielsen,~H.; Winther,~O. Convolutional
  {LSTM} Networks For Subcellular Localization Of Proteins. International
  Conference on Algorithms for Computational Biology. 2015; pp 68--80\relax
\mciteBstWouldAddEndPuncttrue
\mciteSetBstMidEndSepPunct{\mcitedefaultmidpunct}
{\mcitedefaultendpunct}{\mcitedefaultseppunct}\relax
\EndOfBibitem
\bibitem[Almagro~Armenteros \latin{et~al.}(2017)Almagro~Armenteros,
  S{\o}nderby, S{\o}nderby, Nielsen, and Winther]{almagro2017deeploc}
Almagro~Armenteros,~J.~J.; S{\o}nderby,~C.~K.; S{\o}nderby,~S.~K.; Nielsen,~H.;
  Winther,~O. {DeepLoc}: Prediction Of Protein Subcellular Localization Using
  Deep Learning. \emph{Bioinformatics} \textbf{2017}, \emph{33},
  3387--3395\relax
\mciteBstWouldAddEndPuncttrue
\mciteSetBstMidEndSepPunct{\mcitedefaultmidpunct}
{\mcitedefaultendpunct}{\mcitedefaultseppunct}\relax
\EndOfBibitem
\bibitem[Botlani \latin{et~al.}(2018)Botlani, Siddiqui, and
  Varma]{botlani2018machine}
Botlani,~M.; Siddiqui,~A.; Varma,~S. Machine Learning Approaches To Evaluate
  Correlation Patterns In Allosteric Signaling: A Case Study Of The {PDZ2}
  Domain. \emph{J. Chem. Phys.} \textbf{2018}, \emph{148}, 241726\relax
\mciteBstWouldAddEndPuncttrue
\mciteSetBstMidEndSepPunct{\mcitedefaultmidpunct}
{\mcitedefaultendpunct}{\mcitedefaultseppunct}\relax
\EndOfBibitem
\bibitem[Boninsegna \latin{et~al.}(2018)Boninsegna, N{\"u}ske, and
  Clementi]{boninsegna2018sparse}
Boninsegna,~L.; N{\"u}ske,~F.; Clementi,~C. Sparse Learning Of Stochastic
  Dynamical Equations. \emph{J. Chem. Phys.} \textbf{2018}, \emph{148},
  241723\relax
\mciteBstWouldAddEndPuncttrue
\mciteSetBstMidEndSepPunct{\mcitedefaultmidpunct}
{\mcitedefaultendpunct}{\mcitedefaultseppunct}\relax
\EndOfBibitem
\bibitem[Senior \latin{et~al.}(2020)Senior, Evans, Jumper, Kirkpatrick, Sifre,
  Green, Qin, {\v{Z}}{\'\i}dek, Nelson, Bridgland, \latin{et~al.}
  others]{senior2020improved}
Senior,~A.~W.; Evans,~R.; Jumper,~J.; Kirkpatrick,~J.; Sifre,~L.; Green,~T.;
  Qin,~C.; {\v{Z}}{\'\i}dek,~A.; Nelson,~A.~W.; Bridgland,~A., \latin{et~al.}
  Improved Protein Structure Prediction Using Potentials From Deep Learning.
  \emph{Nature} \textbf{2020}, \emph{577}, 706--710\relax
\mciteBstWouldAddEndPuncttrue
\mciteSetBstMidEndSepPunct{\mcitedefaultmidpunct}
{\mcitedefaultendpunct}{\mcitedefaultseppunct}\relax
\EndOfBibitem
\bibitem[Scherer \latin{et~al.}(2015)Scherer, Trendelkamp-Schroer, Paul,
  P{\'e}rez-Hern{\'a}ndez, Hoffmann, Plattner, Wehmeyer, Prinz, and
  No{\'e}]{scherer2015pyemma}
Scherer,~M.~K.; Trendelkamp-Schroer,~B.; Paul,~F.; P{\'e}rez-Hern{\'a}ndez,~G.;
  Hoffmann,~M.; Plattner,~N.; Wehmeyer,~C.; Prinz,~J.-H.; No{\'e},~F. {PyEMMA}
  2: A Software Package For Estimation, Validation, And Analysis Of Markov
  Models. \emph{J. Chem. Theory Comput.} \textbf{2015}, \emph{11},
  5525--5542\relax
\mciteBstWouldAddEndPuncttrue
\mciteSetBstMidEndSepPunct{\mcitedefaultmidpunct}
{\mcitedefaultendpunct}{\mcitedefaultseppunct}\relax
\EndOfBibitem
\bibitem[Mardt \latin{et~al.}(2018)Mardt, Pasquali, Wu, and
  No{\'e}]{mardt2018vampnets}
Mardt,~A.; Pasquali,~L.; Wu,~H.; No{\'e},~F. {VAMPnets} For Deep Learning Of
  Molecular Kinetics. \emph{Nat. Comm.} \textbf{2018}, \emph{9}, 1--11\relax
\mciteBstWouldAddEndPuncttrue
\mciteSetBstMidEndSepPunct{\mcitedefaultmidpunct}
{\mcitedefaultendpunct}{\mcitedefaultseppunct}\relax
\EndOfBibitem
\bibitem[Wehmeyer and No{\'e}(2018)Wehmeyer, and No{\'e}]{wehmeyer2018time}
Wehmeyer,~C.; No{\'e},~F. Time-lagged Autoencoders: Deep Learning Of Slow
  Collective Variables For Molecular Kinetics. \emph{J. Chem. Phys.}
  \textbf{2018}, \emph{148}, 241703\relax
\mciteBstWouldAddEndPuncttrue
\mciteSetBstMidEndSepPunct{\mcitedefaultmidpunct}
{\mcitedefaultendpunct}{\mcitedefaultseppunct}\relax
\EndOfBibitem
\bibitem[Wu \latin{et~al.}(2018)Wu, Mardt, Pasquali, and Noe]{wu2018deep}
Wu,~H.; Mardt,~A.; Pasquali,~L.; Noe,~F. deep Generative Markov State Models.
  Adv. Neural. Inf. Process. Syst. 2018; pp 3975--3984\relax
\mciteBstWouldAddEndPuncttrue
\mciteSetBstMidEndSepPunct{\mcitedefaultmidpunct}
{\mcitedefaultendpunct}{\mcitedefaultseppunct}\relax
\EndOfBibitem
\bibitem[Chen \latin{et~al.}(2019)Chen, Sidky, and Ferguson]{chen2019nonlinear}
Chen,~W.; Sidky,~H.; Ferguson,~A.~L. Nonlinear Discovery Of Slow Molecular
  Modes Using State-free Reversible {VAMPnets}. \emph{J. Chem. Phys.}
  \textbf{2019}, \emph{150}, 214114\relax
\mciteBstWouldAddEndPuncttrue
\mciteSetBstMidEndSepPunct{\mcitedefaultmidpunct}
{\mcitedefaultendpunct}{\mcitedefaultseppunct}\relax
\EndOfBibitem
\bibitem[Klus \latin{et~al.}(2019)Klus, Husic, Mollenhauer, and
  No{\'e}]{klus2019kernel}
Klus,~S.; Husic,~B.~E.; Mollenhauer,~M.; No{\'e},~F. Kernel Methods For
  Detecting Coherent Structures In Dynamical Data. \emph{Chaos} \textbf{2019},
  \emph{29}, 123112\relax
\mciteBstWouldAddEndPuncttrue
\mciteSetBstMidEndSepPunct{\mcitedefaultmidpunct}
{\mcitedefaultendpunct}{\mcitedefaultseppunct}\relax
\EndOfBibitem
\bibitem[Sidky \latin{et~al.}(2019)Sidky, Chen, and Ferguson]{sidky2019high}
Sidky,~H.; Chen,~W.; Ferguson,~A.~L. High-resolution Markov State Models For
  The Dynamics Of Trp-cage Miniprotein Constructed Over Slow Folding Modes
  Identified By State-free Reversible {VAMPnets}. \emph{J. Phys. Chem. B}
  \textbf{2019}, \emph{123}, 7999--8009\relax
\mciteBstWouldAddEndPuncttrue
\mciteSetBstMidEndSepPunct{\mcitedefaultmidpunct}
{\mcitedefaultendpunct}{\mcitedefaultseppunct}\relax
\EndOfBibitem
\bibitem[Chen \latin{et~al.}(2018)Chen, Chen, Pinamonti, and
  Clementi]{chen2018learning}
Chen,~J.; Chen,~J.; Pinamonti,~G.; Clementi,~C. Learning Effective Molecular
  Models From Experimental Observables. \emph{J. Chem. Theory Comput.}
  \textbf{2018}, \emph{14}, 3849--3858\relax
\mciteBstWouldAddEndPuncttrue
\mciteSetBstMidEndSepPunct{\mcitedefaultmidpunct}
{\mcitedefaultendpunct}{\mcitedefaultseppunct}\relax
\EndOfBibitem
\bibitem[Wang \latin{et~al.}(2019)Wang, Olsson, Wehmeyer, P{\'e}rez, Charron,
  De~Fabritiis, No{\'e}, and Clementi]{wang2019machine}
Wang,~J.; Olsson,~S.; Wehmeyer,~C.; P{\'e}rez,~A.; Charron,~N.~E.;
  De~Fabritiis,~G.; No{\'e},~F.; Clementi,~C. Machine Learning Of
  Coarse-grained Molecular Dynamics Force Fields. \emph{ACS Cent. Sci.}
  \textbf{2019}, \emph{5}, 755--767\relax
\mciteBstWouldAddEndPuncttrue
\mciteSetBstMidEndSepPunct{\mcitedefaultmidpunct}
{\mcitedefaultendpunct}{\mcitedefaultseppunct}\relax
\EndOfBibitem
\bibitem[N{\"u}ske \latin{et~al.}(2019)N{\"u}ske, Boninsegna, and
  Clementi]{nuske2019coarse}
N{\"u}ske,~F.; Boninsegna,~L.; Clementi,~C. Coarse-graining Molecular Systems
  By Spectral Matching. \emph{J. Chem. Phys.} \textbf{2019}, \emph{151},
  044116\relax
\mciteBstWouldAddEndPuncttrue
\mciteSetBstMidEndSepPunct{\mcitedefaultmidpunct}
{\mcitedefaultendpunct}{\mcitedefaultseppunct}\relax
\EndOfBibitem
\bibitem[Wang \latin{et~al.}(2020)Wang, Chmiela, M{\"u}ller, No{\'e}, and
  Clementi]{wang2020ensemble}
Wang,~J.; Chmiela,~S.; M{\"u}ller,~K.-R.; No{\'e},~F.; Clementi,~C. Ensemble
  Learning Of Coarse-grained Molecular Dynamics Force Fields With A Kernel
  Approach. \emph{J. Chem. Phys.} \textbf{2020}, \emph{152}, 194106\relax
\mciteBstWouldAddEndPuncttrue
\mciteSetBstMidEndSepPunct{\mcitedefaultmidpunct}
{\mcitedefaultendpunct}{\mcitedefaultseppunct}\relax
\EndOfBibitem
\bibitem[No{\'e} \latin{et~al.}(2020)No{\'e}, Tkatchenko, M{\"u}ller, and
  Clementi]{noe2020machinemolsim}
No{\'e},~F.; Tkatchenko,~A.; M{\"u}ller,~K.-R.; Clementi,~C. Machine Learning
  For Molecular Simulation. \emph{Annu. Rev. Phys. Chem.} \textbf{2020},
  \emph{71}, 361--390\relax
\mciteBstWouldAddEndPuncttrue
\mciteSetBstMidEndSepPunct{\mcitedefaultmidpunct}
{\mcitedefaultendpunct}{\mcitedefaultseppunct}\relax
\EndOfBibitem
\bibitem[Sch{\"u}tt \latin{et~al.}(2019)Sch{\"u}tt, Gastegger, Tkatchenko,
  M{\"u}ller, and Maurer]{schutt2019unifying}
Sch{\"u}tt,~K.; Gastegger,~M.; Tkatchenko,~A.; M{\"u}ller,~K.-R.; Maurer,~R.~J.
  Unifying Machine Learning And Quantum Chemistry With A Deep Neural Network
  For Molecular Wavefunctions. \emph{Nat. Comm.} \textbf{2019}, \emph{10},
  1--10\relax
\mciteBstWouldAddEndPuncttrue
\mciteSetBstMidEndSepPunct{\mcitedefaultmidpunct}
{\mcitedefaultendpunct}{\mcitedefaultseppunct}\relax
\EndOfBibitem
\bibitem[Gastegger \latin{et~al.}(2020)Gastegger, McSloy, Luya, Sch{\"u}tt, and
  Maurer]{gastegger2020deep}
Gastegger,~M.; McSloy,~A.; Luya,~M.; Sch{\"u}tt,~K.; Maurer,~R. A Deep Neural
  Network For Molecular Wave Functions In Quasi-atomic Minimal Basis
  Representation. \emph{J. Chem. Phys.} \textbf{2020}, \emph{153}, 044123\relax
\mciteBstWouldAddEndPuncttrue
\mciteSetBstMidEndSepPunct{\mcitedefaultmidpunct}
{\mcitedefaultendpunct}{\mcitedefaultseppunct}\relax
\EndOfBibitem
\bibitem[St{\"o}hr \latin{et~al.}(2020)St{\"o}hr, Sandonas, and
  Tkatchenko]{stohr2020accurate}
St{\"o}hr,~M.; Sandonas,~L.~M.; Tkatchenko,~A. Accurate Many-body Repulsive
  Potentials For Density-functional Tight-binding From Deep Tensor Neural
  Networks. \emph{J. Phys. Chem. Lett.} \textbf{2020}, \emph{11},
  6835--6843\relax
\mciteBstWouldAddEndPuncttrue
\mciteSetBstMidEndSepPunct{\mcitedefaultmidpunct}
{\mcitedefaultendpunct}{\mcitedefaultseppunct}\relax
\EndOfBibitem
\bibitem[Snyder \latin{et~al.}(2012)Snyder, Rupp, Hansen, M{\"u}ller, and
  Burke]{snyder2012finding}
Snyder,~J.~C.; Rupp,~M.; Hansen,~K.; M{\"u}ller,~K.-R.; Burke,~K. Finding
  Density Functionals With Machine Learning. \emph{Phys. Rev. Lett.}
  \textbf{2012}, \emph{108}, 253002\relax
\mciteBstWouldAddEndPuncttrue
\mciteSetBstMidEndSepPunct{\mcitedefaultmidpunct}
{\mcitedefaultendpunct}{\mcitedefaultseppunct}\relax
\EndOfBibitem
\bibitem[Brockherde \latin{et~al.}(2017)Brockherde, Vogt, Li, Tuckerman, Burke,
  and M{\"u}ller]{brockherde2017bypassing}
Brockherde,~F.; Vogt,~L.; Li,~L.; Tuckerman,~M.~E.; Burke,~K.;
  M{\"u}ller,~K.-R. Bypassing The Kohn-sham Equations With Machine Learning.
  \emph{Nat. Comm.} \textbf{2017}, \emph{8}, 1--10\relax
\mciteBstWouldAddEndPuncttrue
\mciteSetBstMidEndSepPunct{\mcitedefaultmidpunct}
{\mcitedefaultendpunct}{\mcitedefaultseppunct}\relax
\EndOfBibitem
\bibitem[Bogojeski \latin{et~al.}(2020)Bogojeski, Vogt-Maranto, Tuckerman,
  M{\"u}ller, and Burke]{bogojeski2019density}
Bogojeski,~M.; Vogt-Maranto,~L.; Tuckerman,~M.~E.; M{\"u}ller,~K.-R.; Burke,~K.
  Quantum Chemical Accuracy From Density Functional Approximations Via Machine
  Learning. \emph{Nat. Comm.} \textbf{2020}, \emph{11}, 1--11\relax
\mciteBstWouldAddEndPuncttrue
\mciteSetBstMidEndSepPunct{\mcitedefaultmidpunct}
{\mcitedefaultendpunct}{\mcitedefaultseppunct}\relax
\EndOfBibitem
\bibitem[Hermann \latin{et~al.}(2019)Hermann, Sch{\"a}tzle, and
  No{\'e}]{hermann2019deep}
Hermann,~J.; Sch{\"a}tzle,~Z.; No{\'e},~F. Deep Neural Network Solution Of The
  Electronic Schr\"odinger Equation. \emph{arXiv preprint arXiv:1909.08423}
  \textbf{2019}, \relax
\mciteBstWouldAddEndPunctfalse
\mciteSetBstMidEndSepPunct{\mcitedefaultmidpunct}
{}{\mcitedefaultseppunct}\relax
\EndOfBibitem
\bibitem[Carleo and Troyer(2017)Carleo, and Troyer]{carleo2017solving}
Carleo,~G.; Troyer,~M. Solving The Quantum Many-body Problem With Artificial
  Neural Networks. \emph{Science} \textbf{2017}, \emph{355}, 602--606\relax
\mciteBstWouldAddEndPuncttrue
\mciteSetBstMidEndSepPunct{\mcitedefaultmidpunct}
{\mcitedefaultendpunct}{\mcitedefaultseppunct}\relax
\EndOfBibitem
\bibitem[Gebauer \latin{et~al.}(2018)Gebauer, Gastegger, and
  Sch{\"u}tt]{gebauer2018generating}
Gebauer,~N.~W.; Gastegger,~M.; Sch{\"u}tt,~K.~T. Generating Equilibrium
  Molecules With Deep Neural Networks. NeurIPS 2018 Workshop on Machine
  Learning for Molecules and Materials. 2018\relax
\mciteBstWouldAddEndPuncttrue
\mciteSetBstMidEndSepPunct{\mcitedefaultmidpunct}
{\mcitedefaultendpunct}{\mcitedefaultseppunct}\relax
\EndOfBibitem
\bibitem[Gebauer \latin{et~al.}(2019)Gebauer, Gastegger, and
  Sch{\"u}tt]{gebauer2019symmetry}
Gebauer,~N.; Gastegger,~M.; Sch{\"u}tt,~K. Symmetry-adapted Generation Of 3d
  Point Sets For The Targeted Discovery Of Molecules. Adv. Neural. Inf.
  Process. Syst. 2019; pp 7566--7578\relax
\mciteBstWouldAddEndPuncttrue
\mciteSetBstMidEndSepPunct{\mcitedefaultmidpunct}
{\mcitedefaultendpunct}{\mcitedefaultseppunct}\relax
\EndOfBibitem
\bibitem[Hoffmann and No{\'e}(2019)Hoffmann, and
  No{\'e}]{hoffmann2019generating}
Hoffmann,~M.; No{\'e},~F. Generating Valid Euclidean Distance Matrices.
  \emph{arXiv preprint arXiv:1910.03131} \textbf{2019}, \relax
\mciteBstWouldAddEndPunctfalse
\mciteSetBstMidEndSepPunct{\mcitedefaultmidpunct}
{}{\mcitedefaultseppunct}\relax
\EndOfBibitem
\bibitem[Winter \latin{et~al.}(2019)Winter, Montanari, Steffen, Briem, No{\'e},
  and Clevert]{winter2019efficient}
Winter,~R.; Montanari,~F.; Steffen,~A.; Briem,~H.; No{\'e},~F.; Clevert,~D.-A.
  Efficient Multi-objective Molecular Optimization In A Continuous Latent
  Space. \emph{Chem. Sci.} \textbf{2019}, \emph{10}, 8016--8024\relax
\mciteBstWouldAddEndPuncttrue
\mciteSetBstMidEndSepPunct{\mcitedefaultmidpunct}
{\mcitedefaultendpunct}{\mcitedefaultseppunct}\relax
\EndOfBibitem
\bibitem[Simm \latin{et~al.}(2020)Simm, Pinsler, and
  Hern{\'a}ndez-Lobato]{simm2020reinforcement}
Simm,~G.~N.; Pinsler,~R.; Hern{\'a}ndez-Lobato,~J.~M. Reinforcement Learning
  For Molecular Design Guided By Quantum Mechanics. \emph{arXiv preprint
  arXiv:2002.07717} \textbf{2020}, \relax
\mciteBstWouldAddEndPunctfalse
\mciteSetBstMidEndSepPunct{\mcitedefaultmidpunct}
{}{\mcitedefaultseppunct}\relax
\EndOfBibitem
\bibitem[Strieth-Kalthoff \latin{et~al.}(2020)Strieth-Kalthoff, Sandfort,
  Segler, and Glorius]{strieth2020machine}
Strieth-Kalthoff,~F.; Sandfort,~F.; Segler,~M.~H.; Glorius,~F. Machine Learning
  The Ropes: Principles, Applications And Directions In Synthetic Chemistry.
  \emph{Chem. Soc. Rev.} \textbf{2020}, \emph{49}, 6154--6168\relax
\mciteBstWouldAddEndPuncttrue
\mciteSetBstMidEndSepPunct{\mcitedefaultmidpunct}
{\mcitedefaultendpunct}{\mcitedefaultseppunct}\relax
\EndOfBibitem
\bibitem[Chmiela \latin{et~al.}(2018)Chmiela, Sauceda, M{\"u}ller, and
  Tkatchenko]{chmiela2018}
Chmiela,~S.; Sauceda,~H.~E.; M{\"u}ller,~K.-R.; Tkatchenko,~A. Towards Exact
  Molecular Dynamics Simulations With Machine-learned Force Fields. \emph{Nat.
  Commun.} \textbf{2018}, \emph{9}, 3887\relax
\mciteBstWouldAddEndPuncttrue
\mciteSetBstMidEndSepPunct{\mcitedefaultmidpunct}
{\mcitedefaultendpunct}{\mcitedefaultseppunct}\relax
\EndOfBibitem
\bibitem[Sauceda \latin{et~al.}(2019)Sauceda, Chmiela, Poltavsky, M{\"u}ller,
  and Tkatchenko]{sauceda2019molecular}
Sauceda,~H.~E.; Chmiela,~S.; Poltavsky,~I.; M{\"u}ller,~K.-R.; Tkatchenko,~A.
  Molecular Force Fields With Gradient-domain Machine Learning: Construction
  And Application To Dynamics Of Small Molecules With Coupled Cluster Forces.
  \emph{J. Chem. Phys.} \textbf{2019}, \emph{150}, 114102\relax
\mciteBstWouldAddEndPuncttrue
\mciteSetBstMidEndSepPunct{\mcitedefaultmidpunct}
{\mcitedefaultendpunct}{\mcitedefaultseppunct}\relax
\EndOfBibitem
\bibitem[Gastegger \latin{et~al.}(2017)Gastegger, Behler, and
  Marquetand]{gastegger2017machine}
Gastegger,~M.; Behler,~J.; Marquetand,~P. Machine Learning Molecular Dynamics
  For The Simulation Of Infrared Spectra. \emph{Chem. Sci.} \textbf{2017},
  \emph{8}, 6924--6935\relax
\mciteBstWouldAddEndPuncttrue
\mciteSetBstMidEndSepPunct{\mcitedefaultmidpunct}
{\mcitedefaultendpunct}{\mcitedefaultseppunct}\relax
\EndOfBibitem
\bibitem[Christensen \latin{et~al.}(2019)Christensen, Faber, and von
  Lilienfeld]{christensen2019operators}
Christensen,~A.~S.; Faber,~F.~A.; von Lilienfeld,~O.~A. Operators In Quantum
  Machine Learning: Response Properties In Chemical Space. \emph{J. Chem.
  Phys.} \textbf{2019}, \emph{150}, 064105\relax
\mciteBstWouldAddEndPuncttrue
\mciteSetBstMidEndSepPunct{\mcitedefaultmidpunct}
{\mcitedefaultendpunct}{\mcitedefaultseppunct}\relax
\EndOfBibitem
\bibitem[K{\"a}ser \latin{et~al.}(2020)K{\"a}ser, Unke, and
  Meuwly]{kaser2020reactive}
K{\"a}ser,~S.; Unke,~O.; Meuwly,~M. Reactive Dynamics And Spectroscopy Of
  Hydrogen Transfer From Neural Network-based Reactive Potential Energy
  Surfaces. \emph{New J. Phys.} \textbf{2020}, \emph{22}, 55002\relax
\mciteBstWouldAddEndPuncttrue
\mciteSetBstMidEndSepPunct{\mcitedefaultmidpunct}
{\mcitedefaultendpunct}{\mcitedefaultseppunct}\relax
\EndOfBibitem
\bibitem[Agnihotri(2014)]{agnihotri2014computational}
Agnihotri,~N. Computational Studies Of Charge Transfer In Organic Solar
  Photovoltaic Cells: A Review. \emph{J. Photochem. Photobiol. C}
  \textbf{2014}, \emph{18}, 18--31\relax
\mciteBstWouldAddEndPuncttrue
\mciteSetBstMidEndSepPunct{\mcitedefaultmidpunct}
{\mcitedefaultendpunct}{\mcitedefaultseppunct}\relax
\EndOfBibitem
\bibitem[Payne \latin{et~al.}(1986)Payne, Joannopoulos, Allan, Teter, and
  Vanderbilt]{payne1986molecular}
Payne,~M.; Joannopoulos,~J.; Allan,~D.; Teter,~M.; Vanderbilt,~D.~H. Molecular
  Dynamics And Ab Initio Total Energy Calculations. \emph{Phys. Rev. Lett.}
  \textbf{1986}, \emph{56}, 2656\relax
\mciteBstWouldAddEndPuncttrue
\mciteSetBstMidEndSepPunct{\mcitedefaultmidpunct}
{\mcitedefaultendpunct}{\mcitedefaultseppunct}\relax
\EndOfBibitem
\bibitem[Adcock and McCammon(2006)Adcock, and McCammon]{adcock2006molecular}
Adcock,~S.~A.; McCammon,~J.~A. Molecular Dynamics: Survey Of Methods For
  Simulating The Activity Of Proteins. \emph{Chem. Rev.} \textbf{2006},
  \emph{106}, 1589--1615\relax
\mciteBstWouldAddEndPuncttrue
\mciteSetBstMidEndSepPunct{\mcitedefaultmidpunct}
{\mcitedefaultendpunct}{\mcitedefaultseppunct}\relax
\EndOfBibitem
\bibitem[Paquet and Viktor(2015)Paquet, and Viktor]{paquet2015molecular}
Paquet,~E.; Viktor,~H.~L. Molecular Dynamics, Monte Carlo Simulations, And
  Langevin Dynamics: A Computational Review. \emph{Biomed Res. Int.}
  \textbf{2015}, \emph{2015}\relax
\mciteBstWouldAddEndPuncttrue
\mciteSetBstMidEndSepPunct{\mcitedefaultmidpunct}
{\mcitedefaultendpunct}{\mcitedefaultseppunct}\relax
\EndOfBibitem
\bibitem[Schr{\"o}dinger(1926)]{schrodinger1926undulatory}
Schr{\"o}dinger,~E. An Undulatory Theory Of The Mechanics Of Atoms And
  Molecules. \emph{Phys. Rev.} \textbf{1926}, \emph{28}, 1049--1070\relax
\mciteBstWouldAddEndPuncttrue
\mciteSetBstMidEndSepPunct{\mcitedefaultmidpunct}
{\mcitedefaultendpunct}{\mcitedefaultseppunct}\relax
\EndOfBibitem
\bibitem[Born and Oppenheimer(1927)Born, and
  Oppenheimer]{born1927quantentheorie}
Born,~M.; Oppenheimer,~R. Zur Quantentheorie der Molekeln. \emph{Ann. Phys.}
  \textbf{1927}, \emph{389}, 457--484\relax
\mciteBstWouldAddEndPuncttrue
\mciteSetBstMidEndSepPunct{\mcitedefaultmidpunct}
{\mcitedefaultendpunct}{\mcitedefaultseppunct}\relax
\EndOfBibitem
\bibitem[Meng \latin{et~al.}(2011)Meng, Zhang, Mezei, and
  Cui]{meng2011molecular}
Meng,~X.-Y.; Zhang,~H.-X.; Mezei,~M.; Cui,~M. Molecular Docking: A Powerful
  Approach For Structure-based Drug Discovery. \emph{Curr. Comput. Aided Drug
  Des.} \textbf{2011}, \emph{7}, 146--157\relax
\mciteBstWouldAddEndPuncttrue
\mciteSetBstMidEndSepPunct{\mcitedefaultmidpunct}
{\mcitedefaultendpunct}{\mcitedefaultseppunct}\relax
\EndOfBibitem
\bibitem[Tuckerman \latin{et~al.}(1993)Tuckerman, Berne, Martyna, and
  Klein]{tuckerman1993efficient}
Tuckerman,~M.~E.; Berne,~B.~J.; Martyna,~G.~J.; Klein,~M.~L. Efficient
  Molecular Dynamics And Hybrid Monte Carlo Algorithms For Path Integrals.
  \emph{J. Chem. Phys.} \textbf{1993}, \emph{99}, 2796--2808\relax
\mciteBstWouldAddEndPuncttrue
\mciteSetBstMidEndSepPunct{\mcitedefaultmidpunct}
{\mcitedefaultendpunct}{\mcitedefaultseppunct}\relax
\EndOfBibitem
\bibitem[Chandler and Wolynes(1981)Chandler, and
  Wolynes]{chandler1981exploiting}
Chandler,~D.; Wolynes,~P.~G. Exploiting The Isomorphism Between Quantum Theory
  And Classical Statistical Mechanics Of Polyatomic Fluids. \emph{J. Chem.
  Phys.} \textbf{1981}, \emph{74}, 4078--4095\relax
\mciteBstWouldAddEndPuncttrue
\mciteSetBstMidEndSepPunct{\mcitedefaultmidpunct}
{\mcitedefaultendpunct}{\mcitedefaultseppunct}\relax
\EndOfBibitem
\bibitem[Habershon \latin{et~al.}(2013)Habershon, Manolopoulos, Markland, and
  Miller~III]{habershon2013ring}
Habershon,~S.; Manolopoulos,~D.~E.; Markland,~T.~E.; Miller~III,~T.~F.
  Ring-polymer Molecular Dynamics: Quantum Effects In Chemical Dynamics From
  Classical Trajectories In An Extended Phase Space. \emph{Annu. Rev. Phys.
  Chem.} \textbf{2013}, \emph{64}, 387--413\relax
\mciteBstWouldAddEndPuncttrue
\mciteSetBstMidEndSepPunct{\mcitedefaultmidpunct}
{\mcitedefaultendpunct}{\mcitedefaultseppunct}\relax
\EndOfBibitem
\bibitem[Poltavsky and Tkatchenko(2016)Poltavsky, and
  Tkatchenko]{poltavsky2016modeling}
Poltavsky,~I.; Tkatchenko,~A. Modeling Quantum Nuclei With Perturbed Path
  Integral Molecular Dynamics. \emph{Chem. Sci.} \textbf{2016}, \emph{7},
  1368--1372\relax
\mciteBstWouldAddEndPuncttrue
\mciteSetBstMidEndSepPunct{\mcitedefaultmidpunct}
{\mcitedefaultendpunct}{\mcitedefaultseppunct}\relax
\EndOfBibitem
\bibitem[Verlet(1967)]{verlet1967computer}
Verlet,~L. Computer ``Experiments'' On Classical Fluids. I. Thermodynamical
  Properties Of Lennard-jones Molecules. \emph{Phys. Rev.} \textbf{1967},
  \emph{159}, 98\relax
\mciteBstWouldAddEndPuncttrue
\mciteSetBstMidEndSepPunct{\mcitedefaultmidpunct}
{\mcitedefaultendpunct}{\mcitedefaultseppunct}\relax
\EndOfBibitem
\bibitem[Noether(1918)]{noether1918invarianten}
Noether,~E. Invarianten Beliebiger Differentialausdr{\"u}cke. \emph{G{\"o}tt.
  Nachr., mathematisch-physikalische Klasse} \textbf{1918}, \emph{1918},
  37--44\relax
\mciteBstWouldAddEndPuncttrue
\mciteSetBstMidEndSepPunct{\mcitedefaultmidpunct}
{\mcitedefaultendpunct}{\mcitedefaultseppunct}\relax
\EndOfBibitem
\bibitem[Shannon(1949)]{shannon1949communication}
Shannon,~C.~E. Communication In The Presence Of Noise. \emph{Proceedings of the
  IRE} \textbf{1949}, \emph{37}, 10--21\relax
\mciteBstWouldAddEndPuncttrue
\mciteSetBstMidEndSepPunct{\mcitedefaultmidpunct}
{\mcitedefaultendpunct}{\mcitedefaultseppunct}\relax
\EndOfBibitem
\bibitem[Schaffer(2015)]{schaffer2015not}
Schaffer,~J. What Not To Multiply Without Necessity. \emph{Australas. J.
  Philos.} \textbf{2015}, \emph{93}, 644--664\relax
\mciteBstWouldAddEndPuncttrue
\mciteSetBstMidEndSepPunct{\mcitedefaultmidpunct}
{\mcitedefaultendpunct}{\mcitedefaultseppunct}\relax
\EndOfBibitem
\bibitem[Geman \latin{et~al.}(1992)Geman, Bienenstock, and
  Doursat]{geman1992neural}
Geman,~S.; Bienenstock,~E.; Doursat,~R. Neural Networks And The Bias/Variance
  Dilemma. \emph{Neural Comput.} \textbf{1992}, \emph{4}, 1--58\relax
\mciteBstWouldAddEndPuncttrue
\mciteSetBstMidEndSepPunct{\mcitedefaultmidpunct}
{\mcitedefaultendpunct}{\mcitedefaultseppunct}\relax
\EndOfBibitem
\bibitem[M{\"u}ller \latin{et~al.}(2001)M{\"u}ller, Mika, R{\"a}tsch, Tsuda,
  and Sch{\"o}lkopf]{muller2001introduction}
M{\"u}ller,~K.-R.; Mika,~S.; R{\"a}tsch,~G.; Tsuda,~K.; Sch{\"o}lkopf,~B. An
  Introduction To Kernel-based Learning Algorithms. \emph{IEEE Trans. Neural
  Netw.} \textbf{2001}, \emph{12}, 181--201\relax
\mciteBstWouldAddEndPuncttrue
\mciteSetBstMidEndSepPunct{\mcitedefaultmidpunct}
{\mcitedefaultendpunct}{\mcitedefaultseppunct}\relax
\EndOfBibitem
\bibitem[Sch{\"o}lkopf \latin{et~al.}(1997)Sch{\"o}lkopf, Smola, and
  M{\"u}ller]{scholkopf1997kernel}
Sch{\"o}lkopf,~B.; Smola,~A.; M{\"u}ller,~K.-R. Kernel Principal Component
  Analysis. International Conference on Artificial Neural Networks. 1997; pp
  583--588\relax
\mciteBstWouldAddEndPuncttrue
\mciteSetBstMidEndSepPunct{\mcitedefaultmidpunct}
{\mcitedefaultendpunct}{\mcitedefaultseppunct}\relax
\EndOfBibitem
\bibitem[Sch{\"o}lkopf \latin{et~al.}(1998)Sch{\"o}lkopf, Smola, and
  M{\"u}ller]{scholkopf1998nonlinear}
Sch{\"o}lkopf,~B.; Smola,~A.; M{\"u}ller,~K.-R. Nonlinear Component Analysis As
  A Kernel Eigenvalue Problem. \emph{Neural Comput.} \textbf{1998}, \emph{10},
  1299--1319\relax
\mciteBstWouldAddEndPuncttrue
\mciteSetBstMidEndSepPunct{\mcitedefaultmidpunct}
{\mcitedefaultendpunct}{\mcitedefaultseppunct}\relax
\EndOfBibitem
\bibitem[Sch{\"o}lkopf \latin{et~al.}(1999)Sch{\"o}lkopf, Mika, Burges,
  Knirsch, M{\"u}ller, R{\"a}tsch, and Smola]{scholkopf1999input}
Sch{\"o}lkopf,~B.; Mika,~S.; Burges,~C.~J.; Knirsch,~P.; M{\"u}ller,~K.-R.;
  R{\"a}tsch,~G.; Smola,~A.~J. Input Space Versus Feature Space In Kernel-based
  Methods. \emph{IEEE Trans. Neural Netw.} \textbf{1999}, \emph{10},
  1000--1017\relax
\mciteBstWouldAddEndPuncttrue
\mciteSetBstMidEndSepPunct{\mcitedefaultmidpunct}
{\mcitedefaultendpunct}{\mcitedefaultseppunct}\relax
\EndOfBibitem
\bibitem[Bishop(1995)]{bishop1995neural}
Bishop,~C.~M. \emph{Neural Networks for Pattern Recognition}; Oxford university
  press, 1995\relax
\mciteBstWouldAddEndPuncttrue
\mciteSetBstMidEndSepPunct{\mcitedefaultmidpunct}
{\mcitedefaultendpunct}{\mcitedefaultseppunct}\relax
\EndOfBibitem
\bibitem[M{\"u}ller \latin{et~al.}(1997)M{\"u}ller, Smola, R{\"a}tsch,
  Sch{\"o}lkopf, Kohlmorgen, and Vapnik]{muller1997predicting}
M{\"u}ller,~K.-R.; Smola,~A.~J.; R{\"a}tsch,~G.; Sch{\"o}lkopf,~B.;
  Kohlmorgen,~J.; Vapnik,~V. Predicting Time Series With Support Vector
  Machines. International Conference on Artificial Neural Networks. 1997; pp
  999--1004\relax
\mciteBstWouldAddEndPuncttrue
\mciteSetBstMidEndSepPunct{\mcitedefaultmidpunct}
{\mcitedefaultendpunct}{\mcitedefaultseppunct}\relax
\EndOfBibitem
\bibitem[Boser \latin{et~al.}(1992)Boser, Guyon, and Vapnik]{boser1992training}
Boser,~B.~E.; Guyon,~I.~M.; Vapnik,~V.~N. A Training Algorithm For Optimal
  Margin Classifiers. Proceedings of the Fifth Annual Workshop on Computational
  Learning Theory. 1992; pp 144--152\relax
\mciteBstWouldAddEndPuncttrue
\mciteSetBstMidEndSepPunct{\mcitedefaultmidpunct}
{\mcitedefaultendpunct}{\mcitedefaultseppunct}\relax
\EndOfBibitem
\bibitem[Theodoridis \latin{et~al.}(2008)Theodoridis, Koutroumbas,
  \latin{et~al.} others]{theodoridis2008pattern}
Theodoridis,~S.; Koutroumbas,~K., \latin{et~al.}  \emph{Pattern Recognition};
  Elsevier, 2008\relax
\mciteBstWouldAddEndPuncttrue
\mciteSetBstMidEndSepPunct{\mcitedefaultmidpunct}
{\mcitedefaultendpunct}{\mcitedefaultseppunct}\relax
\EndOfBibitem
\bibitem[Theodoridis(2020)]{theodoridis2020machine}
Theodoridis,~S. \emph{Machine Learning: A Bayesian And Optimization Perspective
  2nd. Ed.}; Academic press, 2020\relax
\mciteBstWouldAddEndPuncttrue
\mciteSetBstMidEndSepPunct{\mcitedefaultmidpunct}
{\mcitedefaultendpunct}{\mcitedefaultseppunct}\relax
\EndOfBibitem
\bibitem[Kamath \latin{et~al.}(2018)Kamath, Vargas-Hern{\'a}ndez, Krems,
  Carrington~Jr, and Manzhos]{kamath2018neural}
Kamath,~A.; Vargas-Hern{\'a}ndez,~R.~A.; Krems,~R.~V.; Carrington~Jr,~T.;
  Manzhos,~S. Neural Networks vs Gaussian Process Regression For Representing
  Potential Energy Surfaces: A Comparative Study Of Fit Quality And Vibrational
  Spectrum Accuracy. \emph{J. Chem. Phys.} \textbf{2018}, \emph{148},
  241702\relax
\mciteBstWouldAddEndPuncttrue
\mciteSetBstMidEndSepPunct{\mcitedefaultmidpunct}
{\mcitedefaultendpunct}{\mcitedefaultseppunct}\relax
\EndOfBibitem
\bibitem[Wolpert(1996)]{wolpert1996lack}
Wolpert,~D.~H. The Lack Of A Priori Distinctions Between Learning Algorithms.
  \emph{Neural Comput.} \textbf{1996}, \emph{8}, 1341--1390\relax
\mciteBstWouldAddEndPuncttrue
\mciteSetBstMidEndSepPunct{\mcitedefaultmidpunct}
{\mcitedefaultendpunct}{\mcitedefaultseppunct}\relax
\EndOfBibitem
\bibitem[Lee \latin{et~al.}(2017)Lee, Bahri, Novak, Schoenholz, Pennington, and
  Sohl-Dickstein]{lee2017deep}
Lee,~J.; Bahri,~Y.; Novak,~R.; Schoenholz,~S.~S.; Pennington,~J.;
  Sohl-Dickstein,~J. Deep Neural Networks As Gaussian Processes. \emph{arXiv
  preprint arXiv:1711.00165} \textbf{2017}, \relax
\mciteBstWouldAddEndPunctfalse
\mciteSetBstMidEndSepPunct{\mcitedefaultmidpunct}
{}{\mcitedefaultseppunct}\relax
\EndOfBibitem
\bibitem[Matthews \latin{et~al.}(2018)Matthews, Rowland, Hron, Turner, and
  Ghahramani]{matthews2018gaussian}
Matthews,~A. G. d.~G.; Rowland,~M.; Hron,~J.; Turner,~R.~E.; Ghahramani,~Z.
  Gaussian Process Behaviour In Wide Deep Neural Networks. \emph{arXiv preprint
  arXiv:1804.11271} \textbf{2018}, \relax
\mciteBstWouldAddEndPunctfalse
\mciteSetBstMidEndSepPunct{\mcitedefaultmidpunct}
{}{\mcitedefaultseppunct}\relax
\EndOfBibitem
\bibitem[Braun \latin{et~al.}(2008)Braun, Buhmann, and
  M{\"u}ller]{braun2008relevant}
Braun,~M.~L.; Buhmann,~J.~M.; M{\"u}ller,~K.-R. On Relevant Dimensions In
  Kernel Feature Spaces. \emph{J. Mach. Learn. Res.} \textbf{2008}, \emph{9},
  1875--1908\relax
\mciteBstWouldAddEndPuncttrue
\mciteSetBstMidEndSepPunct{\mcitedefaultmidpunct}
{\mcitedefaultendpunct}{\mcitedefaultseppunct}\relax
\EndOfBibitem
\bibitem[Montavon \latin{et~al.}(2011)Montavon, Braun, and
  M{\"u}ller]{montavon2011kernel}
Montavon,~G.; Braun,~M.~L.; M{\"u}ller,~K.-R. Kernel Analysis of Deep Networks.
  \emph{J. Mach. Learn. Res.} \textbf{2011}, \emph{12}\relax
\mciteBstWouldAddEndPuncttrue
\mciteSetBstMidEndSepPunct{\mcitedefaultmidpunct}
{\mcitedefaultendpunct}{\mcitedefaultseppunct}\relax
\EndOfBibitem
\bibitem[Chmiela \latin{et~al.}(2017)Chmiela, Tkatchenko, Sauceda, Poltavsky,
  Sch{\"u}tt, and M{\"u}ller]{chmiela2017}
Chmiela,~S.; Tkatchenko,~A.; Sauceda,~H.~E.; Poltavsky,~I.; Sch{\"u}tt,~K.~T.;
  M{\"u}ller,~K.-R. Machine Learning Of Accurate Energy-conserving Molecular
  Force Fields. \emph{Sci. Adv.} \textbf{2017}, \emph{3}, e1603015\relax
\mciteBstWouldAddEndPuncttrue
\mciteSetBstMidEndSepPunct{\mcitedefaultmidpunct}
{\mcitedefaultendpunct}{\mcitedefaultseppunct}\relax
\EndOfBibitem
\bibitem[Faber \latin{et~al.}(2018)Faber, Christensen, Huang, and von
  Lilienfeld]{faber2018alchemical}
Faber,~F.~A.; Christensen,~A.~S.; Huang,~B.; von Lilienfeld,~O.~A. Alchemical
  And Structural Distribution Based Representation For Universal Quantum
  Machine Learning. \emph{J. Chem. Phys.} \textbf{2018}, \emph{148},
  241717\relax
\mciteBstWouldAddEndPuncttrue
\mciteSetBstMidEndSepPunct{\mcitedefaultmidpunct}
{\mcitedefaultendpunct}{\mcitedefaultseppunct}\relax
\EndOfBibitem
\bibitem[Christensen \latin{et~al.}(2020)Christensen, Bratholm, Faber, and
  Anatole~von Lilienfeld]{christensen2020fchl}
Christensen,~A.~S.; Bratholm,~L.~A.; Faber,~F.~A.; Anatole~von Lilienfeld,~O.
  FCHL Revisited: Faster And More Accurate Quantum Machine Learning. \emph{J.
  Chem. Phys.} \textbf{2020}, \emph{152}, 044107\relax
\mciteBstWouldAddEndPuncttrue
\mciteSetBstMidEndSepPunct{\mcitedefaultmidpunct}
{\mcitedefaultendpunct}{\mcitedefaultseppunct}\relax
\EndOfBibitem
\bibitem[Unke and Meuwly(2019)Unke, and Meuwly]{unke2019}
Unke,~O.~T.; Meuwly,~M. Physnet: A Neural Network For Predicting Energies,
  Forces, Dipole Moments And Partial Charges. \emph{J. Chem. Theory Comput.}
  \textbf{2019}, \emph{15}, 3678--3693\relax
\mciteBstWouldAddEndPuncttrue
\mciteSetBstMidEndSepPunct{\mcitedefaultmidpunct}
{\mcitedefaultendpunct}{\mcitedefaultseppunct}\relax
\EndOfBibitem
\bibitem[Sch{\"u}tt \latin{et~al.}(2017)Sch{\"u}tt, Kindermans, Sauceda,
  Chmiela, Tkatchenko, and M{\"u}ller]{schutt2017schnet}
Sch{\"u}tt,~K.; Kindermans,~P.-J.; Sauceda,~H.~E.; Chmiela,~S.; Tkatchenko,~A.;
  M{\"u}ller,~K.-R. {SchNet}: {A} Continuous-filter Convolutional Neural
  Network For Modeling Quantum Interactions. Adv. Neural. Inf. Process. Syst.
  2017; pp 991--1001\relax
\mciteBstWouldAddEndPuncttrue
\mciteSetBstMidEndSepPunct{\mcitedefaultmidpunct}
{\mcitedefaultendpunct}{\mcitedefaultseppunct}\relax
\EndOfBibitem
\bibitem[Klicpera \latin{et~al.}(2020)Klicpera, Gro{\ss}, and
  G{\"u}nnemann]{klicpera_dimenet_2020}
Klicpera,~J.; Gro{\ss},~J.; G{\"u}nnemann,~S. Directional Message Passing For
  Molecular Graphs. International Conference on Learning Representations
  (ICLR). 2020\relax
\mciteBstWouldAddEndPuncttrue
\mciteSetBstMidEndSepPunct{\mcitedefaultmidpunct}
{\mcitedefaultendpunct}{\mcitedefaultseppunct}\relax
\EndOfBibitem
\bibitem[Zhang \latin{et~al.}(2019)Zhang, Hu, and Jiang]{zhang2019embedded}
Zhang,~Y.; Hu,~C.; Jiang,~B. Embedded Atom Neural Network Potentials: Efficient
  And Accurate Machine Learning With A Physically Inspired Representation.
  \emph{J. Phys. Chem. Lett.} \textbf{2019}, \emph{10}, 4962--4967\relax
\mciteBstWouldAddEndPuncttrue
\mciteSetBstMidEndSepPunct{\mcitedefaultmidpunct}
{\mcitedefaultendpunct}{\mcitedefaultseppunct}\relax
\EndOfBibitem
\bibitem[Zhang \latin{et~al.}(2018)Zhang, Han, Wang, Car, and
  Weinan]{zhang2018deep}
Zhang,~L.; Han,~J.; Wang,~H.; Car,~R.; Weinan,~E. Deep Potential Molecular
  Dynamics: A Scalable Model With The Accuracy Of Quantum Mechanics.
  \emph{Phys. Rev. Lett.} \textbf{2018}, \emph{120}, 143001\relax
\mciteBstWouldAddEndPuncttrue
\mciteSetBstMidEndSepPunct{\mcitedefaultmidpunct}
{\mcitedefaultendpunct}{\mcitedefaultseppunct}\relax
\EndOfBibitem
\bibitem[Zhang \latin{et~al.}(2018)Zhang, Han, Wang, Saidi, Car, and
  Weinan]{zhang2018end}
Zhang,~L.; Han,~J.; Wang,~H.; Saidi,~W.; Car,~R.; Weinan,~E. End-to-end
  Symmetry Preserving Inter-atomic Potential Energy Model For Finite And
  Extended Systems. Adv. Neural. Inf. Process. Syst. 2018; pp 4436--4446\relax
\mciteBstWouldAddEndPuncttrue
\mciteSetBstMidEndSepPunct{\mcitedefaultmidpunct}
{\mcitedefaultendpunct}{\mcitedefaultseppunct}\relax
\EndOfBibitem
\bibitem[Behler and Parrinello(2007)Behler, and
  Parrinello]{behler2007generalized}
Behler,~J.; Parrinello,~M. Generalized Neural-Network Representation of
  High-Dimensional Potential-Energy Surfaces. \emph{Phys. Rev. Lett.}
  \textbf{2007}, \emph{98}, 146401\relax
\mciteBstWouldAddEndPuncttrue
\mciteSetBstMidEndSepPunct{\mcitedefaultmidpunct}
{\mcitedefaultendpunct}{\mcitedefaultseppunct}\relax
\EndOfBibitem
\bibitem[Lubbers \latin{et~al.}(2018)Lubbers, Smith, and
  Barros]{lubbers2018hierarchical}
Lubbers,~N.; Smith,~J.~S.; Barros,~K. Hierarchical Modeling Of Molecular
  Energies Using A Deep Neural Network. \emph{J. Chem. Phys.} \textbf{2018},
  \emph{148}, 241715\relax
\mciteBstWouldAddEndPuncttrue
\mciteSetBstMidEndSepPunct{\mcitedefaultmidpunct}
{\mcitedefaultendpunct}{\mcitedefaultseppunct}\relax
\EndOfBibitem
\bibitem[Behler(2011)]{behler2011atom}
Behler,~J. Atom-Centered Symmetry Functions for Constructing High-Dimensional
  Neural Network Potentials. \emph{J. Chem. Phys.} \textbf{2011}, \emph{134},
  074106\relax
\mciteBstWouldAddEndPuncttrue
\mciteSetBstMidEndSepPunct{\mcitedefaultmidpunct}
{\mcitedefaultendpunct}{\mcitedefaultseppunct}\relax
\EndOfBibitem
\bibitem[Bart{\'o}k \latin{et~al.}(2013)Bart{\'o}k, Kondor, and
  Cs{\'a}nyi]{bartok2013representing}
Bart{\'o}k,~A.~P.; Kondor,~R.; Cs{\'a}nyi,~G. On Representing Chemical
  Environments. \emph{Phys. Rev. B} \textbf{2013}, \emph{87}, 184115\relax
\mciteBstWouldAddEndPuncttrue
\mciteSetBstMidEndSepPunct{\mcitedefaultmidpunct}
{\mcitedefaultendpunct}{\mcitedefaultseppunct}\relax
\EndOfBibitem
\bibitem[Sch{\"u}tt \latin{et~al.}(2014)Sch{\"u}tt, Glawe, Brockherde, Sanna,
  M{\"u}ller, and Gross]{schutt2014represent}
Sch{\"u}tt,~K.; Glawe,~H.; Brockherde,~F.; Sanna,~A.; M{\"u}ller,~K.; Gross,~E.
  How To Represent Crystal Structures For Machine Learning: Towards Fast
  Prediction Of Electronic Properties. \emph{Phys. Rev. B} \textbf{2014},
  \emph{89}, 205118\relax
\mciteBstWouldAddEndPuncttrue
\mciteSetBstMidEndSepPunct{\mcitedefaultmidpunct}
{\mcitedefaultendpunct}{\mcitedefaultseppunct}\relax
\EndOfBibitem
\bibitem[Faber \latin{et~al.}(2015)Faber, Lindmaa, von Lilienfeld, and
  Armiento]{faber2015crystal}
Faber,~F.; Lindmaa,~A.; von Lilienfeld,~O.~A.; Armiento,~R. Crystal Structure
  Representations For Machine Learning Models Of Formation Energies. \emph{Int.
  J. Quantum Chem.} \textbf{2015}, \emph{115}, 1094--1101\relax
\mciteBstWouldAddEndPuncttrue
\mciteSetBstMidEndSepPunct{\mcitedefaultmidpunct}
{\mcitedefaultendpunct}{\mcitedefaultseppunct}\relax
\EndOfBibitem
\bibitem[Faber \latin{et~al.}(2016)Faber, Lindmaa, Von~Lilienfeld, and
  Armiento]{faber2016machine}
Faber,~F.~A.; Lindmaa,~A.; Von~Lilienfeld,~O.~A.; Armiento,~R. Machine Learning
  Energies of 2 Million Elpasolite (ABC\$\_2\$D\$\_6\$) Crystals. \emph{Phys.
  Rev. Lett.} \textbf{2016}, \emph{117}, 135502\relax
\mciteBstWouldAddEndPuncttrue
\mciteSetBstMidEndSepPunct{\mcitedefaultmidpunct}
{\mcitedefaultendpunct}{\mcitedefaultseppunct}\relax
\EndOfBibitem
\bibitem[Wahba(1990)]{wahba1990spline}
Wahba,~G. \emph{Spline Models For Observational Data}; Siam, 1990;
  Vol.~59\relax
\mciteBstWouldAddEndPuncttrue
\mciteSetBstMidEndSepPunct{\mcitedefaultmidpunct}
{\mcitedefaultendpunct}{\mcitedefaultseppunct}\relax
\EndOfBibitem
\bibitem[Sch{\"o}lkopf \latin{et~al.}(2001)Sch{\"o}lkopf, Herbrich, and
  Smola]{scholkopf2001generalized}
Sch{\"o}lkopf,~B.; Herbrich,~R.; Smola,~A.~J. A Generalized Representer
  Theorem. International Conference on Computational Learning Theory. 2001; pp
  416--426\relax
\mciteBstWouldAddEndPuncttrue
\mciteSetBstMidEndSepPunct{\mcitedefaultmidpunct}
{\mcitedefaultendpunct}{\mcitedefaultseppunct}\relax
\EndOfBibitem
\bibitem[Argyriou \latin{et~al.}(2009)Argyriou, Micchelli, and
  Pontil]{argyriou2009there}
Argyriou,~A.; Micchelli,~C.~A.; Pontil,~M. When Is There A Representer Theorem?
  Vector Versus Matrix Regularizers. \emph{J. Mach. Learn. Res.} \textbf{2009},
  \emph{10}, 2507--2529\relax
\mciteBstWouldAddEndPuncttrue
\mciteSetBstMidEndSepPunct{\mcitedefaultmidpunct}
{\mcitedefaultendpunct}{\mcitedefaultseppunct}\relax
\EndOfBibitem
\bibitem[Berlinet and Thomas-Agnan(2011)Berlinet, and
  Thomas-Agnan]{berlinet2011reproducing}
Berlinet,~A.; Thomas-Agnan,~C. \emph{Reproducing Kernel Hilbert Spaces in
  Probability and Statistics}; Springer Science \& Business Media Dordrecht,
  2011\relax
\mciteBstWouldAddEndPuncttrue
\mciteSetBstMidEndSepPunct{\mcitedefaultmidpunct}
{\mcitedefaultendpunct}{\mcitedefaultseppunct}\relax
\EndOfBibitem
\bibitem[Sch{\"o}lkopf and Smola(2002)Sch{\"o}lkopf, and
  Smola]{scholkopf2002learning}
Sch{\"o}lkopf,~B.; Smola,~A.~J. \emph{Learning With Kernels: Support Vector
  Machines, Regularization, Optimization, And Beyond}; MIT press, 2002\relax
\mciteBstWouldAddEndPuncttrue
\mciteSetBstMidEndSepPunct{\mcitedefaultmidpunct}
{\mcitedefaultendpunct}{\mcitedefaultseppunct}\relax
\EndOfBibitem
\bibitem[Rasmussen(2003)]{rasmussen2003gaussian}
Rasmussen,~C.~E. \emph{Summer School on Machine Learning}; 2003; pp
  63--71\relax
\mciteBstWouldAddEndPuncttrue
\mciteSetBstMidEndSepPunct{\mcitedefaultmidpunct}
{\mcitedefaultendpunct}{\mcitedefaultseppunct}\relax
\EndOfBibitem
\bibitem[Murphy(2012)]{murphy2012machine}
Murphy,~K.~P. \emph{Machine Learning: A Probabilistic Perspective}; MIT press,
  2012\relax
\mciteBstWouldAddEndPuncttrue
\mciteSetBstMidEndSepPunct{\mcitedefaultmidpunct}
{\mcitedefaultendpunct}{\mcitedefaultseppunct}\relax
\EndOfBibitem
\bibitem[Smola \latin{et~al.}(1998)Smola, Sch{\"o}lkopf, and
  M{\"u}ller]{smola1998connection}
Smola,~A.~J.; Sch{\"o}lkopf,~B.; M{\"u}ller,~K.-R. The Connection Between
  Regularization Operators And Support Vector Kernels. \emph{Neural Netw.}
  \textbf{1998}, \emph{11}, 637--649\relax
\mciteBstWouldAddEndPuncttrue
\mciteSetBstMidEndSepPunct{\mcitedefaultmidpunct}
{\mcitedefaultendpunct}{\mcitedefaultseppunct}\relax
\EndOfBibitem
\bibitem[Micchelli \latin{et~al.}(2006)Micchelli, Xu, and
  Zhang]{micchelli2006universal}
Micchelli,~C.~A.; Xu,~Y.; Zhang,~H. Universal Kernels. \emph{J. Mach. Learn.
  Res.} \textbf{2006}, \emph{7}, 2651--2667\relax
\mciteBstWouldAddEndPuncttrue
\mciteSetBstMidEndSepPunct{\mcitedefaultmidpunct}
{\mcitedefaultendpunct}{\mcitedefaultseppunct}\relax
\EndOfBibitem
\bibitem[Huang and Von~Lilienfeld(2016)Huang, and
  Von~Lilienfeld]{huang2016communication}
Huang,~B.; Von~Lilienfeld,~O.~A. Communication: Understanding Molecular
  Representations In Machine Learning: The Role Of Uniqueness And Target
  Similarity. \emph{J. Chem. Phys.} \textbf{2016}, \emph{145}, 161102\relax
\mciteBstWouldAddEndPuncttrue
\mciteSetBstMidEndSepPunct{\mcitedefaultmidpunct}
{\mcitedefaultendpunct}{\mcitedefaultseppunct}\relax
\EndOfBibitem
\bibitem[Hofmann \latin{et~al.}(2008)Hofmann, Sch{\"o}lkopf, and
  Smola]{hofmann2008kernel}
Hofmann,~T.; Sch{\"o}lkopf,~B.; Smola,~A.~J. Kernel Methods In Machine
  Learning. \emph{Ann. Stat.} \textbf{2008}, 1171--1220\relax
\mciteBstWouldAddEndPuncttrue
\mciteSetBstMidEndSepPunct{\mcitedefaultmidpunct}
{\mcitedefaultendpunct}{\mcitedefaultseppunct}\relax
\EndOfBibitem
\bibitem[Golub and Van~Loan(2012)Golub, and Van~Loan]{golub2012matrix}
Golub,~G.~H.; Van~Loan,~C.~F. \emph{Matrix Computations}; JHU Press Baltimore,
  2012; Vol.~3\relax
\mciteBstWouldAddEndPuncttrue
\mciteSetBstMidEndSepPunct{\mcitedefaultmidpunct}
{\mcitedefaultendpunct}{\mcitedefaultseppunct}\relax
\EndOfBibitem
\bibitem[Raykar and Duraiswami(2007)Raykar, and Duraiswami]{raykar2007fast}
Raykar,~V.~C.; Duraiswami,~R. Fast Large Scale Gaussian Process Regression
  Using Approximate Matrix-vector Products. Learning workshop. 2007\relax
\mciteBstWouldAddEndPuncttrue
\mciteSetBstMidEndSepPunct{\mcitedefaultmidpunct}
{\mcitedefaultendpunct}{\mcitedefaultseppunct}\relax
\EndOfBibitem
\bibitem[Williams and Seeger(2001)Williams, and Seeger]{williams2001using}
Williams,~C.~K.; Seeger,~M. Using The Nystr{\"o}m Method To Speed Up Kernel
  Machines. Adv. Neural. Inf. Process. Syst. 2001; pp 682--688\relax
\mciteBstWouldAddEndPuncttrue
\mciteSetBstMidEndSepPunct{\mcitedefaultmidpunct}
{\mcitedefaultendpunct}{\mcitedefaultseppunct}\relax
\EndOfBibitem
\bibitem[Qui{\~n}onero-Candela and Rasmussen(2005)Qui{\~n}onero-Candela, and
  Rasmussen]{quinonero2005unifying}
Qui{\~n}onero-Candela,~J.; Rasmussen,~C.~E. A Unifying View Of Sparse
  Approximate Gaussian Process Regression. \emph{J. Mach. Learn. Res.}
  \textbf{2005}, \emph{6}, 1939--1959\relax
\mciteBstWouldAddEndPuncttrue
\mciteSetBstMidEndSepPunct{\mcitedefaultmidpunct}
{\mcitedefaultendpunct}{\mcitedefaultseppunct}\relax
\EndOfBibitem
\bibitem[Snelson and Ghahramani(2006)Snelson, and
  Ghahramani]{snelson2006sparse}
Snelson,~E.; Ghahramani,~Z. Sparse Gaussian Processes Using Pseudo-inputs. Adv.
  Neural. Inf. Process. Syst. 2006; pp 1257--1264\relax
\mciteBstWouldAddEndPuncttrue
\mciteSetBstMidEndSepPunct{\mcitedefaultmidpunct}
{\mcitedefaultendpunct}{\mcitedefaultseppunct}\relax
\EndOfBibitem
\bibitem[Rahimi and Recht(2008)Rahimi, and Recht]{rahimi2008random}
Rahimi,~A.; Recht,~B. Random Features For Large-scale Kernel Machines. Adv.
  Neural. Inf. Process. Syst. 2008; pp 1177--1184\relax
\mciteBstWouldAddEndPuncttrue
\mciteSetBstMidEndSepPunct{\mcitedefaultmidpunct}
{\mcitedefaultendpunct}{\mcitedefaultseppunct}\relax
\EndOfBibitem
\bibitem[Rudi \latin{et~al.}(2017)Rudi, Carratino, and Rosasco]{rudi2017falkon}
Rudi,~A.; Carratino,~L.; Rosasco,~L. Falkon: An Optimal Large Scale Kernel
  Method. Adv. Neural. Inf. Process. Syst. 2017; pp 3888--3898\relax
\mciteBstWouldAddEndPuncttrue
\mciteSetBstMidEndSepPunct{\mcitedefaultmidpunct}
{\mcitedefaultendpunct}{\mcitedefaultseppunct}\relax
\EndOfBibitem
\bibitem[Moore(1920)]{moore1920reciprocal}
Moore,~E.~H. On The Reciprocal Of The General Algebraic Matrix. \emph{Bull. Am.
  Math. Soc.} \textbf{1920}, \emph{26}, 394--395\relax
\mciteBstWouldAddEndPuncttrue
\mciteSetBstMidEndSepPunct{\mcitedefaultmidpunct}
{\mcitedefaultendpunct}{\mcitedefaultseppunct}\relax
\EndOfBibitem
\bibitem[Penrose(1955)]{penrose1955generalized}
Penrose,~R. A Generalized Inverse For Matrices. Math. Proc. Cambridge. 1955; pp
  406--413\relax
\mciteBstWouldAddEndPuncttrue
\mciteSetBstMidEndSepPunct{\mcitedefaultmidpunct}
{\mcitedefaultendpunct}{\mcitedefaultseppunct}\relax
\EndOfBibitem
\bibitem[Cutajar \latin{et~al.}(2016)Cutajar, Osborne, Cunningham, and
  Filippone]{cutajar2016preconditioning}
Cutajar,~K.; Osborne,~M.; Cunningham,~J.; Filippone,~M. Preconditioning Kernel
  Matrices. International Conference on Machine Learning. 2016; pp
  2529--2538\relax
\mciteBstWouldAddEndPuncttrue
\mciteSetBstMidEndSepPunct{\mcitedefaultmidpunct}
{\mcitedefaultendpunct}{\mcitedefaultseppunct}\relax
\EndOfBibitem
\bibitem[Tikhonov \latin{et~al.}(1977)Tikhonov, Arsenin, and
  John]{tikhonov1977solutions}
Tikhonov,~A.~N.; Arsenin,~V.~I.; John,~F. \emph{Solutions of Ill-Posed
  Problems}; Winston Washington, DC, 1977; Vol.~14\relax
\mciteBstWouldAddEndPuncttrue
\mciteSetBstMidEndSepPunct{\mcitedefaultmidpunct}
{\mcitedefaultendpunct}{\mcitedefaultseppunct}\relax
\EndOfBibitem
\bibitem[McCulloch and Pitts(1943)McCulloch, and Pitts]{mcculloch1943logical}
McCulloch,~W.~S.; Pitts,~W. A Logical Calculus of the Ideas Immanent in Nervous
  Activity. \emph{Bull. Math. Biophys.} \textbf{1943}, \emph{5}, 115--133\relax
\mciteBstWouldAddEndPuncttrue
\mciteSetBstMidEndSepPunct{\mcitedefaultmidpunct}
{\mcitedefaultendpunct}{\mcitedefaultseppunct}\relax
\EndOfBibitem
\bibitem[Kohonen(1988)]{kohonen1988introduction}
Kohonen,~T. An Introduction to Neural Computing. \emph{Neural Netw.}
  \textbf{1988}, \emph{1}, 3--16\relax
\mciteBstWouldAddEndPuncttrue
\mciteSetBstMidEndSepPunct{\mcitedefaultmidpunct}
{\mcitedefaultendpunct}{\mcitedefaultseppunct}\relax
\EndOfBibitem
\bibitem[Abdi(1994)]{abdi1994neural}
Abdi,~H. A Neural Network Primer. \emph{J. Biol. Syst.} \textbf{1994},
  \emph{2}, 247--281\relax
\mciteBstWouldAddEndPuncttrue
\mciteSetBstMidEndSepPunct{\mcitedefaultmidpunct}
{\mcitedefaultendpunct}{\mcitedefaultseppunct}\relax
\EndOfBibitem
\bibitem[Clark(1999)]{clark1999neural}
Clark,~J.~W. \emph{Scientific Applications of Neural Nets}; Springer, 1999; pp
  1--96\relax
\mciteBstWouldAddEndPuncttrue
\mciteSetBstMidEndSepPunct{\mcitedefaultmidpunct}
{\mcitedefaultendpunct}{\mcitedefaultseppunct}\relax
\EndOfBibitem
\bibitem[Ripley(2007)]{ripley2007pattern}
Ripley,~B.~D. \emph{Pattern Recognition and Neural Networks}; Cambridge
  university press, 2007\relax
\mciteBstWouldAddEndPuncttrue
\mciteSetBstMidEndSepPunct{\mcitedefaultmidpunct}
{\mcitedefaultendpunct}{\mcitedefaultseppunct}\relax
\EndOfBibitem
\bibitem[Haykin(2009)]{haykin2009neural}
Haykin,~S.~S. \emph{Neural Networks and Learning Machines}; Pearson Upper
  Saddle River, NJ, USA:, 2009; Vol.~3\relax
\mciteBstWouldAddEndPuncttrue
\mciteSetBstMidEndSepPunct{\mcitedefaultmidpunct}
{\mcitedefaultendpunct}{\mcitedefaultseppunct}\relax
\EndOfBibitem
\bibitem[LeCun \latin{et~al.}(2012)LeCun, Bottou, Orr, and
  M{\"u}ller]{lecun2012efficient}
LeCun,~Y.~A.; Bottou,~L.; Orr,~G.~B.; M{\"u}ller,~K.-R. \emph{Neural networks:
  Tricks of the trade}; Springer LNCS 7700, 2012; pp 9--48\relax
\mciteBstWouldAddEndPuncttrue
\mciteSetBstMidEndSepPunct{\mcitedefaultmidpunct}
{\mcitedefaultendpunct}{\mcitedefaultseppunct}\relax
\EndOfBibitem
\bibitem[Cybenko(1989)]{gybenko1989approximation}
Cybenko,~G. Approximation By Superposition Of Sigmoidal Functions. \emph{Math.
  Control Signals Syst.} \textbf{1989}, \emph{2}, 303--314\relax
\mciteBstWouldAddEndPuncttrue
\mciteSetBstMidEndSepPunct{\mcitedefaultmidpunct}
{\mcitedefaultendpunct}{\mcitedefaultseppunct}\relax
\EndOfBibitem
\bibitem[Hornik(1991)]{hornik1991approximation}
Hornik,~K. Approximation Capabilities of Multilayer Feedforward Networks.
  \emph{Neural Netw.} \textbf{1991}, \emph{4}, 251--257\relax
\mciteBstWouldAddEndPuncttrue
\mciteSetBstMidEndSepPunct{\mcitedefaultmidpunct}
{\mcitedefaultendpunct}{\mcitedefaultseppunct}\relax
\EndOfBibitem
\bibitem[Eldan and Shamir(2016)Eldan, and Shamir]{eldan2016power}
Eldan,~R.; Shamir,~O. The Power of Depth for Feedforward Neural Networks.
  Conference on Learning Theory. 2016; pp 907--940\relax
\mciteBstWouldAddEndPuncttrue
\mciteSetBstMidEndSepPunct{\mcitedefaultmidpunct}
{\mcitedefaultendpunct}{\mcitedefaultseppunct}\relax
\EndOfBibitem
\bibitem[Cohen \latin{et~al.}(2016)Cohen, Sharir, and
  Shashua]{cohen2016expressive}
Cohen,~N.; Sharir,~O.; Shashua,~A. On The Expressive Power Of Deep Learning: A
  Tensor Analysis. Conference On Learning Theory. 2016; pp 698--728\relax
\mciteBstWouldAddEndPuncttrue
\mciteSetBstMidEndSepPunct{\mcitedefaultmidpunct}
{\mcitedefaultendpunct}{\mcitedefaultseppunct}\relax
\EndOfBibitem
\bibitem[Telgarsky(2016)]{telgarsky2016benefits}
Telgarsky,~M. Benefits Of Depth In Neural Networks. Conference On Learning
  Theory. 2016; pp 1517--1539\relax
\mciteBstWouldAddEndPuncttrue
\mciteSetBstMidEndSepPunct{\mcitedefaultmidpunct}
{\mcitedefaultendpunct}{\mcitedefaultseppunct}\relax
\EndOfBibitem
\bibitem[Lu \latin{et~al.}(2017)Lu, Pu, Wang, Hu, and Wang]{lu2017expressive}
Lu,~Z.; Pu,~H.; Wang,~F.; Hu,~Z.; Wang,~L. The Expressive Power Of Neural
  Networks: A View From The Width. \emph{Adv. Neural. Inf. Process. Syst.}
  \textbf{2017}, \emph{30}, 6231--6239\relax
\mciteBstWouldAddEndPuncttrue
\mciteSetBstMidEndSepPunct{\mcitedefaultmidpunct}
{\mcitedefaultendpunct}{\mcitedefaultseppunct}\relax
\EndOfBibitem
\bibitem[Montavon \latin{et~al.}(2012)Montavon, Orr, and
  M{\"u}ller]{montavon2012neural}
Montavon,~G.; Orr,~G.; M{\"u}ller,~K.-R. \emph{Neural Networks: Tricks Of The
  Trade}; Springer LNCS 7700, 2012; Vol.~2\relax
\mciteBstWouldAddEndPuncttrue
\mciteSetBstMidEndSepPunct{\mcitedefaultmidpunct}
{\mcitedefaultendpunct}{\mcitedefaultseppunct}\relax
\EndOfBibitem
\bibitem[Snoek \latin{et~al.}(2012)Snoek, Larochelle, and
  Adams]{snoek2012practical}
Snoek,~J.; Larochelle,~H.; Adams,~R.~P. Practical Bayesian Optimization Of
  Machine Learning Algorithms. \emph{Adv. Neural. Inf. Process. Syst.}
  \textbf{2012}, \emph{25}, 2951--2959\relax
\mciteBstWouldAddEndPuncttrue
\mciteSetBstMidEndSepPunct{\mcitedefaultmidpunct}
{\mcitedefaultendpunct}{\mcitedefaultseppunct}\relax
\EndOfBibitem
\bibitem[Hastie \latin{et~al.}(2009)Hastie, Tibshirani, and
  Friedman]{hastie2009elements}
Hastie,~T.; Tibshirani,~R.; Friedman,~J. \emph{The Elements Of Statistical
  Learning: Data Mining, Inference, And Prediction}; Springer Science \&
  Business Media, 2009\relax
\mciteBstWouldAddEndPuncttrue
\mciteSetBstMidEndSepPunct{\mcitedefaultmidpunct}
{\mcitedefaultendpunct}{\mcitedefaultseppunct}\relax
\EndOfBibitem
\bibitem[Sch{\"u}tt \latin{et~al.}(2017)Sch{\"u}tt, Arbabzadah, Chmiela,
  M{\"u}ller, and Tkatchenko]{schutt2017quantum}
Sch{\"u}tt,~K.~T.; Arbabzadah,~F.; Chmiela,~S.; M{\"u}ller,~K.~R.;
  Tkatchenko,~A. Quantum-Chemical Insights from Deep Tensor Neural Networks.
  \emph{Nat. Commun.} \textbf{2017}, \emph{8}, 13890\relax
\mciteBstWouldAddEndPuncttrue
\mciteSetBstMidEndSepPunct{\mcitedefaultmidpunct}
{\mcitedefaultendpunct}{\mcitedefaultseppunct}\relax
\EndOfBibitem
\bibitem[Sauceda \latin{et~al.}(2020)Sauceda, Vassilev-Galindo, Chmiela,
  M\"{u}ller, and Tkatchenko]{sauceda2020}
Sauceda,~H.~E.; Vassilev-Galindo,~V.; Chmiela,~S.; M\"{u}ller,~K.-R.;
  Tkatchenko,~A. Dynamical Strengthening Of Covalent And Non-covalent Molecular
  Interactions By Nuclear Quantum Effects At Finite Temperature. \emph{arXiv
  preprint arXiv:2006.10578} \textbf{2020}, \relax
\mciteBstWouldAddEndPunctfalse
\mciteSetBstMidEndSepPunct{\mcitedefaultmidpunct}
{}{\mcitedefaultseppunct}\relax
\EndOfBibitem
\bibitem[Anselmi \latin{et~al.}(2016)Anselmi, Rosasco, and
  Poggio]{anselmi2016invariance}
Anselmi,~F.; Rosasco,~L.; Poggio,~T. On Invariance And Selectivity In
  Representation Learning. \emph{Information and Inference: A Journal of the
  IMA} \textbf{2016}, \emph{5}, 134--158\relax
\mciteBstWouldAddEndPuncttrue
\mciteSetBstMidEndSepPunct{\mcitedefaultmidpunct}
{\mcitedefaultendpunct}{\mcitedefaultseppunct}\relax
\EndOfBibitem
\bibitem[Hellman(1937)]{hellman1937einfuhrung}
Hellman,~H. \emph{Einf{\"u}hrung In Die Quantenchemie}; Leipzig: F. Deuticke,
  1937; Vol.~0\relax
\mciteBstWouldAddEndPuncttrue
\mciteSetBstMidEndSepPunct{\mcitedefaultmidpunct}
{\mcitedefaultendpunct}{\mcitedefaultseppunct}\relax
\EndOfBibitem
\bibitem[Feynman(1939)]{feynman1939forces}
Feynman,~R.~P. Forces In Molecules. \emph{Phys. Rev.} \textbf{1939}, \emph{56},
  340\relax
\mciteBstWouldAddEndPuncttrue
\mciteSetBstMidEndSepPunct{\mcitedefaultmidpunct}
{\mcitedefaultendpunct}{\mcitedefaultseppunct}\relax
\EndOfBibitem
\bibitem[Chmiela \latin{et~al.}(2019)Chmiela, Sauceda, Poltavsky, M{\"u}ller,
  and Tkatchenko]{chmiela2019}
Chmiela,~S.; Sauceda,~H.~E.; Poltavsky,~I.; M{\"u}ller,~K.-R.; Tkatchenko,~A.
  {sGDML}: {C}Onstructing Accurate And Data Efficient Molecular Force Fields
  Using Machine Learning. \emph{Comput. Phys. Commun.} \textbf{2019},
  \emph{240}, 38--45\relax
\mciteBstWouldAddEndPuncttrue
\mciteSetBstMidEndSepPunct{\mcitedefaultmidpunct}
{\mcitedefaultendpunct}{\mcitedefaultseppunct}\relax
\EndOfBibitem
\bibitem[Adamo and Barone(1999)Adamo, and Barone]{adamo1999toward}
Adamo,~C.; Barone,~V. Toward Reliable Density Functional Methods Without
  Adjustable Parameters: The {PBE0} Model. \emph{J. Chem. Phys.} \textbf{1999},
  \emph{110}, 6158--6170\relax
\mciteBstWouldAddEndPuncttrue
\mciteSetBstMidEndSepPunct{\mcitedefaultmidpunct}
{\mcitedefaultendpunct}{\mcitedefaultseppunct}\relax
\EndOfBibitem
\bibitem[Perdew \latin{et~al.}(1996)Perdew, Burke, and Ernzerhof]{PBE1996}
Perdew,~J.~P.; Burke,~K.; Ernzerhof,~M. Generalized Gradient Approximation Made
  Simple. \emph{Phys. Rev. Lett.} \textbf{1996}, \emph{77}, 3865--3868\relax
\mciteBstWouldAddEndPuncttrue
\mciteSetBstMidEndSepPunct{\mcitedefaultmidpunct}
{\mcitedefaultendpunct}{\mcitedefaultseppunct}\relax
\EndOfBibitem
\bibitem[Tkatchenko and Scheffler(2009)Tkatchenko, and
  Scheffler]{tkatchenko2009accurate}
Tkatchenko,~A.; Scheffler,~M. Accurate Molecular Van Der Waals Interactions
  From Ground-state Electron Density And Free-atom Reference Data. \emph{Phys.
  Rev. Lett.} \textbf{2009}, \emph{102}, 073005\relax
\mciteBstWouldAddEndPuncttrue
\mciteSetBstMidEndSepPunct{\mcitedefaultmidpunct}
{\mcitedefaultendpunct}{\mcitedefaultseppunct}\relax
\EndOfBibitem
\bibitem[Montavon \latin{et~al.}(2012)Montavon, Hansen, Fazli, Rupp, Biegler,
  Ziehe, Tkatchenko, Lilienfeld, and M{\"u}ller]{montavon2012learning}
Montavon,~G.; Hansen,~K.; Fazli,~S.; Rupp,~M.; Biegler,~F.; Ziehe,~A.;
  Tkatchenko,~A.; Lilienfeld,~A.~V.; M{\"u}ller,~K.-R. Learning Invariant
  Representations Of Molecules For Atomization Energy Prediction. Adv. Neural.
  Inf. Process. Syst. 2012; pp 440--448\relax
\mciteBstWouldAddEndPuncttrue
\mciteSetBstMidEndSepPunct{\mcitedefaultmidpunct}
{\mcitedefaultendpunct}{\mcitedefaultseppunct}\relax
\EndOfBibitem
\bibitem[Huo and Rupp(2017)Huo, and Rupp]{huo2017unified}
Huo,~H.; Rupp,~M. Unified Representation Of Molecules And Crystals For Machine
  Learning. \emph{arXiv preprint arXiv:1704.06439} \textbf{2017}, \relax
\mciteBstWouldAddEndPunctfalse
\mciteSetBstMidEndSepPunct{\mcitedefaultmidpunct}
{}{\mcitedefaultseppunct}\relax
\EndOfBibitem
\bibitem[Hirn \latin{et~al.}(2017)Hirn, Mallat, and Poilvert]{hirn2017wavelet}
Hirn,~M.; Mallat,~S.; Poilvert,~N. Wavelet Scattering Regression Of Quantum
  Chemical Energies. \emph{Multiscale Modeling \& Simulation} \textbf{2017},
  \emph{15}, 827--863\relax
\mciteBstWouldAddEndPuncttrue
\mciteSetBstMidEndSepPunct{\mcitedefaultmidpunct}
{\mcitedefaultendpunct}{\mcitedefaultseppunct}\relax
\EndOfBibitem
\bibitem[Eickenberg \latin{et~al.}(2017)Eickenberg, Exarchakis, Hirn, and
  Mallat]{eickenberg2017solid}
Eickenberg,~M.; Exarchakis,~G.; Hirn,~M.; Mallat,~S. Solid Harmonic Wavelet
  Scattering: Predicting Quantum Molecular Energy From Invariant Descriptors Of
  3d Electronic Densities. Adv. Neural. Inf. Process. Syst. 2017; pp
  6540--6549\relax
\mciteBstWouldAddEndPuncttrue
\mciteSetBstMidEndSepPunct{\mcitedefaultmidpunct}
{\mcitedefaultendpunct}{\mcitedefaultseppunct}\relax
\EndOfBibitem
\bibitem[Kriege \latin{et~al.}(2016)Kriege, Giscard, and
  Wilson]{kriege2016valid}
Kriege,~N.~M.; Giscard,~P.-L.; Wilson,~R. On Valid Optimal Assignment Kernels
  And Applications To Graph Classification. Adv. Neural. Inf. Process. Syst.
  2016; pp 1623--1631\relax
\mciteBstWouldAddEndPuncttrue
\mciteSetBstMidEndSepPunct{\mcitedefaultmidpunct}
{\mcitedefaultendpunct}{\mcitedefaultseppunct}\relax
\EndOfBibitem
\bibitem[Vert(2008)]{Vert2008}
Vert,~J. The Optimal Assignment Kernel Is Not Positive Definite. \emph{arXiv
  preprint arXiv:0801.4061} \textbf{2008}, \emph{abs/0801.4061}\relax
\mciteBstWouldAddEndPuncttrue
\mciteSetBstMidEndSepPunct{\mcitedefaultmidpunct}
{\mcitedefaultendpunct}{\mcitedefaultseppunct}\relax
\EndOfBibitem
\bibitem[Pachauri \latin{et~al.}(2013)Pachauri, Kondor, and
  Singh]{pachauri2013solving}
Pachauri,~D.; Kondor,~R.; Singh,~V. \emph{Adv. Neural. Inf. Process. Syst.};
  2013; pp 1860--1868\relax
\mciteBstWouldAddEndPuncttrue
\mciteSetBstMidEndSepPunct{\mcitedefaultmidpunct}
{\mcitedefaultendpunct}{\mcitedefaultseppunct}\relax
\EndOfBibitem
\bibitem[Bart{\'o}k and Cs{\'a}nyi(2015)Bart{\'o}k, and
  Cs{\'a}nyi]{bartok2015g}
Bart{\'o}k,~A.~P.; Cs{\'a}nyi,~G. Gaussian Approximation Potentials: A Brief
  Tutorial Introduction. \emph{Int. J. Quantum Chem.} \textbf{2015},
  \emph{115}, 1051--1057\relax
\mciteBstWouldAddEndPuncttrue
\mciteSetBstMidEndSepPunct{\mcitedefaultmidpunct}
{\mcitedefaultendpunct}{\mcitedefaultseppunct}\relax
\EndOfBibitem
\bibitem[De \latin{et~al.}(2016)De, Bart{\'o}k, Cs{\'a}nyi, and
  Ceriotti]{de2016comparing}
De,~S.; Bart{\'o}k,~A.~P.; Cs{\'a}nyi,~G.; Ceriotti,~M. Comparing Molecules And
  Solids Across Structural And Alchemical Space. \emph{Phys. Chem. Chem. Phys.}
  \textbf{2016}, \emph{18}, 13754--13769\relax
\mciteBstWouldAddEndPuncttrue
\mciteSetBstMidEndSepPunct{\mcitedefaultmidpunct}
{\mcitedefaultendpunct}{\mcitedefaultseppunct}\relax
\EndOfBibitem
\bibitem[Umeyama(1988)]{Umeyama1988}
Umeyama,~S. An Eigendecomposition Approach To Weighted Graph Matching Problems.
  \emph{IEEE Trans. Pattern Anal. Mach. Intell.} \textbf{1988}, \emph{10},
  695--703\relax
\mciteBstWouldAddEndPuncttrue
\mciteSetBstMidEndSepPunct{\mcitedefaultmidpunct}
{\mcitedefaultendpunct}{\mcitedefaultseppunct}\relax
\EndOfBibitem
\bibitem[Bart{\'o}k \latin{et~al.}(2010)Bart{\'o}k, Payne, Kondor, and
  Cs{\'a}nyi]{bartok2010gaussian}
Bart{\'o}k,~A.~P.; Payne,~M.~C.; Kondor,~R.; Cs{\'a}nyi,~G. Gaussian
  Approximation Potentials: The Accuracy Of Quantum Mechanics, Without The
  Electrons. \emph{Phys. Rev. Lett.} \textbf{2010}, \emph{104}, 136403\relax
\mciteBstWouldAddEndPuncttrue
\mciteSetBstMidEndSepPunct{\mcitedefaultmidpunct}
{\mcitedefaultendpunct}{\mcitedefaultseppunct}\relax
\EndOfBibitem
\bibitem[Bart{\'o}k \latin{et~al.}(2017)Bart{\'o}k, De, Poelking, Bernstein,
  Kermode, Cs{\'a}nyi, and Ceriotti]{bartok2017machine}
Bart{\'o}k,~A.~P.; De,~S.; Poelking,~C.; Bernstein,~N.; Kermode,~J.~R.;
  Cs{\'a}nyi,~G.; Ceriotti,~M. Machine Learning Unifies The Modeling Of
  Materials And Molecules. \emph{Sci. Adv.} \textbf{2017}, \emph{3},
  e1701816\relax
\mciteBstWouldAddEndPuncttrue
\mciteSetBstMidEndSepPunct{\mcitedefaultmidpunct}
{\mcitedefaultendpunct}{\mcitedefaultseppunct}\relax
\EndOfBibitem
\bibitem[Grisafi \latin{et~al.}(2018)Grisafi, Wilkins, Cs{\'a}nyi, and
  Ceriotti]{grisafi2018symmetry}
Grisafi,~A.; Wilkins,~D.~M.; Cs{\'a}nyi,~G.; Ceriotti,~M. Symmetry-adapted
  Machine Learning For Tensorial Properties Of Atomistic Systems. \emph{Phys.
  Rev. Lett.} \textbf{2018}, \emph{120}, 036002\relax
\mciteBstWouldAddEndPuncttrue
\mciteSetBstMidEndSepPunct{\mcitedefaultmidpunct}
{\mcitedefaultendpunct}{\mcitedefaultseppunct}\relax
\EndOfBibitem
\bibitem[Cs{\'a}nyi \latin{et~al.}(2020)Cs{\'a}nyi, Willatt, and
  Ceriotti]{csanyi2020machine}
Cs{\'a}nyi,~G.; Willatt,~M.~J.; Ceriotti,~M. \emph{Machine Learning Meets
  Quantum Physics}; Springer, 2020; pp 99--127\relax
\mciteBstWouldAddEndPuncttrue
\mciteSetBstMidEndSepPunct{\mcitedefaultmidpunct}
{\mcitedefaultendpunct}{\mcitedefaultseppunct}\relax
\EndOfBibitem
\bibitem[Blank \latin{et~al.}(1995)Blank, Brown, Calhoun, and
  Doren]{blank1995neural}
Blank,~T.~B.; Brown,~S.~D.; Calhoun,~A.~W.; Doren,~D.~J. Neural Network Models
  Of Potential Energy Surfaces. \emph{J. Chem. Phys.} \textbf{1995},
  \emph{103}, 4129--4137\relax
\mciteBstWouldAddEndPuncttrue
\mciteSetBstMidEndSepPunct{\mcitedefaultmidpunct}
{\mcitedefaultendpunct}{\mcitedefaultseppunct}\relax
\EndOfBibitem
\bibitem[Brown \latin{et~al.}(1996)Brown, Gibbs, and Clary]{brown1996combining}
Brown,~D.~F.; Gibbs,~M.~N.; Clary,~D.~C. Combining \textit{Ab Initio}
  Computations, Neural Networks, And Diffusion Monte Carlo: An Efficient Method
  To Treat Weakly Bound Molecules. \emph{J. Chem. Phys.} \textbf{1996},
  \emph{105}, 7597--7604\relax
\mciteBstWouldAddEndPuncttrue
\mciteSetBstMidEndSepPunct{\mcitedefaultmidpunct}
{\mcitedefaultendpunct}{\mcitedefaultseppunct}\relax
\EndOfBibitem
\bibitem[Tafeit \latin{et~al.}(1996)Tafeit, Estelberger, Horejsi, Moeller,
  Oettl, Vrecko, and Reibnegger]{tafeit1996neural}
Tafeit,~E.; Estelberger,~W.; Horejsi,~R.; Moeller,~R.; Oettl,~K.; Vrecko,~K.;
  Reibnegger,~G. Neural Networks As A Tool For Compact Representation Of
  \textit{ab Initio} Molecular Potential Energy Surfaces. \emph{J. Mol. Graph.}
  \textbf{1996}, \emph{14}, 12--18\relax
\mciteBstWouldAddEndPuncttrue
\mciteSetBstMidEndSepPunct{\mcitedefaultmidpunct}
{\mcitedefaultendpunct}{\mcitedefaultseppunct}\relax
\EndOfBibitem
\bibitem[No \latin{et~al.}(1997)No, Chang, Kim, Jhon, and
  Scheraga]{no1997description}
No,~K.~T.; Chang,~B.~H.; Kim,~S.~Y.; Jhon,~M.~S.; Scheraga,~H.~A. Description
  Of The Potential Energy Surface Of The Water Dimer With An Artificial Neural
  Network. \emph{Chem. Phys. Lett.} \textbf{1997}, \emph{271}, 152--156\relax
\mciteBstWouldAddEndPuncttrue
\mciteSetBstMidEndSepPunct{\mcitedefaultmidpunct}
{\mcitedefaultendpunct}{\mcitedefaultseppunct}\relax
\EndOfBibitem
\bibitem[Prudente and Neto(1998)Prudente, and Neto]{prudente1998fitting}
Prudente,~F.~V.; Neto,~J.~S. The Fitting Of Potential Energy Surfaces Using
  Neural Networks. Application To The Study Of The Photodissociation Processes.
  \emph{Chem. Phys. Lett.} \textbf{1998}, \emph{287}, 585--589\relax
\mciteBstWouldAddEndPuncttrue
\mciteSetBstMidEndSepPunct{\mcitedefaultmidpunct}
{\mcitedefaultendpunct}{\mcitedefaultseppunct}\relax
\EndOfBibitem
\bibitem[Manzhos and Carrington~Jr(2006)Manzhos, and
  Carrington~Jr]{manzhos2006random}
Manzhos,~S.; Carrington~Jr,~T. A Random-Sampling High Dimensional Model
  Representation Neural Network for Building Potential Energy Surfaces.
  \emph{J. Chem. Phys.} \textbf{2006}, \emph{125}, 084109\relax
\mciteBstWouldAddEndPuncttrue
\mciteSetBstMidEndSepPunct{\mcitedefaultmidpunct}
{\mcitedefaultendpunct}{\mcitedefaultseppunct}\relax
\EndOfBibitem
\bibitem[Manzhos and Carrington~Jr(2007)Manzhos, and
  Carrington~Jr]{manzhos2007using}
Manzhos,~S.; Carrington~Jr,~T. Using Redundant Coordinates to Represent
  Potential Energy Surfaces with Lower-Dimensional Functions. \emph{J. Chem.
  Phys.} \textbf{2007}, \emph{127}, 014103\relax
\mciteBstWouldAddEndPuncttrue
\mciteSetBstMidEndSepPunct{\mcitedefaultmidpunct}
{\mcitedefaultendpunct}{\mcitedefaultseppunct}\relax
\EndOfBibitem
\bibitem[Malshe \latin{et~al.}(2009)Malshe, Narulkar, Raff, Hagan, Bukkapatnam,
  Agrawal, and Komanduri]{malshe2009development}
Malshe,~M.; Narulkar,~R.; Raff,~L.; Hagan,~M.; Bukkapatnam,~S.; Agrawal,~P.;
  Komanduri,~R. Development of Generalized Potential-Energy Surfaces using
  Many-Body Expansions, Neural Networks, and Moiety Energy Approximations.
  \emph{J. Chem. Phys.} \textbf{2009}, \emph{130}, 184102\relax
\mciteBstWouldAddEndPuncttrue
\mciteSetBstMidEndSepPunct{\mcitedefaultmidpunct}
{\mcitedefaultendpunct}{\mcitedefaultseppunct}\relax
\EndOfBibitem
\bibitem[Khorshidi and Peterson(2016)Khorshidi, and Peterson]{khorshidi2016amp}
Khorshidi,~A.; Peterson,~A.~A. Amp: A Modular Approach To Machine Learning In
  Atomistic Simulations. \emph{Comput. Phys. Commun.} \textbf{2016},
  \emph{207}, 310--324\relax
\mciteBstWouldAddEndPuncttrue
\mciteSetBstMidEndSepPunct{\mcitedefaultmidpunct}
{\mcitedefaultendpunct}{\mcitedefaultseppunct}\relax
\EndOfBibitem
\bibitem[Artrith \latin{et~al.}(2017)Artrith, Urban, and
  Ceder]{artrith2017efficient}
Artrith,~N.; Urban,~A.; Ceder,~G. Efficient And Accurate Machine-learning
  Interpolation Of Atomic Energies In Compositions With Many Species.
  \emph{Phys. Rev. B} \textbf{2017}, \emph{96}, 014112\relax
\mciteBstWouldAddEndPuncttrue
\mciteSetBstMidEndSepPunct{\mcitedefaultmidpunct}
{\mcitedefaultendpunct}{\mcitedefaultseppunct}\relax
\EndOfBibitem
\bibitem[Unke and Meuwly(2018)Unke, and Meuwly]{unke2018reactive}
Unke,~O.~T.; Meuwly,~M. A Reactive, Scalable, And Transferable Model For
  Molecular Energies From A Neural Network Approach Based On Local Information.
  \emph{J. Chem. Phys.} \textbf{2018}, \emph{148}, 241708\relax
\mciteBstWouldAddEndPuncttrue
\mciteSetBstMidEndSepPunct{\mcitedefaultmidpunct}
{\mcitedefaultendpunct}{\mcitedefaultseppunct}\relax
\EndOfBibitem
\bibitem[Smith \latin{et~al.}(2017)Smith, Isayev, and Roitberg]{smith2017ani}
Smith,~J.~S.; Isayev,~O.; Roitberg,~A.~E. {ANI-1}: An Extensible Neural Network
  Potential With {DFT} Accuracy At Force Field Computational Cost. \emph{Chem.
  Sci.} \textbf{2017}, \emph{8}, 3192--3203\relax
\mciteBstWouldAddEndPuncttrue
\mciteSetBstMidEndSepPunct{\mcitedefaultmidpunct}
{\mcitedefaultendpunct}{\mcitedefaultseppunct}\relax
\EndOfBibitem
\bibitem[Yao \latin{et~al.}(2018)Yao, Herr, Toth, Mckintyre, and
  Parkhill]{yao2018tensormol}
Yao,~K.; Herr,~J.~E.; Toth,~D.~W.; Mckintyre,~R.; Parkhill,~J. The
  {TensorMol-0.1} Model Chemistry: A Neural Network Augmented With Long-range
  Physics. \emph{Chem. Sci.} \textbf{2018}, \emph{9}, 2261--2269\relax
\mciteBstWouldAddEndPuncttrue
\mciteSetBstMidEndSepPunct{\mcitedefaultmidpunct}
{\mcitedefaultendpunct}{\mcitedefaultseppunct}\relax
\EndOfBibitem
\bibitem[Duvenaud \latin{et~al.}(2015)Duvenaud, Maclaurin, Iparraguirre,
  Bombarell, Hirzel, Aspuru-Guzik, and Adams]{duvenaud2015convolutional}
Duvenaud,~D.~K.; Maclaurin,~D.; Iparraguirre,~J.; Bombarell,~R.; Hirzel,~T.;
  Aspuru-Guzik,~A.; Adams,~R.~P. Convolutional Networks On Graphs For Learning
  Molecular Fingerprints. Adv. Neural. Inf. Process. Syst. 2015; pp
  2224--2232\relax
\mciteBstWouldAddEndPuncttrue
\mciteSetBstMidEndSepPunct{\mcitedefaultmidpunct}
{\mcitedefaultendpunct}{\mcitedefaultseppunct}\relax
\EndOfBibitem
\bibitem[Kearnes \latin{et~al.}(2016)Kearnes, McCloskey, Berndl, Pande, and
  Riley]{kearnes2016molecular}
Kearnes,~S.; McCloskey,~K.; Berndl,~M.; Pande,~V.; Riley,~P. Molecular Graph
  Convolutions: Moving Beyond Fingerprints. \emph{J. Comput. Aided Mol. Des.}
  \textbf{2016}, \emph{30}, 595--608\relax
\mciteBstWouldAddEndPuncttrue
\mciteSetBstMidEndSepPunct{\mcitedefaultmidpunct}
{\mcitedefaultendpunct}{\mcitedefaultseppunct}\relax
\EndOfBibitem
\bibitem[Gilmer \latin{et~al.}(2017)Gilmer, Schoenholz, Riley, Vinyals, and
  Dahl]{gilmer2017neural}
Gilmer,~J.; Schoenholz,~S.~S.; Riley,~P.~F.; Vinyals,~O.; Dahl,~G.~E. Neural
  Message Passing for Quantum Chemistry. International Conference on Machine
  Learning. 2017; pp 1263--1272\relax
\mciteBstWouldAddEndPuncttrue
\mciteSetBstMidEndSepPunct{\mcitedefaultmidpunct}
{\mcitedefaultendpunct}{\mcitedefaultseppunct}\relax
\EndOfBibitem
\bibitem[Scarselli \latin{et~al.}(2008)Scarselli, Gori, Tsoi, Hagenbuchner, and
  Monfardini]{scarselli2008graph}
Scarselli,~F.; Gori,~M.; Tsoi,~A.~C.; Hagenbuchner,~M.; Monfardini,~G. The
  Graph Neural Network Model. \emph{IEEE Trans. Neural Netw.} \textbf{2008},
  \emph{20}, 61--80\relax
\mciteBstWouldAddEndPuncttrue
\mciteSetBstMidEndSepPunct{\mcitedefaultmidpunct}
{\mcitedefaultendpunct}{\mcitedefaultseppunct}\relax
\EndOfBibitem
\bibitem[Sch{\"u}tt \latin{et~al.}(2018)Sch{\"u}tt, Gastegger, Tkatchenko, and
  M{\"u}ller]{schutt2018quantum}
Sch{\"u}tt,~K.~T.; Gastegger,~M.; Tkatchenko,~A.; M{\"u}ller,~K.-R.
  Quantum-chemical Insights From Interpretable Atomistic Neural Networks.
  \emph{arXiv preprint arXiv:1806.10349} \textbf{2018}, \relax
\mciteBstWouldAddEndPunctfalse
\mciteSetBstMidEndSepPunct{\mcitedefaultmidpunct}
{}{\mcitedefaultseppunct}\relax
\EndOfBibitem
\bibitem[Hy \latin{et~al.}(2018)Hy, Trivedi, Pan, Anderson, and
  Kondor]{hy2018predicting}
Hy,~T.~S.; Trivedi,~S.; Pan,~H.; Anderson,~B.~M.; Kondor,~R. Predicting
  Molecular Properties With Covariant Compositional Networks. \emph{J. Chem.
  Phys.} \textbf{2018}, \emph{148}, 241745\relax
\mciteBstWouldAddEndPuncttrue
\mciteSetBstMidEndSepPunct{\mcitedefaultmidpunct}
{\mcitedefaultendpunct}{\mcitedefaultseppunct}\relax
\EndOfBibitem
\bibitem[Anderson \latin{et~al.}(2019)Anderson, Hy, and
  Kondor]{anderson2019cormorant}
Anderson,~B.; Hy,~T.~S.; Kondor,~R. Cormorant: Covariant Molecular Neural
  Networks. Adv. Neural. Inf. Process. Syst. 2019; pp 14537--14546\relax
\mciteBstWouldAddEndPuncttrue
\mciteSetBstMidEndSepPunct{\mcitedefaultmidpunct}
{\mcitedefaultendpunct}{\mcitedefaultseppunct}\relax
\EndOfBibitem
\bibitem[Weiler \latin{et~al.}(2018)Weiler, Geiger, Welling, Boomsma, and
  Cohen]{weiler20183d}
Weiler,~M.; Geiger,~M.; Welling,~M.; Boomsma,~W.; Cohen,~T.~S. 3d Steerable
  {CNNs}: Learning Rotationally Equivariant Features In Volumetric Data. Adv.
  Neural. Inf. Process. Syst. 2018; pp 10381--10392\relax
\mciteBstWouldAddEndPuncttrue
\mciteSetBstMidEndSepPunct{\mcitedefaultmidpunct}
{\mcitedefaultendpunct}{\mcitedefaultseppunct}\relax
\EndOfBibitem
\bibitem[Nair and Hinton(2010)Nair, and Hinton]{nair2010rectified}
Nair,~V.; Hinton,~G.~E. Rectified Linear Units Improve Restricted Boltzmann
  Machines. Int. Conf. Mach. Learn. 2010\relax
\mciteBstWouldAddEndPuncttrue
\mciteSetBstMidEndSepPunct{\mcitedefaultmidpunct}
{\mcitedefaultendpunct}{\mcitedefaultseppunct}\relax
\EndOfBibitem
\bibitem[Behler(2015)]{behler2015constructing}
Behler,~J. Constructing High-dimensional Neural Network Potentials: A Tutorial
  Review. \emph{Int. J. Quantum Chem.} \textbf{2015}, \emph{115},
  1032--1050\relax
\mciteBstWouldAddEndPuncttrue
\mciteSetBstMidEndSepPunct{\mcitedefaultmidpunct}
{\mcitedefaultendpunct}{\mcitedefaultseppunct}\relax
\EndOfBibitem
\bibitem[Gastegger \latin{et~al.}(2018)Gastegger, Schwiedrzik, Bittermann,
  Berzsenyi, and Marquetand]{gastegger2018wacsf}
Gastegger,~M.; Schwiedrzik,~L.; Bittermann,~M.; Berzsenyi,~F.; Marquetand,~P.
  {wACSF} -- Weighted Atom-centered Symmetry Functions As Descriptors In
  Machine Learning Potentials. \emph{J. Chem. Phys.} \textbf{2018}, \emph{148},
  241709\relax
\mciteBstWouldAddEndPuncttrue
\mciteSetBstMidEndSepPunct{\mcitedefaultmidpunct}
{\mcitedefaultendpunct}{\mcitedefaultseppunct}\relax
\EndOfBibitem
\bibitem[Kondor and Lafferty(2002)Kondor, and Lafferty]{kondor2002diffusion}
Kondor,~R.~I.; Lafferty,~J. Diffusion Kernels On Graphs And Other Discrete
  Structures. Proceedings of the 19th international conference on machine
  learning. 2002; pp 315--22\relax
\mciteBstWouldAddEndPuncttrue
\mciteSetBstMidEndSepPunct{\mcitedefaultmidpunct}
{\mcitedefaultendpunct}{\mcitedefaultseppunct}\relax
\EndOfBibitem
\bibitem[Vinyals \latin{et~al.}(2015)Vinyals, Bengio, and
  Kudlur]{vinyals2015order}
Vinyals,~O.; Bengio,~S.; Kudlur,~M. Order Matters: Sequence To Sequence For
  Sets. \emph{arXiv preprint arXiv:1511.06391} \textbf{2015}, \relax
\mciteBstWouldAddEndPunctfalse
\mciteSetBstMidEndSepPunct{\mcitedefaultmidpunct}
{}{\mcitedefaultseppunct}\relax
\EndOfBibitem
\bibitem[Grimme \latin{et~al.}(2010)Grimme, Antony, Ehrlich, and
  Krieg]{grimme2010consistent}
Grimme,~S.; Antony,~J.; Ehrlich,~S.; Krieg,~H. A Consistent And Accurate Ab
  Initio Parametrization Of Density Functional Dispersion Correction {(DFT-D)}
  For The 94 Elements {H-Pu}. \emph{J. Chem. Phys.} \textbf{2010}, \emph{132},
  154104\relax
\mciteBstWouldAddEndPuncttrue
\mciteSetBstMidEndSepPunct{\mcitedefaultmidpunct}
{\mcitedefaultendpunct}{\mcitedefaultseppunct}\relax
\EndOfBibitem
\bibitem[Unke and Meuwly(2019)Unke, and Meuwly]{unke2019sn2}
Unke,~O.; Meuwly,~M. S$_{\mathrm{N}}$2 reactions data set.
  \url{http://doi.org/10.5281/zenodo.2605341}, 2019\relax
\mciteBstWouldAddEndPuncttrue
\mciteSetBstMidEndSepPunct{\mcitedefaultmidpunct}
{\mcitedefaultendpunct}{\mcitedefaultseppunct}\relax
\EndOfBibitem
\bibitem[Monticelli and Tieleman(2013)Monticelli, and
  Tieleman]{monticelli2013force}
Monticelli,~L.; Tieleman,~D.~P. \emph{Biomolecular simulations}; Springer,
  2013; pp 197--213\relax
\mciteBstWouldAddEndPuncttrue
\mciteSetBstMidEndSepPunct{\mcitedefaultmidpunct}
{\mcitedefaultendpunct}{\mcitedefaultseppunct}\relax
\EndOfBibitem
\bibitem[Friesner(2005)]{friesner2005ab}
Friesner,~R.~A. Ab Initio Quantum Chemistry: Methodology And Applications.
  \emph{Proc. Natl. Acad. Sci. USA} \textbf{2005}, \emph{102}, 6648--6653\relax
\mciteBstWouldAddEndPuncttrue
\mciteSetBstMidEndSepPunct{\mcitedefaultmidpunct}
{\mcitedefaultendpunct}{\mcitedefaultseppunct}\relax
\EndOfBibitem
\bibitem[Brickel \latin{et~al.}(2019)Brickel, Das, Unke, Turan, and
  Meuwly]{brickel2019reactive}
Brickel,~S.; Das,~A.~K.; Unke,~O.~T.; Turan,~H.~T.; Meuwly,~M. Reactive
  Molecular Dynamics For The {[Cl--CH3--Br]-} Reaction In The Gas Phase And In
  Solution: A Comparative Study Using Empirical And Neural Network Force
  Fields. \emph{Electron. Struct.} \textbf{2019}, \emph{1}, 024002\relax
\mciteBstWouldAddEndPuncttrue
\mciteSetBstMidEndSepPunct{\mcitedefaultmidpunct}
{\mcitedefaultendpunct}{\mcitedefaultseppunct}\relax
\EndOfBibitem
\bibitem[Sauceda \latin{et~al.}(2020)Sauceda, Gastegger, Chmiela, M\"uller, and
  Tkatchenko]{sauceda2020molecular}
Sauceda,~H.~E.; Gastegger,~M.; Chmiela,~S.; M\"uller,~K.-R.; Tkatchenko,~A.
  Molecular Force Fields With Gradient-domain Machine Learning (gdml):
  Comparison And Synergies With Classical Force Fields. \emph{J. Chem. Phys.}
  \textbf{2020}, \emph{153}, 124109\relax
\mciteBstWouldAddEndPuncttrue
\mciteSetBstMidEndSepPunct{\mcitedefaultmidpunct}
{\mcitedefaultendpunct}{\mcitedefaultseppunct}\relax
\EndOfBibitem
\bibitem[Szalay \latin{et~al.}(2012)Szalay, Muller, Gidofalvi, Lischka, and
  Shepard]{szalay2012multiconfiguration}
Szalay,~P.~G.; Muller,~T.; Gidofalvi,~G.; Lischka,~H.; Shepard,~R.
  Multiconfiguration Self-consistent Field And Multireference Configuration
  Interaction Methods And Applications. \emph{Chem. Rev.} \textbf{2012},
  \emph{112}, 108--181\relax
\mciteBstWouldAddEndPuncttrue
\mciteSetBstMidEndSepPunct{\mcitedefaultmidpunct}
{\mcitedefaultendpunct}{\mcitedefaultseppunct}\relax
\EndOfBibitem
\bibitem[Jia \latin{et~al.}(2020)Jia, Wang, Chen, Lu, Liu, Lin, Car, Zhang,
  \latin{et~al.} others]{jia2020pushing}
Jia,~W.; Wang,~H.; Chen,~M.; Lu,~D.; Liu,~J.; Lin,~L.; Car,~R.; Zhang,~L.,
  \latin{et~al.}  Pushing The Limit Of Molecular Dynamics With Ab Initio
  Accuracy To 100 Million Atoms With Machine Learning. \emph{arXiv preprint
  arXiv:2005.00223} \textbf{2020}, \relax
\mciteBstWouldAddEndPunctfalse
\mciteSetBstMidEndSepPunct{\mcitedefaultmidpunct}
{}{\mcitedefaultseppunct}\relax
\EndOfBibitem
\bibitem[Sanders and Saxe(2017)Sanders, and Saxe]{sanders2017garbage}
Sanders,~H.; Saxe,~J. Garbage In, Garbage Out: How Purportedly Great Ml Models
  Can Be Screwed Up By Bad Data. Proceedings of Blackhat. 2017\relax
\mciteBstWouldAddEndPuncttrue
\mciteSetBstMidEndSepPunct{\mcitedefaultmidpunct}
{\mcitedefaultendpunct}{\mcitedefaultseppunct}\relax
\EndOfBibitem
\bibitem[Blum \latin{et~al.}(2009)Blum, Gehrke, Hanke, Havu, Havu, Ren, Reuter,
  and Scheffler]{FHIaims2009}
Blum,~V.; Gehrke,~R.; Hanke,~F.; Havu,~P.; Havu,~V.; Ren,~X.; Reuter,~K.;
  Scheffler,~M. Ab Initio Molecular Simulations With Numeric Atom-centered
  Orbitals. \emph{Comput. Phys. Commun.} \textbf{2009}, \emph{180},
  2175--2196\relax
\mciteBstWouldAddEndPuncttrue
\mciteSetBstMidEndSepPunct{\mcitedefaultmidpunct}
{\mcitedefaultendpunct}{\mcitedefaultseppunct}\relax
\EndOfBibitem
\bibitem[Cs{\'a}nyi \latin{et~al.}(2004)Cs{\'a}nyi, Albaret, Payne, and
  De~Vita]{csanyi2004learn}
Cs{\'a}nyi,~G.; Albaret,~T.; Payne,~M.; De~Vita,~A. ``Learn On The Fly'': A
  Hybrid Classical And Quantum-mechanical Molecular Dynamics Simulation.
  \emph{Phys. Rev. Lett.} \textbf{2004}, \emph{93}, 175503\relax
\mciteBstWouldAddEndPuncttrue
\mciteSetBstMidEndSepPunct{\mcitedefaultmidpunct}
{\mcitedefaultendpunct}{\mcitedefaultseppunct}\relax
\EndOfBibitem
\bibitem[Seung \latin{et~al.}(1992)Seung, Opper, and
  Sompolinsky]{seung1992query}
Seung,~H.~S.; Opper,~M.; Sompolinsky,~H. Query By Committee. Proceedings of the
  Fifth Annual Workshop on Computational Learning Theory. 1992; pp
  287--294\relax
\mciteBstWouldAddEndPuncttrue
\mciteSetBstMidEndSepPunct{\mcitedefaultmidpunct}
{\mcitedefaultendpunct}{\mcitedefaultseppunct}\relax
\EndOfBibitem
\bibitem[Morawietz \latin{et~al.}(2012)Morawietz, Sharma, and
  Behler]{morawietz2012neural}
Morawietz,~T.; Sharma,~V.; Behler,~J. A Neural Network Potential-energy Surface
  For The Water Dimer Based On Environment-dependent Atomic Energies And
  Charges. \emph{J. Chem. Phys.} \textbf{2012}, \emph{136}, 064103\relax
\mciteBstWouldAddEndPuncttrue
\mciteSetBstMidEndSepPunct{\mcitedefaultmidpunct}
{\mcitedefaultendpunct}{\mcitedefaultseppunct}\relax
\EndOfBibitem
\bibitem[Smith \latin{et~al.}(2018)Smith, Nebgen, Lubbers, Isayev, and
  Roitberg]{smith2018less}
Smith,~J.~S.; Nebgen,~B.; Lubbers,~N.; Isayev,~O.; Roitberg,~A.~E. Less Is
  More: Sampling Chemical Space With Active Learning. \emph{J. Chem. Phys.}
  \textbf{2018}, \emph{148}, 241733\relax
\mciteBstWouldAddEndPuncttrue
\mciteSetBstMidEndSepPunct{\mcitedefaultmidpunct}
{\mcitedefaultendpunct}{\mcitedefaultseppunct}\relax
\EndOfBibitem
\bibitem[Srivastava \latin{et~al.}(2014)Srivastava, Hinton, Krizhevsky,
  Sutskever, and Salakhutdinov]{srivastava2014dropout}
Srivastava,~N.; Hinton,~G.; Krizhevsky,~A.; Sutskever,~I.; Salakhutdinov,~R.
  Dropout: A Simple Way To Prevent Neural Networks From Overfitting. \emph{J.
  Mach. Learn. Res.} \textbf{2014}, \emph{15}, 1929--1958\relax
\mciteBstWouldAddEndPuncttrue
\mciteSetBstMidEndSepPunct{\mcitedefaultmidpunct}
{\mcitedefaultendpunct}{\mcitedefaultseppunct}\relax
\EndOfBibitem
\bibitem[Gal and Ghahramani(2016)Gal, and Ghahramani]{gal2016dropout}
Gal,~Y.; Ghahramani,~Z. Dropout As A Bayesian Approximation: Representing Model
  Uncertainty In Deep Learning. International Conference on Machine Learning.
  2016; pp 1050--1059\relax
\mciteBstWouldAddEndPuncttrue
\mciteSetBstMidEndSepPunct{\mcitedefaultmidpunct}
{\mcitedefaultendpunct}{\mcitedefaultseppunct}\relax
\EndOfBibitem
\bibitem[Li \latin{et~al.}(2015)Li, Kermode, and De~Vita]{li2015molecular}
Li,~Z.; Kermode,~J.~R.; De~Vita,~A. Molecular Dynamics With On-the-fly Machine
  Learning Of Quantum-mechanical Forces. \emph{Phys. Rev. Lett.} \textbf{2015},
  \emph{114}, 096405\relax
\mciteBstWouldAddEndPuncttrue
\mciteSetBstMidEndSepPunct{\mcitedefaultmidpunct}
{\mcitedefaultendpunct}{\mcitedefaultseppunct}\relax
\EndOfBibitem
\bibitem[Gastegger and Marquetand(2020)Gastegger, and
  Marquetand]{gastegger2020molecular}
Gastegger,~M.; Marquetand,~P. \emph{Machine Learning Meets Quantum Physics};
  Springer, 2020; pp 233--252\relax
\mciteBstWouldAddEndPuncttrue
\mciteSetBstMidEndSepPunct{\mcitedefaultmidpunct}
{\mcitedefaultendpunct}{\mcitedefaultseppunct}\relax
\EndOfBibitem
\bibitem[Shapeev \latin{et~al.}(2020)Shapeev, Gubaev, Tsymbalov, and
  Podryabinkin]{shapeev2020active}
Shapeev,~A.; Gubaev,~K.; Tsymbalov,~E.; Podryabinkin,~E. \emph{Machine Learning
  Meets Quantum Physics}; Springer, 2020; pp 309--329\relax
\mciteBstWouldAddEndPuncttrue
\mciteSetBstMidEndSepPunct{\mcitedefaultmidpunct}
{\mcitedefaultendpunct}{\mcitedefaultseppunct}\relax
\EndOfBibitem
\bibitem[Barducci \latin{et~al.}(2011)Barducci, Bonomi, and
  Parrinello]{barducci2011metadynamics}
Barducci,~A.; Bonomi,~M.; Parrinello,~M. Metadynamics. \emph{Wiley Interdiscip.
  Rev. Comput. Mol. Sci.} \textbf{2011}, \emph{1}, 826--843\relax
\mciteBstWouldAddEndPuncttrue
\mciteSetBstMidEndSepPunct{\mcitedefaultmidpunct}
{\mcitedefaultendpunct}{\mcitedefaultseppunct}\relax
\EndOfBibitem
\bibitem[Herr \latin{et~al.}(2018)Herr, Yao, McIntyre, Toth, and
  Parkhill]{herr2018metadynamics}
Herr,~J.~E.; Yao,~K.; McIntyre,~R.; Toth,~D.~W.; Parkhill,~J. Metadynamics For
  Training Neural Network Model Chemistries: A Competitive Assessment. \emph{J.
  Chem. Phys.} \textbf{2018}, \emph{148}, 241710\relax
\mciteBstWouldAddEndPuncttrue
\mciteSetBstMidEndSepPunct{\mcitedefaultmidpunct}
{\mcitedefaultendpunct}{\mcitedefaultseppunct}\relax
\EndOfBibitem
\bibitem[Sugiyama \latin{et~al.}(2007)Sugiyama, Krauledat, and
  M{\"u}ller]{sugiyama2007covariate}
Sugiyama,~M.; Krauledat,~M.; M{\"u}ller,~K.-R. Covariate Shift Adaptation By
  Importance Weighted Cross Validation. \emph{J. Mach. Learn. Res.}
  \textbf{2007}, \emph{8}, 985--1005\relax
\mciteBstWouldAddEndPuncttrue
\mciteSetBstMidEndSepPunct{\mcitedefaultmidpunct}
{\mcitedefaultendpunct}{\mcitedefaultseppunct}\relax
\EndOfBibitem
\bibitem[Lemm \latin{et~al.}(2011)Lemm, Blankertz, Dickhaus, and
  M{\"u}ller]{lemm2011introduction}
Lemm,~S.; Blankertz,~B.; Dickhaus,~T.; M{\"u}ller,~K.-R. Introduction To
  Machine Learning For Brain Imaging. \emph{Neuroimage} \textbf{2011},
  \emph{56}, 387--399\relax
\mciteBstWouldAddEndPuncttrue
\mciteSetBstMidEndSepPunct{\mcitedefaultmidpunct}
{\mcitedefaultendpunct}{\mcitedefaultseppunct}\relax
\EndOfBibitem
\bibitem[Nesterov(1983)]{nesterov1983method}
Nesterov,~Y.~E. A Method For Solving The Convex Programming Problem With
  Convergence Rate $\mathcal{O}(1/k^2)$. Proc. USSR Acad. Sci. 1983; pp
  543--547\relax
\mciteBstWouldAddEndPuncttrue
\mciteSetBstMidEndSepPunct{\mcitedefaultmidpunct}
{\mcitedefaultendpunct}{\mcitedefaultseppunct}\relax
\EndOfBibitem
\bibitem[Qian(1999)]{qian1999momentum}
Qian,~N. On The Momentum Term In Gradient Descent Learning Algorithms.
  \emph{Neural Netw.} \textbf{1999}, \emph{12}, 145--151\relax
\mciteBstWouldAddEndPuncttrue
\mciteSetBstMidEndSepPunct{\mcitedefaultmidpunct}
{\mcitedefaultendpunct}{\mcitedefaultseppunct}\relax
\EndOfBibitem
\bibitem[Duchi \latin{et~al.}(2011)Duchi, Hazan, and Singer]{duchi2011adaptive}
Duchi,~J.; Hazan,~E.; Singer,~Y. Adaptive Subgradient Methods For Online
  Learning And Stochastic Optimization. \emph{J. Mach. Learn. Res.}
  \textbf{2011}, \emph{12}\relax
\mciteBstWouldAddEndPuncttrue
\mciteSetBstMidEndSepPunct{\mcitedefaultmidpunct}
{\mcitedefaultendpunct}{\mcitedefaultseppunct}\relax
\EndOfBibitem
\bibitem[Zeiler(2012)]{zeiler2012adadelta}
Zeiler,~M.~D. Adadelta: An Adaptive Learning Rate Method. \emph{arXiv preprint
  arXiv:1212.5701} \textbf{2012}, \relax
\mciteBstWouldAddEndPunctfalse
\mciteSetBstMidEndSepPunct{\mcitedefaultmidpunct}
{}{\mcitedefaultseppunct}\relax
\EndOfBibitem
\bibitem[Ruder(2016)]{ruder2016overview}
Ruder,~S. An Overview Of Gradient Descent Optimization Algorithms. \emph{arXiv
  preprint arXiv:1609.04747} \textbf{2016}, \relax
\mciteBstWouldAddEndPunctfalse
\mciteSetBstMidEndSepPunct{\mcitedefaultmidpunct}
{}{\mcitedefaultseppunct}\relax
\EndOfBibitem
\bibitem[Kingma and Ba(2015)Kingma, and Ba]{kingma2015adam}
Kingma,~D.~P.; Ba,~J. Adam: A Method For Stochastic Optimization. International
  Conference on Learning Representations. 2015; pp 1--13\relax
\mciteBstWouldAddEndPuncttrue
\mciteSetBstMidEndSepPunct{\mcitedefaultmidpunct}
{\mcitedefaultendpunct}{\mcitedefaultseppunct}\relax
\EndOfBibitem
\bibitem[Huber(1992)]{huber1992robust}
Huber,~P.~J. \emph{Breakthroughs in statistics}; Springer, 1992; pp
  492--518\relax
\mciteBstWouldAddEndPuncttrue
\mciteSetBstMidEndSepPunct{\mcitedefaultmidpunct}
{\mcitedefaultendpunct}{\mcitedefaultseppunct}\relax
\EndOfBibitem
\bibitem[Barron(2019)]{barron2019general}
Barron,~J.~T. A General And Adaptive Robust Loss Function. Proceedings of the
  IEEE Conference on Computer Vision and Pattern Recognition. 2019; pp
  4331--4339\relax
\mciteBstWouldAddEndPuncttrue
\mciteSetBstMidEndSepPunct{\mcitedefaultmidpunct}
{\mcitedefaultendpunct}{\mcitedefaultseppunct}\relax
\EndOfBibitem
\bibitem[Chmiela(2019)]{chmiela2019towards}
Chmiela,~S. \emph{Towards Exact Molecular Dynamics Simulations With Invariant
  Machine-learned Models}; Technische Universit{\"a}t Berlin (Germany),
  2019\relax
\mciteBstWouldAddEndPuncttrue
\mciteSetBstMidEndSepPunct{\mcitedefaultmidpunct}
{\mcitedefaultendpunct}{\mcitedefaultseppunct}\relax
\EndOfBibitem
\bibitem[Christensen and von Lilienfeld(2020)Christensen, and von
  Lilienfeld]{christensen2020on}
Christensen,~A.~S.; von Lilienfeld,~O.~A. On The Role Of Gradients For Machine
  Learning Of Molecular Energies And Forces. \emph{Machine Learning: Science
  and Technology} \textbf{2020}, \emph{1}, 045018\relax
\mciteBstWouldAddEndPuncttrue
\mciteSetBstMidEndSepPunct{\mcitedefaultmidpunct}
{\mcitedefaultendpunct}{\mcitedefaultseppunct}\relax
\EndOfBibitem
\bibitem[Meyer \latin{et~al.}(2020)Meyer, Weichselbaum, and
  Hauser]{meyer2020machine}
Meyer,~R.; Weichselbaum,~M.; Hauser,~A.~W. Machine Learning Approaches Toward
  Orbital-free Density Functional Theory: Simultaneous Training On The Kinetic
  Energy Density Functional And Its Functional Derivative. \emph{J. Chem.
  Theory Comput.} \textbf{2020}, \emph{16}, 5685--5694\relax
\mciteBstWouldAddEndPuncttrue
\mciteSetBstMidEndSepPunct{\mcitedefaultmidpunct}
{\mcitedefaultendpunct}{\mcitedefaultseppunct}\relax
\EndOfBibitem
\bibitem[Bergstra and Bengio(2012)Bergstra, and Bengio]{bergstra2012random}
Bergstra,~J.; Bengio,~Y. Random Search For Hyper-parameter Optimization.
  \emph{J. Mach. Learn. Res.} \textbf{2012}, \emph{13}, 281--305\relax
\mciteBstWouldAddEndPuncttrue
\mciteSetBstMidEndSepPunct{\mcitedefaultmidpunct}
{\mcitedefaultendpunct}{\mcitedefaultseppunct}\relax
\EndOfBibitem
\bibitem[Prechelt(1998)]{prechelt1998early}
Prechelt,~L. \emph{Neural Networks: Tricks of the trade}; Springer, 1998; pp
  55--69\relax
\mciteBstWouldAddEndPuncttrue
\mciteSetBstMidEndSepPunct{\mcitedefaultmidpunct}
{\mcitedefaultendpunct}{\mcitedefaultseppunct}\relax
\EndOfBibitem
\bibitem[Sch{\"u}tt \latin{et~al.}(2018)Sch{\"u}tt, Kessel, Gastegger, Nicoli,
  Tkatchenko, and M\"uller]{schutt2018schnetpack}
Sch{\"u}tt,~K.; Kessel,~P.; Gastegger,~M.; Nicoli,~K.; Tkatchenko,~A.;
  M\"uller,~K.-R. {SchNetPack}: A Deep Learning Toolbox For Atomistic Systems.
  \emph{J. Chem. Theory Comput.} \textbf{2018}, \emph{15}, 448--455\relax
\mciteBstWouldAddEndPuncttrue
\mciteSetBstMidEndSepPunct{\mcitedefaultmidpunct}
{\mcitedefaultendpunct}{\mcitedefaultseppunct}\relax
\EndOfBibitem
\bibitem[Larsen \latin{et~al.}(2017)Larsen, Mortensen, Blomqvist, Castelli,
  Christensen, Du\l{}ak, Friis, Groves, Hammer, Hargus, Hermes, Jennings,
  Jensen, Kermode, Kitchin, Kolsbjerg, Kubal, Kaasbjerg, Lysgaard, Maronsson,
  Maxson, Olsen, Pastewka, Peterson, Rostgaard, Schi\o{}tz, Sch\"{u}tt,
  Strange, Thygesen, Vegge, Vilhelmsen, Walter, Zeng, and Jacobsen]{ase17}
Larsen,~A.~H. \latin{et~al.}  The Atomic Simulation Environment -- A Python
  Library For Working With Atoms. \emph{J. Phys. Condens. Matter}
  \textbf{2017}, \emph{29}, 273002\relax
\mciteBstWouldAddEndPuncttrue
\mciteSetBstMidEndSepPunct{\mcitedefaultmidpunct}
{\mcitedefaultendpunct}{\mcitedefaultseppunct}\relax
\EndOfBibitem
\bibitem[Kapil \latin{et~al.}(2018)Kapil, Rossi, Marsalek, Petraglia, Litman,
  Spura, Cheng, Cuzzocrea, Meißner, Wilkins, Juda, Bienvenue, Fang, Kessler,
  Poltavsky, Vandenbrande, Wieme, Corminboeuf, K\"{u}hne, avnd Thomas
  E.~Markland, Richardson, Tkatchenko, Tribello, Speybroeck, and
  Ceriotti]{ipiv2}
Kapil,~V. \latin{et~al.}  {i-PI} 2.0: A Universal Force Engine For Advanced
  Molecular Simulations. \emph{Comput. Phys. Commun.} \textbf{2018},
  \emph{236}, 214--223\relax
\mciteBstWouldAddEndPuncttrue
\mciteSetBstMidEndSepPunct{\mcitedefaultmidpunct}
{\mcitedefaultendpunct}{\mcitedefaultseppunct}\relax
\EndOfBibitem
\bibitem[Paszke \latin{et~al.}(2019)Paszke, Gross, Massa, Lerer, Bradbury,
  Chanan, Killeen, Lin, Gimelshein, Antiga, \latin{et~al.}
  others]{paszke2019pytorch}
Paszke,~A.; Gross,~S.; Massa,~F.; Lerer,~A.; Bradbury,~J.; Chanan,~G.;
  Killeen,~T.; Lin,~Z.; Gimelshein,~N.; Antiga,~L., \latin{et~al.}  {PyTorch}:
  An Imperative Style, High-performance Deep Learning Library. Adv. Neural.
  Inf. Process. Syst. 2019; pp 8026--8037\relax
\mciteBstWouldAddEndPuncttrue
\mciteSetBstMidEndSepPunct{\mcitedefaultmidpunct}
{\mcitedefaultendpunct}{\mcitedefaultseppunct}\relax
\EndOfBibitem
\bibitem[Ramakrishnan \latin{et~al.}(2014)Ramakrishnan, Dral, Rupp, and
  Von~Lilienfeld]{ramakrishnan2014quantum}
Ramakrishnan,~R.; Dral,~P.~O.; Rupp,~M.; Von~Lilienfeld,~O.~A. Quantum
  Chemistry Structures And Properties Of 134 Kilo Molecules. \emph{Sci. Data}
  \textbf{2014}, \emph{1}, 1--7\relax
\mciteBstWouldAddEndPuncttrue
\mciteSetBstMidEndSepPunct{\mcitedefaultmidpunct}
{\mcitedefaultendpunct}{\mcitedefaultseppunct}\relax
\EndOfBibitem
\bibitem[Abadi \latin{et~al.}(2016)Abadi, Barham, Chen, Chen, Davis, Dean,
  Devin, Ghemawat, Irving, Isard, \latin{et~al.} others]{abadi2016tensorflow}
Abadi,~M.; Barham,~P.; Chen,~J.; Chen,~Z.; Davis,~A.; Dean,~J.; Devin,~M.;
  Ghemawat,~S.; Irving,~G.; Isard,~M., \latin{et~al.}  Tensorflow: A System For
  Large-scale Machine Learning. 12th $\{$USENIX$\}$ symposium on operating
  systems design and implementation ($\{$OSDI$\}$ 16). 2016; pp 265--283\relax
\mciteBstWouldAddEndPuncttrue
\mciteSetBstMidEndSepPunct{\mcitedefaultmidpunct}
{\mcitedefaultendpunct}{\mcitedefaultseppunct}\relax
\EndOfBibitem
\bibitem[Plimpton(1993)]{plimpton1993fast}
Plimpton,~S. \emph{Fast Parallel Algorithms For Short-range Molecular
  Dynamics}; 1993\relax
\mciteBstWouldAddEndPuncttrue
\mciteSetBstMidEndSepPunct{\mcitedefaultmidpunct}
{\mcitedefaultendpunct}{\mcitedefaultseppunct}\relax
\EndOfBibitem
\bibitem[Wang \latin{et~al.}(2018)Wang, Zhang, Han, and Weinan]{wang2018deepmd}
Wang,~H.; Zhang,~L.; Han,~J.; Weinan,~E. {DeePMD-kit}: A Deep Learning Package
  For Many-body Potential Energy Representation And Molecular Dynamics.
  \emph{Comput. Phys. Commun.} \textbf{2018}, \emph{228}, 178--184\relax
\mciteBstWouldAddEndPuncttrue
\mciteSetBstMidEndSepPunct{\mcitedefaultmidpunct}
{\mcitedefaultendpunct}{\mcitedefaultseppunct}\relax
\EndOfBibitem
\bibitem[Himanen \latin{et~al.}(2020)Himanen, J{\"a}ger, Morooka,
  Federici~Canova, Ranawat, Gao, Rinke, and Foster]{dscribe}
Himanen,~L.; J{\"a}ger,~M. O.~J.; Morooka,~E.~V.; Federici~Canova,~F.;
  Ranawat,~Y.~S.; Gao,~D.~Z.; Rinke,~P.; Foster,~A.~S. {{DScribe}: Library Of
  Descriptors For Machine Learning In Materials Science}. \emph{Comput. Phys.
  Commun.} \textbf{2020}, \emph{247}, 106949\relax
\mciteBstWouldAddEndPuncttrue
\mciteSetBstMidEndSepPunct{\mcitedefaultmidpunct}
{\mcitedefaultendpunct}{\mcitedefaultseppunct}\relax
\EndOfBibitem
\bibitem[Christensen \latin{et~al.}(2017)Christensen, Faber, Huang, Bratholm,
  Tkatchenko, M\"uller, and von Lilienfeld]{christensen2017qml}
Christensen,~A.; Faber,~F.; Huang,~B.; Bratholm,~L.; Tkatchenko,~A.;
  M\"uller,~K.-R.; von Lilienfeld,~O. {QML}: A Python Toolkit For Quantum
  Machine Learning. \emph{URL https://github.com/qmlcode/qml} \textbf{2017},
  \relax
\mciteBstWouldAddEndPunctfalse
\mciteSetBstMidEndSepPunct{\mcitedefaultmidpunct}
{}{\mcitedefaultseppunct}\relax
\EndOfBibitem
\bibitem[Unke and Meuwly(2017)Unke, and Meuwly]{unke2017toolkit}
Unke,~O.~T.; Meuwly,~M. Toolkit For The Construction Of Reproducing
  Kernel-based Representations Of Data: Application To Multidimensional
  Potential Energy Surfaces. \emph{J. Chem. Inf. Model.} \textbf{2017},
  \emph{57}, 1923--1931\relax
\mciteBstWouldAddEndPuncttrue
\mciteSetBstMidEndSepPunct{\mcitedefaultmidpunct}
{\mcitedefaultendpunct}{\mcitedefaultseppunct}\relax
\EndOfBibitem
\bibitem[Boltzmann(1898)]{boltzmann2017vorlesungen}
Boltzmann,~L. \emph{Vorlesungen {\"u}ber Gastheorie: 2. Teil}; Leipzig: J. A.
  Barth, 1898\relax
\mciteBstWouldAddEndPuncttrue
\mciteSetBstMidEndSepPunct{\mcitedefaultmidpunct}
{\mcitedefaultendpunct}{\mcitedefaultseppunct}\relax
\EndOfBibitem
\bibitem[Botu \latin{et~al.}(2017)Botu, Batra, Chapman, and
  Ramprasad]{botu2017machine}
Botu,~V.; Batra,~R.; Chapman,~J.; Ramprasad,~R. Machine Learning Force Fields:
  Construction, Validation, And Outlook. \emph{J. Phys. Chem. C} \textbf{2017},
  \emph{121}, 511--522\relax
\mciteBstWouldAddEndPuncttrue
\mciteSetBstMidEndSepPunct{\mcitedefaultmidpunct}
{\mcitedefaultendpunct}{\mcitedefaultseppunct}\relax
\EndOfBibitem
\bibitem[Behler(2017)]{behler2017first}
Behler,~J. First Principles Neural Network Potentials For Reactive Simulations
  Of Large Molecular And Condensed Systems. \emph{Angew. Chem. Int. Ed.}
  \textbf{2017}, \emph{56}, 12828--12840\relax
\mciteBstWouldAddEndPuncttrue
\mciteSetBstMidEndSepPunct{\mcitedefaultmidpunct}
{\mcitedefaultendpunct}{\mcitedefaultseppunct}\relax
\EndOfBibitem
\bibitem[Deringer \latin{et~al.}(2019)Deringer, Caro, and
  Cs{\'a}nyi]{deringer2019machine}
Deringer,~V.~L.; Caro,~M.~A.; Cs{\'a}nyi,~G. Machine Learning Interatomic
  Potentials As Emerging Tools For Materials Science. \emph{Adv. Mater.}
  \textbf{2019}, \emph{31}, 1902765\relax
\mciteBstWouldAddEndPuncttrue
\mciteSetBstMidEndSepPunct{\mcitedefaultmidpunct}
{\mcitedefaultendpunct}{\mcitedefaultseppunct}\relax
\EndOfBibitem
\bibitem[Morawietz \latin{et~al.}(2016)Morawietz, Singraber, Dellago, and
  Behler]{morawietz2016van}
Morawietz,~T.; Singraber,~A.; Dellago,~C.; Behler,~J. How Van Der Waals
  Interactions Determine The Unique Properties Of Water. \emph{Proc. Natl.
  Acad. Sci. U.S.A.} \textbf{2016}, \emph{113}, 8368--8373\relax
\mciteBstWouldAddEndPuncttrue
\mciteSetBstMidEndSepPunct{\mcitedefaultmidpunct}
{\mcitedefaultendpunct}{\mcitedefaultseppunct}\relax
\EndOfBibitem
\bibitem[Andrade \latin{et~al.}(2020)Andrade, Ko, Zhang, Car, and
  Selloni]{andrade2020free}
Andrade,~M. F.~C.; Ko,~H.-Y.; Zhang,~L.; Car,~R.; Selloni,~A. Free Energy Of
  Proton Transfer At The Water--{TiO} 2 Interface From Ab Initio Deep Potential
  Molecular Dynamics. \emph{Chem. Sci.} \textbf{2020}, \emph{11},
  2335--2341\relax
\mciteBstWouldAddEndPuncttrue
\mciteSetBstMidEndSepPunct{\mcitedefaultmidpunct}
{\mcitedefaultendpunct}{\mcitedefaultseppunct}\relax
\EndOfBibitem
\bibitem[Deringer and Cs{\'a}nyi(2017)Deringer, and
  Cs{\'a}nyi]{deringer2017machine}
Deringer,~V.~L.; Cs{\'a}nyi,~G. Machine Learning Based Interatomic Potential
  For Amorphous Carbon. \emph{Phys. Rev. B} \textbf{2017}, \emph{95},
  094203\relax
\mciteBstWouldAddEndPuncttrue
\mciteSetBstMidEndSepPunct{\mcitedefaultmidpunct}
{\mcitedefaultendpunct}{\mcitedefaultseppunct}\relax
\EndOfBibitem
\bibitem[Behler \latin{et~al.}(2008)Behler, Marto{\v{n}}{\'a}k, Donadio, and
  Parrinello]{behler2008pressure}
Behler,~J.; Marto{\v{n}}{\'a}k,~R.; Donadio,~D.; Parrinello,~M.
  Pressure-induced Phase Transitions In Silicon Studied By Neural Network-based
  Metadynamics Simulations. \emph{Phys. Status Solidi B} \textbf{2008},
  \emph{245}, 2618--2629\relax
\mciteBstWouldAddEndPuncttrue
\mciteSetBstMidEndSepPunct{\mcitedefaultmidpunct}
{\mcitedefaultendpunct}{\mcitedefaultseppunct}\relax
\EndOfBibitem
\bibitem[Bart{\'o}k \latin{et~al.}(2018)Bart{\'o}k, Kermode, Bernstein, and
  Cs{\'a}nyi]{bartok2018machine}
Bart{\'o}k,~A.~P.; Kermode,~J.; Bernstein,~N.; Cs{\'a}nyi,~G. Machine Learning
  A General-purpose Interatomic Potential For Silicon. \emph{Phys. Rev. X}
  \textbf{2018}, \emph{8}, 041048\relax
\mciteBstWouldAddEndPuncttrue
\mciteSetBstMidEndSepPunct{\mcitedefaultmidpunct}
{\mcitedefaultendpunct}{\mcitedefaultseppunct}\relax
\EndOfBibitem
\bibitem[Deringer \latin{et~al.}(2018)Deringer, Bernstein, Bart{\'o}k, Cliffe,
  Kerber, Marbella, Grey, Elliott, and Cs{\'a}nyi]{deringer2018realistic}
Deringer,~V.~L.; Bernstein,~N.; Bart{\'o}k,~A.~P.; Cliffe,~M.~J.;
  Kerber,~R.~N.; Marbella,~L.~E.; Grey,~C.~P.; Elliott,~S.~R.; Cs{\'a}nyi,~G.
  Realistic Atomistic Structure Of Amorphous Silicon From
  Machine-learning-driven Molecular Dynamics. \emph{J. Phys. Chem. Lett.}
  \textbf{2018}, \emph{9}, 2879--2885\relax
\mciteBstWouldAddEndPuncttrue
\mciteSetBstMidEndSepPunct{\mcitedefaultmidpunct}
{\mcitedefaultendpunct}{\mcitedefaultseppunct}\relax
\EndOfBibitem
\bibitem[Bonati and Parrinello(2018)Bonati, and Parrinello]{bonati2018silicon}
Bonati,~L.; Parrinello,~M. Silicon Liquid Structure And Crystal Nucleation From
  Ab Initio Deep Metadynamics. \emph{Phys. Rev. Lett.} \textbf{2018},
  \emph{121}, 265701\relax
\mciteBstWouldAddEndPuncttrue
\mciteSetBstMidEndSepPunct{\mcitedefaultmidpunct}
{\mcitedefaultendpunct}{\mcitedefaultseppunct}\relax
\EndOfBibitem
\bibitem[Unke \latin{et~al.}(2016)Unke, Castro-Palacio, Bemish, and
  Meuwly]{unke2016collision}
Unke,~O.~T.; Castro-Palacio,~J.~C.; Bemish,~R.~J.; Meuwly,~M. Collision-induced
  Rotational Excitation In {N$_2^+$($^2\Sigma_g^+$, $v=0$)--Ar}: Comparison Of
  Computations And Experiment. \emph{J. Chem. Phys.} \textbf{2016}, \emph{144},
  224307\relax
\mciteBstWouldAddEndPuncttrue
\mciteSetBstMidEndSepPunct{\mcitedefaultmidpunct}
{\mcitedefaultendpunct}{\mcitedefaultseppunct}\relax
\EndOfBibitem
\bibitem[Denis-Alpizar \latin{et~al.}(2017)Denis-Alpizar, Unke, Bemish, and
  Meuwly]{denis2017quantum}
Denis-Alpizar,~O.; Unke,~O.~T.; Bemish,~R.~J.; Meuwly,~M. Quantum And
  Quasiclassical Trajectory Studies Of Rotational Relaxation In {Ar--N$_2^+$}
  Collisions. \emph{Phys. Chem. Chem. Phys.} \textbf{2017}, \emph{19},
  27945--27951\relax
\mciteBstWouldAddEndPuncttrue
\mciteSetBstMidEndSepPunct{\mcitedefaultmidpunct}
{\mcitedefaultendpunct}{\mcitedefaultseppunct}\relax
\EndOfBibitem
\bibitem[Lu \latin{et~al.}(2020)Lu, Li, and Guo]{lu2020comprehensive}
Lu,~D.; Li,~J.; Guo,~H. Comprehensive Investigations Of The
  {Cl+CH$_3$OH$\rightarrow$HCl+CH$_3$O/CH$_2$OH} Reaction: Validation Of
  Experiment And Dynamic Insights. \emph{CCS Chem.} \textbf{2020}, \emph{2},
  882--894\relax
\mciteBstWouldAddEndPuncttrue
\mciteSetBstMidEndSepPunct{\mcitedefaultmidpunct}
{\mcitedefaultendpunct}{\mcitedefaultseppunct}\relax
\EndOfBibitem
\bibitem[Sweeny \latin{et~al.}(2020)Sweeny, Pan, Kassem, Sawyer, Ard, Shuman,
  Viggiano, Brickel, Unke, Upadhyay, \latin{et~al.} others]{sweeny2020thermal}
Sweeny,~B.~C.; Pan,~H.; Kassem,~A.; Sawyer,~J.~C.; Ard,~S.~G.; Shuman,~N.~S.;
  Viggiano,~A.~A.; Brickel,~S.; Unke,~O.~T.; Upadhyay,~M., \latin{et~al.}
  Thermal Activation Of Methane By {MgO+}: Temperature Dependent Kinetics,
  Reactive Molecular Dynamics Simulations And Statistical Modeling. \emph{Phys.
  Chem. Chem. Phys.} \textbf{2020}, \emph{22}, 8913--8923\relax
\mciteBstWouldAddEndPuncttrue
\mciteSetBstMidEndSepPunct{\mcitedefaultmidpunct}
{\mcitedefaultendpunct}{\mcitedefaultseppunct}\relax
\EndOfBibitem
\bibitem[K{\"a}ser \latin{et~al.}(2020)K{\"a}ser, Unke, and
  Meuwly]{kaser2020isomerization}
K{\"a}ser,~S.; Unke,~O.~T.; Meuwly,~M. Isomerization And Decomposition
  Reactions Of Acetaldehyde Relevant To Atmospheric Processes From Dynamics
  Simulations On Neural Network-based Potential Energy Surfaces. \emph{J. Chem.
  Phys.} \textbf{2020}, \emph{152}, 214304\relax
\mciteBstWouldAddEndPuncttrue
\mciteSetBstMidEndSepPunct{\mcitedefaultmidpunct}
{\mcitedefaultendpunct}{\mcitedefaultseppunct}\relax
\EndOfBibitem
\bibitem[Rivero \latin{et~al.}(2019)Rivero, Unke, Meuwly, and
  Willitsch]{rivero2019reactive}
Rivero,~U.; Unke,~O.~T.; Meuwly,~M.; Willitsch,~S. Reactive Atomistic
  Simulations Of Diels-alder Reactions: The Importance Of Molecular Rotations.
  \emph{J. Chem. Phys.} \textbf{2019}, \emph{151}, 104301\relax
\mciteBstWouldAddEndPuncttrue
\mciteSetBstMidEndSepPunct{\mcitedefaultmidpunct}
{\mcitedefaultendpunct}{\mcitedefaultseppunct}\relax
\EndOfBibitem
\bibitem[Liu \latin{et~al.}(2018)Liu, Zhou, Zhou, Zhang, Luo, Guo, and
  Jiang]{liu2018constructing}
Liu,~Q.; Zhou,~X.; Zhou,~L.; Zhang,~Y.; Luo,~X.; Guo,~H.; Jiang,~B.
  Constructing High-dimensional Neural Network Potential Energy Surfaces For
  Gas--surface Scattering And Reactions. \emph{J. Phys. Chem. C} \textbf{2018},
  \emph{122}, 1761--1769\relax
\mciteBstWouldAddEndPuncttrue
\mciteSetBstMidEndSepPunct{\mcitedefaultmidpunct}
{\mcitedefaultendpunct}{\mcitedefaultseppunct}\relax
\EndOfBibitem
\bibitem[Sch\"utt \latin{et~al.}(2018)Sch\"utt, Sauceda, Kindermans,
  Tkatchenko, and M\"uller]{schutt2018}
Sch\"utt,~K.~T.; Sauceda,~H.~E.; Kindermans,~P.-J.; Tkatchenko,~A.;
  M\"uller,~K.-R. {SchNet} -- {A} Deep Learning Architecture For Molecules And
  Materials. \emph{J. Chem. Phys.} \textbf{2018}, \emph{148}, 241722\relax
\mciteBstWouldAddEndPuncttrue
\mciteSetBstMidEndSepPunct{\mcitedefaultmidpunct}
{\mcitedefaultendpunct}{\mcitedefaultseppunct}\relax
\EndOfBibitem
\bibitem[Hellstr\"om \latin{et~al.}(2018)Hellstr\"om, Ceriotti, and
  Behler]{hellstrom2018nuclear}
Hellstr\"om,~M.; Ceriotti,~M.; Behler,~J. Nuclear Quantum Effects In Sodium
  Hydroxide Solutions From Neural Network Molecular Dynamics Simulations.
  \emph{J. Phys. Chem. B} \textbf{2018}, \emph{122}, 10158--10171\relax
\mciteBstWouldAddEndPuncttrue
\mciteSetBstMidEndSepPunct{\mcitedefaultmidpunct}
{\mcitedefaultendpunct}{\mcitedefaultseppunct}\relax
\EndOfBibitem
\bibitem[Chen \latin{et~al.}(2018)Chen, Liu, Fang, Dral, and Cui]{chen2018deep}
Chen,~W.-K.; Liu,~X.-Y.; Fang,~W.-H.; Dral,~P.~O.; Cui,~G. Deep Learning For
  Nonadiabatic Excited-state Dynamics. \emph{J. Phys. Chem. Lett.}
  \textbf{2018}, \emph{9}, 6702--6708\relax
\mciteBstWouldAddEndPuncttrue
\mciteSetBstMidEndSepPunct{\mcitedefaultmidpunct}
{\mcitedefaultendpunct}{\mcitedefaultseppunct}\relax
\EndOfBibitem
\bibitem[Westermayr \latin{et~al.}(2019)Westermayr, Gastegger, Menger, Mai,
  Gonz{\'a}lez, and Marquetand]{westermayr2019machine}
Westermayr,~J.; Gastegger,~M.; Menger,~M.~F.; Mai,~S.; Gonz{\'a}lez,~L.;
  Marquetand,~P. Machine Learning Enables Long Time Scale Molecular
  Photodynamics Simulations. \emph{Chem. Sci.} \textbf{2019}, \emph{10},
  8100--8107\relax
\mciteBstWouldAddEndPuncttrue
\mciteSetBstMidEndSepPunct{\mcitedefaultmidpunct}
{\mcitedefaultendpunct}{\mcitedefaultseppunct}\relax
\EndOfBibitem
\bibitem[Westermayr \latin{et~al.}(2020)Westermayr, Gastegger, and
  Marquetand]{westermayr2020combining}
Westermayr,~J.; Gastegger,~M.; Marquetand,~P. Combining {SchNet} and {SHARC}:
  The {SchNarc} Machine Learning Approach For Excited-state Dynamics. \emph{J.
  Phys. Chem. Lett.} \textbf{2020}, \emph{11}, 3828--3834\relax
\mciteBstWouldAddEndPuncttrue
\mciteSetBstMidEndSepPunct{\mcitedefaultmidpunct}
{\mcitedefaultendpunct}{\mcitedefaultseppunct}\relax
\EndOfBibitem
\bibitem[Raimbault \latin{et~al.}(2019)Raimbault, Grisafi, Ceriotti, and
  Rossi]{raimbault2019using}
Raimbault,~N.; Grisafi,~A.; Ceriotti,~M.; Rossi,~M. Using Gaussian Process
  Regression To Simulate The Vibrational Raman Spectra Of Molecular Crystals.
  \emph{New J. Phys.} \textbf{2019}, \emph{21}, 105001\relax
\mciteBstWouldAddEndPuncttrue
\mciteSetBstMidEndSepPunct{\mcitedefaultmidpunct}
{\mcitedefaultendpunct}{\mcitedefaultseppunct}\relax
\EndOfBibitem
\bibitem[Sommers \latin{et~al.}(2020)Sommers, Andrade, Zhang, Wang, and
  Car]{sommers2020raman}
Sommers,~G.~M.; Andrade,~M. F.~C.; Zhang,~L.; Wang,~H.; Car,~R. Raman Spectrum
  And Polarizability Of Liquid Water From Deep Neural Networks. \emph{Phys.
  Chem. Chem. Phys.} \textbf{2020}, \emph{22}, 10592--10602\relax
\mciteBstWouldAddEndPuncttrue
\mciteSetBstMidEndSepPunct{\mcitedefaultmidpunct}
{\mcitedefaultendpunct}{\mcitedefaultseppunct}\relax
\EndOfBibitem
\bibitem[Hermann \latin{et~al.}(2017)Hermann, DiStasio~Jr, and
  Tkatchenko]{hermann2017first}
Hermann,~J.; DiStasio~Jr,~R.~A.; Tkatchenko,~A. First-principles Models For Van
  Der Waals Interactions In Molecules And Materials: Concepts, Theory, And
  Applications. \emph{Chem. Rev.} \textbf{2017}, \emph{117}, 4714--4758\relax
\mciteBstWouldAddEndPuncttrue
\mciteSetBstMidEndSepPunct{\mcitedefaultmidpunct}
{\mcitedefaultendpunct}{\mcitedefaultseppunct}\relax
\EndOfBibitem
\bibitem[Guillot(2002)]{guillot2002reappraisal}
Guillot,~B. A Reappraisal Of What We Have Learnt During Three Decades Of
  Computer Simulations On Water. \emph{J. Mol. Liq.} \textbf{2002}, \emph{101},
  219--260\relax
\mciteBstWouldAddEndPuncttrue
\mciteSetBstMidEndSepPunct{\mcitedefaultmidpunct}
{\mcitedefaultendpunct}{\mcitedefaultseppunct}\relax
\EndOfBibitem
\bibitem[M{\o}ller and Plesset(1934)M{\o}ller, and Plesset]{moller1934note}
M{\o}ller,~C.; Plesset,~M.~S. Note On An Approximation Treatment For
  Many-electron Systems. \emph{Phys. Rev.} \textbf{1934}, \emph{46}, 618\relax
\mciteBstWouldAddEndPuncttrue
\mciteSetBstMidEndSepPunct{\mcitedefaultmidpunct}
{\mcitedefaultendpunct}{\mcitedefaultseppunct}\relax
\EndOfBibitem
\bibitem[Dunning~Jr(1989)]{dunning1989gaussian}
Dunning~Jr,~T.~H. Gaussian Basis Sets For Use In Correlated Molecular
  Calculations. I. The Atoms Boron Through Neon And Hydrogen. \emph{J. Chem.
  Phys.} \textbf{1989}, \emph{90}, 1007--1023\relax
\mciteBstWouldAddEndPuncttrue
\mciteSetBstMidEndSepPunct{\mcitedefaultmidpunct}
{\mcitedefaultendpunct}{\mcitedefaultseppunct}\relax
\EndOfBibitem
\bibitem[Unke \latin{et~al.}(2019)Unke, Brickel, and Meuwly]{unke2019sampling}
Unke,~O.~T.; Brickel,~S.; Meuwly,~M. Sampling Reactive Regions In Phase Space
  By Following The Minimum Dynamic Path. \emph{J. Chem. Phys.} \textbf{2019},
  \emph{150}, 074107\relax
\mciteBstWouldAddEndPuncttrue
\mciteSetBstMidEndSepPunct{\mcitedefaultmidpunct}
{\mcitedefaultendpunct}{\mcitedefaultseppunct}\relax
\EndOfBibitem
\bibitem[Manzhos and Carrington(2020)Manzhos, and
  Carrington]{manzhos2020neural}
Manzhos,~S.; Carrington,~T. Neural Network Potential Energy Surfaces For Small
  Molecules And Reactions. \emph{Chem. Rev.} \textbf{2020}, \relax
\mciteBstWouldAddEndPunctfalse
\mciteSetBstMidEndSepPunct{\mcitedefaultmidpunct}
{}{\mcitedefaultseppunct}\relax
\EndOfBibitem
\bibitem[Spura \latin{et~al.}(2015)Spura, Elgabarty, and
  Kühne]{Spura_CC-PIMD_PCCP2015}
Spura,~T.; Elgabarty,~H.; Kühne,~T.~D. “On-the-fly” Coupled Cluster
  Path-integral Molecular Dynamics: Impact Of Nuclear Quantum Effects On The
  Protonated Water Dimer. \emph{Phys. Chem. Chem. Phys.} \textbf{2015},
  \emph{17}, 14355--14359\relax
\mciteBstWouldAddEndPuncttrue
\mciteSetBstMidEndSepPunct{\mcitedefaultmidpunct}
{\mcitedefaultendpunct}{\mcitedefaultseppunct}\relax
\EndOfBibitem
\bibitem[Wang \latin{et~al.}(2017)Wang, Fried, and
  Markland]{Wang_Ketosteroid-PIMD_JPCB2017}
Wang,~L.; Fried,~S.~D.; Markland,~T.~E. Proton Network Flexibility Enables
  Robustness And Large Electric Fields In The Ketosteroid Isomerase Active
  Site. \emph{J. Phys. Chem. B} \textbf{2017}, \emph{121}, 9807--9815\relax
\mciteBstWouldAddEndPuncttrue
\mciteSetBstMidEndSepPunct{\mcitedefaultmidpunct}
{\mcitedefaultendpunct}{\mcitedefaultseppunct}\relax
\EndOfBibitem
\bibitem[Litman \latin{et~al.}(2019)Litman, Richardson, Kumagai, and
  Rossi]{Litman_Porphycen-PIMD_JACS2019}
Litman,~Y.; Richardson,~J.~O.; Kumagai,~T.; Rossi,~M. Elucidating The Nuclear
  Quantum Dynamics Of Intramolecular Double Hydrogen Transfer In Porphycene.
  \emph{J. Am. Chem. Soc.} \textbf{2019}, \emph{141}, 2526--2534\relax
\mciteBstWouldAddEndPuncttrue
\mciteSetBstMidEndSepPunct{\mcitedefaultmidpunct}
{\mcitedefaultendpunct}{\mcitedefaultseppunct}\relax
\EndOfBibitem
\bibitem[Schran \latin{et~al.}(2018)Schran, Brieuc, and
  Marx]{Schran_Zundel-PIMD_JCTC2018}
Schran,~C.; Brieuc,~F.; Marx,~D. Converged Colored Noise Path Integral
  Molecular Dynamics Study Of The Zundel Cation Down To Ultralow Temperatures
  At Coupled Cluster Accuracy. \emph{J. Chem. Theory Comput.} \textbf{2018},
  \emph{14}, 5068--5078\relax
\mciteBstWouldAddEndPuncttrue
\mciteSetBstMidEndSepPunct{\mcitedefaultmidpunct}
{\mcitedefaultendpunct}{\mcitedefaultseppunct}\relax
\EndOfBibitem
\bibitem[Heisenberg(1985)]{heisenberg1985anschaulichen}
Heisenberg,~W. \emph{Original Scientific Papers Wissenschaftliche
  Originalarbeiten}; Springer, 1985; pp 478--504\relax
\mciteBstWouldAddEndPuncttrue
\mciteSetBstMidEndSepPunct{\mcitedefaultmidpunct}
{\mcitedefaultendpunct}{\mcitedefaultseppunct}\relax
\EndOfBibitem
\bibitem[Merchant \latin{et~al.}(1973)Merchant, Srivastava, and
  Pandey]{Merchant1973}
Merchant,~H.~D.; Srivastava,~K.~K.; Pandey,~H.~D. Equations Of State And
  Thermal Expansion Of Alkali Halides. \emph{Crit. Rev. Solid State}
  \textbf{1973}, \emph{3}, 451--504\relax
\mciteBstWouldAddEndPuncttrue
\mciteSetBstMidEndSepPunct{\mcitedefaultmidpunct}
{\mcitedefaultendpunct}{\mcitedefaultseppunct}\relax
\EndOfBibitem
\bibitem[Kirchner \latin{et~al.}(2000)Kirchner, Heinke, Hommel, Domagala, and
  Leszczynski]{Kirchner2000}
Kirchner,~V.; Heinke,~H.; Hommel,~D.; Domagala,~J.~Z.; Leszczynski,~M. Thermal
  Expansion Of Bulk And Homoepitaxial {GaN}. \emph{Appl. Phys. Lett.}
  \textbf{2000}, \emph{77}, 1434--1436\relax
\mciteBstWouldAddEndPuncttrue
\mciteSetBstMidEndSepPunct{\mcitedefaultmidpunct}
{\mcitedefaultendpunct}{\mcitedefaultseppunct}\relax
\EndOfBibitem
\bibitem[Jiang \latin{et~al.}(2004)Jiang, Liu, Huang, and
  Hwang]{NanotubeCTE2004}
Jiang,~H.; Liu,~B.; Huang,~Y.; Hwang,~K.~C. Thermal Expansion Of Single Wall
  Carbon Nanotubes. \emph{ASME. J. Eng. Mater. Technol.} \textbf{2004},
  \emph{126}, 265--270\relax
\mciteBstWouldAddEndPuncttrue
\mciteSetBstMidEndSepPunct{\mcitedefaultmidpunct}
{\mcitedefaultendpunct}{\mcitedefaultseppunct}\relax
\EndOfBibitem
\bibitem[Kim \latin{et~al.}(2018)Kim, Hellman, Herriman, Smith, Lin, Shulumba,
  Niedziela, Li, Abernathy, and Fultz]{ThermalExpSi2018}
Kim,~D.~S.; Hellman,~O.; Herriman,~J.; Smith,~H.~L.; Lin,~J. Y.~Y.;
  Shulumba,~N.; Niedziela,~J.~L.; Li,~C.~W.; Abernathy,~D.~L.; Fultz,~B.
  Nuclear Quantum Effect With Pure Anharmonicity And The Anomalous Thermal
  Expansion Of Silicon. \emph{Proc. Natl. Acad. Sci. U.S.A.} \textbf{2018},
  \emph{115}, 1992--1997\relax
\mciteBstWouldAddEndPuncttrue
\mciteSetBstMidEndSepPunct{\mcitedefaultmidpunct}
{\mcitedefaultendpunct}{\mcitedefaultseppunct}\relax
\EndOfBibitem
\bibitem[Hermet \latin{et~al.}(2015)Hermet, Koza, Ritter, Reibel, and
  Viennois]{ThermalExpZnSb2015}
Hermet,~P.; Koza,~M.~M.; Ritter,~C.; Reibel,~C.; Viennois,~R. Origin Of The
  Highly Anisotropic Thermal Expansion Of The Semiconducting Znsb And Relations
  With Its Thermoelectric Applications. \emph{RSC Adv.} \textbf{2015},
  \emph{5}, 87118--87131\relax
\mciteBstWouldAddEndPuncttrue
\mciteSetBstMidEndSepPunct{\mcitedefaultmidpunct}
{\mcitedefaultendpunct}{\mcitedefaultseppunct}\relax
\EndOfBibitem
\bibitem[Poltavsky \latin{et~al.}(2018)Poltavsky, Zheng, Mortazavi, and
  Tkatchenko]{Poltavsky-graphene}
Poltavsky,~I.; Zheng,~L.; Mortazavi,~M.; Tkatchenko,~A. Quantum Tunneling Of
  Thermal Protons Through Pristine Graphene. \emph{J. Chem. Phys.}
  \textbf{2018}, \emph{148}, 204707\relax
\mciteBstWouldAddEndPuncttrue
\mciteSetBstMidEndSepPunct{\mcitedefaultmidpunct}
{\mcitedefaultendpunct}{\mcitedefaultseppunct}\relax
\EndOfBibitem
\bibitem[Markland and Ceriotti(2018)Markland, and Ceriotti]{NQEreview2018}
Markland,~T.~E.; Ceriotti,~M. Nuclear Quantum Effects Enter The Mainstream.
  \emph{Nat. Rev. Chem.} \textbf{2018}, \emph{2}, 0109\relax
\mciteBstWouldAddEndPuncttrue
\mciteSetBstMidEndSepPunct{\mcitedefaultmidpunct}
{\mcitedefaultendpunct}{\mcitedefaultseppunct}\relax
\EndOfBibitem
\bibitem[Freitas \latin{et~al.}(2018)Freitas, Asta, and
  Bulatov]{Freitas_disloc-PIMD_npjCM2018}
Freitas,~R.; Asta,~M.; Bulatov,~V.~V. Quantum Effects On Dislocation Motion
  From Ring-polymer Molecular Dynamics. \emph{npj Computational Materials}
  \textbf{2018}, \emph{4}, 55\relax
\mciteBstWouldAddEndPuncttrue
\mciteSetBstMidEndSepPunct{\mcitedefaultmidpunct}
{\mcitedefaultendpunct}{\mcitedefaultseppunct}\relax
\EndOfBibitem
\bibitem[Poltavsky \latin{et~al.}(2018)Poltavsky, DiStasio, and
  Tkatchenko]{pPI_2018}
Poltavsky,~I.; DiStasio,~R.~A.; Tkatchenko,~A. Perturbed Path Integrals In
  Imaginary Time: Efficiently Modeling Nuclear Quantum Effects In Molecules And
  Materials. \emph{J. Chem. Phys.} \textbf{2018}, \emph{148}, 102325\relax
\mciteBstWouldAddEndPuncttrue
\mciteSetBstMidEndSepPunct{\mcitedefaultmidpunct}
{\mcitedefaultendpunct}{\mcitedefaultseppunct}\relax
\EndOfBibitem
\bibitem[Zhu and Nakamura(1993)Zhu, and Nakamura]{zhu1993two}
Zhu,~C.; Nakamura,~H. The Two-state Linear Curve Crossing Problems Revisited.
  Iii. Analytical Approximations For Stokes Constant And Scattering Matrix:
  Nonadiabatic Tunneling Case. \emph{J. Chem. Phys.} \textbf{1993}, \emph{98},
  6208--6222\relax
\mciteBstWouldAddEndPuncttrue
\mciteSetBstMidEndSepPunct{\mcitedefaultmidpunct}
{\mcitedefaultendpunct}{\mcitedefaultseppunct}\relax
\EndOfBibitem
\bibitem[Westermayr and Marquetand(0)Westermayr, and
  Marquetand]{westermayr2020machine2}
Westermayr,~J.; Marquetand,~P. Machine Learning For Electronically Excited
  States Of Molecules. \emph{Chem. Rev.} \textbf{0}, \emph{0}, null\relax
\mciteBstWouldAddEndPuncttrue
\mciteSetBstMidEndSepPunct{\mcitedefaultmidpunct}
{\mcitedefaultendpunct}{\mcitedefaultseppunct}\relax
\EndOfBibitem
\bibitem[Blum and Reymond(2009)Blum, and Reymond]{blum2009970}
Blum,~L.~C.; Reymond,~J.-L. 970 Million Druglike Small Molecules For Virtual
  Screening In The Chemical Universe Database {GDB-13}. \emph{J. Am. Chem.
  Soc.} \textbf{2009}, \emph{131}, 8732--8733\relax
\mciteBstWouldAddEndPuncttrue
\mciteSetBstMidEndSepPunct{\mcitedefaultmidpunct}
{\mcitedefaultendpunct}{\mcitedefaultseppunct}\relax
\EndOfBibitem
\bibitem[Ruddigkeit \latin{et~al.}(2012)Ruddigkeit, Van~Deursen, Blum, and
  Reymond]{ruddigkeit2012enumeration}
Ruddigkeit,~L.; Van~Deursen,~R.; Blum,~L.~C.; Reymond,~J.-L. Enumeration Of 166
  Billion Organic Small Molecules In The Chemical Universe Database {GDB-17}.
  \emph{J. Chem. Inf. Model.} \textbf{2012}, \emph{52}, 2864--2875\relax
\mciteBstWouldAddEndPuncttrue
\mciteSetBstMidEndSepPunct{\mcitedefaultmidpunct}
{\mcitedefaultendpunct}{\mcitedefaultseppunct}\relax
\EndOfBibitem
\bibitem[Ghasemi \latin{et~al.}(2015)Ghasemi, Hofstetter, Saha, and
  Goedecker]{ghasemi2015interatomic}
Ghasemi,~S.~A.; Hofstetter,~A.; Saha,~S.; Goedecker,~S. Interatomic Potentials
  For Ionic Systems With Density Functional Accuracy Based On Charge Densities
  Obtained By A Neural Network. \emph{Phys. Rev. B} \textbf{2015}, \emph{92},
  045131\relax
\mciteBstWouldAddEndPuncttrue
\mciteSetBstMidEndSepPunct{\mcitedefaultmidpunct}
{\mcitedefaultendpunct}{\mcitedefaultseppunct}\relax
\EndOfBibitem
\bibitem[Ko \latin{et~al.}(2020)Ko, Finkler, Goedecker, and
  Behler]{ko2020fourth}
Ko,~T.~W.; Finkler,~J.~A.; Goedecker,~S.; Behler,~J. A Fourth-Generation
  High-Dimensional Neural Network Potential With Accurate Electrostatics
  Including Non-Local Charge Transfer. \emph{arXiv preprint arXiv:2009.06484}
  \textbf{2020}, \relax
\mciteBstWouldAddEndPunctfalse
\mciteSetBstMidEndSepPunct{\mcitedefaultmidpunct}
{}{\mcitedefaultseppunct}\relax
\EndOfBibitem
\bibitem[Huang and von Lilienfeld(2017)Huang, and von Lilienfeld]{huang2017dna}
Huang,~B.; von Lilienfeld,~O.~A. The ``{DNA}'' Of Chemistry: Scalable Quantum
  Machine Learning With ``amons''. \emph{arXiv preprint arXiv:1707.04146}
  \textbf{2017}, \relax
\mciteBstWouldAddEndPunctfalse
\mciteSetBstMidEndSepPunct{\mcitedefaultmidpunct}
{}{\mcitedefaultseppunct}\relax
\EndOfBibitem
\bibitem[Huang and von Lilienfeld(2020)Huang, and von
  Lilienfeld]{huang2020quantum}
Huang,~B.; von Lilienfeld,~O.~A. Quantum Machine Learning Using
  Atom-in-molecule-based Fragments Selected On The Fly. \emph{Nat. Chem.}
  \textbf{2020}, \emph{12}, 945–--951\relax
\mciteBstWouldAddEndPuncttrue
\mciteSetBstMidEndSepPunct{\mcitedefaultmidpunct}
{\mcitedefaultendpunct}{\mcitedefaultseppunct}\relax
\EndOfBibitem
\bibitem[Dewar and Thiel(1977)Dewar, and Thiel]{dewar1977ground}
Dewar,~M.~J.; Thiel,~W. Ground States Of Molecules. 38. The {MNDO} Method.
  Approximations And Parameters. \emph{J. Am. Chem. Soc.} \textbf{1977},
  \emph{99}, 4899--4907\relax
\mciteBstWouldAddEndPuncttrue
\mciteSetBstMidEndSepPunct{\mcitedefaultmidpunct}
{\mcitedefaultendpunct}{\mcitedefaultseppunct}\relax
\EndOfBibitem
\bibitem[Hirshfeld(1977)]{hirshfeld1977bonded}
Hirshfeld,~F.~L. Bonded-atom Fragments For Describing Molecular Charge
  Densities. \emph{Theor. Chim. Acta} \textbf{1977}, \emph{44}, 129--138\relax
\mciteBstWouldAddEndPuncttrue
\mciteSetBstMidEndSepPunct{\mcitedefaultmidpunct}
{\mcitedefaultendpunct}{\mcitedefaultseppunct}\relax
\EndOfBibitem
\bibitem[Wilkins \latin{et~al.}(2019)Wilkins, Grisafi, Yang, Lao, DiStasio, and
  Ceriotti]{wilkins2019accurate}
Wilkins,~D.~M.; Grisafi,~A.; Yang,~Y.; Lao,~K.~U.; DiStasio,~R.~A.;
  Ceriotti,~M. Accurate Molecular Polarizabilities With Coupled Cluster Theory
  and Machine Learning. \emph{Proc. Natl. Acad. Sci. USA} \textbf{2019},
  \emph{116}, 3401--3406\relax
\mciteBstWouldAddEndPuncttrue
\mciteSetBstMidEndSepPunct{\mcitedefaultmidpunct}
{\mcitedefaultendpunct}{\mcitedefaultseppunct}\relax
\EndOfBibitem
\bibitem[Morawietz and Behler(2013)Morawietz, and Behler]{morawietz2013density}
Morawietz,~T.; Behler,~J. A Density-Functional Theory-Based Neural Network
  Potential for Water Clusters Including van der {Waals} Corrections. \emph{J.
  Phys. Chem. A} \textbf{2013}, \emph{117}, 7356--7366\relax
\mciteBstWouldAddEndPuncttrue
\mciteSetBstMidEndSepPunct{\mcitedefaultmidpunct}
{\mcitedefaultendpunct}{\mcitedefaultseppunct}\relax
\EndOfBibitem
\bibitem[Uteva \latin{et~al.}(2017)Uteva, Graham, Wilkinson, and
  Wheatley]{uteva2017interpolation}
Uteva,~E.; Graham,~R.~S.; Wilkinson,~R.~D.; Wheatley,~R.~J. Interpolation Of
  Intermolecular Potentials Using Gaussian Processes. \emph{J. Chem. Phys.}
  \textbf{2017}, \emph{147}, 161706\relax
\mciteBstWouldAddEndPuncttrue
\mciteSetBstMidEndSepPunct{\mcitedefaultmidpunct}
{\mcitedefaultendpunct}{\mcitedefaultseppunct}\relax
\EndOfBibitem
\bibitem[Li \latin{et~al.}(2018)Li, Collins, Tanha, Gordon, and
  Yaron]{li2018density}
Li,~H.; Collins,~C.; Tanha,~M.; Gordon,~G.~J.; Yaron,~D.~J. A Density
  Functional Tight Binding Layer For Deep Learning Of Chemical Hamiltonians.
  \emph{J. Chem. Theory Comput.} \textbf{2018}, \emph{14}, 5764--5776\relax
\mciteBstWouldAddEndPuncttrue
\mciteSetBstMidEndSepPunct{\mcitedefaultmidpunct}
{\mcitedefaultendpunct}{\mcitedefaultseppunct}\relax
\EndOfBibitem
\bibitem[Zubatyuk \latin{et~al.}(2019)Zubatyuk, Nebgen, Lubbers, Smith,
  Zubatyuk, Zhou, Koh, Barros, Isayev, and Tretiak]{zubatyuk2019machine}
Zubatyuk,~T.; Nebgen,~B.; Lubbers,~N.; Smith,~J.~S.; Zubatyuk,~R.; Zhou,~G.;
  Koh,~C.; Barros,~K.; Isayev,~O.; Tretiak,~S. Machine Learned H\"uckel Theory:
  Interfacing Physics and Deep Neural Networks. \emph{arXiv preprint
  arXiv:1909.12963} \textbf{2019}, \relax
\mciteBstWouldAddEndPunctfalse
\mciteSetBstMidEndSepPunct{\mcitedefaultmidpunct}
{}{\mcitedefaultseppunct}\relax
\EndOfBibitem
\bibitem[Lahey and Rowley(2020)Lahey, and Rowley]{lahey2020simulating}
Lahey,~S.-L.~J.; Rowley,~C.~N. Simulating Protein--Ligand Binding With Neural
  Network Potentials. \emph{Chem. Sci.} \textbf{2020}, \emph{11},
  2362--2368\relax
\mciteBstWouldAddEndPuncttrue
\mciteSetBstMidEndSepPunct{\mcitedefaultmidpunct}
{\mcitedefaultendpunct}{\mcitedefaultseppunct}\relax
\EndOfBibitem
\bibitem[Gastegger \latin{et~al.}(2020)Gastegger, Sch{\"u}tt, and
  M{\"u}ller]{gastegger2020machine}
Gastegger,~M.; Sch{\"u}tt,~K.~T.; M{\"u}ller,~K.-R. Machine Learning Of Solvent
  Effects On Molecular Spectra And Reactions. \emph{arXiv preprint
  arXiv:2010.14942} \textbf{2020}, \relax
\mciteBstWouldAddEndPunctfalse
\mciteSetBstMidEndSepPunct{\mcitedefaultmidpunct}
{}{\mcitedefaultseppunct}\relax
\EndOfBibitem
\bibitem[Zhang \latin{et~al.}(2018)Zhang, Shen, and Yang]{zhang2018solvation}
Zhang,~P.; Shen,~L.; Yang,~W. Solvation free energy calculations with quantum
  mechanics/molecular mechanics and machine learning models. \emph{J. Phys.
  Chem. B} \textbf{2018}, \emph{123}, 901--908\relax
\mciteBstWouldAddEndPuncttrue
\mciteSetBstMidEndSepPunct{\mcitedefaultmidpunct}
{\mcitedefaultendpunct}{\mcitedefaultseppunct}\relax
\EndOfBibitem
\bibitem[B{\"o}selt \latin{et~al.}(2020)B{\"o}selt, Th{\"u}rlemann, and
  Riniker]{boselt2020machine}
B{\"o}selt,~L.; Th{\"u}rlemann,~M.; Riniker,~S. Machine Learning In {QM/MM}
  Molecular Dynamics Simulations Of Condensed-Phase Systems. \emph{arXiv
  preprint arXiv:2010.11610} \textbf{2020}, \relax
\mciteBstWouldAddEndPunctfalse
\mciteSetBstMidEndSepPunct{\mcitedefaultmidpunct}
{}{\mcitedefaultseppunct}\relax
\EndOfBibitem
\bibitem[Nilsson(1982)]{nilsson1982principles}
Nilsson,~N.~J. \emph{Principles of Artificial Intelligence}; Springer Science
  \& Business Media, 1982\relax
\mciteBstWouldAddEndPuncttrue
\mciteSetBstMidEndSepPunct{\mcitedefaultmidpunct}
{\mcitedefaultendpunct}{\mcitedefaultseppunct}\relax
\EndOfBibitem
\bibitem[LeCun \latin{et~al.}(2015)LeCun, Bengio, and Hinton]{lecun2015deep}
LeCun,~Y.; Bengio,~Y.; Hinton,~G. Deep Learning. \emph{Nature} \textbf{2015},
  \emph{521}, 436--444\relax
\mciteBstWouldAddEndPuncttrue
\mciteSetBstMidEndSepPunct{\mcitedefaultmidpunct}
{\mcitedefaultendpunct}{\mcitedefaultseppunct}\relax
\EndOfBibitem
\bibitem[Schmidhuber(2015)]{schmidhuber2015deep}
Schmidhuber,~J. Deep Learning In Neural Networks: An Overview. \emph{Neural
  Netw.} \textbf{2015}, \emph{61}, 85--117\relax
\mciteBstWouldAddEndPuncttrue
\mciteSetBstMidEndSepPunct{\mcitedefaultmidpunct}
{\mcitedefaultendpunct}{\mcitedefaultseppunct}\relax
\EndOfBibitem
\bibitem[Goodfellow \latin{et~al.}(2016)Goodfellow, Bengio, and
  Courville]{goodfellow2016deep}
Goodfellow,~I.; Bengio,~Y.; Courville,~A. \emph{Deep Learning}; MIT press
  Cambridge, 2016\relax
\mciteBstWouldAddEndPuncttrue
\mciteSetBstMidEndSepPunct{\mcitedefaultmidpunct}
{\mcitedefaultendpunct}{\mcitedefaultseppunct}\relax
\EndOfBibitem
\bibitem[Cortes and Vapnik(1995)Cortes, and Vapnik]{cortes1995support}
Cortes,~C.; Vapnik,~V. Support-vector Networks. \emph{Machine learning}
  \textbf{1995}, \emph{20}, 273--297\relax
\mciteBstWouldAddEndPuncttrue
\mciteSetBstMidEndSepPunct{\mcitedefaultmidpunct}
{\mcitedefaultendpunct}{\mcitedefaultseppunct}\relax
\EndOfBibitem
\bibitem[Vapnik(1995)]{vapnik1995nature}
Vapnik,~V. \emph{The Nature Of Statistical Learning Theory}; Springer,
  1995\relax
\mciteBstWouldAddEndPuncttrue
\mciteSetBstMidEndSepPunct{\mcitedefaultmidpunct}
{\mcitedefaultendpunct}{\mcitedefaultseppunct}\relax
\EndOfBibitem
\bibitem[Lapuschkin \latin{et~al.}(2019)Lapuschkin, W{\"a}ldchen, Binder,
  Montavon, Samek, and M{\"u}ller]{lapuschkin2019unmasking}
Lapuschkin,~S.; W{\"a}ldchen,~S.; Binder,~A.; Montavon,~G.; Samek,~W.;
  M{\"u}ller,~K.-R. Unmasking Clever Hans Predictors And Assessing What
  Machines Really Learn. \emph{Nat. Comm.} \textbf{2019}, \emph{10}, 1096\relax
\mciteBstWouldAddEndPuncttrue
\mciteSetBstMidEndSepPunct{\mcitedefaultmidpunct}
{\mcitedefaultendpunct}{\mcitedefaultseppunct}\relax
\EndOfBibitem
\bibitem[Sauceda \latin{et~al.}(2020)Sauceda, Chmiela, Poltavsky, M{\"u}ller,
  and Tkatchenko]{sauceda2020construction}
Sauceda,~H.~E.; Chmiela,~S.; Poltavsky,~I.; M{\"u}ller,~K.-R.; Tkatchenko,~A.
  \emph{Machine Learning Meets Quantum Physics}; Springer, 2020; pp
  277--307\relax
\mciteBstWouldAddEndPuncttrue
\mciteSetBstMidEndSepPunct{\mcitedefaultmidpunct}
{\mcitedefaultendpunct}{\mcitedefaultseppunct}\relax
\EndOfBibitem
\end{mcitethebibliography}

\end{document}